\documentclass[11pt,twoside,nofootinbib,smartcite]{mitthesis}
\usepackage{lgrind,graphicx,amssymb,amsmath,url,mathrsfs,appendix,rotating} 
\def\etal{{et al. }}
\def\ie{{i.e. }}
\def\eg{{e.g. }}

\def\ben{\begin{equation}}
\def\een{\end{equation}}
\def\beq#1{\begin{equation}\label{#1}}
\def\eeq{\end{equation}}
\def\bena{\begin{eqnarray}}
\def\eena{\end{eqnarray}}
\def\beqa#1{\begin{eqnarray}\label{#1}}
\def\eeqa{\end{eqnarray}}
\def\Eq#1{Eq.~(\ref{#1})}
\def\eq#1{Eq.~(\ref{#1})}
\def\Eqnopar#1{Eq.~\ref{#1}}
\def\Fig#1{Figure~\ref{#1}}
\def\fig#1{Figure~\ref{#1}}
\def\Sec#1{Section~\ref{#1}}
\def\x{{\bf x}}
\def\I{{\bf I}}
\def\J{{\bf J}}
\def\R{{\bf R}}

\def\spose#1{\hbox to 0pt{#1\hss}}
\def\simlt{\mathrel{\spose{\lower 3pt\hbox{$\mathchar"218$}}
     \raise 2.0pt\hbox{$\mathchar"13C$}}}
\def\simgt{\mathrel{\spose{\lower 3pt\hbox{$\mathchar"218$}}
     \raise 2.0pt\hbox{$\mathchar"13E$}}}
\def\simpropto{\mathrel{\spose{\lower 3pt\hbox{$\mathchar"218$}}
     \raise 2.0pt\hbox{$\propto$}}}
\def\lesssim{\simlt}
\def\gtrsim{\simgt}

\newcommand{\milliK}{{\rm mK}}

\newcommand{\Mpc}{{\rm Mpc}}
\newcommand{\perMpc}{\ensuremath{\Mpc^{-1}}}

\newcommand{\eV}{{\rm eV}}

\newcommand{\Ob}{\ensuremath{\Omega_{\rm b}}}
\newcommand{\Oc}{\ensuremath{\Omega_{\rm cdm}}}
\newcommand{\Ok}{\ensuremath{\Omega_{\rm k}}}
\newcommand{\Ol}{\ensuremath{\Omega_\Lambda}}
\newcommand{\Om}{\ensuremath{\Omega_{\rm m}}}

\newcommand{\On}{\ensuremath{\Omega_\nu}}
\newcommand{\ob}{\ensuremath{\Omega_{\rm b} h^2}}
\newcommand{\ocdm}{\ensuremath{\Omega_{\rm cdm} h^2}}

\newcommand{\om}{\ensuremath{\Omega_{\rm m} h^2}}
\newcommand{\on}{\ensuremath{\Omega_\nu h^2}}

\newcommand{\fn}{\ensuremath{f_\nu}}
\newcommand{\ns}{\ensuremath{n_{\rm s}}}
\newcommand{\nt}{\ensuremath{n_{\rm t}}}
\newcommand{\al}{\ensuremath{\alpha}}
\newcommand{\Ot}{\ensuremath{\Omega_{\rm tot}}}
\newcommand{\As}{\ensuremath{A_{\rm s}}}
\newcommand{\At}{\ensuremath{A_{\rm t}}}
\newcommand{\mnu}{\ensuremath{m_\nu}}
\newcommand{\xH}{\ensuremath{\bar{x}_{\rm H}}}
\newcommand{\xI}{\ensuremath{\bar{x}_{\rm i}}}
\newcommand{\Ts}{\ensuremath{T_S}}
\newcommand{\Tcmb}{\ensuremath{T_{\rm CMB}}}

\newcommand{\PDT}{\ensuremath{P_{\Delta T}}}
\newcommand{\Pxx}{\ensuremath{P_{\rm xx}}}
\newcommand{\Pxd}{\ensuremath{P_{{\rm x}\delta}}}
\newcommand{\Pdd}{\ensuremath{P_{\delta\delta}}}
\newcommand{\sPdd}{\ensuremath{\mathscr{P}_{\delta\delta}}}
\newcommand{\sPxd}{\ensuremath{\mathscr{P}_{x\delta}}}
\newcommand{\sPxx}{\ensuremath{\mathscr{P}_{xx}}}
\newcommand{\Pmuzero}{\ensuremath{P_{\mu^0}}}
\newcommand{\Pmutwo}{\ensuremath{P_{\mu^2}}}
\newcommand{\Pmufour}{\ensuremath{P_{\mu^4}}}
\newcommand{\Pmusix}{\ensuremath{P_{\mu^6}}}
\newcommand{\bsqxx}{\ensuremath{b^2_{\rm xx}}}
\newcommand{\Rxx}{\ensuremath{R_{\rm xx}}}
\newcommand{\alxx}{\ensuremath{\al_{\rm xx}}}
\newcommand{\gaxx}{\ensuremath{\gamma_{\rm xx}}}
\newcommand{\bsqxd}{\ensuremath{b^2_{{\rm x}\delta}}}
\newcommand{\Rxd}{\ensuremath{R_{{\rm x}\delta}}}
\newcommand{\alxd}{\ensuremath{\al_{{\rm x}\delta}}}

\newcommand{\kmax}{\ensuremath{k_{\rm max}}}
\newcommand{\kmin}{\ensuremath{k_{\rm min}}}

\newcommand{\onesig}{\ensuremath{1\sigma}}

\newcommand{\kpar}{\ensuremath{k_{\parallel}}}
\newcommand{\uperp}{\ensuremath{{\bf u}_{\perp}}}
\newcommand{\upar}{\ensuremath{u_{\parallel}}}
\newcommand{\bfk}{{\bf k}}

\newcommand{\bfx}{{\bf x}}
\newcommand{\bfu}{{\bf u}}
\newcommand{\bfV}{{\bf V}}
\newcommand{\bfy}{{\bf y}}
\newcommand{\bfF}{{\bf F}}
\def\ocdm{\omega_{\rm c}}

\def\percent{\%}
\def\Mnu{M_\nu}
\def\zeq{{z_{\rm eq}}}
\def\zrec{{z_{\rm rec}}}
\def\zion{{z_{\rm ion}}}
\def\zacc{{z_{\rm acc}}}
\def\teq{t_{\rm eq}}
\def\trec{t_{\rm req}}
\def\tion{t_{\rm ion}}
\def\tacc{t_{\rm acc}}
\def\tnow{t_{\rm now}}

\def\ng{n_\gamma}

\def\rhob{\rho_{\rm b}}
\def\rhoc{\rho_{\rm c}}

\def\rhon{\rho_\nu}

\def\rhom{\rho_{\rm m}}

\def\rhohalo{\rho_{\rm halo}}

\def\xib{\xi_{\rm b}}
\def\xic{\xi_{\rm c}}

\def\xin{\xi_\nu}

\def\rhocr{\rho_{\rm cr}}

\newcommand{\Tab}[1]{Table~\ref{#1}}         
\newcommand{\bem}{\begin{enumerate}}
\newcommand{\eem}{\end{enumerate}}
\newcommand{\bit}{\begin{itemize}}
\newcommand{\eit}{\end{itemize}}
\newcommand{\la}{\lesssim}
\newcommand{\nodata}{\ensuremath{\cdots}}
\newcommand{\mnras}{Mon.~Not.~Roy.~Astron.~Soc. }
\newcommand{\apj}{Astrophys.~J.}

\newcommand{\Mpl}{M_{\rm{pl}}}

\def\bit{\begin{itemize}}
\def\eit{\end{itemize}}

\begin{document}

%
%
%
%
%
%
%
\title{Constraining Gravitational and Cosmological Parameters with Astrophysical Data}

\author{Yi Mao}
\department{Department of Physics}
\degree{Doctor of Philosophy}
\degreemonth{June}
\degreeyear{2008}
\thesisdate{May 23, 2008}  


\supervisor{Max Tegmark}{Associate Professor of Physics}
\supervisor{Alan H. Guth}{Victor F. Weisskopf Professor of Physics}

\chairman{Thomas J. Greytak}{Associate Department Head for Education}

\maketitle



\cleardoublepage
\setcounter{savepage}{\thepage}
\begin{abstractpage}
%
%
%

We use astrophysical data to shed light on fundamental physics by constraining parametrized theoretical cosmological and gravitational models.

Gravitational parameters are those constants that parametrize possible departures from Einstein's general theory of relativity (GR).
We develop a general framework to describe torsion in the spacetime around the Earth, and show that certain observables of the Gravity Probe B (GPB) experiment can be computed in this framework.  We examine a toy model showing how a specific theory in this framework can be constrained by GPB data.
We also search for viable theories of gravity where the Ricci scalar $R$ in the Lagrangian is replaced by an arbitrary function $f(R)$.
Making use of the equivalence between such theories and scalar-tensor gravity, we find that models can be made consistent with solar system
constraints either by giving the scalar a high mass or by exploiting the so-called Chameleon Effect.
We explore observational constraints from the  late-time cosmic acceleration, big bang nucleosynthesis and inflation.

Cosmology can successfully describe the evolution of our universe using six or more adjustable cosmological parameters.
There is growing interest in using 3-dimensional neutral hydrogen mapping with the redshifted 21 cm line as a cosmological probe.
We quantify how the precision with which cosmological parameters can be measured depends on a broad range of
assumptions.  We present an accurate and robust method for measuring cosmological
parameters that exploits the fact that the ionization power spectra are rather smooth functions that can be accurately fit by $7$ phenomenological parameters.
We find that a future square kilometer array optimized for 21 cm tomography could have great potential,
improving the sensitivity to spatial curvature and neutrino masses by up to two orders of magnitude, to $\Delta\Omega_k\approx 0.0002$ and
$\Delta m_\nu\approx 0.007$ eV, and giving a $4\sigma$ detection of the spectral index running predicted by the simplest inflation models.

\end{abstractpage}


\cleardoublepage

\section*{Acknowledgments}

The work presented in this thesis, and my time at MIT over the last six years, has benefited from the contributions and support of many individuals.  At MIT, I would 
like to thank my advisors, Max Tegmark and Alan H. Guth, my thesis committee, 
Scott A. Hughes and Erotokritos Katsavounidis, and Hong Liu, Jackie Hewitt, Iain Stewart, Miguel Morales, Serkan Cabi, Thomas Faulkner, 
along with Scott Morley, Joyce Berggren, Charles Suggs and Omri Schwarz.  
I would also like to thank my collaborators Matias Zaldarriaga, Matthew McQuinn and Oliver Zahn at the Center for Astrophysics at Harvard University, and 
Emory F. Bunn at University of Richmond.  The particle-theory and astro grads have been a constant source of support and entertainment, especially Molly Swanson, Qudsia Ejaz, Mark Hertzberg, Onur Ozcan, Dacheng Lin, Adrian Liu, and in earlier times, Judd Bowman and Ying Liu.  I would like to give my special thanks to my dad Zhenzhong Mao, my wife Yi Zheng and my sister Su Jiang, for their constant support and encouragement without which I could not accomplish the course of study.


\pagestyle{plain}
\tableofcontents
\newpage
\listoffigures
\newpage
\listoftables

\chapter{Introduction}

Study of gravitation and cosmology has a long history, tracing back to antiquity when a number of Greek philosophers attempted to summarize and explain observations from the natural world, and has now evolved into two successful and flourishing areas.  
Since Einstein's general theory of relativity (GR) was first proposed about ninety years ago, it has emerged as the hands-down most popular candidate for the laws governing gravitation.   
Moreover, during the past decade a cosmological concordance model, in which the cosmic matter budget consists of about 5\% ordinary matter (baryons), 
30\% cold dark matter and 65\% dark energy, 
has emerged in good agreement with all cosmological data, including 
the cosmic microwave background observations, galaxy surveys, type Ia supernovae, gravitational lensing and the Lyman-$\alpha$ forest.  

Why is gravitation on an equal footing with cosmology in this thesis?  This is because they are   closely related subjects: 
gravitation is the theoretical foundation of cosmology, and cosmology can test gravity on the scale of the universe.   
Gravitation has influenced cosmology right from the start:  
the modern Big Bang cosmology began with two historical discoveries, the Hubble diagram and the Friedmann equation.  As an application of GR, the latter predicted the possibility of an expanding universe.  
In recent years, attempts have been made to explain away the dark energy and/or dark matter by modifying GR.  So-called $f(R)$-gravity  
\cite{Amendola:2006kh,Bean:2006up,Capozziello:2006dj,Carroll:2003wy,Cembranos:2005fi,
Chiba:2003ir,Chiba:2006jp,Clifton:2005aj,Dick:2003dw,Dolgov:2003px,Erickcek:2006vf, 
Navarro:2006mw, Nojiri:2006gh,Olmo:2005hc,PerezBergliaffa:2006ni,Song:2006ej,Zhang:2005vt}, which generalizes the gravitational Lagrangian to contain a function of the curvature $R$, can potentially explain the late-time cosmic acceleration without dark energy, or provide the inflaton field in the early universe.  
DGP gravity \cite{Dvali:2000hr}, named after its inventors Dvali, Gabadadze and Porrati, adopts a radical approach that assumes that a 3-dimensional brane is embedded in a 5-dimensional spacetime, and also claims that it can reproduce the cosmic acceleration of dark energy.  
The approach of Modified Newtonian Dynamics (MOND) \cite{Milgrom:1983ca}, in particular 
the relativistic version -- Bekenstein's tensor-vector-scalar (TeVeS) theory \cite{Bekenstein:2004ne} --  
purports to explain galaxy rotation curves without dark matter.

Turning to how cosmology has influenced gravitation, 
the cosmological concordance model assumes that the expansion and structure formation of the universe are governed by equations derived from GR, mostly to linear order.  
It is therefore not a surprise that modified theories of gravity can imprint their signatures on the expansion history and the density perturbations of the universe.  
Recent research in this direction has undergone rapid progresses towards the so-called Parametrized Post-Friedmann formalism \cite{Bertschinger:2006aw,Hu:2008zd,Hu:2007pj} that can in principle use the avalanche of cosmology data to test gravity on scales up to the cosmic horizon.  

In a nutshell, gravitation and cosmology are united.  To test both --- which is the subject of this thesis --- we will generalize the standard models of both gravitation and cosmology, such that our ignorance can be parametrized by a few constants, and constrain those constants with astrophysical data.

\section{Testing gravity}

\subsection{Was Einstein right?}

\begin{table}
\noindent 
\footnotesize{
\begin{center}
\begin{tabular}{p{1.2cm}p{3cm}p{0.9cm}p{3.4cm}p{1.5cm}p{3.2cm}}
\hline
Parameter & Meaning & GR value & Effect & Limits  & Remarks \\ \hline
$\gamma-1$ & Amount of spacetime & 0 & (i) Shapiro time delay & $2.3\times 10^{-5}$  & Cassini tracking \\ \cline{4-6}
 & curvature produced  & & (ii) Light deflection & $3\times 10^{-4}$ & VLBI \\ \cline{4-6}
 & by unit rest mass   & & (iii) Geodetic precession & $1.1\times 10^{-4}$ (anticipated) & Gravity Probe B \\  \hline
$\beta-1$ & Amount of ``non-linearity'' in the & 0 & (i) Perihelion shift & $3\times 10^{-3}$  & $J_2=10^{-7}$ from helioseismology \\ \cline{4-6}
& superposition law for gravity & & (ii) Nordtvedt effect & $5\times 10^{-4}$ & \\ \hline
$\xi$ & Preferred-location effects & 0 & Earth tides & $10^{-3}$  & Gravimeter data \\ \hline
$\alpha_1$ & Preferred-frame & 0 & Orbital polarization & $10^{-4}$ & Lunar laser ranging \\ \cline{4-6}
$\alpha_2$ & effects  & 0 & Solar spin precession & $4\times 10^{-7}$ & Alignment of Sun and ecliptic \\ \cline{4-6}
$\alpha_3$ &   & 0 & Pulsar acceleration & $2\times 10^{-20}$ & Pulsar $\dot{P}$ statistics \\ \hline
$\zeta_1$ & Violation of  & 0 & -- & $2\times 10^{-2}$ & Combined PPN bounds \\ \cline{4-6}
$\zeta_2$ & conservation  & 0 & Binary motion & $4\times 10^{-5}$ & $\ddot{P}_p$ for PSR 1913+16 \\ \cline{4-6}
$\zeta_3$ & of total momentum & 0 & Newton's 3rd law & $10^{-8}$ & Lunar acceleration \\ \cline{4-6}
$\zeta_4$ &  & 0 & -- & -- & Not independent \\ \hline
\end{tabular}
\caption[PPN parameters, their significance and experimental bounds]{The PPN parameters, their significance and experimental bounds.  Contents of this table except for the GPB  entry are taken from Table 2 of \cite{Will:2005va} and Table 1 of \cite{Will:2005yc}.  }
\label{tab:PPN}
\end{center}
}
\end{table}

Einstein's GR has been hailed as one of the greatest intellectual successes of human beings.  This reputation is a consequence of both its elegant structure and its impressive agreement with a host of experimental tests.  In Relativity, space is not merely a stage where events unfold over time as it was in Newtonian mechanics.  Rather, space-time is a unified entity which has a life of its own: it can be sufficiently curved to form an event horizon around a black hole, it can propagate ripples in ``empty'' space with the speed of light, and it can expand the universe with the driving force given by the matter and energy inside it.  

In GR and most generalized gravity theories, spacetime is, in mathematical language, a manifold whose geometry is dictated by the metric, a tensor defined at each location.  GR contains two essential ingredients that experiments can test.  The first is how spacetime influences test particles (particles small enough that they do not significantly change  the spacetime around them) such as photons or astrophysical objects.  In the absence of non-gravitational forces, test particles move along geodesics, which are generalized straight lines in GR.  I term experiments that use a planet or a light ray to probe the metric around a massive gravitating object (Sun or Earth) as {\it geometric tests} of GR.  Geometric tests have included the well-known weak-field solar system tests: Mercury's perihelion shift, light bending, gravitational redshift, Shapiro time delay and lunar laser-ranging.  An on-going satellite experiment, Gravity Probe B, that measures the spin precession of free-falling gyroscopes, falls into this category too, since the spin precession is a response to the spacetime metric that arises from both the mass (geodetic precession) and the rotation (frame-dragging effect) of Earth.  
Additionally, the LAGEOS \cite{Ciufolini:2004rq} experiment has directly confirmed the frame-dragging effect on the orbits of test particles around the rotating Earth.  

In the weak-field regime, there exists a mature formalism, the so-called Parametrized Post-Newtonian (PPN) formalism, that parametrizes departures from GR in terms of a few constants under certain reasonable assumptions.  Thus the success of GR in this regime can be fully described by the observational constraints on these PPN parameters near their GR predictions, as summarized in Table \ref{tab:PPN}.  

The second part of GR describes how matter, in the form of masses, pressures and sometimes shears, curves the spacetime around them.  Specifically the metric and matter are related by the Einstein field equation (EFE).  I term any experiments that test EFE {\it dynamical tests}.  The linearized EFE predicts the existence of gravitational waves, i.e. propagation of tensorial gravitational perturbations.  
The upcoming experiments LIGO and LISA are direct probes of gravitational waves from black holes or the primordial universe.  An indirect probe is the observation of the damping of binary orbits due to gravitational radiation reaction in the binary pulsars.  
Upcoming observations of strong-field binary compact objects and black-hole accretion will be exquisite dynamical tests of GR.  In addition, as mentioned above, cosmology can test the EFE, since the cosmic expansion and structure formation are determined by the zeroth and linear order EFE.

\subsection{Generalizing GR}

As mentioned, since GR is the foundation of both modern gravitation and cosmology theory, testing GR is of central interest in the science community.  
Two popular approaches have been taken to test GR in a broader framework.  
The first approach is to generalize GR in a model-independent fashion by making a few assumptions that are valid in a certain limit, and parametrize any possible extensions of GR by a few constants.  For example, the PPN formalism \cite{PPN1,PPN2,PPN3} parametrizes theories of gravity in the weak-field limit with 10 PPN parameters (see Table \ref{tab:PPN}) that can be constrained by solar-system test data.  The developing Parametrized Post-Friedmann (PPF) formalism, as a second example, parametrizes the cosmology of modified gravity theories to linear order in cosmic density perturbations and may end up with a few PPF parameters too, that may be constrained by cosmological experiments in the future.  The second approach to testing GR is to follow the debate strategy: if we can rule out all modified theories of gravity that we can think of, then GR becomes more trustworthy.  Arguably the most beautiful aspect of GR is that it geometrizes gravitation.  Consequently, there are at least three general methods that can generalize GR, corresponding to different geometries.  

The first method is to introduce extra dynamical degrees of freedom in the same geometry as GR.  The geometry where GR is defined is the so-called Riemann spacetime, that is completely specified by the metric $g_{\mu\nu}(x)$, a tensor at each spacetime position.  In the Riemann spacetime, a free-falling particle moves along a covariantly ``constant'' velocity curve, in the sense that the 4D velocity vector $dx^{\mu}/d\tau$ has vanishing covariant derivative ($D\ldots/D\tau$), because the change in the absolute differentiation ($d\ldots/d\tau$) is compensated for by a term involving the so-called \emph{connection} that characterizes a curved manifold and defines the spacetime \emph{curvature} $R$.  
The connection and curvature are not free in the Riemann spacetime --- they are defined in terms of the metric and its 1st and/or 2nd derivatives.  The dynamics of Einstein's GR is given by the simple action, ${\sl S} = (1/2\kappa)\int d^4x \sqrt{-g}\,R$, from which EFE is derived.  Here $\kappa = 8\pi G/c^4$, where $G$ is Newton's gravitational constant and $c$ is the speed of light.  The factor $\sqrt{-g}$ is inserted so that $d^4x \sqrt{-g}$ is covariant (i.e. unchanged under arbitrary coordinate transformations).  
The outstanding simplicity of GR is that it contains no free parameters, given that $G$ is fixed by the inverse-square law in the Newtonian limit.  
To generalize GR, however, one trades off the simplicity for the generalization.  
For example, one can take the action to contain, in principle, \emph{arbitrary} functions of the curvature, i.e. ${\sl S} = (1/2\kappa) \int d^4x \sqrt{-g}\,f(R)$ --- which defines so-called $f(R)$ gravity.  
A new scalar field $\phi$ with arbitrary potential and couplings to the metric can also be introduced into the action --- this is so-called scalar-tensor gravity.  
In fact, $f(R)$ gravity is equivalent to a special class of scalar-tensor gravity theories.  
Additionally, the action can include even more fields (vector fields plus scalar fields), as in the so-called TeVeS (standing for ``tensor-vector-scalar'') gravity, a relativistic version of MOND.  

The second method to generalize GR is to generalize the geometry such that the emergent degrees of freedom in the spacetime manifold are dynamic variables.  The simplest extension to Riemann spacetime is the so-called Riemann-Cartan spacetime with nonzero \emph{torsion}.  In a nutshell, torsion is the antisymmetric part of the connection mentioned earlier -- in Riemann spacetime the connection is constrained to be symmetric, so allowing for non-zero torsion relaxes this constraint.  
The geometry of Riemann-Cartan spacetime is pinned down by the metric and torsion 
-- the so-called $U_4$ torsion theory is established in terms of these two pieces.  
Just as Riemann spacetime is a special case of Riemann-Cartan spacetime with zero torsion, there is an exotic brother of the Riemann spacetime, so-called Weitzenb\"ock spacetime, that is characterized by zero total curvature.  That means that gravitation in the Weitzenb\"ock spacetime is carried only by torsion, e.g. in the Hayashi-Shirafuji theory \cite{HS1} and teleparallel gravity \cite{DeAndrade,aldrovandi}.  It is even possible to extend the geometry more generally than the Riemann-Cartan spacetime, as illustrated by \Fig{fig:spaces}, and use more spacetime degrees of freedom to gravitate differently.  

The third method is to generalize the \emph{dimensionality} of the spacetime.  Spacetime with extra dimensions was first considered by Kaluza in 1919 and Klein in 1926.  Despite the failure of their old theories, modern versions of Kaluza-Klein theory continue to attract attention.  A typical example is the above-mentioned DGP theory which exploits the perspective that the ordinary world is a (3+1)-D brane to which electromagnetism, the strong and the weak forces are confined, with gravitation extending into the (4+1)-D bulk.

Theories in all of the above three categories might explain away dark matter or dark energy, or may be of exotic phenomenological interest.  
In this thesis, we will focus on the first two categories, in particular $f(R)$ gravity and torsion theory, and give further introductions in more details below.

\subsection{$f(R)$ gravity}

There are two important classes of $f(R)$ theories: massive $f(R)$ theories and $f(R)$ dark energy (DE) theories.  Interestingly, both classes were motivated by two accelerating eras in the universe.  Massive $f(R)$ theories, namely polynomials $f(R)=-2\Lambda+R+aR^2+bR^3\ldots$, contain higher order corrections that dominate over the linear GR Lagrangian $f(R)=-2\Lambda+R$ in the \emph{early} universe, as the curvature was presumably larger in the past.  More subtly, an $f(R)$ theory is equivalent to a scalar-tensor gravity theory, and in the massive $f(R)$ case, the emergent scalar field can roll down the emergent potential, which drives inflation at early times.  
In contrast, the $f(R)$-DE branch, exemplified by $R^m\,(m<0)$, is motivated by explaining dark energy that causes the \emph{late-time} cosmic acceleration.  Naively, since $R$ is small at late times, negative powers of $R$ dominate over the linear GR Lagrangian, and the emergent scalar field can have negative pressure, thus driving the late-time acceleration and explaining dark energy.    

However, the archetypal $f(R)$-DE model, $f(R)=R-\mu^4/R$ for $\mu\approx H_0$ \cite{Carroll:2001bv}, where $H_0$ is the Hubble constant at today, suffers from serious problems.  First, the theory does not pass solar system tests \cite{Chiba:2003ir,Dick:2003dw,Dolgov:2003px,Erickcek:2006vf}.  Although the Schwarzschild metric can naively solve the field equations for this theory, it can be shown that it is not the solution that satisfies the correct boundary conditions.  In fact, it has been shown that the solution that satisfies both the field equations and the correct boundary conditions has the PPN parameter $\gamma=1/2$, so this theory is ruled out by, e.g., Shapiro time delay, and deflection of light.  Second, the cosmology for this theory is inconsistent with observation when non-relativistic matter is present \cite{Amendola:2006kh}.  

Does this mean that $f(R)$-DE theories are dead?  The answer is no.  In Chapter 3, we exploit the so-called Chameleon effect, which uses non-linear effects from a very specific singular form of the potential to hide the scalar field from current tests of gravity.  In other words, the Chameleon $f(R)$-DE models are still consistent with both solar system tests and the late-time cosmic acceleration.  We will constrain the gravitational parameters that parametrize the departure from GR in the Chameleon $f(R)$-DE models, using solar system tests and cosmological tests in Chapter 3.

\subsection{Torsion theories}

\begin{figure}
\centering
\includegraphics[width=0.7\textwidth]{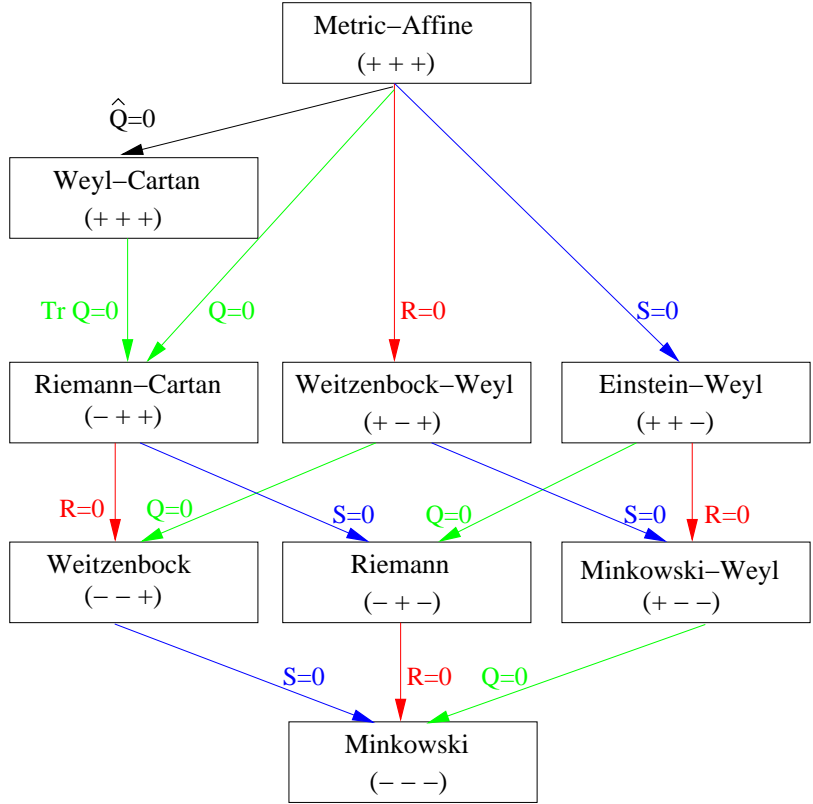}
\caption[Classification of spaces (Q,R,S) and the reduction flow]{\footnotesize%
Classification of spaces (Q,R,S) and the reduction flow.  Metric-Affine spacetime is a manifold endowed with Lorentzian
metric and linear affine connection without any restrictions.  All spaces below it except the Weyl-Cartan space are special cases obtained
from it by imposing three
types of constraints: vanishing non-metricity tensor $Q_{\mu\nu\rho}$ ($Q$ for short), 
vanishing Riemann curvature tensor
$R_{\mu\nu\rho\sigma}$ ($R$ for short), or vanishing torsion tensor 
$S_{\mu\nu}^{\phantom{01}\rho}$ ($S$ for short).  A plus
sign in a parenthesis indicates a non-vanishing quantity from the set $(Q,R,S)$, and a minus sign a vanishing quantity.  
For example, Riemann
spacetime $(-+-)$ means that $Q=S=0$ but $R\ne 0$.  Weyl-Cartan space is a Metric-Affine space with vanishing 
``tracefree nonmetricity'' $\hat{Q}_{\mu\nu\rho}$ ($\hat{Q}$ for short), 
defined by $\hat{Q}_{\mu\nu\rho}\equiv Q_{\mu\nu\rho}-\frac{1}{4} ({\rm tr}\>Q)_\mu g_{\nu\rho}$.  
The trace of the nonmetricity is defined by $({\rm tr}\>Q)_{\mu}\equiv g^{\nu\rho}Q_{\mu\nu\rho}$; thus $\hat{Q}$ is automatically trace free, i.e., $({\rm tr}\hat{Q})_{\mu}=0$.  
Subsets of this classification scheme are shown in Fig.~2 of \cite{Hehl:1994ue}, Fig.~1 of \cite{Puetzfeld:2004yg} 
and Fig.~5 of \cite{Gronwald:1995em}.  Among the terms, {\it Einstein-Weyl, Weitzenb\"ock} and  {\it Minkowski} spaces are standard, 
{\it Metric-Affine, Weyl-Cartan, Riemann-Cartan} and {\it Riemann} spaces follow \cite{Hehl:1994ue}, and we here introduce the 
terms {\it Weitzenb\"ock-Weyl} and {\it Minkowski-Weyl} space by symmetry.}
\label{fig:spaces}
\end{figure}

\begin{table}
\noindent
\footnotesize{
\begin{center}
\begin{tabular}{p{2.2cm}p{2cm}p{1cm}p{2.5cm}p{1cm}|p{4cm}}
\hline
Theory & Dynamical DOF & Vacuum & Source & Ref. & Notes \\ \hline\hline

$U_4$ theory & $g_{\mu\nu}$, $S_{\mu\nu}^{\phantom{12}\rho}$ & N & Spin & \cite{hehl} & \\ \hline
Pagels theory & $O(5)$ gauge fields $\omega_{\mu}^{\phantom{1}AB}$  & N & Spin & \cite{Pagels:1983pq} & An $O(5)$ gauge theory of gravity \\ \hline 
Metric-affine gravity & General gauge fields & P & Spin & \cite{Hehl:1994ue} & A gauge theory of gravity in the metric-affine space\\ \hline
Stelle-West & $SO(3,2)$ gauge fields $\omega_{\mu}^{\phantom{1}AB}$ & P & Spin, Gradient of the Higgs field & \cite{Stelle:1979aj} & A $SO(3,2)$ gauge theory of gravity spontaneously broken to $SO(3,1)$ \\ \hline 
Hayashi-Shirafuji & Tetrads $e^{\,k}_{\,\phantom{1}\mu}$ & P & Spin, Mass, Rotation & \cite{HS1} & A theory in Weitzenb\"ock space\\ \hline 
Einstein-Hayashi-Shirafuji & Tetrads $e^{\,k}_{\,\phantom{1}\mu}$ & P & Spin, Mass, Rotation & \cite{Mao:2006bb} & A class of theories in Riemann-Cartan space \\ \hline
Teleparallel gravity & Tetrads $e^{\,k}_{\,\phantom{1}\mu}$ & P & Spin, Mass, Rotation & \cite{DeAndrade,aldrovandi} & A theory in Weitzenb\"ock space \\ \hline 

\end{tabular}
\caption[A short list of torsion theories of gravity]{A short list of torsion theories of gravity.  The ``DOF'' in the second column is short for ``degrees of freedom''.  In the column \emph{Vacuum}, ``N'' refers to non-propagating torsion in the vacuum while ``P'' means propagating torsion.  In the column \emph{Source}, ``spin'' refers to intrinsic spin, ``mass'' means non-rotating mass, and ``rotation'' means rotational angular momentum.}
\label{tab:torsion-theory}
\end{center}
}
\end{table}

As illustrated in \Fig{fig:spaces}, for the most general manifold with a metric $g$ and a connection $\Gamma$, departures from Minkowski
space are characterized by three geometrical entities: non-metricity ($Q$), curvature ($R$) and torsion ($S$), defined as follows:
\bena
Q_{\mu\nu\rho}&\equiv&\nabla_{\mu} g_{\nu\rho}\,,\\
R^{\rho}_{\phantom{1}\lambda\nu\mu} &\equiv& \Gamma^{\rho}_{\phantom{1}\mu\lambda,\nu}-\Gamma^{\rho}_{\phantom{1}\nu\lambda,\mu}+\Gamma^\rho_{\phantom{1}\nu\alpha}\Gamma^\alpha_{\phantom{1}\mu\lambda}-\Gamma^\rho_{\phantom{1}\mu\alpha}\Gamma^\alpha_{\phantom{1}\nu\lambda}\,,\\
S_{\mu\nu}^{\phantom{\mu\nu}\rho}&\equiv&\frac{1}{2}(\Gamma^{\rho}_{\phantom{\rho}\mu\nu}-\Gamma^{\rho}_{\phantom{\rho}\nu\mu})\,.
\eena
In GR, spacetime is distorted only by curvature, restricting non-metricity and torsion to vanish.  In Riemann-Cartan spacetime, gravitation is manifested in the terms of nonzero torsion as well as curvature.  There have been many attempts to construct gravitational theories involving torsion, as shown in Table \ref{tab:torsion-theory}.  However, testing torsion in the solar system was not a popular idea in the old-fashioned theories for the following two reasons.  First, in some torsion theories, e.g. $U_4$ theory \cite{hehl} and Pagels theory \cite{Pagels:1983pq}, the field equations for the torsion are in algebraic \footnote{For example, in $U_4$ theory, both the affine connection and the metric are independent dynamical variables with respect to which the action is differentiated in order to get the field equation.  Consequently, the second field equation is like (the first one for the metric is similar to the Einstein field equation):
\[ {\rm modified\, torsion} = (8\pi G/c^4) \times {\rm spin\, angular\, momentum}\,. \] 
Since there is no spin in the vacuum, torsion must be identically zero.
}
rather than differential form, which means that torsion cannot propagate into the vacuum.  Second, it is well-entrenched folklore that the source of the torsion can only be the intrinsic spin of elementary particles, not the mass or the rotational angular momentum of a massive object.  That means that, even if torsion could propagate and exist in the vacuum, since the net spin polarizations of the Sun and Earth are negligible, the torsion should still vanish in the solar system.  This second assumption also implies that even if torsion were present in the solar system, 
it would only affect particles with intrinsic spin (e.g. a gyroscope with net magnetic polarization) \cite{pereira1,cognola,Hojman,Hojman2,kopczynski,NSH,sy1,sy2}, while having no influence on the precession of a gyroscope without nuclear spin \cite{NSH,sy1,sy2} such as a gyroscope in Gravity Probe B.  The upshot is that any torsion theory that makes either of two above assumptions has already found itself in a position that can never be ruled out by solar system tests.  Such a torsion theory can have noticeable effects only in extreme situations, e.g. near neutron stars with net magnetic polarizations.  

Whether torsion does or does not satisfy these pessimistic assumptions depends on what the Lagrangian is, which is of course one of the things that should be tested experimentally rather than assumed. Taken at face value, the Hayashi-Shirafuji Lagrangian \cite{HS1}  and teleparallel gravity 
provide an explicit counterexample to both assumptions, with even a static massive body generating a torsion field.  
They show that one cannot dismiss out of hand the possibility that mass and angular momentum sources non-local torsion 
(see also Table \ref{tab:torsion-theory}).  
In Chapter 2, we show that gyroscope experiments such as Gravity Probe B are perfect for testing torsion in non-conventional torsion theories in which torsion can be sourced by rotational angular momentum and can affect the precession of a gyroscope.  

After the work in Chapter 2 \cite{Mao:2006bb} was published, it has both generated interest \cite{Aldrovandi:2008sd,Alexander:2008wi,Capozziello:2007tj,Janssen:2007yu,Mantz:2008hm,Pereira:2007rb,Puetzfeld:2007ye} and drawn criticism \cite{Flanagan:2007dc,Hehl:private,Hehl:2007bn}.  The controversies are on two levels.  On the technical level, in \cite{Mao:2006bb} we developed as an illustrative example a family of tetrad theories, the so-called Einstein-Hayashi-Shirafuji (EHS) Lagrangian, in Riemann-Cartan space which linearly interpolates between GR and the Hayashi-Shirafuji theory.  After we submitted the first version of \cite{Mao:2006bb}, Flanagan and Rosenthal \cite{Flanagan:2007dc} pointed out that the EHS Lagrangian has serious defects.  More specifically, in order for the EHS Lagrangian to be a consistent theory (i.e. ghost-free and having well-posed initial value formulation), the parameters of the EHS Lagrangian need to be carefully pre-selected, and in addition the torsion tensor needs to be minimally coupled to matter.  Satisfying these requirements, however, results in a theory that violates the ``action equals reaction'' principle.  Ultimately, then, the EHS Lagrangian does not yield a consistent theory that is capable of predicting a detectable torsion signal for gyroscope experiments.  It is worth noting, however, that Flanagan and Rosenthal paper \cite{Flanagan:2007dc} leaves open the possibility that there may be other viable Lagrangians in the same class (where spinning objects generate and feel propagating torsion).  The EHS Lagrangian should therefore not be viewed as a viable physical model, but as a pedagogical toy model giving concrete illustrations of the various effects and constraints that we discuss.  

On the level of perspectives, Hehl \cite{Hehl:private} argued that orbital angular momentum density is not a tensor in the field theory since the orbital angular momentum depends on the reference point and the point where momentum acts.  Therefore the orbital (and rotational) angular momentum cannot be the source of torsion.  In addition, using the multipole exansion method and conservation laws from Noether's theorem, Puetzfeld and Obukhov \cite{Puetzfeld:2007ye} argued that non-Riemannian spacetime can only be detected by test particles with intrinsic spins.  Their arguments altogether imply that there must be zero torsion in the solar system (no source), and that the GPB gyroscopes, since they have no net polarization, cannot register any signal due to torsion (no coupling).  From our point of view, however, the questions of torsion source and coupling have not yet been rigorously settled.  The spirit behind our work in \cite{Mao:2006bb} is that 
the answers to these difficult questions can and should be tested experimentally, and that it never hurts to place experimental constraints on an effect even if there are theoretical reasons that favor its non-existence.  
The history of science is full of theoretically unexpected discoveries.    
An example is the discovery of high temperature superconductivity in ceramic compounds containing copper-oxide planes: only in metals and metal alloys that had been cooled below 23 K had superconductivity been observed before the mid-1980s, but in 1986 Bednorz and M\"uller \cite{muller-bednorz} discovered that the lanthanum copper oxide, which is an insulator, becomes a superconductor with a high transition temperature of 36 K when it is doped with barium.  
In the same spirit, we feel that it is valuable to constrain the torsion parameters using the GPB data, despite the non-existence arguments mentioned above.

\section{Cosmology and 21cm tomography}

\subsection{Cosmological parameters}

Thanks to the spectacular technological advancements in circuits and computers, modern cosmologists are fortunate to live in the era of precision cosmology.  Using the avalanche of astrophysical data from CMB experiments, large scale galaxy surveys, Type IA supernovae, Ly$\alpha$ forest, gravitational lensing and future probes (e.g. 21cm tomography), cosmologists can constrain cosmological parameters to unprecedented accuracies, and in the future may even be able to measure cosmological  functions in addition to parameters.  In this section, we will give an overview of cosmological parameters, also summarized in Table \ref{tab:cospar}.

Just like there is a concordance theory --- GR --- in the area of gravitation, there is a concordance model --- the standard cosmological model with inflation --- in cosmology, successfully parametrized in terms of merely six cosmological parameters.  The standard cosmological model is based on the following assumptions:
\begin{enumerate}
\item On large scales, the universe is spatially homogeneous and isotropic (i.e. invariant under translation and rotation) and density fluctuations are small.  

\item The correct gravitational theory is GR.  

\item The universe consists of ordinary baryonic matter, cold non-baryonic dark matter, dark energy, and electromagnetic and neutrino background radiation.

\item The primordial density fluctuations are seeded during an inflationary epoch in the early universe.  
 
\end{enumerate}
By the first assumption, the intimidating non-linear partial differential equations of GR can be accurately solved by using Taylor expansions to linear order in the density fluctuations.  Thus, the full description of cosmology consists of two parts: zeroth order (ignoring fluctuations) and linear order (concerning perturbations). 

\begin{table}
\noindent 
{\tiny
\begin{center}
\caption[Cosmological parameters measured from WMAP and SDSS LRG data]{ 
Cosmological parameters measured from WMAP and SDSS LRG data.  Error bars are $1\sigma$.  This table is adapted from Table 2 of \cite{Tegmark:2006az}.}
\label{tab:cospar}

\begin{tabular}{|l|l|l|p{5.5cm}|}
\hline
Parameter		&Value	 &Meaning			&Definition\\
\hline
\multicolumn{3}{|l|}{Matter budget parameters:}&\\
\cline{1-4}
$        \Omega_{\rm{tot}}$      &$ 1.003^{+ 0.010}_{- 0.009}$                                          &Total density/critical density 	&$\Ot=\Om+\Ol=1-\Ok$\\
$           \Omega_\Lambda$      &$ 0.761^{+ 0.017}_{- 0.018}$                                          &Dark energy density parameter  	&$\Ol\approx h^{-2}\rho_\Lambda(1.88\times 10^{-26}$kg$/$m$^3)$\\
$                 \omega_b$      &$ 0.0222^{+ 0.0007}_{- 0.0007}$                                       &Baryon density		 	&$\ob=\Ob h^2 \approx \rho_b/(1.88\times 10^{-26}$kg$/$m$^3)$\\
$                 \omega_c$      &$ 0.1050^{+ 0.0041}_{- 0.0040}$                                       &Cold dark matter density 	 	&$\ocdm=\Oc h^2 \approx \rho_c/(1.88\times 10^{-26}$kg$/$m$^3)$\\
$               \omega_\nu$      &$< 0.010 \>(95\percent)$                                              &Massive neutrino density		&$\on=\On h^2 \approx \rho_\nu/(1.88\times 10^{-26}$kg$/$m$^3)$\\
$                        w$      &$-0.941^{+ 0.087}_{- 0.101}$                                          &Dark energy equation of state		&$p_\Lambda/\rho_\Lambda$ (approximated as constant)\\
\hline
\multicolumn{3}{|l|}{Seed fluctuation parameters:}&\\
\cline{1-4}
$                      A_s$      &$ 0.690^{+ 0.045}_{- 0.044}$                                          &Scalar fluctuation amplitude		&Primordial scalar power at $k=0.05$/Mpc\\
$                        r$      &$< 0.30 \>(95\percent)$                                               &Tensor-to-scalar ratio		&Tensor-to-scalar power ratio at $k=0.05$/Mpc\\
$                      n_s$      &$ 0.953^{+ 0.016}_{- 0.016}$                                          &Scalar spectral index			&Primordial spectral index at $k=0.05$/Mpc\\
$                    n_t+1$      &$ 0.9861^{+ 0.0096}_{- 0.0142}$                                       &Tensor spectral index			&$\nt=-r/8$ assumed\\
$                   \alpha$      &$-0.040^{+ 0.027}_{- 0.027}$                                          &Running of spectral index		&$\alpha=d\ns/d\ln k$ (approximated as constant)\\
\hline
\multicolumn{3}{|l|}{Nuisance parameters:}&\\
\cline{1-4}
$                     \tau$      &$ 0.087^{+ 0.028}_{- 0.030}$                                          &Reionization optical depth		&\\
$                        b$      &$ 1.896^{+ 0.074}_{- 0.069}$                                          &Galaxy bias factor			&$b=[P_{\rm galaxy}(k)/P(k)]^{1/2}$ on large scales, where $P(k)$ refers to today.\\
\hline\hline
\multicolumn{3}{|l|}{Other popular parameters (determined by those above):}&\\
\cline{1-4}
$                        h$      &$ 0.730^{+ 0.019}_{- 0.019}$                                          &Hubble parameter			&$h = \sqrt{(\ob+\ocdm+\on)/(\Ot-\Ol)}$\\
$                 \Omega_m$      &$ 0.239^{+ 0.018}_{- 0.017}$                                          &Matter density/critical density	&$\Om=\Ot-\Ol$\\
$                 \Omega_b$      &$ 0.0416^{+ 0.0019}_{- 0.0018}$                                       &Baryon density/critical density 	&$\Ob=\ob/h^2$\\
$                 \Omega_c$      &$ 0.197^{+ 0.016}_{- 0.015}$                                          &CDM density/critical density 		&$\Oc=\ocdm/h^2$\\
$               \Omega_\nu$      &$< 0.024 \>(95\percent)$                                              &Neutrino density/critical density 	&$\On=\on/h^2$\\
$                 \Omega_k$      &$-0.0030^{+ 0.0095}_{- 0.0102}$                                       &Spatial curvature			&$\Ok=1-\Ot$\\
$                 \omega_m$      &$ 0.1272^{+ 0.0044}_{- 0.0043}$                                       &Matter density			&$\om=\ob+\ocdm+\on = \Om h^2$\\
$                    f_\nu$      &$< 0.090 \>(95\percent)$                                              &Dark matter neutrino fraction		&$\fn=\rho_\nu/\rho_d$\\
$                      A_t$      &$< 0.21 \>(95\percent)$                                               &Tensor fluctuation amplitude		&$\At=r\As$\\
$                    M_\nu$      &$< 0.94 \>(95\percent)$ eV                                            &Sum of neutrino masses 		&$\Mnu\approx(94.4\>{\rm eV})\times\on$~~~\cite{KolbTurnerBook}\\
$                 A_{.002}$      &$ 0.801^{+ 0.042}_{- 0.043}$                                          &WMAP3 normalization parameter		&$\As$ scaled to $k=0.002$/Mpc: $A_{.002} = 25^{1-\ns}\As$ if $\al=0$\\
$                 r_{.002}$      &$< 0.33 \>(95\percent)$                                               &Tensor-to-scalar ratio (WMAP3)	&Tensor-to-scalar power ratio at $k=0.002$/Mpc\\
$                 \sigma_8$      &$ 0.756^{+ 0.035}_{- 0.035}$                                          &Density fluctuation amplitude		&$\sigma_8=\{4\pi\int_0^\infty [{3\over x^3}(\sin x-x\cos x)]^2 P(k) {k^2 dk\over(2\pi)^3}\}^{1/2}$, $x\equiv k\times 8h^{-1}$Mpc\\
$   \sigma_8\Omega_m^{0.6}$      &$ 0.320^{+ 0.024}_{- 0.023}$                                          &Velocity fluctuation amplitude&\\
\hline
\multicolumn{3}{|l|}{Cosmic history parameters:}&\\
\cline{1-4}
$              z_{\rm{eq}}$      &$3057^{+105}_{-102}$                                                  &Matter-radiation Equality redshift	&$z_{\rm eq}\approx 24074\om - 1$\\
$             z_{\rm{rec}}$      &$1090.25^{+ 0.93}_{- 0.91}$                                           &Recombination redshift		&$z_{\rm rec}(\om,\ob)$ given by eq.~(18) of \cite{Hu04}\\
$             z_{\rm{ion}}$      &$11.1^{+ 2.2}_{- 2.7}$                                                &Reionization redshift (abrupt)	&$\zion\approx 92 (0.03h\tau/\ob)^{2/3}\Om^{1/3}$ (assuming abrupt reionization; \cite{reion})\\
$             z_{\rm{acc}}$      &$ 0.855^{+ 0.059}_{- 0.059}$                                          &Acceleration redshift			&$\zacc=[(-3w-1)\Ol/\Om]^{-1/3w}-1$ if $w<-1/3$\\
$              t_{\rm{eq}}$      &$ 0.0634^{+ 0.0045}_{- 0.0041}$ Myr                                   &Matter-radiation Equality time 	&$\teq \approx$($9.778$ Gyr)$\times h^{-1}\int_{\zeq}^\infty [H_0/H(z)(1+z)]dz$~~~\cite{KolbTurnerBook}\\
$             t_{\rm{rec}}$      &$ 0.3856^{+ 0.0040}_{- 0.0040}$ Myr                                   &Recombination time 			&$\trec\approx$($9.778$ Gyr)$\times h^{-1}\int_{\zrec}^\infty [H_0/H(z)(1+z)]dz$~~~\cite{KolbTurnerBook}\\
$             t_{\rm{ion}}$      &$ 0.43^{+ 0.20}_{- 0.10}$ Gyr                                         &Reionization time			&$\tion\approx$($9.778$ Gyr)$\times h^{-1}\int_{\zion}^\infty [H_0/H(z)(1+z)]dz$~~~\cite{KolbTurnerBook}\\
$             t_{\rm{acc}}$      &$ 6.74^{+ 0.25}_{- 0.24}$ Gyr                                         &Acceleration time 			&$\tacc\approx$($9.778$ Gyr)$\times h^{-1}\int_{\zacc}^\infty [H_0/H(z)(1+z)]dz$~~~\cite{KolbTurnerBook}\\
$             t_{\rm{now}}$      &$13.76^{+ 0.15}_{- 0.15}$ Gyr                                         &Age of Universe now			&$\tnow\approx$($9.778$ Gyr)$\times h^{-1}\int_0^\infty [H_0/H(z)(1+z)]dz$~~~\cite{KolbTurnerBook}\\
\hline
\multicolumn{3}{|l|}{Fundamental parameters (independent of observing epoch):}&\\
\cline{1-4}
$                        Q$      &$ 1.945^{+ 0.051}_{- 0.053}$ $\times10^{-5}$                          &Primordial fluctuation amplitude	&$Q=\delta_h\approx A_{.002}^{1/2}\times 59.2384\mu$K$/T_{\rm CMB}$\\
$                   \kappa$      &$ 1.3^{+ 3.7}_{- 4.3}$ $\times10^{-61}$                               &Dimensionless spatial curvature \cite{Q}&$\kappa=(\hbar c/k_B T_{\rm CMB} a)^2 k$\\ 
$             \rho_\Lambda$      &$ 1.48^{+ 0.11}_{- 0.11}$ $\times10^{-123}\rho_{\rm{Pl}}$             &Dark energy density			&$\rho_\Lambda\approx h^2\Ol\times(1.88\times 10^{-26}$kg$/$m$^3)$\\
$         \rho_{\rm{halo}}$      &$ 6.6^{+ 1.2}_{- 1.0}$ $\times10^{-123}\rho_{\rm{Pl}}$                &Halo formation density		&$\rhohalo=18\pi^2 Q^3\xi^4$\\
$                      \xi$      &$ 3.26^{+ 0.11}_{- 0.11}$ eV                                          &Matter mass per photon 		&$\xi =\rhom/\ng$\\
$                    \xi_b$      &$ 0.569^{+ 0.018}_{- 0.018}$ eV                                       &Baryon mass per photon 		&$\xib=\rhob/\ng$\\
$                    \xi_c$      &$ 2.69^{+ 0.11}_{- 0.10}$ eV                                          &CDM mass per photon 			&$\xic=\rhoc/\ng$\\
$                  \xi_\nu$      &$< 0.26 \>(95\percent)$ eV                                            &Neutrino mass per photon 		&$\xin=\rhon/\ng$\\
$                     \eta$      &$ 6.06^{+ 0.20}_{- 0.19}$ $\times10^{-10}$                            &Baryon/photon ratio			&$\eta=n_b/n_g=\xib/m_p$\\
$                A_\Lambda$      &$2077^{+135}_{-125}$                                                  &Expansion during matter domination	&$(1+\zeq)(\Om/\Ol)^{1/3}$ \cite{tegmark05}\\  
$      \sigma^*_{\rm{gal}}$      &$ 0.561^{+ 0.024}_{- 0.023}$ $\times10^{-3}$                          &Seed amplitude on galaxy scale	&Like $\sigma_8$ but on galactic ($M=10^{12} M_\odot$) scale early on\\
\hline
\end{tabular}
\end{center}     
} 
\end{table}

\subsubsection{Zeroth order: the cosmic expansion}

To zeroth order, the metric for a spatially homogeneous and isotropic universe is completely specified by the so-called Friedmann-Robertson-Walker(FRW) line element, 
\ben
ds^2=-c^2 dt^2+a(t)^2 \left(\frac{dr^2}{1-kr^2}+r^2d\theta^2+r^2\sin^2\theta d\phi^2 \right)\,,
\een
which has only one free function $a(t)$, describing the expansion of the universe over time, and one free parameter $k$, the curvature of the 3D space.  The Hubble parameter is 
defined as $H(z)\equiv d\ln a/dt$  where the redshift is $1+z\equiv a(t_{\rm today})/a(t)$.  The Hubble parameter is both more closely related to observations, and determined by the Friedmann equation 
\ben
H(z)^2=\frac{8\pi G}{3}\rho(z)-k\frac{c^2}{a^2}\,,
\een
obtained by applying the EFE to the FRW metric and a perfect fluid with density $\rho$ and pressure $p$.  Here $G$ is Newton's gravitational constant.  The Hubble parameter today is usually written $H_0=100h\,{\rm km \,s^{-1}\, Mpc^{-1}}$ where $h$ is a unitless number parametrizing our ignorance.  The measured value is $h=0.73\pm 0.02$ from WMAP+SDSS data \cite{Tegmark:2006az}.

Cosmological parameters and their measured values are summarized in Table \ref{tab:cospar}.  A critical density $\rhocr\equiv 3H_0^2/8\pi G$ can be defined such that a universe with total current density equal to $\rhocr$ is flat ($k=0$).  The matter budget of the universe can be quantified by dimensionless parameters as follows: total matter density $\Omega_m\equiv \rho_{m,0}/\rhocr$, baryonic matter density $\Omega_b \equiv \rho_{b,0}/\rhocr$, dark matter density $\Omega_d\equiv \rho_{d,0}/\rhocr$, massive neutrino density $\Omega_\nu\equiv \rho_{\nu,0}/\rhocr$, electromagnetic radiation density $\Omega_r\equiv \rho_{r,0}/\rhocr$, and spatial curvature $\Omega_k\equiv -k\,c^2/H_0^2$.  The subscript ``0'' denotes the value at the present epoch.  The simplest model for dark energy is a cosmological constant (c.c.) $\Lambda$, or the vacuum energy, corresponding to the parameter $\Omega_\Lambda\equiv \Lambda /3H_0^2$.  A popular approach to generalizing the c.c. is to assume that the equation of state for dark energy $w\equiv p_\Lambda / \rho_\Lambda$ is constant.  

These parameters are not all independent, e.g. $\Omega_\Lambda+\Omega_m+\Omega_k=1$ ($\Omega_r$ is negligible) and $\Omega_m=\Omega_b+\Omega_d$.  The mathematically equivalent quantities more closely related to observations are $\omega_b\equiv \Omega_b h^2$, $\omega_d\equiv \Omega_d h^2$, $\omega_m\equiv \Omega_m h^2$, $\omega_\nu\equiv \Omega_\nu h^2$, dark matter neutrino fraction $f_\nu\equiv \Omega_\nu/\Omega_d$, and sum of neutrino masses $m_\nu \approx (94.4\,{\rm eV})\times \Omega_\nu h^2$, since these quantities are simply proportional to the corresponding densities.  
The energy density $\rho$ of these components have simple dependences on redshift: $\rho_m(z)=\rho_{m,0} (1+z)^3$, $\rho_\Lambda(z)=\rho_{\Lambda,0}$, $\rho_k(z)=\rho_{k,0} (1+z)^2$, and $\rho_r(z)=\rho_{r,0}(1+z)^4$.  Thus, the Friedmann equation relates the Hubble parameter to these unitless matter budget parameters,
\ben
H(z)=H_0\sqrt{\Omega_\Lambda+\Omega_m (1+z)^3 + \Omega_r (1+z)^4 + \Omega_k (1+z)^2}\,.
\een  

\subsubsection{First order: the density fluctuations}
To linear order, perturbations come in two important types: gravitational waves and density fluctuations.  The former propagate with the speed of light without growing in amplitude.  The latter, however, can get amplified by gravitational instability, and are therefore responsible for structure formation.  Density fluctuations are so far observationally consistent with 
having uncorrelated Gaussian-distributed amplitudes.  It is therefore sufficient to use a single function, the so-called power spectrum $P(k,z)$ which gives the variance of the fluctuations as a function of wavenumber $k$ and redshift $z$, to characterize the first-order density perturbations.  In principle, $P(k,z)$ can be computed by solving linearized EFE that involves fluctuations in the metric, energy density, pressure, and sometimes shear.  In general, $P(k,z)$ depends on three things: 
\begin{enumerate}
\item The cosmic matter budget
\item The seed fluctuations in the early universe
\item Galaxy formation, including reionization, bias, etc.
\end{enumerate}

In the currently most popular scenario, a large and almost constant energy density stored in a scalar field caused an exponentially rapid expansion $a(t)\sim e^{Ht}$ at perhaps $t\lesssim 10^{-34}$ seconds during a period known as inflation.  The theory of inflation can successfully predict negligible spatial curvature ($\Omega_k=0$), and solve the horizon problem that the last scattering surface was naively out of causal contact in the non-inflationary standard model while the cosmic microwave background radiation (CMBR) is highly spatially homogeneous and isotropic ($\delta T/T\sim 10^{-5}$).  Furthermore, inflation can stunningly explain where seed density fluctuations were created: 
microscopic quantum fluctuations in the aftermath of the Big Bang were stretched to enormous scales during the inflationary epoch.  After inflation ended, these seed fluctuations grew into the observed galaxies and galaxy clustering patterns by gravitational instability.  The theory of inflation generically predicts almost Gaussian-distributed primordial fluctuations and a nearly scale invariant ($n_s\approx 1$) adiabatic scalar power spectrum with subdominant gravitational waves.  In typical inflation models, the initial power spectrum can be written in the approximate form  
\ben
P_\Phi(k) = A_s (k/k_{\rm fid})^{n_s} \label{eqn:ps}
\een
for the fluctuations in the gravitational potential.  Here $A_s$ is the scalar fluctuation amplitude, and $n_s$ the scalar spectral index, at $k_{\rm fid}=0.05\,{\rm Mpc}^{-1}$.  The minimal set of cosmological parameters approximates $n_s$ to be constant.  In a conservative extension, $n_s(k)$ runs linearly in $\ln(k)$, i.e. 
\ben
n_s(k)=n_s(k_{\rm fid})+\alpha \ln(k/k_{\rm fid})\,,
\een
where $\alpha$, the logarithmic running of the tilt, is approximated as a constant.  
In addition to scalar perturbations, the tensor perturbations, related to subdominant gravitational waves, were seeded with the initial power spectrum written in the same form as \Eq{eqn:ps} except for $A_s$ and $n_s$ replaced by the tensor fluctuation amplitude $A_t$ and the tensor spectral index $n_t+1$, respectively.  A quantity more closely related by observations is the tensor-to-scalar ratio $r\equiv A_t/A_s$.  

When seed fluctuations grow into stars, galaxies and galaxy clustering patterns, a number of complicated astrophysical processes are triggered by the structure formation and may influence the clumpiness.  For example, during the Epoch of Reionization ($6\lesssim z\lesssim 20$), the newly-formed Pop-III stars emitted Ly$\alpha$ photons, and x-rays that re-ionized neutral hydrogen atoms in the inter-galactic medium.  Some microwave background photons that have propagated during billions of years from the distant last scattering surface were scattered from the intervening free electrons, generating more anisotropies in the CMBR through the so-called Sunyaev-Zeldovich effect.  As a consequence, the CMB power spectrum is sensitive to an integrated quantity known as the reionization optical depth $\tau$.  The Epoch of Reionization is one of the most poorly understood epochs in the cosmic evolution and is therefore of particular interest to cosmologists and astrophysicists.  

In addition to reionization, the power spectrum of density fluctuations for galaxies or gas depends on the linear bias $b$.  Ordinary baryonic matter cannot gravitate enough to form the observed clumpy structure such as galaxies.  In the currently most popular scenario, instead, the observed galaxies trace dark matter halos.  As a result, the observed power spectrum from galaxy surveys should be closely related to the real matter power spectrum.  A simple widely used model is that $P_{\rm galaxy}=b^2 P_{\rm mass}$ on large scales.  

\subsubsection{$\Lambda$CDM model}

As discussed above, the power spectrum $P(k,z)$ of density fluctuations depends on the cosmic matter budget, the seed fluctuations and nuisance astrophysical parameters.  It is striking that the concordance model can fit everything with a fairly small number of cosmological parameters.  In this model, the cosmic matter budget consists of about 5\% ordinary matter, 30\% cold dark matter, $\lesssim 0.1\%$ hot dark matter (neutrinos) and 65\% dark energy.  The minimal model space, so-called vanilla set, is parametrized by  $(\Omega_\Lambda,\omega_b,\omega_c, A_s, n_s, \tau, b)$, setting $\Omega_k=\omega_\nu=\alpha=r=n_t=0$ and $w=-1$.  We show a comprehensive set of cosmological parameters in Table \ref{tab:cospar}.

\subsection{A brief history of the universe}

\subsubsection{Cosmic plasma}

According to the Big Bang theory, the early universe was filled with hot plasma whose contents evolved over time through a series of phase transitions.  
In the very early universe, 
the particle constituents were all types of particles in the Standard Model (SM) of particle physics, unidentified dark matter (DM) particles from some extended model of particle physics beyond the SM (e.g. lightest supersymmetric particle and/or axions), and an equal amount of all corresponding anti-particles.  The universe cooled as it expanded.  When the thermal energy of the cosmic plasma dropped roughly below the rest energy of DM particles, DM particles froze out (at $t\sim 10^{-10}\,{\rm seconds}$ for typical WIMPs) and have not been in thermal equilibrium with other constituents since.  DM particles eventually became an almost collisionless and cold (non-relativistic) component that constitutes about 20\% of the cosmic matter budget at the present day.  

As the cosmic temperature kept decreasing, the symmetry between baryons and anti-baryons was broken at $t\approx 10^{-4}\,{\rm seconds}$.  The tiny asymmetry at the level of $10^{-9}$ was followed by matter-antimatter annihilation, forming protons that constitute about 4\% of the cosmic matter budget at the present day.  This is a hypothetical process known as baryogenesis.  After the baryogenesis, the cosmic hot soup was a cauldron of protons, electrons and photons, and a smattering of other particles (e.g. hot neutrinos).  

When the universe was cooled to below about 1 MeV -- the mass difference between a neutron and a proton -- neutrons froze out at $t\approx 2\,{\rm minutes}$ as weak interactions like $p+e^- \leftrightarrow n+\nu_e$ ceased.  Subsequently, protons and neutrons combined to form light element such as deuterium ($\phantom{}^2 {\rm H}$ or ${\rm D}$), tritium ($\phantom{}^3 {\rm H}$) and helium ($\phantom{}^3 {\rm He}$ and $\phantom{}^4 {\rm He}$) in a process known as big bang nucleosynthesis (BBN).  For example, deuterium forms via $p+n\to {\rm D}+\gamma$;  then ${\rm D}+{\rm D}\to n+\phantom{}^3 {\rm He}$, after which $\phantom{}^3 {\rm He} + {\rm D} \to p + \phantom{}^4 {\rm He}$.  The helium nucleus ($\phantom{}^4 {\rm He}$) is the most stable among light elements, and after BBN, about 75\% of baryons in 
the universe are hydrogen nuclei (i.e. protons), while nearly 25\% are helium nuclei.  

The freely-moving electrons tightly coupled to photons via Compton scattering and electrons to protons and other nuclei via Coulomb scattering, keeping the cosmic plasma in equilibrium.  All components except for photons were in the form of ions until temperatures fell to $3,000$ Kelvin, when protons and electrons combined to form electrically neutral hydrogen atoms --- a process known as recombination.  The photons at that temperature were no longer energetic enough to re-ionize significant amounts of electrons.  The Compton scattering process therefore ended, decoupling the photons from the matter.  Thus, the cosmos become almost transparent to photons, releasing the microwave background.  The gas temperature continues to drop as the universe expands, so 
one might expect that the cosmic gas would still be cold and neutral today.  

Surprisingly, it is not.  To understand why, we take a detour and first review how galaxies form, and come back to this question subsequently.  

\begin{figure}
\centering
\includegraphics[width=1\textwidth]{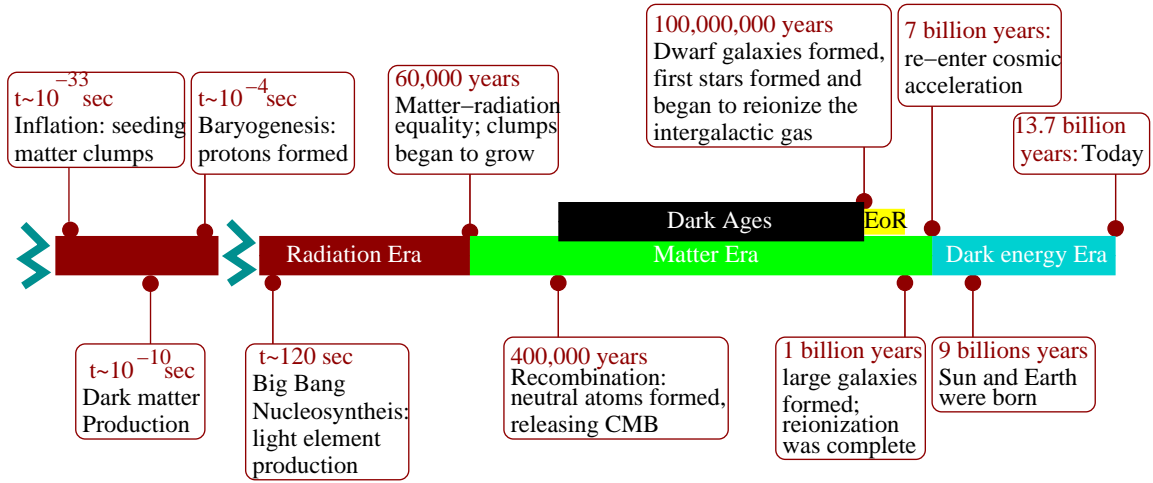}
\caption[Cosmic time line]{Cosmic time line: a brief history of the universe from the aftermath of the big bang to the present day.}
\label{fig:cosmohistory}
\end{figure}

\subsubsection{Galaxy formation}

According to the current most popular scenario, 
at $t\lesssim 10^{-33}\,{\rm seconds}$, the universe underwent a period of inflation.  The cosmic inflation stretched the universe by $\gtrsim$ 55 e-foldings, i.e. a lattice grid was more than $e^{55}\approx 10^{24}$ times larger than itself before inflation, making the universe extremely flat.  After inflation, the universe was approximately spatially homogeneous and isotropic because particles at any two largely separated points that otherwise could by no means have causal contact without inflation may actually be in close together during the inflation, and equilibrate their temperatures through the exchange of force carriers that would have had time to propagate back and forth between them.  

Cosmic inflation also created seed fluctuations at the level of one part in a hundred thousand in the early universe.  After the end of inflation, however, the universe was dominated by radiation, i.e. ultra-relativistic particles that moved fast enough to keep these primordial density fluctuations from growing.  Fortunately, the energy density of radiation dropped more rapidly than matter density as the universe expanded; quantitatively, $\rho\propto a(t)^{-4}$ for radiation and $\rho\propto a(t)^{-3}$ for matter.  Consequently, at $t\approx 60,000$ years, matter became the dominant component of the universe, and the fluctuations began to grow due to gravitational instability --- which means that a region that started slightly denser than average pulled itself together by its own gravity.  More specifically, the denser region initially expanded with the whole universe, but its extra gravity slowed its expansion down, turned it around and eventually made the region collapse on itself to form a bound object such as a galaxy.  

\subsubsection{Reionization: cosmic plasma revisited}

Now we come back to the question: is the present universe filled with mostly neutral hydrogen atoms?  Although the terrestrial world is composed of atoms, the intergalactic medium hosts the majority of ordinary matter in the form of plasma.  Conclusive evidence comes from two types of observations.  The Wilkinson Microwave Anisotropy Probe (WMAP) and other experiments have confirmed that the CMBR is slightly polarized (in so-called EE modes).  Since only free electrons (and not neutral hydrogen atoms) scatter and polarize this radiation, the amount of polarization observed on large angular scales suggests that the neutral gas was reionized into plasma as early as a few hundred million years after our big bang.  Independent confirmation of cosmic reionization come from the observed spectra of the distant quasars that indicates that reionization should be complete by a billion years after the big bang.  

The details of cosmic reionization are still a blank page that needs to be filled by upcoming observations.  There are, however, some plausible pictures that reside in the minds of theorists.  In the current models, the oldest galaxies are dwarf galaxies that started to form at a cosmic age of a few hundred million years.  Larger galaxies such as the Milky Way were latecomers that were born from the gradual coalescence of many dwarf galaxies.  Stars were created when the gas in embryonic galaxies got cool and fragmented.  The first generation of stars, so-called Pop-III stars, triggered the nuclear fusion of hydrogen and released energy in the form of ultraviolet photons in amounts a million times larger than the energy needed to ionize the same mass of neutral gas ($13.6\,{\rm eV}$ for each hydrogen atom).  The emitted ultraviolet photons leaked into the intergalactic medium, broke the neutral hydrogen atoms back down into their constituent protons and electrons, and created an expanding bubble of ionized gas.  As new galaxies took root, more bubbles appeared, overlapped and eventually filled all of intergalactic space.

Some researchers conjecture that black holes rather than stars may have caused cosmic reionization.  Like stars, black holes arise from galaxies.  Particles that plummeted into black holes emitted x-rays in an amount of energy 10 million times larger than the ionization energy of the same amount of hydrogen.  The mechanisms of reionization by massive stars or black holes can be distinguished by observing the boundaries of ionized bubbles in upcoming experiments.   Ultraviolet photons emitted by massive stars were easily absorbed by the neutral gas, while x-rays from black holes can penetrate deeply into the intergalactic medium, so, black holes are associated with fuzzier bubble boundaries.

\subsubsection{Dark Ages}

Both reionization models predict that the cosmic reionization started to take shape after the first galaxies formed at $t\approx 100,000,000\,{\rm years}$.  Between the release of the microwave background at $t\approx 400,000\,{\rm years}$ and the formation of first galaxies, however, there is a tremendous gap!  During these so-called Dark Ages (DA), the universe was dark since ordinary matter was in the form of neutral atoms that were not hot enough to radiate light.  Since the cosmic matter was transparent to the microwave background photons, the CMB photons no longer traced the distribution of matter.  However, the DA were not a boring period.  In fact, the DA are an interesting embryonic interlude between the seeding of density fluctuations and the birth of first galaxies: within the inky blackness, the primordial matter clumps grew by their extra gravity and eventually collapsed on themselves into galaxies.  The secret of galaxy formation is hidden in the DA.  

But how can we probe a period that was by its very nature practically dark?  Fortunately, even cold hydrogen atoms emit feeble light with a wavelength of 21 centimeters.  Below we describe how observations of the 21cm line are emerging as a promising probe of the epoch of reionization (EoR) and the Dark Ages.  

\subsection{21cm line: spin temperature}

In quantum mechanics, particles carry an intrinsic angular momentum known as spin.  For example, a particle with spin $1/2$ such as a proton or electron can have its angular momentum vector point either ``up'' or ``down''.  In a hydrogen atom, the interaction between the spins of the nucleus (the proton) and the electron splits the ground state into two hyperfine states, i.e., the triplet states of parallel spins and the singlet state of anti-parallel spins.  The anti-parallel spin state has lower energy than the parallel spin state, and the transition between them corresponds to the emission or absorption of a photon with the wavelength of 21 centimeters.  For the 21cm transition, the so-called spin temperature $T_s$ quantifies the fraction of atoms in each of the two states: the ratio of number densities is 
\ben
\frac{n_1}{n_0}\equiv\frac{g_1}{g_0}e^{-E_{10}/k_B T_s}=3e^{-T_*/T_s}\,.
\een 
Here the subscripts 1 and 0 denote the parallel and the anti-parallel spin state, respectively.  $n_i$ is the number density of atoms in the $i$-th state, and $g_i$ is the statistical weight ($g_1$=3 and $g_0$=1), $E_{10}=5.9\times 10^{-6}\,{\rm eV}=\hbar c/21\,{\rm cm}$ is the energy splitting, and $T_* = E_{10}/k_B = 0.068\,{\rm K}$ is the equivalent temperature.  

21cm observations aim to compare lines of sight through intergalactic hydrogen gas to hypothetical sightlines without gas and with clear views of CMB.  Thus, one should observe emission lines if $T_s > T_{\rm cmb}$, or absorption lines if $T_s <  T_{\rm cmb}$.  Here $T_{\rm cmb} (z) = 2.73 (1+z)\,{\rm K}$ is the CMB temperature at redshift $z$.  

There are three competing mechanisms that drive $T_s$: (1) absorption of CMB photons; (2) collisions with other hydrogen atoms, free electrons and protons; and (3) scattering of ultraviolet photons.  For the first mechanism, the process of absorption of microwave background photons tends to equilibrate the spin temperature with the CMB temperature.  
For the second, spin exchange due to collisions is efficient when gas density is large.  The third mechanism, also known as the Wouthuysen-Field mechanism, involves transitions from one $1s$ hyperfine state to the first excited state $2P$ and then down to the other $1s$ hyperfine state with different spin orientation, which couples the 21cm excitation to the ultraviolet radiation field.  

\begin{figure}[t]
\centering
\includegraphics[width=0.9\textwidth]{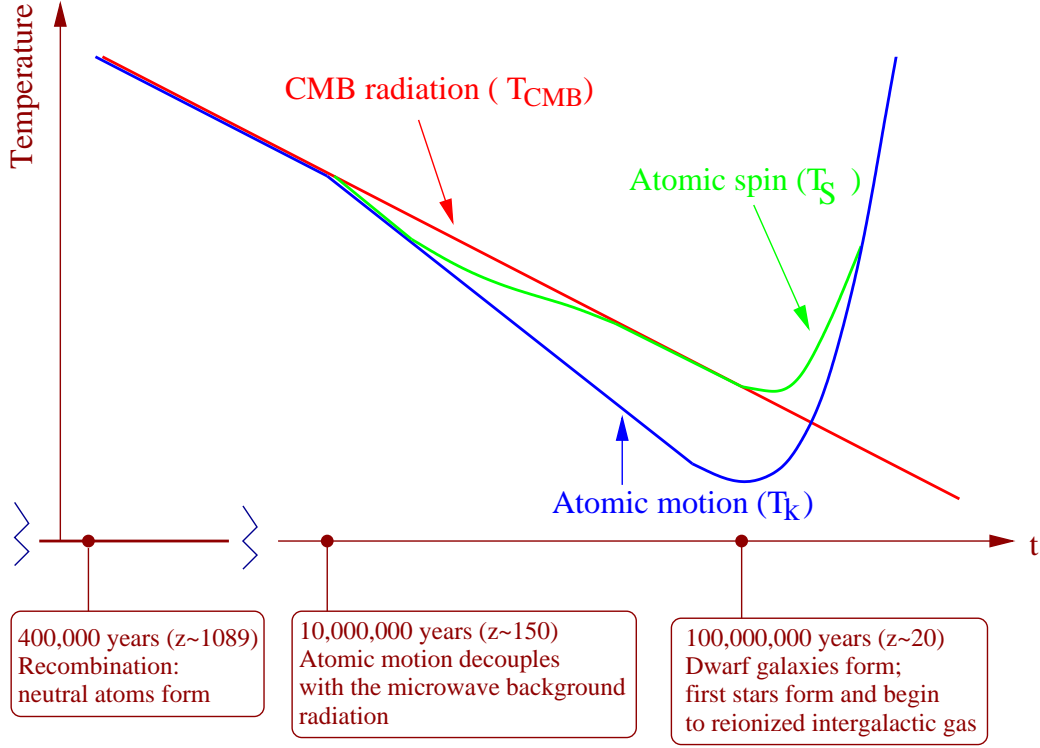}
\caption[Thermal history of the intergalactic gas]{Thermal history of the intergalactic gas}
\label{fig:IGM-history}
\end{figure}

The global history of the intergalactic gas is defined by three temperatures: the spin temperature $T_s$ (a measure of the spin excitation) as defined above, the kinetic temperature $T_k$ of the intergalactic gas (a measure of atomic motions), and the microwave background temperature $T_{\rm cmb}$ (a measure of the energy of background photons).  These temperatures can approach or deviate from one another, depending on which physical processes are dominant.  In a three-way relation (see Figure \ref{fig:IGM-history}), after an initial period when three temperatures are all equal, spin temperature first traces the kinetic temperature, then the background temperature, and eventually the kinetic temperature again.

Initially after the CMB is released, although the neutral atoms are transparent to the background photons, free electrons left over from recombination mediate the exchange of energy between background photons and atoms via Compton scattering (between photons and free electrons) and Coulomb scattering (between free electrons and hydrogen nuclei).  The kinetic temperature therefore tracks the CMB temperature, and also the spin temperature due to collisions between hydrogen atoms.  Observations of this period will therefore find neither emission or absorption of 21cm lines against the microwave background.

The first transition took place when the universe was about 10 million years old.  As the universe expanded, the gas was diluted and cooled, and the free electron mediation eventually became so inefficient that the atomic motions decoupled from the background radiation at a redshift of about 150 and underwent adiabatic cooling at a more rapid rate ($T_k \propto a(t)^{-2}$) than the cooling of the CMB ($T_{\rm cmb} \propto a(t)^{-1}$). 
In this phase, the spin temperature matched the kinetic temperature due to collisions, and neutral gas absorbed the background photons.  

When the universe was close to a hundred million years old, a second transition occurred.   As the gas continued to expand, collisions between atoms became infrequent, and made the coupling between kinetic temperature and spin temperature inefficient.  
As a consequence, the spin excitation reached equilibrium with the background photons again.  Thus, we cannot observe gas from this period.

After the first stars and quasars lit up, a third transition occurred.  The intergalactic gas was heated up by ultraviolet photons, x-rays or shocks from galaxies.  In addition, spin exchange through the scattering of ultraviolet photons became important, coupling spin temperature back to approximately the kinetic temperature.  Since flipping the spins takes much less energy than ionizing atoms, neutral gas began to glow in 21cm radiation well before becoming  ionized.  Finally, as the hydrogen became fully reionized, the 21cm emission faded away.

\subsection{21cm cosmology}

The three-way relation between $T_s$, $T_k$ and $T_{\rm cmb}$ determines whether absorption or emission lines, or neither, of 21cm signals can be detected against the microwave background.  However, the observed quantities from 21cm experiments are not these temperatures.  In this section we will describe how to extract cosmological information from the 21cm signal.

The observable in 21cm experiments is the difference between the observed 21 cm brightness temperature at the redshifted frequency $\nu$ and the CMB temperature $\Tcmb$, given by \cite{field59a}
\ben
T_b(\bfx) = \frac{3 c^3 h A_{10} n_H(\bfx) [\Ts(\bfx)-\Tcmb]}{32\pi k_B \nu_0^2 \Ts(\bfx) (1+z)^2 (dv_{\parallel}/dr)}\,,
\een
where $n_H$ is number density of the neutral hydrogen gas, and
$A_{10}\approx 2.85\times 10^{-15}$s$^{-1}$ is the spontaneous decay rate of 21cm excitation. The factor
$dv_{\parallel}/dr$ is the gradient of the physical velocity along the line of sight ($r$ is the
comoving distance), which is $H(z)/(1+z)$  on average (\ie for no peculiar velocity).  Here $H(z)$ is the Hubble parameter at redshift $z$. 

21cm experiments can measure the statistical properties (such as power spectrum) of  brightness temperature and even map $T_s$.  The brightness temperature is determined by four quantities --- hydrogen mass density, spin temperature, neutral fraction, and peculiar velocity.  Among them, only fluctuations in the hydrogen mass density can be used to test cosmological models, and how to disentangle density fluctuations from other quantities remains an open question.  Although fluctuations in $T_s$ are poorly understood, this complication can be circumvented using the fact that the factor $(T_s-\Tcmb)/T_s$ is saturated to be unity when $T_s \gg \Tcmb$.  This condition is usually satisfied, 
since the gas should be heated enough by ultraviolet photons, x-rays and shocks before and during the reionization.  Consequently, the fluctuations in the factor $(T_s-\Tcmb)/T_s$ can be neglected. 

Fluctuations in neutral fraction are important during reionization, and are unfortunately also poorly understood.  In order to effectively use 21cm lines as probes of cosmology, two solutions 
to the problem of how to disentangle matter density fluctuations from 
the fluctuations in neutral fraction 
have been proposed in the past.
 
\begin{enumerate}

\item Since flipping the spin takes much less energy than ionizing an atom, it is plausible that there exists a pre-reionization period in which both $T_s\gg \Tcmb$ and $x_{\rm HI}=1$ hold.  In this period, the matter power spectrum dominates the total power spectrum.

\item As long as density fluctuations are much smaller than unity on large scales, linear perturbation theory will be valid, so the peculiar velocity can be used to decompose the total 21cm power spectrum into parts with different dependencies on $\mu$ caused by the so-called redshift space distortion, where $\mu=\hat{\bfk}\cdot\hat{{\bf n}}$ is the cosine of the angle between the Fourier vector $\bfk$ and the line of sight.  Only the forth moment in the total power spectrum, i.e. a term containing $\mu^4$, depends on the matter power spectrum alone, and all other moments are contaminated by power spectra related to fluctuations in neutral fraction.  One can in principle separate the $\mu^4$ term from the contaminated terms, and use only it to constrain cosmology.  

\end{enumerate}
In Chapter 4, we will develop a third method that exploits the smoothness of the nuisance power spectra and parametrizes them in terms of seven constants at each redshift.  Thus, the combination of cosmological parameters and nuisance parameters completely dictate total power spectrum.  This so-called MID method turns out to be as effective as the simplest methods for long-term 21cm experiments, but more accurate.

\subsection{Prospects of 21cm tomography}

By observing 21cm signal from a broad range of epochs, neutral intergalactic hydrogen gas can be mapped by upcoming 21cm experiments.  This approach, known as 21cm tomography, is emerging as one of the most promising cosmological probes for the next decades, since it encodes a wealth of information about cosmology, arguably even more than the microwave background.  The reasons behind this optimistic perspective are as follows.

First, mapping of neutral hydrogen can be done over a broad range of frequencies corresponding to different redshifts, and is therefore three-dimensional, with the third dimension along the line of sight.  In contrast, the two-dimensional microwave background is a map of anisotropy of radiations in the sky from the last scattering surface, a narrow spherical shell at the epoch of recombination.  The 3D mapping, in principle, measures a much larger number of modes than 2D mapping, and therefore has the potential to measure the matter power spectrum and cosmological parameters with less sample variance.  

Second, the range of 21cm tomography goes from the dark ages to the epoch of reionization, which is almost a complete time line of galaxy formation.  Mapping of neutral hydrogen along this time line provides an observational view of how primordial density fluctuations evolved to form galaxies, a picture that has hitherto only existed in theorists' minds.  

Third, 21cm tomography contains information not only about the matter density fluctuations that seeded galaxies, but also on the effects that the galaxies, after their formation, had on their surroundings, e.g. reionization and heating of the intergalactic gas, etc.  Separating physics (matter power spectrum) from astrophysics (ionization power spectrum, power spectrum of spin temperature fluctuations) can be used not only to constrain cosmology, but to learn about astrophysical process.

Last but not the least, 21cm tomography can shed light on testing fundamental particle physics and gravitational physics.  During the dark ages, the spin temperature traces the kinetic temperature by collisions of neutral atoms or microwave background temperature by absorption of CMB photons.  Since no complicated astrophysics (e.g. reionization) takes effect during the dark ages, the dark ages are a well controlled cosmic laboratory.  Non-standard particle physics models may have unusual decay of dark matter which imprints a signature on the dark ages.  Also, many modified gravitational theories can be distinguished by their predictions for galaxy formation.

However, observers will have to overcome a great deal of challenges.  Firstly, 
the redshifted 21cm signals fall in the low-frequency radio band, from $1.5\,{\rm m}$ to $30\,{\rm m}$.  Thus, low-frequency radio broadcasts on Earth must be filtered out.  In fact, most 21cm experiments (except LOFAR) have chosen their sites at low-population spots.  Secondly, thermal noise is approximately proportional to the wavelength to roughly the 2.6 power, because of synchrotron radio from our own galaxy.  Noise at the ultra low frequency side will therefore overwhelm the signal from the dark ages, making observation of the dark ages technically unrealistic with the upcoming first generation of 21cm experiments.  Even at the higher frequencies corresponding to the epoch of reionization, 
synchrotron foreground is about four orders of magnitude more intense than the cosmic signal.  Fortunately, the foreground spectra are smooth functions of wavelength and may vary slowly, allowing them to be accurately subtracted out.

To detect the 21cm signal, four first generation observatories --- the Murchison Widefield Array (MWA) \cite{MWA}, the 21 Centimeter Array (21CMA) \cite{21CMA}, the Low Frequency Array (LOFAR) \cite{LOFAR} and the Precision Array to Probe Epoch of Reionization (PAPER) \cite{paper} --- are currently under development.  The next generation observatory, Square Kilometre Array (SKA) \cite{SKA}, is in the fund-raising and design stage.  Furthermore, 21cm tomography optimized square kilometer array known as the Fast Fourier Transform Telescope (FFTT) \cite{FFTT}, which has been forecast to be capable of extremely accurate cosmological parameters measurements, has been proposed.  The next two decades should be a golden age for 21cm tomography, both observationally and theoretically.

\section{Road map}

The rest of this thesis is organized as follows.  In Chapter 2, we parametrize the torsion field around Earth, derive the precession rate of GPB gyroscopes in terms of the above-mentioned model-independent parameters, and constrain the torsion parameters with the ongoing GPB experiment together with other solar system tests.  We also present the EHS theory as a toy model of an angular-momentum coupled torsion theory, and constrain the EHS parameters with the same set of experiments.  The work in Chapter 2 has been published in Physical Review D  \cite{Mao:2006bb}.  
In Chapter 3, after a review of the equivalence of $f(R)$ theories with scalar tensor theories, we explore the Chameleon model and massive theories, respectively, focusing on observational constraints.  The work in Chapter 3 has been published in Physical Review D \cite{Faulkner:2006ub}.  
In Chapter 4, we explain  the assumptions that affect the forecast of cosmological parameter measurements with 21cm tomography, and also present a new method for modeling the ionization power spectra.  We quantify how the cosmological parameter measurement accuracy depends on each assumption, derive simple analytic approximations of these relations, and discuss the relative
importance of these assumptions and implications for experimental design.  
The work in Chapter 4 has been accepted for publication in Physical Review D \cite{Mao:2008ug}.  In Chapter 5, we conclude and discuss possible extensions to the work in the thesis.  

The contributions to the work in this thesis are as follows.  For \cite{Mao:2006bb}, 
I carried out all detailed calculations and plots.  Max Tegmark initially suggested the idea of constraining torsion with GPB, and he and Alan Guth were extensively involved in the discussion of results.  Serkan Cabi contributed to the discussion of generalized gravitational theories.  For \cite{Faulkner:2006ub}, Tom Faulkner carried out all detailed calculations and plots.  I checked and corrected preliminary results.  Max Tegmark initially suggested the idea of constraining viable $f(R)$ theories, and he and Ted Bunn were extensively involved in the discussion of results.  For \cite{Mao:2008ug}, I did the bulk of the work, including writing analysis software, inventing the MID model parametrization of nuisance power spectra, performing the calculations and consistency checks.  Max Tegmark initially suggested the idea of investigating how forecasts of 21cm tomography depend on various assumptions and was extensively involved in discussions of the results as the project progressed.  Matt McQuinn contributed his radiative transfer simulation results, and he, Matias Zaldarriaga and Oliver Zahn also participated in detailed discussions of results and strategy.  Oliver Zahn also helped with consistency checks of the Fisher matrix results.

\chapter{Constraining torsion with Gravity Probe B}

\section{Introduction}

Einstein's General Theory of Relativity (GR) has emerged as the hands down most popular candidate for a relativistic theory of 
gravitation, owing both to its elegant structure and to its impressive agreement with a host of experimental tests 
since it was first proposed about ninety years ago \cite{Will:2005yc,Will:1993ns,Will:2005va}. 
Yet it remains worthwhile to subject GR to further tests whenever possible, since these can either build further confidence in the theory or
uncover new physics. Early efforts in this regard focused on weak-field solar system tests, and efforts to test GR have since been extended
to probe stronger gravitational fields involved in binary compact objects, black hole accretion and cosmology \cite{Hulse:1974eb,Weisberg:2002qg,Weisberg:2004hi,Champion:2004hc,Peters:1963ux,Stairs:1997kz,Stairs:1999dv,Hotan:2004ua,Hotan:2004ub,vanStraten:2001zk,Armitage:2004ga,Chakrabarti:2004uz,Merloni:2002gx,Menou:2001ga,Peldan:1993hi,Sotiriou:2006hs,Sotiriou:2006qn,Sotiriou:2005cd,Akbar:2006er,Koivisto:2006ie,Amarzguioui:2005zq,Hwang:2001pu,Meng:2003en,Esposito-Farese:1999pa,Esposito-Farese:2004cc,Damour:1996xx,Puetzfeld:2004yg,Kasper:1994xv,Biswas:1999fa,Mukherjee:2005zt,Beesham:1987gd,Mahato:2006gi}.

\subsection{Generalizing general relativity}

The arguably most beautiful aspect of GR is that it 
geometrizes gravitation, with Minkowski spacetime being deformed by the matter (and energy) inside it.
As illustrated in \Fig{fig:spaces}, for the most general manifold with a metric $g$ and a connection $\Gamma$, departures from Minkowski
space are characterized by three geometrical entities: non-metricity ($Q$), curvature ($R$) and torsion ($S$), defined as follows:
\bena
Q_{\mu\nu\rho}&\equiv&\nabla_{\mu} g_{\nu\rho}\,,\label{eqn:nonmetricity}\\
R^{\rho}_{\phantom{1}\lambda\nu\mu} &\equiv& \Gamma^{\rho}_{\phantom{1}\mu\lambda,\nu}-\Gamma^{\rho}_{\phantom{1}\nu\lambda,\mu}+\Gamma^\rho_{\phantom{1}\nu\alpha}\Gamma^\alpha_{\phantom{1}\mu\lambda}-\Gamma^\rho_{\phantom{1}\mu\alpha}\Gamma^\alpha_{\phantom{1}\nu\lambda}\,,\\
S_{\mu\nu}^{\phantom{\mu\nu}\rho}&\equiv&\frac{1}{2}(\Gamma^{\rho}_{\phantom{\rho}\mu\nu}-\Gamma^{\rho}_{\phantom{\rho}\nu\mu})\,.\label{TorsionDefEq}
\eena
GR is the special case where the non-metricity and torsion are assumed to vanish identically ($Q=S=0$, i.e., Riemann spacetime), 
which determines the connection in terms of the metric and leaves 
the metric as the only dynamical entity. 
However, as \Fig{fig:spaces} illustrates, this is by no means the only possibility, and many alternative geometric gravity
theories have been discussed in the literature \cite{Blagojevic:book,hammond,Gronwald:1995em,Hehl:1997bz,DeAndrade,aldrovandi,hehl,nester,watanabe,capozziello,Gasperini:1986mv,Shapiro:2001rz,Sotiriou:2006qn,Baekler:2006de,Poltorak:2004tz,Mielke:2004gg,Kleyn:2004yj,Vassiliev:2003dk,Minkevich:2003it,Obukhov:2002tm,Tresguerres:1995un,King:2000ha,Gronwald:1997bx,Hehl:1999sb,Poberii:1994rz,Hehl:1994ue,Lord:1978qz,Hehl:1976my,Rajaraman:2003st,Vollick:2003ic,Chiba:2003ir,Gruver:2001tt,Berthias:1993aa,Lam:2002ve,Saa:1993fx,Antoniadis:1992ep,Bytsenko:1993qn,Fabbri:2006xq,Carroll:1994dq} corresponding to 
alternative deforming geometries where other subsets of $(Q,R,S)$ vanish. Embedding GR in a broader parametrized class of theories allowing 
non-vanishing torsion and non-metricity, and experimentally constraining these parameters would provide a natural generalization of
the highly successful parametrized post-Newtonian (PPN) program for GR testing, which assumes vanishing torsion \cite{Will:2005yc,Will:1993ns,Will:2005va}.

For the purposes of this chapter, a particularly interesting generalization of Riemann spacetime is Riemann-Cartan Spacetime (also known as $U_4$),
which retains $Q=0$ but is characterized by non-vanishing torsion. In $U_4$,
torsion can be dynamical and consequently play a role in gravitation alongside the metric. 
Note that gravitation theories including torsion retain what are often regarded as the most beautiful aspects of
General Relativity, \ie general covariance and the idea that ``gravity is geometry''. 
Torsion is just as geometrical an entity as curvature, and torsion theories can be consistent with the Weak Equivalence 
Principle (WEP).

\subsection{Why torsion testing is timely}

Experimental searches for torsion have so far been rather limited \cite{hammond}, in part because most published torsion theories predict
a negligible amount of torsion in the solar system. First of all, many torsion Lagrangians imply that torsion is related to its source 
via an algebraic equation rather than via a differential equation, so that (as opposed to curvature), torsion must vanish in vacuum. 
Second, even within the subset of torsion theories where torsion propagates and can exist in vacuum, it is usually assumed that it
couples only to intrinsic spin, not to rotational angular momentum \cite{hehl,sy1,sy2}, and is therefore negligibly small far from extreme objects
such as neutron stars. This second assumption also implies that even if torsion were present in the solar system, 
it would only affect particles with intrinsic spin (\eg\ a
gyroscope with net magnetic polarization) \cite{sy1,sy2,NSH,Hojman,Hojman2,cognola,kopczynski,pereira1}, while having no influence on the precession of a gyroscope 
without nuclear spin
\cite{sy1,sy2,NSH} such as a gyroscope in Gravity Probe B.  

Whether torsion does or does not satisfy these pessimistic assumptions depends on what the Lagrangian is, which is of course one of the
things that should be tested experimentally rather than assumed. 
Taken at face value, the Hayashi-Shirafuji Lagrangian \cite{HS1} 
provides an explicit counterexample to both assumptions, with even a static massive body generating a torsion field --- indeed, 
such a strong one that the gravitational forces are due entirely to torsion, not to curvature.
As another illustrative example, we will develop in
\Sec{subsec:linearinterpol} a family of tetrad theories in Riemann-Cartan space which linearly interpolate
between GR and the Hayashi-Shirafuji theory.  
Although these particular Lagrangeans come with important caveats to which we return below (see also \cite{Flanagan:2007dc}),
they show that one cannot dismiss out of hand the possibility that angular momentum sources non-local torsion
(see also Table \ref{tab:torsion-theory}).
Note that the proof\cite{sy1,sy2,NSH} of the oft-repeated assertion that a
gyroscope without nuclear spin cannot feel torsion crucially relies on the assumption that orbital angular momentum cannot be
the source of torsion. This proof is therefore not generally applicable in the context of non-standard torsion theories. 

More generally, in the spirit of action$=$reaction, if a (non-rotating or rotating) mass like a planet can generate torsion, then a gyroscope
without nuclear spin could be expected feel torsion, 
so the question of whether a non-standard gravitational Lagrangian causes 
torsion in the solar system is one which can and should be addressed experimentally.

This experimental question is timely because the Stanford-led
gyroscope satellite experiment, Gravity Probe B\footnote{http://einstein.stanford.edu/} (GPB), was launched in April 2004 and has successfully been taking 
data.
Preliminary GPB results, released in April 2007, have confirmed the geodetic precession to better than 1\%, 
and the full results, which are highly relevant to this chapter, are due to be released soon. 
GPB contains a set of four extremely spherical gyroscopes and flies in a circular polar orbit with altitude 640 kilometers, and we will show that it has the potential
to severely constrain a broad class of previously allowed torsion theories.
GPB was intended to test the GR prediction \cite{schiff,Will:2002ma,Adler:1999yt,biemond:2004,Ashby:1990,Barker:1970zr} that a gyroscope in this orbit precesses about 6,614.4 milli-arcseconds per year around its orbital angular momentum vector 
(geodetic precession) and about 40.9 milli-arcseconds per year about Earth's angular momentum vector (frame-dragging)\footnote{These numerical
precession rates are taken from the GPB website.}.  Most impressively, GPB should convincingly observe the frame-dragging effect, an 
effect of the off-diagonal metric elements that originate from the rotation of Earth. Of particular interest to us is that GPB can reach a
precision of $0.005\%$ for the geodetic precession, which as we will see enables precision discrimination\footnote{GPB 
also has potential for constraining other GR extensions \cite{Moffat:2004cv} than those we consider in this chapter.}  
between GR and a class of torsion theories.  

\subsection{How this chapter is organized}

In general, torsion has 24 independent components, each being a function of time and position. Fortunately, symmetry arguments and a perturbative expansion
will allow us to greatly simplify the possible form of any torsion field of Earth, a nearly
spherical slowly rotating massive object. We will show that the most general possibility can be elegantly parametrized by merely seven 
numerical constants to be constrained experimentally. We then derive the effect of torsion on 
the precession rate of a gyroscope in Earth orbit and work out how the anomalous precession that GPB would register depends on these seven parameters.

The rest of this chapter is organized as follows.  In Section \ref{sec:u4}, we review the basics of Riemann-Cartan spacetime. In \Sec{sec:param}, we derive
the results of parametrizing the torsion field around Earth. 
In \Sec{sec:eom}, we discuss the equation of motion for the precession of a gyroscope and the world-line of its center of mass.  
We use the results to calculate the instantaneous precession rate in \Sec{sec:instan-precess}, 
and then analyze the Fourier moments for the particular orbit of GPB in \Sec{sec:mom-ana}.  
In \Sec{sec:general-torsion-constraints}, we show that GPB can constrain two linear combinations of 
the seven torsion parameters, given the constraints on the PPN parameters $\gamma$ and $\alpha_1$ from other solar system tests. 
To make our discussion less abstract, we study Hayashi-Shirafuji torsion gravity as an 
explicit illustrative example of an alternative gravitational theory that can be tested within our framework.
In \Sec{sec:counterexample}, we review the basics of Weitzenb\"ock spacetime and
Hayashi-Shirafuji theory, and then give the torsion-equivalent of the linearized Kerr solution.  
In \Sec{subsec:linearinterpol}, we generalize 
the Hayashi-Shirafuji theory to a two-parameter family of gravity theories, 
which we will term Einstein-Hayashi-Shirafuji (EHS) theories, interpolating between torsion-free GR and the Hayashi-Shirafuji 
maximal torsion theory. In \Sec{sec:constrain-torsion}, we apply the precession rate results to the EHS theories and discuss the observational 
constraints that GPB, alongside other solar system tests, will be able to place on the parameter space of the family of EHS theories.   
We conclude in \Sec{sec:conclusion}.  Technical details of torsion parametrization (\ie \Sec{sec:param}) are given in 
Appendices \ref{appendix:spher-symm} \& \ref{appendix:axisymm}. Derivation of solar system tests are given in Appendix \ref{appendix:solar-tests}. 
We also demonstrate in Appendix \ref{appendix:photon-mass} that current ground-based experimental upper bounds on the photon mass do not place more
stringent constraints on the torsion parameters $t_1$ or $t_2$ than GPB will.

After the first version of the paper \cite{Mao:2006bb} involving the work in this chapter was submitted, Flanagan and Rosenthal showed that the Einstein-Hayashi-Shirafuji Lagrangian 
has serious defects \cite{Flanagan:2007dc}, while leaving open the possibility that there may be other viable Lagrangians
in the same class (where spinning objects generate and feel propagating torsion).
The EHS Lagrangian should therefore not be viewed as a viable physical model, but as a pedagogical toy model 
giving concrete illustrations of the various effects and constraints that we discuss.
 
Throughout this chapter, we use natural gravitational units where $c=G=1$.  Unless we explicitly state otherwise, a Greek letter denotes
an index running from 0 to 3 and a Latin letter an index from 1 to 3. We use the metric signature convention $(- + + +)$.

\section{Riemann-Cartan spacetime}\label{sec:u4}

We review the basics of Riemann-Cartan spacetime only briefly here, and refer the interested reader to 
Hehl {\etal} \cite{hehl} for a more comprehensive discussion of spacetime with torsion. 
Riemann-Cartan spacetime is a connected $C^{\infty}$ four-dimensional
manifold endowed with metric $g_{\mu\nu}$ of Lorentzian signature and an affine 
connection $\Gamma^{\mu}_{\phantom{\mu}\nu\rho}$
such that the non-metricity defined by \Eq{eqn:nonmetricity} with respect to the full connection identically vanishes. 
In other words, the connection in Riemann-Cartan spacetime may have torsion, but it must still be 
compatible with the metric ($g_{\mu\nu;\lambda}=0$).
The covariant derivative of a vector is given by
\bena
\nabla_{\mu}V^{\nu} &=& \partial_{\mu}V^{\nu} + \Gamma^{\nu}_{\phantom{1}\mu\rho}V^{\rho}\,,\label{eqn:aff-conn-def1}\\
\nabla_{\mu}V_{\nu} &=& \partial_{\mu}V_{\nu} - \Gamma^{\rho}_{\phantom{1}\mu\nu}V_{\rho}\,,\label{eqn:aff-conn-def2}
\eena
where the first of the lower indices on $\Gamma^{\lambda}_{\phantom{1}\mu\sigma}$ always corresponds to the index on $\nabla_\mu$.

The full connection has 64 independent components.  The condition of vanishing non-metricity $\nabla_{\mu} g_{\nu\rho} = 0$ 
gives 40 constraints, and the remaining 24 components are the degrees of freedom of the torsion tensor.  

In the more familiar case of Riemann spacetime, the two conditions $S_{\mu\nu}^{\phantom{\mu\nu}\rho}=0$ and $Q_{\mu\nu\rho}=0$ imply that
the connection must be the so-called Levi-Civita connection (Christoffel symbol), uniquely determined by the metric as 
\begin{equation} 
\left\{ \begin{array}{c} \rho \\ \mu\nu \end{array} \right\} =\frac{1}{2} g^{\rho\lambda}(\partial_{\mu}g_{\nu\lambda}+\partial_\nu g_{\mu\lambda}-\partial_{\lambda}g_{\mu\nu})\,.
\end{equation}
In the more general case when torsion is present, the connection must depart from the Levi-Civita connection in order to be metric-compatible 
($\nabla_{\mu} g_{\nu\rho} = 0$), 
and this departure is (up to a historical minus sign) called the \emph{contorsion}, defined as 
\beq{eqn:fullconn1}
K_{\mu\nu}^{\phantom{\mu\nu}\rho} \equiv  \left\{ \begin{array}{c} \rho \\ \mu\nu \end{array} \right\} - \Gamma^{\rho}_{\phantom{\rho}\mu\nu} \,.
\eeq
Using the fact that the torsion is the part of the connection that is antisymmetric in the first two indices (Eq.~\ref{TorsionDefEq}), one readily shows that 
\begin{equation}
K_{\mu\nu}^{\phantom{\mu\nu}\rho} =  -S_{\mu\nu}^{\phantom{\mu\nu}\rho}-S^\rho_{\phantom{\rho}\nu\mu}-S_{\phantom{\rho}\mu\nu}^{\rho}\,.
\end{equation}
In Riemann-Cartan spacetime, the metric is used to raise or lower the indices as usual. 

The curvature tensor is defined as usual, in terms of the full connection rather than the Levi-Civita connection:
\beq{eqn:curvature}
R^{\rho}_{\phantom{1}\lambda\nu\mu}= \partial_{\nu}\Gamma^{\rho}_{\phantom{1}\mu\lambda}-\partial_{\mu}\Gamma^{\rho}_{\phantom{1}\nu\lambda}+\Gamma^{\rho}_{\phantom{1}\nu\alpha}\Gamma^{\alpha}_{\phantom{1}\mu\lambda}-\Gamma^{\rho}_{\phantom{1}\mu\alpha}\Gamma^{\alpha}_{\phantom{1}\nu\lambda}\,.
\eeq
As in Riemann spacetime, one can prove that $R^{\rho}_{\phantom{1}\lambda\nu\mu}$ is a tensor by showing that for any vector $V^{\mu}$,
\ben
\nabla_{[\nu}\nabla_{\mu]}V^{\rho}=\frac{1}{2}R^{\rho}_{\phantom{1}\lambda\nu\mu}V^{\lambda}-S_{\nu\mu}^{\phantom{12}\alpha}\nabla_{\alpha}V^{\rho}\,.
\een
The Ricci tensor and Ricci scalar are defined by contraction the Riemann tensor just as in Riemann spacetime.

\section{Parametrization of the Torsion and Connection}\label{sec:param}

The torsion tensor has twenty-four independent components since it is antisymmetric in its first two
indices.  However, its form can be greatly simplified by the fact that Earth is well approximated as a uniformly
rotating spherical object. 
Throughout this chapter, we will therefore Taylor expand all quantities with respect to the dimensionless mass parameter
\beq{lambdaM}
\varepsilon_m\equiv {m\over r}, 
\eeq
 and the dimensionless angular momentum parameter 
\beq{lambdaDefEq}
\varepsilon_a\equiv {a\over r}, 
\eeq
where $a\equiv J/m$ is the specific angular momentum , which has units of length, and $r$ is the distance of the field point from the central gravitating body.  Here $m$ and $J$ are Earth's mass and rotational angular momentum, respectively.  
Since Earth is slowly rotating ($\varepsilon_a\ll 1$), we will only need to keep track of zeroth and first order terms in $\varepsilon_a$.
We will also Taylor expand with respect to $\varepsilon_m$ to first order, since we are interested in objects with orbital radii vastly exceeding
Earth's Schwarzschild radius ($\varepsilon_m\ll 1$).\footnote{These two approximations $\varepsilon_m \ll 1$ and $\varepsilon_a \ll 1$ are highly
accurate for the GPB satellite in an Earth orbit with altitude about 640 kilometers:  $\varepsilon_m \simeq 6.3 \times 10^{-10}$ and
$\varepsilon_a \simeq 5.6 \times 10^{-7}$.}  All calculations will be to first order in $\varepsilon_m$, because to zeroth order in
$\varepsilon_m$, \ie in Minkowski spacetime, there is no torsion.  Consequently, we use the terms ``zeroth order'' and ``first order'' below
with respect to the expansion in $\varepsilon_a$.

We start by studying in section \ref{subsec:spher-symm} the zeroth order part: the static,
spherically and parity symmetric case where Earth's rotation is ignored.  The first correction will be treated in
section \ref{subsection:axisymm}: the stationary and spherically axisymmetric contribution caused by Earth's
rotation.  For each case, we start by giving the symmetry constraints that apply for \emph{any} quantity. We then give the most general 
parametrization of torsion and connection that is consistent with these symmetries, as derived in the appendices.  
The Kerr-like torsion solution of Hayashi-Shirafuji Lagrangian given in \Sec{sec:counterexample} is an explicit
example within this parametrized class. In \Sec{sec:instan-precess}, we will apply these results to the precession
of a gyroscope around Earth. 

\subsection{Zeroth order: the static, spherically and parity symmetric case}\label{subsec:spher-symm}

This is the order at which Earth's slow rotation is neglected ($\varepsilon_a=0$). For this, three convenient
coordinate systems are often employed -- isotropic rectangular coordinates, isotropic spherical coordinates, and standard
spherical coordinates. In the following, we will find it most convenient to work in isotropic rectangular coordinates 
to set up and solve the problem, and then transform the result to standard spherical coordinates.

\subsubsection{Symmetry Principles}\label{subsubsec:general-setup1}

Tetrad spaces with spherical symmetry have been studied by Robertson \cite{robertson} and Hayashi and
Shirafuji \cite{HS1}.  Our approach in this section essentially follows their work.

Given spherical symmetry, one can naturally find a class of isotropic rectangular coordinates $(t,x,y,z)$. Consider
a general quantity $\mathcal{O}(x)$ that may bear upper and lower indices. It may or may not be a tensor. In
either case, its transformation law $\mathcal{O}(x)\to\mathcal{O}\,'(x')$ under the general coordinate transformation
$x\to x'$ should be given. By definition, a quantity $\mathcal{O}$ is static, spherically and parity symmetric if it
has the \emph{formal functional invariance} 
\[
\mathcal{O}\,'(x') = \mathcal{O}(x')
\]
under the following coordinate transformations (note that $\mathcal{O}(x')$ denotes the original function 
$\mathcal{O}(x)$ evaluated at the coordinates $x'$):
\begin{enumerate}

\item{Time translation:} $t\to t'\equiv t+t_0$ where $t_0$ is an arbitrary constant.

\item{Time reversal:} $t\to t'\equiv -t$.

\item{Continuous rotation and space inversion:}
\beq{xform:spherical}
\x\to\x'\equiv\R\x\,,
\eeq
where $\R$ is any $3\times 3$ constant orthogonal ($\R^t\R=\I$) matrix.  
Note that the parity symmetry allows $\R$ to be an improper rotation.

\end{enumerate}

\subsubsection{Parametrization of torsion}\label{subsubsec:param-tor}

It can be shown (see Appendix A) that, under the above conditions, there are only two independent components
of the torsion tensor.  The non-zero torsion components can be parametrized in isotropic rectangular coordinates
as follows: 
\bena
S_{0i}^{\phantom{0i}0} &=& t_1\frac{m}{2r^3}x^i\, ,\label{eqn:t1}\\
S_{jk}^{\phantom{0i}i} &=& t_2\frac{m}{2r^3}(x^j \delta_{ki}-x^k \delta_{ji})\, ,\label{eqn:t2}
\eena
where $t_1$ and $t_2$ are dimensionless constants.  
It is of course only the two combinations $t_1 m$ and $t_2 m$ that correspond to the physical parameters;
we have chosen to introduce a third redundant quantity 
$m$ here, with units of mass, to keep $t_1$ and $t_2$
dimensionless. Below we will see that in the context of specific torsion Lagrangians,
$m$ can be naturally identified with the mass of the object generating the torsion, up to a numerical factor close to unity.

We call $t_1$ the ``anomalous geodetic torsion'' and $t_2$ the ``normal geodetic torsion'', because both
will contribute to the geodetic spin precession of a gyroscope, the former ``anomalously'' and the latter
``regularly'', as will become clear in \Sec{sec:instan-precess} and \ref{sec:mom-ana}.

\subsubsection{Torsion and connection in standard spherical coordinates}

In spherical coordinates, the torsion tensor has the following non-vanishing components:
\beq{eqn:torsion-spher-symm1}
S_{tr}^{\phantom{01}t}(r) = t_1\frac{m}{2r^2}\quad ,\quad S_{r\theta}^{\phantom{12}\theta}(r)=S_{r\phi}^{\phantom{13}\phi}(r) = t_2\frac{m}{2r^2}\, ,
\eeq
where $t_1$ and $t_2$ are the same torsion constants as defined above.

The above parametrization of torsion was derived in isotropic coordinates, but it is also valid in other 
spherical coordinates as far as the linear perturbation around the Minkowski spacetime is concerned.  
The decomposition formula (\Eqnopar{eqn:fullconn1}), derived from $\nabla_{\mu} g_{\nu\rho}=0$, enables one 
to calculate the full connection exactly. However, for that purpose the coordinates with a metric must be specified.  
In general, a spherically symmetric coordinate system has the line element \cite{MTW-Weinberg-Islam:book}
\[
\mathrm{d}s^2 = -h(r)\mathrm{d}t^2+f(r)\mathrm{d}r^2+\alpha(r)r^2\left[\mathrm{d}\theta^2+\sin^2\theta\mathrm{d}\phi^2\right]\,.
\]
There is freedom to rescale the radius, so-called isotropic spherical coordinates corresponding to the
choice $\alpha(r)=f(r)$. Throughout this chapter, we make the common choice $\alpha(r)=1$, where $r$ can be interpreted as
$(2\pi)^{-1}$ times the circumference of a circle.
To linear order, 
\begin{eqnarray*}
h(r) &=& 1 + \mathcal{H}\frac{m}{r}\, ,\label{eqn:h}\\
f(r) &=& 1 + \mathcal{F}\frac{m}{r}\, ,\label{eqn:f}
\end{eqnarray*}
where $\mathcal{H}$ and $\mathcal{F}$ are dimensionless constants.

It is straightforward to show that, in the linear regime, the most general
connection that is static, spherically and parity symmetric in Riemann-Cartan spacetime with standard spherical coordinates is as follows:
\beqa{eqn2:conn}
\Gamma^{t}_{\phantom{0}tr} &=& \left(t_1-\frac{\mathcal{H}}{2}\right)\frac{m}{r^2}\,, \nonumber\\
\Gamma^{t}_{\phantom{0}rt} &=& -\frac{\mathcal{H}}{2}\frac{m}{r^2}\,, \nonumber\\
\Gamma^{r}_{\phantom{0}tt} &=& \left(t_1-\frac{\mathcal{H}}{2}\right)\frac{m}{r^2}\,, \nonumber\\
\Gamma^{r}_{\phantom{0}rr} &=& -\frac{\mathcal{F}}{2}\frac{m}{r^2}\,, \nonumber\\
\Gamma^{r}_{\phantom{0}\theta\theta} &=& -r+(\mathcal{F}+t_2)m \,, \\
\Gamma^{r}_{\phantom{0}\phi\phi} &=& -r\sin^2\theta+(\mathcal{F}+t_2)m\sin ^2 \theta \,, \nonumber\\
\Gamma^{\theta}_{\phantom{0}r\theta} &=& \Gamma^{\phi}_{\phantom{0}r\phi}\quad = \quad \frac{1}{r}\,, \nonumber\\
\Gamma^{\theta}_{\phantom{0}\theta r} &=& \Gamma^{\phi}_{\phantom{0}\phi r}\quad = \quad \frac{1}{r} - t_2\frac{m}{r^2} \,, \nonumber\\
\Gamma^{\theta}_{\phantom{0}\phi\phi} &=& -\sin \theta \cos \theta \,, \nonumber\\
\Gamma^{\phi}_{\phantom{0}\theta\phi} &=& \Gamma^{\phi}_{\phantom{0}\phi\theta}\quad = \quad \cot \theta \,. \nonumber
\eeqa
By ``the most general'' we mean that any other connections are related to the one in \Eq{eqn2:conn} by the nonlinear coordinate transformation law
\beq{eqn:gamma-law}
\Gamma\,'^{\mu}_{\phantom{\rho}\nu\lambda}(x')= \frac{\partial x\,'^{\mu}}{\partial x^{\alpha}}\frac{\partial x^{\beta}}{\partial x\,'^{\nu}}\frac{\partial x^{\gamma}}{\partial x\,'^{\lambda}}\Gamma ^{\alpha}_{\phantom{\rho}\beta\gamma}(x)+\frac{\partial x\,'^{\mu}}{\partial x^{\alpha}}\frac{\partial ^2 x^{\alpha}}{\partial x\,'^{\nu}\partial x\,'^{\lambda}}\,.
\eeq

Note that the terms independent of metric and torsion merely reflect the spherical coordinate system and do not represent
a deformation of spacetime --- in other words, the special case 
$t_1=t_2=\mathcal{H}=-\mathcal{F}=0$ corresponds to the connection for Minkowski spacetime.
The case $t_1=t_2=0$ and $\mathcal{H}=-\mathcal{F}=-2$ corresponds to the standard connection for Schwarzschild spacetime in the linear regime ($r\gg m$).

\subsection{First-order: stationary, spherically axisymmetric case}\label{subsection:axisymm}

The terms added at this order are due to Earth's rotation. 
Roughly speaking, ``{\it spherically axisymmetric}'' refers to the property
that a system is spherically symmetric except for symmetries broken by an angular momentum vector.  The rigorous mathematical definition is given in \Sec{subsubsec:prob-setup-axisym}.  Subtleties related to coordinate system choices at this order fortunately do not matter in the $\varepsilon_m \ll 1$ and $\varepsilon_a \ll 1$ limit that we are interested in.

\subsubsection{Symmetry Principles}
\label{subsubsec:prob-setup-axisym}

Suppose we have a field configuration which depends explicitly on the angular momentum $\J$ of the central spinning body.  We can
denote the fields generically as $\mathcal{O}(x|\J)$, which is a function of coordinates $x$ and the value of the angular momentum
vector $\J$.  We assume that the underlying laws of physics are symmetric under rotations, parity, time translation, and time
reversal, so that the field configurations for various values of $\J$ can be related to each other. Specifically, we assume that $\J$
rotates as a vector, reverses under time-reversal, and is invariant under time translation and parity.  It is then possible to define
transformations for the field configurations, $\mathcal{O}(x|\J)\to\mathcal{O}\,'(x'|\J)$, for these same symmetry operations.  Here
$\mathcal{O}\,'(x'|\J)$ denotes the transform of the field configuration that was specified by $\J$ before the transformation;
$\mathcal{O}$ may or may not be a tensor, but its transformation properties are assumed to be specified. The symmetries of the
underlying laws of physics then imply that the configurations $\mathcal{O}(x|\J)$ are stationary and {\it spherically axisymmetric} in
the sense that the transformed configuration is identical to the configuration that one would compute by transforming $\J \to \J'$. 
That is,
\[  \mathcal{O}\,'(x'|\J)=\mathcal{O}(x'|\J')  \]
under the following coordinate transformations:

\begin{enumerate}

\item{time translation:} $t\to t'\equiv t+t_0$ where $t_0$ is an arbitrary constant.

\item{Time reversal:} $t\to t'\equiv -t$.

\item{Continuous rotation and space inversion:} $\x \to \x'\equiv \R (\x)\,$, \ie $\x'$ is related to $\x$ by any proper or improper rotation.

\end{enumerate}
  
Below we will simplify the problem by keeping track only of terms linear in $J/r^2=\varepsilon_m \varepsilon_a$.

\subsubsection{Parametrization of metric}
\label{subsubsec:metric-axisy} 

With these symmetries, it can be shown that the first-order contribution to the metric is
\ben
g_{ti}=g_{it}= \frac{\mathcal{G}}{r^2}\epsilon_{ijk}J^j \hat{x}^{k}
\een
in rectangular coordinates $x^\mu = (t,x^i)$, where $\mathcal{G}$ is a constant, or
\ben
g_{t\phi}= g_{\phi t}= \mathcal{G}\frac{J}{r}\sin^2\theta
\een
in spherical coordinates $x^{\mu}=(t,r,\theta,\phi)$ where the polar angle $\theta$ is the angle 
with respect to the rotational angular momentum $\J$.  The details of the 
derivation are given in Appendix \ref{appendix:axisymm}.

\subsubsection{Parametrization of torsion}
\label{subsubsec:torsion-axisy}

In Appendix \ref{appendix:axisymm}, we 
show that, in rectangular coordinates, the first-order correction to the torsion is
\begin{eqnarray*}
S_{ij}^{\phantom{01}t} &=& \frac{f_1}{2r^3}\epsilon_{ijk}J^k+\frac{f_2}{2r^3}J^k \hat{x}^l (\epsilon_{ikl}\hat{x}^j-\epsilon_{jkl}\hat{x}^i)\,,\\
S_{tij} &=& \frac{f_3}{2r^3}\epsilon_{ijk}J^k+\frac{f_4}{2r^3}J^k \hat{x}^l \epsilon_{ikl}\hat{x}^j+\frac{f_5}{2r^3}J^k \hat{x}^l \epsilon_{jkl}\hat{x}^i\,.
\end{eqnarray*}
In spherical coordinates, these first-order torsion terms are
\begin{eqnarray*}
S_{r\phi}^{\phantom{0i}t} &=& w_1\frac{ma}{2r^2}\sin^2\theta\,,\\
S_{\theta\phi}^{\phantom{0i}t} &=& w_2\frac{ma}{2r}\sin\theta\cos\theta\,,\\
S_{t\phi}^{\phantom{0i}r} &=& w_3\frac{ma}{2r^2}\sin^2\theta\,, \\
S_{t\phi}^{\phantom{0i}\theta} &=& w_4\frac{ma}{2r^3}\sin\theta\cos\theta\,,\\
S_{tr}^{\phantom{0i}\phi} &=& w_5 \frac{ma}{2r^4}\,,\\
S_{t\theta}^{\phantom{0i}\phi} &=& -w_4\frac{ma}{2r^3}\cot\theta\,.
\end{eqnarray*}
Here $f_1,\ldots,f_5$ and $w_1,\ldots,w_5$ are constants.  The latter are linear combinations of the former.  
The details of the derivation are given
in Appendix \ref{appendix:axisymm}.  We call $w_1$,\ldots,$w_5$ the ``frame-dragging torsion'', since they will 
contribute the frame-dragging spin precession of a gyroscope as will become clear in \Sec{sec:instan-precess}.

\subsection{Around Earth}

We now summarize the results to linear order. We have computed the parametrization perturbatively
in the dimensionless parameters $\varepsilon_m\equiv m/r$ and $\varepsilon_a\equiv a/r$.
The zeroth order ($\varepsilon_a=0$) solution, where Earth's slow rotation is ignored, is simply the solution around 
a static spherical body, i.e. the case studied in \Sec{subsec:spher-symm}.  
The first order correction, due to Earth's rotation, is stationary and spherically axisymmetric as derived 
in \Sec{subsection:axisymm}. 
A quantity $\mathcal{O}$ to linear order is the sum of these two orders. 
In spherical coordinates, a general line element thus takes the form  
\ben
\mathrm{d}s^2 =  -\left[1+\mathcal{H}\frac{m}{r}\right]\mathrm{d}t^2 + \left[1+\mathcal{F}\frac{m}{r}\right]\mathrm{d}r^2 
+r^2 d\Omega^2 +2\,\mathcal{G}\frac{ma}{r}\sin ^2 \theta\mathrm{d}t\mathrm{d}\phi \, ,\label{eqn:metricaxisy}
\een
where $d\Omega^2=\mathrm{d}\theta^2+\sin ^2 \theta \mathrm{d}\phi^2$.  Here $\mathcal{H}$, $\mathcal{F}$ and $\mathcal{G}$ are
dimensionless constants.  In GR, the Kerr metric \cite{kerr,boyer-lind} at large distance gives the constants
$\mathcal{H}=-\mathcal{F}=\mathcal{G}=-2$.  
The result $\mathcal{G}=-2$ can also be derived more generally as shown by de Sitter \cite{deSitter:1916} and Lense \& Thirring \cite{Lense-Thirring:1918}. 
As above, $J=ma$ denotes the magnitude of Earth's rotational angular
momentum. 

Combining our 0th and 1st order expressions from above for the torsion around Earth, we obtain
\begin{eqnarray}
S_{tr}^{\phantom{01}t} &=& t_1\frac{m}{2r^2}\,,\nonumber\\
S_{r\theta}^{\phantom{12}\theta} &=& S_{r\phi}^{\phantom{13}\phi} = t_2\frac{m}{2r^2}\,,\nonumber\\
S_{r\phi}^{\phantom{0i}t} &=& w_1\frac{ma}{2r^2}\sin^2\theta\,,\nonumber\\
S_{\theta\phi}^{\phantom{0i}t} &=& w_2\frac{ma}{2r}\sin\theta\cos\theta\,,\label{eqn:w2-axi}\\
S_{t\phi}^{\phantom{0i}r} &=& w_3\frac{ma}{2r^2}\sin^2\theta\,,\nonumber \\
S_{t\phi}^{\phantom{0i}\theta} &=& w_4\frac{ma}{2r^3}\sin\theta\cos\theta\,,\nonumber\\
S_{tr}^{\phantom{0i}\phi} &=& w_5 \frac{ma}{2r^4}\,,\nonumber\\
S_{t\theta}^{\phantom{0i}\phi} &=& -w_4\frac{ma}{2r^3}\cot\theta\,.\nonumber
\end{eqnarray}
All other components vanish. Again, $t_1,t_2,w_1$,$w_2$, $w_3$, $w_4$, $w_5$ are dimensionless constants.

The calculation of the corresponding connection is straightforward by virtue of \Eq{eqn:fullconn1}. It is not hard to show that,
to linear order in a Riemann-Cartan spacetime in spherical coordinates, the connection around Earth has the following
non-vanishing components:
\bena
\Gamma^{t}_{\phantom{0}tr} &=& \left(t_1-\frac{\mathcal{H}}{2}\right)\frac{m}{r^2}\,, \nonumber\\
\Gamma^{t}_{\phantom{0}rt} &=& -\frac{\mathcal{H}}{2}\frac{m}{r^2}\,, \nonumber\nonumber\\
\Gamma^{t}_{\phantom{0}r\phi} &=& (3\mathcal{G}+w_1-w_3-w_5)\frac{ma}{2r^2}\sin^2\theta\,,\nonumber\\
\Gamma^{t}_{\phantom{0}\phi r} &=& (3\mathcal{G}-w_1-w_3-w_5)\frac{ma}{2r^2}\sin^2\theta\,,\nonumber\\
\Gamma^{t}_{\phantom{0}\theta\phi} &=& w_2\frac{ma}{2r}\sin\theta\cos\theta\,,\nonumber\\
\Gamma^{t}_{\phantom{0}\phi\theta} &=& -w_2\frac{ma}{2r}\sin\theta\cos\theta\,,\nonumber\\
\Gamma^{r}_{\phantom{0}tt} &=& \left(t_1-\frac{\mathcal{H}}{2}\right)\frac{m}{r^2}\,, \nonumber \\ 
\Gamma^{r}_{\phantom{0}rr} &=& -\frac{\mathcal{F}}{2}\frac{m}{r^2}\,, \nonumber\\
\Gamma^{r}_{\phantom{0}\theta\theta} &=& -r+(\mathcal{F}+t_2)m \,, \label{eqn:full-conn323}\\
\Gamma^{r}_{\phantom{0}\phi\phi} &=& -r\sin^2\theta+(\mathcal{F}+t_2)m\sin ^2 \theta \,, \nonumber\\
\Gamma^{r}_{\phantom{0}t\phi} &=& (\mathcal{G}-w_1+w_3-w_5)\frac{ma}{2r^2}\sin^2\theta\,,\nonumber\\
\Gamma^{r}_{\phantom{0}\phi t} &=& (\mathcal{G}-w_1-w_3-w_5)\frac{ma}{2r^2}\sin^2\theta\,,\nonumber
\eeqa
\bena
\Gamma^{\theta}_{\phantom{0}t\phi} &=& (-2\mathcal{G}-w_2+2w_4)\frac{ma}{2r^3}\sin\theta\cos\theta\,,\nonumber\\
\Gamma^{\theta}_{\phantom{0}\phi t} &=&(-2\mathcal{G}-w_2)\frac{ma}{2r^3}\sin\theta\cos\theta\,,\nonumber\\
\Gamma^{\theta}_{\phantom{0}r\theta} &=& \Gamma^{\phi}_{\phantom{0}r\phi}\; =\; \frac{1}{r} \,,\nonumber\\
\Gamma^{\theta}_{\phantom{0}\theta r} &=& \Gamma^{\phi}_{\phantom{0}\phi r}\; =\; \frac{1}{r}-t_2\frac{m}{r^2} \,,\nonumber\\
\Gamma^{\theta}_{\phantom{0}\phi\phi} &=& -\sin \theta \cos \theta \,,\nonumber\\
\Gamma^{\phi}_{\phantom{0}tr} &=&  (-\mathcal{G}+w_1-w_3+w_5)\frac{ma}{2r^4}\,,\nonumber\\
\Gamma^{\phi}_{\phantom{0}rt} &=&  (-\mathcal{G}+w_1-w_3-w_5)\frac{ma}{2r^4}\,,\nonumber\\
\Gamma^{\phi}_{\phantom{0}t\theta} &=& (2\mathcal{G}+w_2-2w_4)\frac{ma}{2r^3}\cot\theta\,,\nonumber\\
\Gamma^{\phi}_{\phantom{0}\theta t} &=& (2\mathcal{G}+w_2)\frac{ma}{2r^3}\cot\theta\,,\nonumber\\
\Gamma^{\phi}_{\phantom{0}\theta\phi} &=& \Gamma^{\phi}_{\phantom{0}\phi\theta} \;=\; \cot \theta\,.\nonumber
\eeqa

\section{Precession of a gyroscope I: fundamentals}
\label{sec:eom}

\subsection{Rotational angular momentum}

There are two ways to covariantly quantify the angular momentum of a spinning object, in the literature denoted
$S^{\mu}$ and $S^{\mu\nu}$, respectively.  (Despite our overuse of the letter $S$, they can be distinguished by the
number of indices.)  In the rest frame of the center of mass of a gyroscope, the 4-vector $S^{\mu}$ 
is defined as 
\ben
S^{\mu}=(0,\vec{S}_0)\,,
\een
and the 4-tensor $S^{\mu\nu}$ 
is defined to be antisymmetric and have the components
\beq{eqn:cyclic-smunu}
S^{0i}=S^{i0}=0, \qquad  \qquad S^{ij}=\epsilon^{ijk}S_0^{\phantom{1}k}\,,
\eeq
where $i=x,y,z$.  $\quad\vec{S}_0=S_0^{\phantom{1}x} \hat{x} + S_0^{\phantom{1}y} \hat{y} + S_0^{\phantom{1}z} \hat{z}\quad $ is the rotational angular momentum of a gyroscope observed by an observer co-moving with the center of mass of the gyroscope.  The relation between $S^{\mu}$ and $S^{\mu\nu}$ can be written in the local (flat) frame as 
\beq{eqn:ss1}
S^{\mu}=\epsilon^{\,\mu\nu\rho\sigma}u_{\nu}S_{\rho\sigma}\,,
\eeq
where $u^{\mu}=\mathrm{d}x^{\mu}/\mathrm{d}\tau$ is the 4-velocity.  

In curved spacetime, the Levi-Civita symbol is generalized to $\bar{\epsilon}^{\,\mu\nu\rho\sigma}=\epsilon^{\mu\nu\rho\sigma}/\sqrt{-g}$ where $g=\det g_{\mu\nu}$.  It is easy to prove that $\bar{\epsilon}^{\,\mu\nu\rho\sigma}$ is a 4-tensor.  Then \Eq{eqn:ss1} becomes a covariant relation
\beq{eqn:ss2}
S^{\mu}=\bar{\epsilon}^{\,\mu\nu\rho\sigma}u_{\nu}S_{\rho\sigma}\,.
\eeq
In addition, the vanishing of temporal components of $S^{\mu}$ and $S^{\mu\nu}$ can be written as covariant conditions 
as follows:
\bena
S^{\mu}u_{\mu} &=& 0 \,,\label{eqn:condition-smu}\\
S^{\mu\nu}u_{\nu} &=& 0\,.\label{eqn:condition-smunu}
\eena
In the literature \cite{schiff}, \Eq{eqn:condition-smunu} is called Pirani's supplementary condition.  
 
Note, however, that unlike the flat space case, the spatial vectors of $S^\mu$ and $S^{\mu\nu}$ 
(denoted by $\vec{S}$ and $\vec{S}'$ respectively) do not coincide in the curved spacetime. 
The former is the spatial component of the 4-vector $S^\mu$, while the latter is historically 
defined as  $\vec{S}^{\,'\,i} \equiv \epsilon^{ijk}S_{jk}$.  
It follows \Eq{eqn:ss2} that $\vec{S}$ and $\vec{S}'$ differ 
by $\vec{S} = \vec{S}' \left[1+\mathcal{O}(m_E/r)+\mathcal{O}(v^2)\right]$ for a gyroscope moving around Earth.

\subsection{Equation of motion for precession of a gyroscope}
\label{subsec:eom4precession}

To derive the equation of motion for $S^{\mu}$ (or $S^{\mu\nu}$) of a small extended object that may have either
rotational angular momentum or net spin, Papapetrou's method \cite{papapetrou} should be generalized to 
Riemann-Cartan spacetime.  This generalization has been studied by Stoeger \& Yasskin \cite{sy1,sy2} 
as well as Nomura, Shirafuji \& Hayashi \cite{NSH}. The starting point of this method is the Bianchi identity 
or Noether current in a
gravitational theory whose derivation strongly relies on an assumption of what sources torsion. 
Under the common assumption that only intrinsic spin sources torsion, both \cite{sy1,sy2} and
\cite{NSH} drew the conclusion that whereas a particle with net intrinsic spin will precess according to the full
connection, the rotational angular momentum of a gyroscope will \emph{not} feel the background
torsion, \ie it will undergo parallel transport by the Levi-Civita connection along the free-falling orbit --- the
same prediction as in GR.  

These results of \cite{sy1,sy2,NSH} have the simple intuitive interpretation that
if angular momentum is not coupled to torsion, then torsion is not coupled to angular momentum.
In other words, for Lagrangians where the angular momentum of a rotating object cannot generate a torsion field, the 
torsion field cannot affect the angular momentum of a rotating object, in the same spirit as Newton's dictum 
``action $=$ reaction''.  

The Hayashi-Shirafuji theory of gravity, which we will discuss in detail in \Sec{sec:counterexample}, raises an objection to the common assumption that
only intrinsic spin sources torsion, in that in this theory even a non-rotating massive body can generate torsion in the vacuum nearby
\cite{HS1}. This feature also generically holds for teleparallel theories. 
It has been customary to assume that spinless test
particles follow metric geodesics (have their momentum parallel transported by the Levi-Civita connection), \ie, that spinless particles decouple
from the torsion even if it is nonzero. For a certain class of Lagrangians, this can follow from using the conventional variational principle. However, Kleinert and Pelster
\cite{Kleinert:1996yi,Kleinert:1998as} argue that the closure failure of parallelograms in the 
presence of torsion adds an additional term to the
geodesics which causes spinless test particles to follow autoparallel worldlines (have their momentum parallel transported by the full
connection). This scenario thus respects the ``action $=$ reaction'' principle, since a spinless test particle can both generate and feel torsion.
As a natural extension, we explore the possibility that in these theories, a rotating body also generates torsion through its rotational angular momentum,
and the torsion in turn affects the motion of spinning objects such as gyroscopes.

An interesting first-principles derivation of how torsion affects a gyroscope in a specific  theory might involve 
generalizing the matched asymptotic expansion method of \cite{D'Eath1975a,D'Eath1975b}, and match
two generalized Kerr-solutions in the weak-field limit to obtain the gyroscope equation of motion.
Since such a calculation would be way beyond the scope of the present chapter, 
we will simply limit our analysis to exploring some obvious possibilities for laws of motion, based on 
the analogy with spin precession. 

The exact equation of motion for the precession of net spin is model dependent,
depending on the way the matter fields couple to the metric and torsion in the Lagrangian
(see
\cite{sy1,sy2,NSH,Hojman,Hojman2,cognola,kopczynski,pereira1,adamowicz-trautman}).  However, in the linear regime that we are
interested in here, many of the cases reduce to one of the following two equations if there is no external non-gravitational 
force acting on the test particle:
\bena
\frac{\mathrm{D}S^{\mu}}{\mathrm{D}\tau} &=& 0\,,\label{eqn:eom4smu}\\
\textrm{or}\quad \frac{\mathrm{D}S^{\mu\nu}}{\mathrm{D}\tau} &=& 0\,,\label{eqn:eom4smunu}
\eena
where $\mathrm{D}/\mathrm{D}\tau=(\mathrm{d}x^{\mu}/\mathrm{d}\tau)\nabla_{\mu}$ is the covariant differentiation along the world-line  
with respect to the full connection. In other words, the net spin undergoes parallel transport by the full connection along its trajectory.\footnote{If an external non-gravitational force acts on a spinning test particle, it will undergo
Fermi-Walker transport along its world-line.  This situation is beyond the interest of a satellite experiment, so it will be
neglected in the present chapter.}

In analog to the precession of spin, we will work out the implications of the assumption that the rotational angular momentum 
also precesses by parallel transport along the free-fall trajectory using the full connection.

\subsection{World line of the center of mass}

In GR, test particles move along well-defined trajectories -- \emph{geodesics}.  In the presence of torsion, things might be
different.  The idea of \emph{geodesics} originates from two independent concepts: \emph{autoparallels} and \emph{extremals}
\footnote{This terminology follows Hehl {\etal} \cite{hehl}.}.  Autoparallels, or affine geodesics, are curves along which the
velocity vector $\mathrm{d}x^{\mu}/\mathrm{d}\lambda$ is transported parallel to itself by the full connection
$\Gamma^{\rho}_{\phantom{\mu}\mu\nu}$.  With an affine parameter $\lambda$, the geodesic equation is
\begin{equation}\label{eqn:autopa}
\frac{\mathrm{d}^2 x^{\rho}}{\mathrm{d}\lambda ^2}+\Gamma^{\rho}_{\phantom{\mu}(\mu\nu)} \frac{\mathrm{d}x^{\mu}}{\mathrm{d}\lambda} \frac{\mathrm{d}x^{\nu}}{\mathrm{d}\lambda} =0\,.
\end{equation}
Extremals, or metric geodesics, are curves of extremal spacetime interval with respect to the metric $g_{\mu\nu}$.  
Since $\mathrm{d}s=[-g_{\mu\nu}(x)\mathrm{d}x^{\mu}\mathrm{d}x^{\nu}]^{1/2}$ does not depend on the full connection, the geodesic differential equations derived from $\delta \int \mathrm{d}s=0$ state that the 4-vector is parallel transported by the Levi-Civita connection.  That is, with the parameter $\lambda$ properly chosen, 
\begin{equation}\label{eqn:extremal}
\frac{\mathrm{d}^2 x^{\rho}}{\mathrm{d}\lambda ^2}+\left\{ \begin{array}{c} \rho \\ \mu\nu \end{array} \right\} \frac{\mathrm{d}x^{\mu}}{\mathrm{d}\lambda} \frac{\mathrm{d}x^{\nu}}{\mathrm{d}\lambda} =0\,.
\end{equation}
In Riemann spacetime where torsion identically vanishes, Eqs.(\ref{eqn:autopa}) and (\ref{eqn:extremal}) coincide.  In a
Riemann-Cartan spacetime, however, these two curves coincide if and only if the torsion is totally antisymmetric in all three
indices \cite{hehl}.  This is because the symmetric part of the full connection can be written from \Eq{eqn:fullconn1} as follows:
\beq{eqn:symm-conn}
\Gamma^{\rho}_{\phantom{\rho}(\mu\nu)}\equiv\frac{1}{2}(\Gamma^{\rho}_{\phantom{\rho}\mu\nu}+\Gamma^{\rho}_{\phantom{\rho}\nu\mu})= \left\{ \begin{array}{c} \rho \\ \mu\nu \end{array} \right\} + S_{\phantom{\rho}\mu\nu}^{\rho}+S_{\phantom{\rho}\nu\mu}^{\rho}\,.
\eeq

Photons are expected to follow extremal world lines because 
the gauge invariance of the electromagnetic part of the Lagrangian, well established by
numerous experimental upper bounds on the photon mass, prohibits torsion from coupling to the electromagnetic field to lowest order \cite{hehl}. 
As a consequence, the classical path of a light ray is at least to leading order determined by the metric alone as an extremal 
path, or equivalently as an autoparallel curve with respect to the Levi-Civita connection, independent of whether there is torsion.

On the other hand, the trajectory of a rotating test particle is still an open question in theory. 
Papapetrou \cite{papapetrou} claims that, even in GR, a gyroscope will deviate from the metric geodesic, albeit slightly. 
In torsion gravity theories, the equations of motion for the orbital 4-momentum differs more strongly between different 
approaches \cite{hehl,sy2,NSH,Hojman,Hojman2,cognola,kopczynski,pereira1}, and it is an open question to what extent
they are consistent with all classical GR tests (deflection of light rays, gravitational redshift, precession of the
perihelion of Mercury, Shapiro time delay, binary pulsars, etc.).
To bracket the uncertainty, we will examine the two extreme assumption in turn -- that world lines are
autoparallels and extremals, respectively.

Only the autoparallel scheme, not the extremal scheme, is theoretically consistent, for two reasons.  The first reason is based on the equivalence of the two approaches using the two alternative quantities $S^{\mu}$ and $S^{\mu\nu}$ to describe the angular momentum.  The equivalence is automatic in GR.  In a torsion theory, however, \Eq{eqn:eom4smu} and (\ref{eqn:eom4smunu}) can be simultaneously valid only if the trajectory is autoparallel.  This can be seen by taking the covariant differentiation of \Eq{eqn:ss2}. 
Note that $\mathrm{D}\bar{\epsilon}^{\,\mu\nu\rho\sigma}/\mathrm{D}\tau=0$.  One finds
\ben
\bar{\epsilon}^{\,\mu\nu\rho\sigma}\frac{\mathrm{D}u_{\nu}}{\mathrm{D}\tau}S_{\rho\sigma}=0\,.
\een
This equation is satisfied if $\mathrm{D}u_{\nu}/\mathrm{D}\tau=0$, \ie if the gyroscope world line is autoparallel. If an extremal world line is assumed, then
one has to make an \emph{a priori} choice between $S^{\mu}$ and $S^{\mu\nu}$, since the precession rates calculated using the two quantities will differ.

The second reason is that for $S^{\mu}$, the condition $S^{\mu}u_{\mu} = 0$ (\Eq{eqn:condition-smu}) must be satisfied anywhere along the world line.  Taking the covariant differentiation for both sides of \Eq{eqn:condition-smu}, one finds 
\beq{eqn:smu-derived}
S^{\mu}\mathrm{D}u_{\mu}/\mathrm{D}\tau = 0\,,
\eeq
assuming $\mathrm{D}S^{\mu}/\mathrm{D}\tau = 0$.  Obviously, autoparallels are consistent with \Eq{eqn:smu-derived}, while extremals are not.  The same argument applies for $S^{\mu\nu}$, \ie taking the covariant differentiation of both sides of \Eq{eqn:condition-smunu}.

Despite the fact that the extremal scheme is not theoretically consistent in this sense, the inconsistencies are numerically small for the linear regime $m/r \ll 1$.  
They are therefore of interest as an approximate phenomenological prescription that might at some 
time in the future be incorporated into a consistent theory.  We therefore include results also for this case below.

\subsection{Newtonian limit}

In \Sec{sec:param}, we parametrized the metric, torsion and connection of Earth, including an arbitrary parameter $m$ with units of mass.  To give $m$ a physical interpretation, the Newtonian limit of a test particle's orbit should be evaluated.  Obviously, the result depends on whether the autoparallel or extremal scheme is assumed.

In the remainder of this chapter, we denote an arbitrary parameter with units of mass as $m_0$ and the physical mass as $m$.  Metric and torsion parameters in accordance with $m_0$ are denoted with a superscript $(0)$, \ie  $\mathcal{H}^{(0)},\mathcal{F}^{(0)},\mathcal{G}^{(0)},t_1^{(0)},t_2^{(0)},w_1^{(0)}\ldots w_5^{(0)}$.

If an autoparallel world line is assumed, using the parametrization of equations (\ref{eqn:full-conn323}), it can be shown that the equation of motion to lowest order becomes
\beq{eqn:auto}
\frac{\mathrm{d}\vec{v}}{\mathrm{d}t}=-\left[t_1^{(0)}-\frac{\mathcal{H}^{(0)}}{2}\right]\frac{m_0}{r^2}\hat{e}_r\,.
\eeq
Therefore Newton's Second Law interprets the mass of the central gravitating body to be 
\beq{eqn:massauto}
m = \left[t_1^{(0)}-\frac{\mathcal{H}^{(0)}}{2}\right]m_0\,.\quad \textrm{(autoparallel scheme)}
\eeq
However, if $t_1^{(0)}-\mathcal{H}^{(0)}/2=0$, the autoparallel scheme fails totally.

Similarly, for a theory with extremal world-lines, the extremal equation in Newtonian approximation is
\beq{eqn:extrem}
\frac{\mathrm{d}\vec{v}}{\mathrm{d}t}=-\frac{[-\mathcal{H}^{(0)}]}{2}\frac{m_0}{r^2}\hat{e}_r\,.
\eeq
Therefore the physical mass of the body generating the gravity field is 
\beq{eqn:massextrem}
m = -\frac{\mathcal{H}^{(0)}}{2}m_0\,,\quad\textrm{(extremal scheme)}
\eeq
as long as $\mathcal{H}^{(0)}\ne 0$.  For the Schwarzschild metric ($\mathcal{H}^{(0)}=-2$), $m=m_0$.

After re-scaling $m$ from $m_0$, all metric and torsion parameters make the inverse re-scaling, \eg\ $t_1 = t_1^{(0)}(m_0/m)$ since the combination $t_1 m$ is the physical parameters during parametrization of metric and torsion.  This inverse scaling applies to $\mathcal{H}^{(0)},\mathcal{F}^{(0)},\mathcal{G}^{(0)},t_2^{(0)},w_1^{(0)}\ldots w_5^{(0)}$ as well.  A natural consequence of the re-scaling is an identity by definition:
\bena
t_1 - \mathcal{H}/2  & = & 1\,,\quad\textrm{(autoparallel scheme)}\\
\textrm{or}\quad\quad\quad \mathcal{H}    & = & -2 \,,\quad\textrm{(extremal scheme)}
\eena

\section{Precession of a gyroscope II: instantaneous rate}
\label{sec:instan-precess}

We now have the tools to calculate the precession of a gyroscope. Before proceeding, let us summarize the assumptions made so far:
\begin{enumerate}
\item A gyroscope can feel torsion through its rotational angular momentum, and the equation of motion is either 
$\mathrm{D}S^{\mu}/\mathrm{D}\tau=0$ or $\mathrm{D}S^{\mu\nu}/\mathrm{D}\tau=0$.
\item The world line of a gyroscope is either an autoparallel curve or an extremal curve.
\item The torsion and connection around Earth are parametrized by \Eq{eqn:w2-axi} and (\ref{eqn:full-conn323}).  
\end{enumerate}
With these assumptions, the calculation of the precession rate becomes straightforward except for one subtlety described below.  

\subsection{Transformation to the center-of-mass frame}

The precession rate $\mathrm{d}\vec{S}/\mathrm{d}t$ derived from a naive application of the equation of motion $\mathrm{D}S^{\mu}/\mathrm{D}\tau=0$ is the rate
measured by an observer at rest relative to the central gravitating body.  This rate is gauge-dependent and unphysical, 
since it depends on which coordinates the observer uses; for example,
isotropic spherical coordinates and standard spherical coordinates yield different precession rates. The physical observable is the precession rate
$\mathrm{d}\vec{S}_0/\mathrm{d}t$ measured by the observer co-moving with the center of mass of the gyroscope, \ie in the instantaneous local inertial frame.  

The methodology of transforming $\vec{S}$ to $\vec{S}_0$ was first established by Schiff \cite{schiff} in which he used the 4-tensor $S^{\mu\nu}$.  The basic idea
using the 4-vector $S^{\mu}$ is as follows. Since we are interested in the transformation only 
to leading order in $(v/c)^2$ and $m/r$, 
we are allowed to consider the coordinate
transformation and the velocity transformation separately and add them together in the end. We adopt standard spherical coordinates with the line element of
\Eq{eqn:metricaxisy}. The off-diagonal metric element proportional to $ma/r^2$ can be ignored for the purposes of this transformation. Consider a measuring rod in the
rest frame of the central body. It will be elongated by a factor of $(1+\mathcal{F}m/2r)$ in the radial direction measured by the observer in the center-of-mass
frame, but unchanged in the tangential direction.  The 4-vector $S^{\mu}$ transforms as $\mathrm{d}x^{\mu}$; thus its radial component is enlarged by a factor of
$(1+\mathcal{F}m/2r)$ and the tangential components are unchanged.  This can be compactly written in the following form:
\ben
\vec{S}_0=\vec{S}+\mathcal{F}\frac{m}{2r^3}(\vec{S}\cdot\vec{r})\vec{r}\,.
\een
Now consider the velocity transformation to the center-of-mass frame by boosting the observer along the $x$-axis, say, with velocity $v$.  
We have the Lorentz boost from
$S^{\mu}=(S^0,S^x,S^y,S^z)$ to $S_0^{\mu}=\left(S_0\,^0,S_0\,^x,S_0\,^y,S_0\,^z\right)$ as follows:
\bena
S_0\,^0 &=& \gamma(S^0-v\,S^x)\,,\\
S_0\,^x &=& \gamma(S^x-v\,S^0)\,,\\
S_0\,^y &=& S^y\,,\\
S_0\,^z &=& S^z\,,
\eena
where $\gamma=1/\sqrt{1-v^2}\approx 1+v^2/2$.  The condition \mbox{$S^{\mu}u_{\mu}=0$} gives 
\[ S^0 = \vec{v}\cdot\vec{S}=v\,S^x\,,\]
which verifies that $S_0\,^0=0$ in the center-of-mass frame.  The spatial components can be written compactly as 
\ben
\vec{S}_0=\vec{S}-\frac{1}{2}(\vec{S}\cdot\vec{v})\vec{v}\,.
\een
Combining the coordinate transformation and the velocity transformation, we find the following transformation from standard spherical coordinates 
to the center-of-mass frame:
\beq{eqn:xform-smu}
\vec{S}_0=\vec{S}+\mathcal{F}\frac{m}{2r^3}(\vec{S}\cdot\vec{r})\vec{r}-\frac{1}{2}(\vec{S}\cdot\vec{v})\vec{v}\,.
\eeq
The time derivative of \Eq{eqn:xform-smu} will lead to the expression for  \emph{geodetic precession} to leading order , \ie to order $(m/r)v$. 
To complete the discussion of transformations, note that the off-diagonal metric element proportional to $ma/r^2$ could add a term of order $ma/r^2$ 
to \Eq{eqn:xform-smu}, which leads to a precession rate proportional to $(ma/r^2)v$.  Since the leading term of the \emph{frame dragging} 
effect is of the order $ma/r^2$, the leading frame-dragging effect is invariant under these transformations, so we are allowed to ignore the 
off-diagonal metric element in the transformation.

The transformation law obtained using the 4-tensor $S^{\mu\nu}$ is different from using $S^{\mu}$ 
 --- this is not surprising because both descriptions coincide only in the rest frame of the gyroscope's center of mass. Schiff \cite{schiff} gave the transformation law from standard spherical coordinates to the center-of-mass frame, using $S^{\mu\nu}$: 
\ben
\vec{S}_0=\vec{S}'+\mathcal{F}\frac{m}{2r}[\vec{S}'-(\vec{r}/r^2)(\vec{r}\cdot\vec{S}')]-\frac{1}{2}[v^2 \vec{S}'-(\vec{v}\cdot\vec{S}')\vec{v}]\,.\label{eqn:xform}
\een

In taking the time derivative of \Eq{eqn:xform-smu} or (\ref{eqn:xform}), one encounters terms proportional to $\mathrm{d}\vec{v}/\mathrm{d}t$.   \Eq{eqn:auto} or (\ref{eqn:extrem}) should be applied, depending on whether autoparallel  or extremal scheme, respectively, is assumed.

\subsection{Instantaneous rates}

\subsubsection{Autoparallel scheme and using $S^{\mu}$}
\label{subsubsec:smu-auto}

Now we are now ready to calculate the precession rate.  In spherical coordinates $x^{\mu}=(t,r,\theta,\phi)$, we expand the rotational angular momentum 
vector in an orthonormal basis:
\[\vec{S}=S_r\hat{e}_r+S_{\theta}\hat{e}_{\theta}+S_{\phi}\hat{e}_{\phi}\,.\]
In terms of the decomposition coefficients, the 4-vector is
\[ S^{\mu}=(S^0,S^1,S^2,S^3)=(S^0,S_r,S_{\theta}/r,S_{\phi}/r\sin\theta)\,.\]
Applying the equation of motion $\mathrm{D}S^{\mu}/\mathrm{D}\tau=0$, transforming $\vec{S}$ to $\vec{S}_0$ by \Eq{eqn:xform-smu} and taking the time derivative
using autoparallels (Eq.~\ref{eqn:auto}), we obtain the following instantaneous gyroscope precession rate:
\bena
\frac{\mathrm{d}\vec{S}_0}{\mathrm{d}t} &=& \vec{\Omega}\times\vec{S}_0\,, \label{eqn:instn-precession1}\\
\textrm{where}\quad \vec{\Omega}\phantom{G} &=& \vec{\Omega}_G + \vec{\Omega}_F\,,\label{eqn:instn-precession2}\\
\vec{\Omega}_G &=& \left( \frac{\mathcal{F}}{2}-\frac{\mathcal{H}}{4}+t_2+\frac{t_1}{2}\right)\frac{m}{r^3}(\vec{r}\times\vec{v})\,, \label{eqn:GP1}\\
\vec{\Omega}_F &=& \frac{\mathcal{G}I}{r^3}\left[ -\frac{3}{2}(1+\mu_1)(\vec{\omega}_E \cdot \hat{e}_r)\hat{e}_r  +\frac{1}{2}(1+\mu_2)\vec{\omega}_E \right]\,. \label{eqn:FD1}
\eena
Here $I\omega_E=ma$ is the angular momentum of Earth, where $I$ is Earth's moment of inertia about its poles and $\omega_E$ is its angular velocity.  
The new effective torsion constants are defined so that they represent the torsion-induced correction to the GR prediction:
\bena
\mu_1 &\equiv & (w_1-w_2-w_3+2w_4+w_5)/(-3\mathcal{G})\,,\label{eqn:mu1new}\\
\mu_2 &\equiv & (w_1-w_3+w_5)/(-\mathcal{G})\,,
\eena
Since $t_1-\mathcal{H}/2=1$ in the autoparallel scheme, \Eq{eqn:GP1} simplifies to 
\beq{eqn:GP11}
\vec{\Omega}_G = \left( 1+ \mathcal{F}+2t_2\right)\frac{m}{2r^3}(\vec{r}\times\vec{v})\,.
\eeq   

In the literature, the precession due to $\Omega_G$ is called \emph{geodetic precession}, and that due to $\Omega_F$ is called \emph{frame dragging}.  
From \Eq{eqn:GP1}, it is seen that geodetic precession depends on the mass of Earth and not on whether Earth is spinning or not.  It is of order $mv$. The
frame-dragging effect is a unique effect of Earth's rotation and highlights the importance of the GPB experiment, since GPB will be the first to
accurately measure the effect of the off-diagonal metric element that lacks a counterpart in Newtonian gravity.  The frame dragging effect is of order
$ma$, so it is independent of whether the gyroscope is moving or static.  In the presence of torsion, we term $\Omega_G$ the ``generalized geodetic precession'',
and $\Omega_F$ the ``generalized frame-dragging''.

\subsubsection{Extremal scheme and using $S^{\mu}$}

We now repeat the calculation of \Sec{subsubsec:smu-auto}, but assuming an extremal trajectory (\Eqnopar{eqn:extrem}) 
when taking the time derivative of \Eq{eqn:xform-smu}, obtaining the following instantaneous gyroscope precession rate:
\bena
\frac{\mathrm{d}\vec{S}_0}{\mathrm{d}t} &=& \vec{\Omega}\times\vec{S}_0-t_1\frac{m}{r^3}(\vec{S}_0\cdot\vec{v})\vec{r}\,, \label{eqn:instn-precession21}\\
\textrm{where}\quad \vec{\Omega}\phantom{G} &=& \vec{\Omega}_G + \vec{\Omega}_F\,.\nonumber\\
\vec{\Omega}_G &=& \left( \frac{\mathcal{F}}{2}-\frac{\mathcal{H}}{4}+t_2\right)\frac{m}{r^3}(\vec{r}\times\vec{v})\,, \label{eqn:GP-extremal_smu}
\eena
and $\vec{\Omega}_F$ is the same as in \Eq{eqn:FD1}.
Since $\mathcal{H}=-2$ in the extremal scheme, \Eq{eqn:GP-extremal_smu} is simplified to formally coincide with \Eq{eqn:GP11}.

\subsubsection{Extremal scheme and using $S^{\mu\nu}$}

In spherical coordinates, $S^{\mu\nu}$ satisfies
\ben
S^{12}=\frac{1}{r}S^{\,'}_{\phi}\,,\:\: S^{23}=\frac{1}{r^2 \sin\theta}S^{\,'}_{r}\,,\:\: S^{31}=\frac{1}{r\sin\theta}S^{\,'}_{\theta}\,,\label{eqn:ssrel}
\een
where $S^{\,'}_r,\,S^{\,'}_{\theta},\,S^{\,'}_{\phi}$ are the components of $\vec{S}^{\,'}$ in spherical coordinates, i.e. $\vec{S}^{\,'} = S^{\,'}_r\hat{e}_r+S^{\,'}_{\theta}\hat{e}_{\theta}+S^{\,'}_{\phi}\hat{e}_{\phi}\,.$
We now repeat the calculation of \Sec{subsubsec:smu-auto} assuming an extremal trajectory (\Eqnopar{eqn:extrem}) and the $S^{\mu\nu}$-based precession of
\Eq{eqn:xform} when taking the time derivative of \Eq{eqn:xform-smu}, obtaining the following instantaneous gyroscope precession rate:
\bena
\frac{\mathrm{d}\vec{S}_0}{\mathrm{d}t} &=& \vec{\Omega}\times\vec{S}_0+t_1\frac{m}{r^3}\vec{r}\times(\vec{v}\times\vec{S}_0)\,, \label{eqn:instn-precession31}\\
\textrm{where}\quad \vec{\Omega}\phantom{G} &=& \vec{\Omega}_G + \vec{\Omega}_F\,.\nonumber
\eena
$\vec{\Omega}_G$ and $\vec{\Omega}_F$ are the same as in equations (\ref{eqn:GP-extremal_smu}) and (\ref{eqn:FD1}), respectively.

In both cases using extremals, the precession rates have anomalous terms proportional to $t_1$;
see \Eq{eqn:instn-precession21}) and~\ref{eqn:instn-precession31}).  We call these terms the ``anomalous geodetic precession''.  
These anomalies change the angular precession rate of a gyroscope, since their
contributions to $\mathrm{d}\vec{S}_0/\mathrm{d}t$ are not perpendicular to $\vec{S}_0$.  This is a phenomenon that GR does not predict. Meanwhile, $t_2$
contributes to modify only the magnitude and not the direction of $\vec{\Omega}_G$. We therefore term $t_1$ the anomalous geodetic torsion and $t_2$ the normal
geodetic torsion.  The torsion functions $w_1$,\ldots,$w_5$ contribute to the generalized frame-dragging effect via the two 
combinations $\mu_1$ and $\mu_2$, and we therefore term them ``frame-dragging torsions''.  

\subsubsection{Autoparallel scheme and using $S^{\mu\nu}$}

Repeating the calculation of \Sec{subsubsec:smu-auto}
using the $S^{\mu\nu}$-based precession rule of
\Eq{eqn:xform} gives the exact same instantaneous precession rate as in \Sec{subsubsec:smu-auto}.
This is expected since these two precession rules are equivalent in the autoparallel scheme.

\section{Precession of a gyroscope III: moment analysis}
\label{sec:mom-ana}

\begin{figure}
\centering
\includegraphics[width=0.5\textwidth]{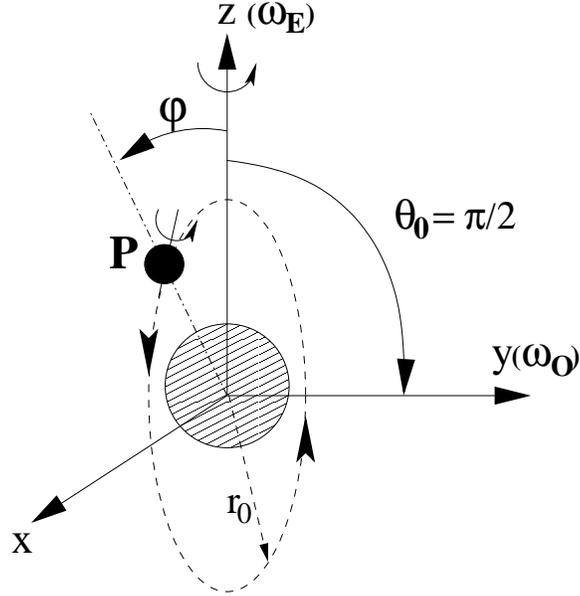}
\caption[Gravity Probe B experiment]{\label{gpb2}\footnotesize%
A Gravity Probe B gyroscope moves around Earth along a circular polar orbit with $\theta_0=\pi/2$. 
$\omega_O$ is its orbital angular velocity and $\omega_E$ is Earth's rotational angular velocity around the $z$-axis.    }
\end{figure}

GPB measures the rotational angular momentum $\vec{S}_0$ of gyroscopes and therefore the precession rate 
$\mathrm{d}\vec{S}_0/\mathrm{d}t$ essentially continuously.  This
provides a wealth of information and deserves careful data analysis. Here we develop a simple but sensitive analysis method based on Fourier transforms.

\subsection{Fourier transforms}

The Gravity Probe B satellite has a circular polar orbit to good approximation\footnote{The actual GPB orbit has an orbital eccentricity of 0.0014 and 
an inclination of $90.007^\circ$ according to the Fact Sheet on the GPB website.  These deviations from the ideal orbit should cause negligible ($\lesssim 10^{-5}$) relative errors in our estimates above.}, \ie the inclination
angle of the orbital angular velocity $\vec{\omega}_O$ with respect
to the Earth's rotation axis (z-axis) is $\theta_0=\pi/2$.  Hence
the orbital plane is perpendicular to the equatorial plane. Let the $y$-axis
point along the vector $\vec{\omega}_O$ and let the $x$-axis be perpendicular
to the $y$-axis in the equatorial plane so that the three axes $\{x,y,z\}$ form a
right-handed coordinate basis as illustrated in \fig{gpb2}. A gyroscope at a point $P$ is marked by
the monotonically increasing angle $\varphi$ with respect to z axis.
The polar angle of the point $P$ can be regarded as a periodic
function of $\varphi$:
\beq{eqn:thetafn}
\theta(\varphi)=\left\{\begin{array}{lcl}\varphi &, & 0\le\varphi\le\pi\\ 
2\pi-\varphi&, & \pi\le\varphi\le 2\pi \end{array}\right.
\eeq 
So for a particular circular polar orbit, $\mathrm{d}\vec{S}_0/\mathrm{d}t(\vec{r},\vec{v})$ can be regarded as a periodic function of $\varphi$, 
where $r_0$ is the fixed radius, allowing us to write $\mathrm{d}\vec{S}_0/\mathrm{d}t(\vec{r},\vec{v})\equiv\mathrm{d}\vec{S}_0/\mathrm{d}t(\varphi)$.

Now define the Fourier \emph{moments} of the precession rate as 
\bena
\vec{a}_0 &=&
\frac{1}{2\pi}\int_0^{2\pi}\frac{\mathrm{d}\vec{S}_0}{\mathrm{d}t}(\varphi)\mathrm{d}\varphi
=\left<\frac{\mathrm{d}\vec{S}_0}{\mathrm{d}t}(\varphi)\right>,\\
\vec{a}_n &=& \frac{1}{2\pi}\int_0^{2\pi}\frac{\mathrm{d}\vec{S}_0}{\mathrm{d}t}(\varphi)\cos
 n\varphi \mathrm{d}\varphi\,,\\
\vec{b}_n &=& \frac{1}{2\pi}\int_0^{2\pi}\frac{\mathrm{d}\vec{S}_0}{\mathrm{d}t}(\varphi)\sin
 n\varphi \mathrm{d}\varphi\,,
\eena
where $n=1,2,\ldots$, so that we can write 
\beq{eqn:dsdt-fourier}
\frac{\mathrm{d}\vec{S}_0}{\mathrm{d}t}(\varphi)=\vec{a}_0+2\sum_{n=1}^{\infty}(\vec{a}_n\cos
n\varphi+\vec{b}_n\sin n\varphi)\,.
\eeq

\subsection{Average precession}
\label{subsec:ave_precession}

We now write equations (\ref{eqn:instn-precession1}), (\ref{eqn:instn-precession2}), (\ref{eqn:FD1}), (\ref{eqn:GP11}), (\ref{eqn:instn-precession21}) and (\ref{eqn:instn-precession31})
 explicitly in terms of $\varphi$ and perform the Fourier transforms.  The average precession in the three calculation schemes above can be compactly written as follows:
\ben
\vec{a}_0  \equiv  \left<\frac{\mathrm{d}\vec{S}_0}{\mathrm{d}t}(\varphi)\right> = \vec{\Omega}_{\textrm{eff}}\times \vec{S}_0\label{eqn:moment1}\,.
\een

The angular precession rate is
\ben
\vec{\Omega}_{\textrm{eff}} = b_t \frac{3m}{2r_0}\vec{\omega}_O+b_\mu\frac{I}{2r_0^3}\vec{\omega}_E\,,\label{eqn:moment2}
\een
where $\vec{\omega}_O=\omega_O\hat{y}$ is the orbital angular velocity and $\vec{\omega}_E=\omega_E \hat{z}$ is the rotational angular velocity of Earth. 
Here the ``biases'' relative to the GR prediction are defined by
\bena
b_t &\equiv & \frac{1}{3}(1+\mathcal{F}+2t_2+|\eta| t_1)\,,\label{eqn:btdef}\\
b_\mu &\equiv & \frac{\mathcal{(-G)}}{2}(1+3\mu_1-2\mu_2)\,,\label{eqn:bmudef}\\
      &=& \frac{\mathcal{(-G)}}{2}[1+(w_1+w_2-w_3-2w_4+w_5)/\mathcal{G}]\,,\nonumber
\eena
where the constant $\eta$ reflects the different assumptions that we have explored, and takes the following values:
\beq{eqn:etadef}
\eta=\left\{ \begin{array}{rl}  0 & \textrm{using autoparallels}\\
                                +1 & \textrm{using $S^{\mu\nu}$ and extremals}\\
                                -1 & \textrm{using $S^{\mu}$ and extremals}
\end{array}\right.
\eeq
From the above formulas, we see that the three schemes give identical results when $t_1=0$.

For comparison, GR predicts the average precession rate
\bena
\vec{a}_0 &\equiv & \left<\frac{\mathrm{d}\vec{S}_0}{\mathrm{d}t}(\varphi)\right> = \vec{\Omega}_{\textrm{eff}}\times \vec{S}_0\label{eqn:moment1GR}\,,\nonumber\\
\textrm{where}\quad\vec{\Omega}_{\textrm{eff}} &=& \frac{3m}{2r_0}\vec{\omega}_O+\frac{I}{2r_0^3}\vec{\omega}_E\,,\label{eqn:moment2GR}
\eena
\ie, $b_t=b_\mu=1$.

It is important to note that torsion contributes to the \emph{average} precession above only via {\it magnitudes} of the precession rates, 
leaving the precession axes intact. The geodetic torsion parameters $t_1$ and $t_2$ are degenerate, entering only in the linear combination corresponding 
to the bias $b_t$. 
The frame-dragging torsion parameters $w_1,\ldots,w_5$ are similarly  degenerate, entering only in the linear combination corresponding to the bias $b_\mu$. 
If for technical reasons, the average precession rate is the only quantity that GPB can measure, then only these biases can be constrained.

\subsection{Higher moments}
\label{subsec:moment}

Interestingly, all higher Fourier moments vanish except for $n=2$:
\begin{eqnarray}
\vec{a}_2 &=& \frac{-3\mathcal{G}I\omega_E}{8r_0^3}(1+\mu_1)\hat{z}\times\vec{S}_0 +\eta\, t_1\frac{m}{4r_0}\omega_O(S_0\,^x\hat{z}+S_0\,^z\hat{x}) \,,\nonumber\\
\vec{b}_2 &=& \frac{-3\mathcal{G}I\omega_E}{8r_0^3}(1+\mu_1)\hat{x}\times\vec{S}_0 +\eta\, t_1\frac{m}{4r_0}\omega_O(S_0\,^x\hat{x}-S_0\,^z\hat{z}) \,.\nonumber\\ 
\label{eqn:n2moments}
\end{eqnarray}
Here we use the notation $S_0\,^i\equiv \vec{S}_0\cdot\hat{i}$, where $i$ denotes the $x$, $y$ and $z$ axes.

For comparison, GR predicts the following second moments (moments with $m=1$ and $m>2$ vanish):
\bena
\vec{a}_2 &=& \frac{3I\omega_E}{4r_0^3}\hat{z}\times\vec{S}_0 \,,\\
\vec{b}_2 &=& \frac{3I\omega_E}{4r_0^3}\hat{x}\times\vec{S}_0 \,.
\eena
Technically, it may be difficult to measure these second moments because of the extremely small precession rate per orbit. 
However, if they \emph{could} be measured, they could break the degeneracy between $t_1$ and $t_2$: $|t_1|$ could be measured through the \emph{anomalous} $n=2$
precession moment (the second term in \Eq{eqn:n2moments}). The sign ambiguity of $t_1$ is due to the relative sign difference between the two schemes using
extremals and $S^{\mu\nu}$ versus $S^{\mu}$. The degeneracy between $w_1,\ldots,w_5$ could be alleviated as well, since the linear combination $\mu_1$ (defined in
\Eq{eqn:mu1new}) could be measured through the correction to the \emph{normal} $n=2$ precession moment (the first term in \Eq{eqn:n2moments}).  
By ``anomalous'' or ``normal'', we mean the term whose precession axis has not been or already been, respectively, predicted by GR.
In addition, the anomalous second-moment terms cannot be expressed as the cross product of $\vec{S}_0$ and an angular velocity vector.

\section{Constraining torsion parameters with Gravity Probe B}
\label{sec:general-torsion-constraints}

The parametrized Post-Newtonian (PPN) formalism has over the past decades demonstrated its success as a theoretical framework of testing GR, by
embedding GR in a broader parametrized class of metric theories of gravitation.  This idea can be naturally generalized by introducing more general departures
from GR, \eg\ torsion. For solar system tests, the seven torsion parameters derived in \Sec{sec:param} define the torsion extension of the PPN
parameters, forming a complete set that parametrizes all observable signatures of torsion to lowest order.

However, most of existing solar system tests cannot constrain the torsion degrees of freedom.  Photons are
usually assumed to decouple from the torsion to preserve gauge invariance (we return below to the experimental basis of this), 
in which case tests using electromagnetic signals (\eg\ Shapiro time delay and the deflection of light) can only
constrain the metric, \ie the PPN parameter $\gamma$, as we explicitly calculate in Appendix \ref{appendix:shapiro} and
Appendix \ref{subsec:deflec}.  Naively, one might expect that Mercury's perihelion shift could constrain torsion parameters if Mercury's orbit is an autoparallel curve,
but calculations in Appendix \ref{subsec:advance-perih-auto} and Appendix \ref{subsec:advance-perih-extreme} show that to lowest
order, the perihelion shift is nonetheless only sensitive to the metric. Moreover, PPN calculations
\cite{Will:2005va} show that a complete account of the  perihelion shift must involve second-order parameters in $m/r$ (\eg\ the PPN
parameter $\beta$), which are beyond our first-order parametrization, as well as the first-order ones. We therefore neglect the constraining power of
Mercury's perihelion shift here.  In contrast, the results in \Sec{subsec:ave_precession} show that Gravity Probe B will be very
sensitive to torsion parameters even if only the average precession rates can be measured.  

We may also constrain torsion with experimental upper bounds on the photon mass, since the ``natural'' extension of Maxwell Lagrangian
($\partial_\mu \to \nabla_\mu$ using the full connection) breaks gauge invariance and introduces anomalous electromagnetic forces and a
quadratic term in $A_\mu$ that may be identified with the photon mass. In Appendix \ref{appendix:photon-mass}, we estimate the constraints
on the torsion parameters $t_1$ and $t_2$ from the measured photon mass limits, and show that these ground-based experiments can constrain
$t_1$ or $t_2$ only to a level of the order unity, \ie, not enough to be relevant to this chapter.  

\begin{figure}
\centering
\begin{displaymath}
\begin{array}{cc}
\includegraphics[width=0.5\textwidth]{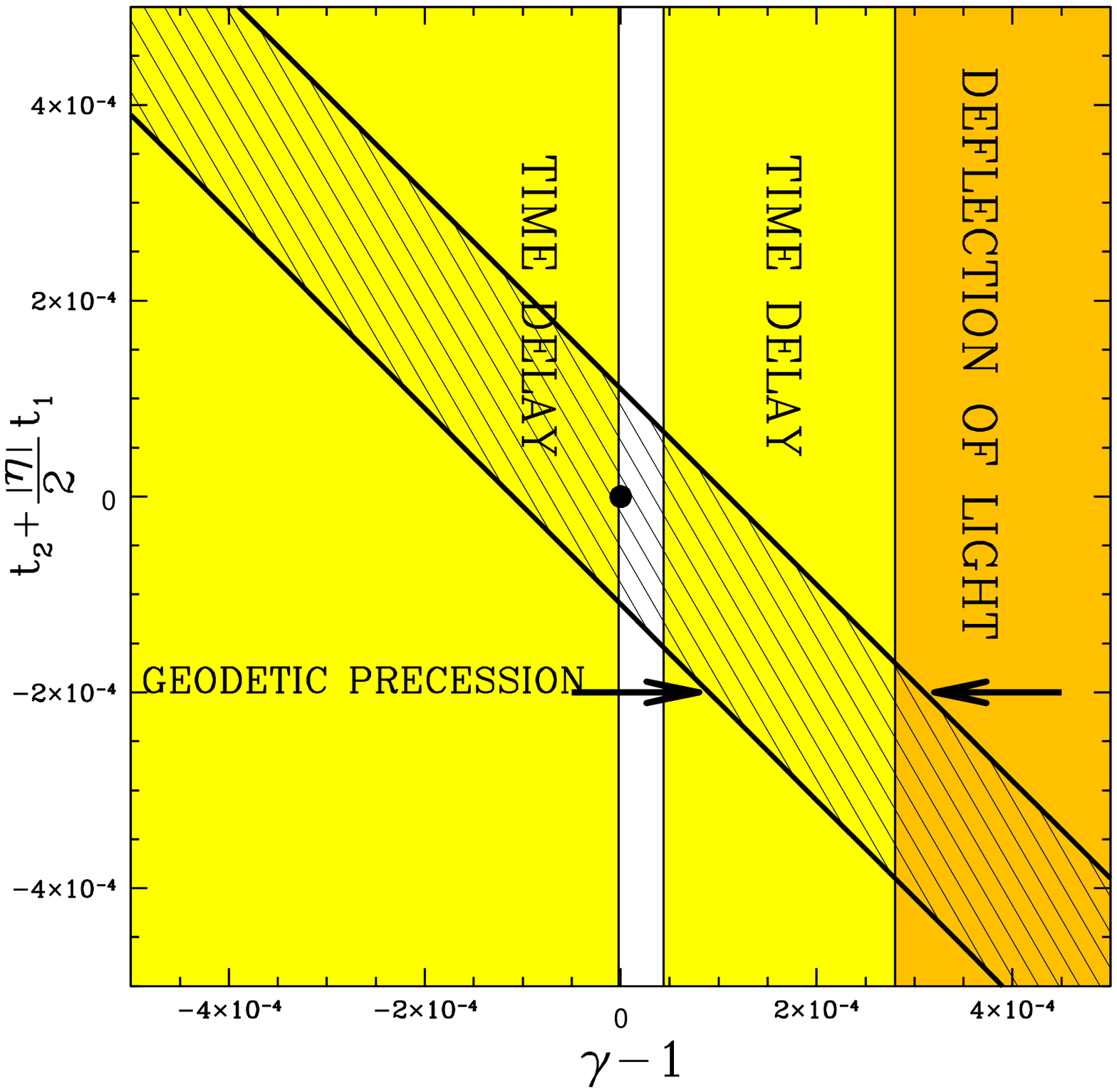} & 
\includegraphics[width=0.5\textwidth]{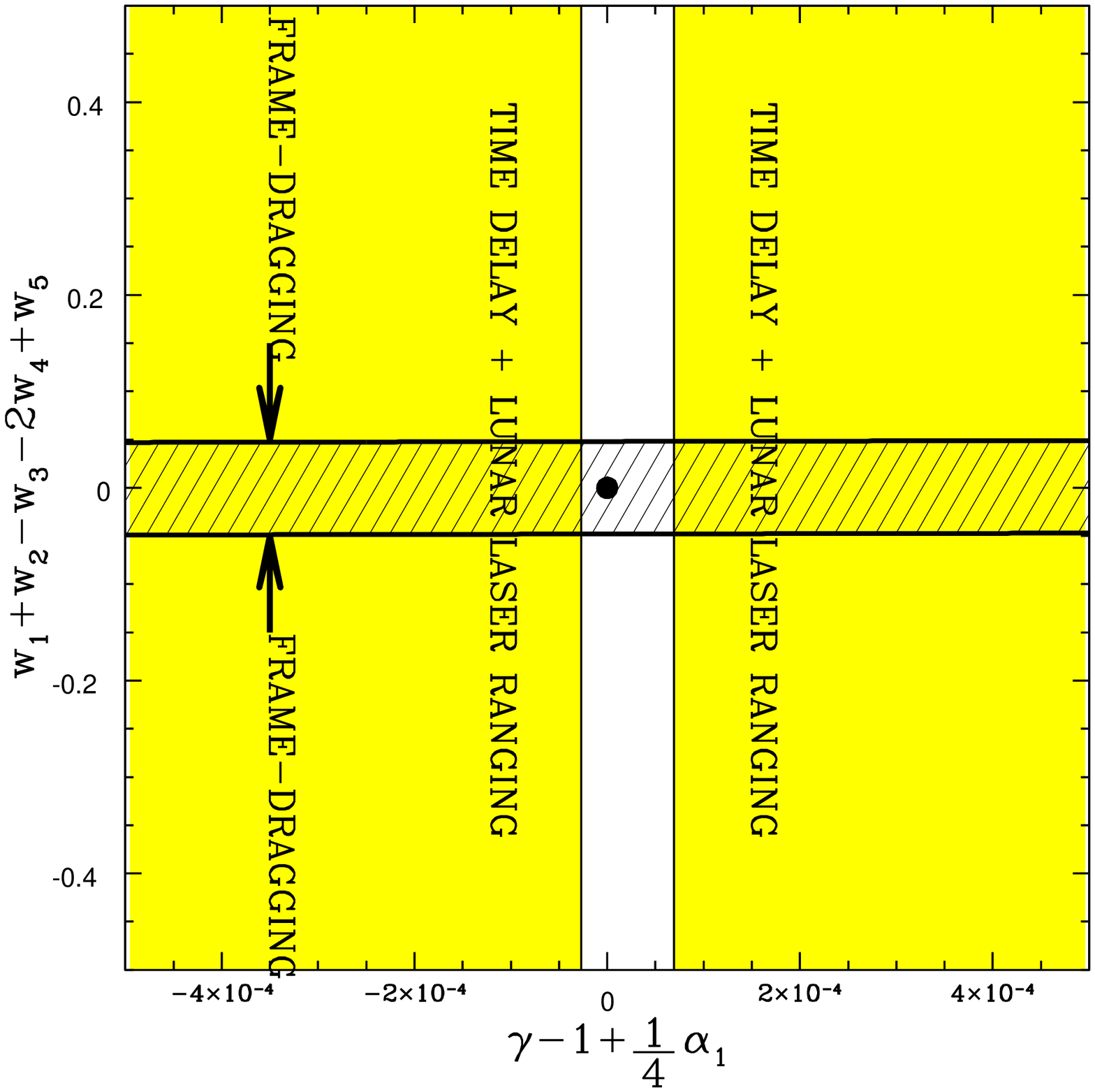} 
\end{array}
\end{displaymath}
\caption[Constraints on the PPN and torsion parameters from solar system tests]{\footnotesize%
Constraints on the PPN parameters ($\gamma$, $\alpha_1$) and torsion parameters 
($t_1$, $t_2$, $w_1$ \ldots $w_5$) from solar system tests.  General Relativity corresponds to the black dot 
($\gamma -1=\alpha_1=$ all torsion parameters $=0$).  
Left panel: the shaded regions in the parameter space have already been ruled out by the deflection of light (orange/grey) 
and Shapiro time delay (yellow/light grey).  Gyroscope experiments are sensitive to torsion parameters.  
If the geodetic precession measured by Gravity Probe B is consistent with GR, this will rule out everything 
outside the hatched region, implying that $-1.5\times 10^{-4} <  t_2+\frac{|\eta|}{2}t_1 < 1.1\times 10^{-4}$
(assuming a target angle accuracy of 0.5 milli-arcseconds).  The unpublished preliminary results of Gravity Probe B have 
confirmed the geodetic precession to better than 1\%, giving a constraint
 $| t_2+\frac{|\eta|}{2}t_1 | \lesssim 0.01$.  
   Right panel: the shaded regions in the parameter space have already been ruled out by Shapiro time delay combined with lunar laser ranging experiment (yellow/light grey).  Lunar laser ranging constrains $|\alpha_1|< 10^{-4}$ \cite{Will:2005va}.
  If the frame-dragging effect measured by Gravity Probe B is consistent with GR, this will rule out everything outside the hatched region, implying that $|w_1+w_2-w_3-2w_4+w_5| < 4.8\times 10^{-2}$.}
\label{fig:extendPPN-constraint}
\end{figure}

In Appendix \ref{appendix:solar-tests}, we confront solar system tests with the predictions from GR generalized with our torsion parameters.
In general, it is natural to assume that all
metric parameters take the same form as in PPN formalism \footnote{This may not be completely true in some particular theories, \eg\
$\mathcal{H} \ne -2$ in Einstein-Hayashi-Shirafuji theories in the autoparallel scheme, shown in Table \ref{table:EHS1}.}, \ie
\cite{Will:2005va}
\bena
\mathcal{H} &=& -2\,,\label{eqn:H=-2}\\
\mathcal{F} &=& 2\gamma\,,\\
\mathcal{G} &=& -(1+\gamma+\frac{1}{4}\alpha_1)\,. 
\eena
Therefore, Shapiro time delay and the deflection of light share the same multiplicative bias factor $(\mathcal{F}-\mathcal{H})/4 = (1+\gamma)/2$ relative to the GR prediction.  
The analogous bias for gravitational redshift is unity since $(\Delta\nu/\nu)/(\Delta\nu/\nu)^{(GR)}=-\mathcal{H}/2=1$. 
In contrast, both the geodetic precession and the frame-dragging effect have a non-trivial multiplicative bias in Eqs.(\ref{eqn:btdef}) and (\ref{eqn:bmudef}):
\bena
b_t &=& \frac{1}{3}(1+2\gamma)+\frac{1}{3}(2t_2+|\eta|t_1) \,,\\
b_\mu &=& \frac{1}{2}(1+\gamma+\frac{1}{4}\alpha_1)-\frac{1}{4}\left(w_1+w_2-w_3 -2w_4+w_5\right)\,.
\eena
We list the observational constraints that solar system tests can place on the PPN and torsion parameters in Table \ref{tab:general-torsion-constraint}
and plot the constraints in the degenerate parameter spaces in Figure \ref{fig:extendPPN-constraint}.
We see that GPB will 
ultimately
 constrain the linear combination $t_2+\frac{|\eta|}{2}t_1$ (with $\eta$ depending on the parallel transport scheme) 
at the $10^{-4}$ level and the combination $w_1+w_2-w_3-2w_4+w_5$ at the $1\%$ level. 
The unpublished preliminary results of GPB have confirmed the geodetic precession to less than 1\% level.  This imposes a constraint on $|t_2+\frac{|\eta|}{2}t_1|\lesssim 0.01$.  The combination $w_1+w_2-w_3-2w_4+w_5$ cannot be constrained by frame-dragging until GPB will manage to improve the accuracy to the target level of less than 1 milli-arcsecond.  

\begin{table}
\noindent 
\footnotesize{
\begin{center}
\begin{tabular}{|p{1.7cm}||p{3.5cm}|p{6cm}|p{2cm}|}
\hline
Effects  & Torsion Biases  &  Observ. Constraints & Remarks \\ \hline
Shapiro time delay & $\Delta t/\Delta t ^{(GR)}=(1+\gamma)/2$ &  $\gamma-1 = (2.1\pm 2.3)\times 10^{-5}$ & Cassini tracking \cite{Bertotti:2003rm}  \\\hline
Deflection of light & $\delta/\delta^{(GR)}=(1+\gamma)/2 $   & $\gamma-1 = (-1.7\pm 4.5)\times 10^{-4}$ &  VLBI \cite{Shapiro2004} \\\hline
Gravitational redshift & $(\Delta\nu/\nu)/(\Delta\nu/\nu)^{(GR)}=1$ & no constraints &  \\\hline
Geodetic Precession & $\Omega_G/\Omega_G^{(GR)}=b_t$ & $\left|(\gamma-1)+(t_2+\frac{|\eta|}{2}t_1)\right|< 1.1\times 10^{-4}$ & Gravity Probe B \\ \hline
Frame-dragging & $\Omega_F/\Omega_F^{(GR)}=b_\mu$ & $|(\gamma-1+\frac{1}{4}\alpha_1)-\frac{1}{2}(w_1+w_2-w_3-2w_4+w_5) |<0.024$ & Gravity Probe B \\ \hline
\end{tabular}
\caption[Constraints of PPN and torsion parameters with solar system tests]{Constraints of PPN and torsion parameters with solar system tests.  The observational constraints on PPN parameters are taken from Table 4 of \cite{Will:2005va}. 
Unpublished preliminary results of Gravity Probe B have confirmed geodetic precession to better than 1\%, 
giving a constraint $|(\gamma-1)+(t_2+\frac{|\eta|}{2}t_1)| \lesssim 0.01$.  
The full GPB results are yet to be released, so whether the frame dragging
will agree with the GR prediction is not currently known.  The last two rows 
show the limits that would correspond to a GPB result consistent with GR, assuming an angle accuracy of 0.5 milli-arcseconds.}
\label{tab:general-torsion-constraint}
\end{center}
}
\end{table}

\section{Linearized Kerr solution with torsion in Weitzenb\"ock spacetime}
\label{sec:counterexample}

So far, we have used only symmetry principles to derive the most general torsion possible around Earth to lowest order. 
We now turn to the separate question of whether there is any gravitational Lagrangian that actually produces torsion around Earth.
We will show that the answer is yes by exploring the specific example of the 
Hayashi-Shirafuji Lagrangian \cite{HS1} in Weitzenb\"ock spacetime, showing that it populates a certain subset of
the torsion degrees of freedom that we parametrized above and that this torsion mimics the Kerr metric to lowest order even though the Riemann curvature of 
spacetime vanishes.
We begin with a
brief review of Weitzenb\"ock spacetime and the Hayashi-Shirafuji Lagrangian, then give the linearized solution in terms of the seven parameters
$t_1,t_2,w_1,\ldots,w_5$ from above. The solution we will derive is a particular special case of 
what the symmetry principles allow, and is for the particularly simple case where the Riemann curvature vanishes (Weitzenb\"ock spacetime). 
 Later in \Sec{subsec:linearinterpol}, 
we will give a more general Lagrangian producing both torsion and curvature, effectively interpolating between the Weitzenb\"ock case below and standard GR.

We adopt the convention only here in \Sec{sec:counterexample} 
 and \Sec{subsec:linearinterpol} 
that Latin letters are indices for the internal basis, whereas Greek letters are 
spacetime indices, both running from 0 to 3.  

\subsection{Weitzenb\"ock spacetime}

We give a compact review of Weitzenb\"ock spacetime and Hayashi-Shirafuji Lagrangian here and in \Sec{subsec:HS} respectively. 
We refer the interested reader to their original papers \cite{weitzenbock,HS1} for a complete survey of these subjects. 

Weitzenb\"ock spacetime is a Riemann-Cartan spacetime in which the Riemann curvature tensor, defined in \Eq{eqn:curvature}, vanishes identically: 
\beq{eqn:vanishingR}
R^{\rho}_{\phantom{\rho}\lambda\nu\mu}(\Gamma)=0\,.
\eeq
\Fig{fig:spaces} illustrates how Weitzenb\"ock spacetime is related to other spacetimes.

Consider a local coordinate neighborhood of a point $p$ in a Weitzenb\"ock manifold with local coordinates $x^{\mu}$.  Introduce the coordinate basis $\left\{\bar{E}_{\,\mu}\right\}=\{\left(\partial/\partial x^{\mu}\right)_p\}$ and the dual basis $\left\{\bar{E}^{\,\mu}\right\}=\{\left(d x^{\mu}\right)_p\}$.  A vector $\bar{V}$ at $p$ can be written as $\bar{V}=V^{\mu}\bar{E}_{\,\mu}$.  The manifold is equipped with an inner product; the metric is the inner product of the coordinate basis vectors,
\[
g(\bar{E}_{\,\mu},\bar{E}_{\,\nu})=g(\bar{E}_{\,\nu},\bar{E}_{\,\mu})=g_{\mu\nu}\,.
\]
There exists a quadruplet of orthonormal vector fields $\bar{e}_{\,k}(p)$, where $\bar{e}_{\,k}(p)=e_{\,k}^{\,\phantom{1}\mu}(p)\bar{E}_{\,\mu}$, such that
\ben
g(\bar{e}_{\,k},\bar{e}_{\,l})=g_{\mu\nu}e_{\,k}^{\,\phantom{1}\mu}e_{\,l}^{\,\phantom{1}\nu}=\eta_{kl}\,,
\een
where $\eta_{kl}=\mathrm{diag}(-1,1,1,1)$.
There also exists a dual quadruplet of orthonormal vector fields $\bar{e}^{\,k}(p)$, where $\bar{e}^{\,k}(p)=e^{\,k}_{\,\phantom{1}\mu}(p)\bar{E}^{\,\mu}$, such that
\ben
e_{\,k}^{\,\phantom{1}\mu}e^{\,k}_{\,\phantom{1}\nu}=\delta_{\phantom{1}\nu}^{\mu}\,,
\qquad  e_{\,k}^{\,\phantom{1}\mu} e^{\,l}_{\,\phantom{1}\mu}=\delta_{k}^{\phantom{1}l}\,.\label{eqn:bbg1}
\een
This implies that
\ben
\eta_{kl}e^{\,k}_{\,\phantom{1}\mu}e^{\,l}_{\,\phantom{1}\nu}=g_{\mu\nu}\,.\label{eqn:bbg2}
\een
which is often phrased as the $4\times 4$ matrix ${\bf e}$ (a.k.a.~the {\it tetrad} or {\it vierbein}) 
being ``the square root of the metric''.

An alternative definition of Weitzenb\"ock spacetime that is equivalent to that of \Eq{eqn:vanishingR} is the requirement that the Riemann-Cartan spacetime 
admit a quadruplet of linearly independent \emph{parallel vector fields} $e_k^{\phantom{1}\mu}$, defined by\footnote{Note that 
Hayashi and Shirafuji \cite{HS1} adopted a convention where the \mbox{order} of the lower index placement in the connection is opposite 
to that in \Eq{eqn:parallel-vector}.}
\beq{eqn:parallel-vector}
\nabla_{\mu}e_k^{\phantom{1}\nu}=\partial_{\mu}e_k^{\phantom{1}\nu}+\Gamma^{\nu}_{\phantom{1}\mu\lambda}e_k^{\phantom{1}\lambda}=0\,.
\eeq
Solving this equation, one finds that
\beq{eqn:conn-b}
\Gamma^{\lambda}_{\phantom{1}\mu\nu}=e_k^{\phantom{1}\lambda}\partial_{\mu}e^k_{\phantom{1}\nu}\,,
\eeq
and that the torsion tensor
\beq{eqn:tor-b}
S_{\mu\nu}^{\phantom{12}\lambda}=\frac{1}{2}e_k^{\phantom{1}\lambda}(\partial_{\mu}e^k_{\phantom{1}\nu}-\partial_{\nu}e^k_{\phantom{1}\mu})\,.
\eeq
This property of allowing globally parallel basis vector fields was termed ``teleparallelism'' by Einstein, since it allows unambiguous parallel transport,
and formed the foundation of the torsion theory he termed ``new general relativity'' \cite{Einstein:1928-30,Einstein:1930b,moller:1961,Pellegrini:1962,moller:1978,Unzicker:2005in,Obukhov:2004hv,Vargas:1992ab,Vargas:1992ac,Mueller-Hoissen:1983vc,Mielke:1992te,Kreisel:1979kh,Treder:1978vf,Kreisel:1980kb,Pimentel:2004bp,Maluf:2001ef}.

A few additional comments are in order:
\begin{enumerate}

\item It is easy to verify that the first definition of Weitzenb\"ock spacetime (as curvature-free, \ie via \Eq{eqn:vanishingR}) 
follows from the second definition --- one simply uses the the explicit expression for the connection (\Eqnopar{eqn:conn-b}). It is
also straightforward to verify that $\nabla_{\mu}g_{\nu\rho}=0$ using \Eq{eqn:bbg2} and (\ref{eqn:parallel-vector}).

\item \Eq{eqn:conn-b} is form invariant under general (spacetime) coordinate transformations due to the nonlinear transformation law (\Eq{eqn:gamma-law}) of
the connection, provided that $e_k^{\phantom{1}\mu}$ and $e^k_{\phantom{1}\mu}$ transform as a contravariant vector and a covariant vector, respectively.

\item The Weitzenb\"ock spacetime preserves its geometry under \emph{global} proper orthochronous Lorentz transformations, \ie a new equivalent quadruplet of
parallel vector fields $\underline{e}'$ is obtained by a global proper orthochronous Lorentz transformation,
${e'}_k^{\phantom{1}\mu}=\Lambda^l_{\phantom{1}k} e_l^{\phantom{1}\mu}$.  

\end{enumerate}

\subsection{Hayashi-Shirafuji Lagrangian}\label{subsec:HS}

The Hayashi-Shirafuji Lagrangian \cite{HS1} is a gravitational Lagrangian density constructed in the geometry of Weitzenb\"ock
spacetime\footnote{The Hayashi-Shirafuji theory differs from the teleparallel gravity theory decribed in \cite{Arcos:2004zh,Arcos:2005ec}, which is argued to be fully equivalent to GR.}.  It is a Poincar\'e gauge theory in that the parallel vector fields $\underline{e}\,_k$ (rather than the metric or
torsion) are the basic entities with respect to which the action is varied to obtain the gravitational field equations.  

First, note that the torsion tensor in \Eq{eqn:tor-b} is \emph{reducible} under the group of global Lorentz transformation.  
It can be decomposed into three irreducible parts under this Lorentz group \cite{Hayashi-Bregman}\footnote{Note 
that we denote the irreducible parts (\ie  $t_{\lambda\mu\nu},v_{\mu},a_{\mu}$) by 
the same letters as in \cite{HS1}, but that these quantities here are only one half as large as in \cite{HS1}, 
due to different conventions in the definition of torsion.  
Similarly, the quantities $c_1,c_2,c_3$ in \Eq{eqn:HSlag} are four times as large as in \cite{HS1}.},
\ie into parts which do not mix under a global Lorentz transformation:
\bena
t_{\lambda\mu\nu} &=& \frac{1}{2}(S_{\nu\mu\lambda}+S_{\nu\lambda\mu})+\frac{1}{6}(g_{\nu\lambda}v_{\mu}+g_{\nu\mu}v_{\lambda}) -\frac{1}{3}g_{\lambda\mu}v_{\nu}\,,\label{eqn:decomp-torsion-t}\\
v_{\mu}&=& S_{\mu\lambda}^{\phantom{12}\lambda}\,,\\
a_{\mu}&=& \frac{1}{6}\bar{\epsilon}_{\mu\nu\rho\sigma}S^{\sigma\rho\nu}\,,\label{eqn:decomp-torsion-a}
\eena
Here $\bar{\epsilon}_{\mu\nu\rho\sigma}=\sqrt{-g}\epsilon_{\mu\nu\rho\sigma}$ and
$\bar{\epsilon}\,^{\mu\nu\rho\sigma}=\epsilon\,^{\mu\nu\rho\sigma}/\sqrt{-g}$ are 4-tensors, and the Levi-Civita symbol is
normalized such that $\epsilon_{0123}=-1$ and $\epsilon^{0123}=+1$. The tensor $t_{\lambda\mu\nu}$ satisfies
$t_{\lambda\mu\nu}=t_{\mu\lambda\nu}$, $g^{\mu\nu}t_{\lambda\mu\nu}=g^{\lambda\mu}t_{\lambda\mu\nu}=0$, and
$t_{\lambda\mu\nu}+t_{\mu\nu\lambda}+t_{\nu\lambda\mu}=0$. Conversely, the torsion can be written in terms of its 
irreducible parts as 
\beq{eqn:decomp-torsion-combination} S_{\nu\mu\lambda} =
\frac{2}{3}(t_{\lambda\mu\nu}-t_{\lambda\nu\mu})+\frac{1}{3}(g_{\lambda\mu}v_{\nu}-g_{\lambda\nu}v_{\mu})+\bar{\epsilon}_{\lambda\mu\nu\rho}a^{\rho}\,.
\eeq

In order that the field equation be a second-order differential equation in $\underline{e}\,_k$ (so that torsion can propagate),
the Lagrangian is required to be quadratic in the torsion tensor. In addition, the Lagrangian should be invariant under the group of
general coordinate transformations, under the global proper orthochronous Lorentz group, and under parity reversal in the
internal basis ($\underline{e}\,_{0}\to \underline{e}\,_{0},\,\underline{e}\,_{a}\to -\underline{e}\,_{a}$). Hayashi and
Shirafuji suggested the gravitational action of the following form \cite{HS1}:
\beq{eqn:HSlag}
I_G = \int \mathrm{d}^4 x \sqrt{-g}\:[\:\frac{1}{2\kappa}R\left(\{\,\}\right)+c_1\, t^{\lambda\mu\nu}t_{\lambda\mu\nu} +c_2\, v^{\mu}v_{\mu}+c_3\, a^{\mu} a_{\mu}]\,,
\eeq
where $c_1,c_2,c_3$ are three free parameters, 
$R\left(\{\,\}\right)$ is the scalar curvature calculated using the Levi-Civita connection and $\kappa=8\pi G/c^4$.  
The \emph{vacuum} field equations are obtained by varying this action with respect to 
the tetrad
$e^{k}_{\phantom{1}\nu}$ and then multiplying by 
$\eta^{kj} e_{j}^{\phantom{1}\mu}$.  
Note that in Hayashi-Shirafuji theory, the torsion (or equivalently, the connection) is not an independent variable as in some standard torsion theories
\cite{hehl}.  Instead, the torsion is exclusively determined by the tetrad via \Eq{eqn:tor-b}.  The resultant field equation is 
\beq{eqn:HSfieldeq}
\frac{1}{2\kappa}G^{\mu\nu}(\{\,\})+\nabla_{\lambda}F^{\mu\nu\lambda}+v_{\lambda}F^{\mu\nu\lambda}+H^{\mu\nu}-\frac{1}{2}g^{\mu\nu}L_2=0\,.
\eeq
Here the first term denotes the Einstein tensor calculated using the Levi-Civita connection,  
but the field equation receives important non-Riemannian contributions from torsion through the other terms.  The other tensors in \Eq{eqn:HSfieldeq} are defined
as follows:
\bena
F\,^{\mu\nu\lambda} &=& c_1(t^{\mu\nu\lambda}-t^{\mu\lambda\nu})+c_2(g^{\mu\nu}v^{\lambda}-g^{\mu\lambda}v^{\nu}) -\frac{1}{3}c_3\bar{\epsilon}^{\mu\nu\lambda\rho}a_{\rho}\,,\\
H^{\mu\nu} &=& 2S^{\mu\sigma\rho}F_{\rho\sigma}^{\phantom{12}\nu}-S^{\sigma\rho\nu}F^{\mu}_{\phantom{1}\rho\sigma}\,,\label{eqn:eom-HS-Hterm}\\
L_2 &=& c_1\, t^{\lambda\mu\nu}t_{\lambda\mu\nu} +c_2\, v^{\mu}v_{\mu}+c_3\, a^{\mu} a_{\mu}\,.
\eena
Since torsion is the first derivative of the tetrad as per \Eq{eqn:tor-b}, the field equation is a nonlinear second-order differential equation of the tetrad.  
Consequently, the tetrad (hence the torsion) can propagate in the vacuum.

\subsection{Static, spherically and parity symmetric vacuum solution}\label{subsubsec:sph-sym-HS}

Hayashi and Shirafuji derived the exact static, spherically and parity symmetric $R_{\mu\nu\rho\sigma}=0$ 
vacuum solutions for this Lagrangian in \cite{HS1}.  The parallel vector fields take the following form in isotropic 
rectangular coordinates (here Latin letters are spatial indices) \cite{HS1}:
\bena
e_{0}^{\phantom{1}0} &=& \left(1-\frac{m_0}{pr}\right)^{-p/2} \left(1+\frac{m_0}{qr}\right)^{q/2}\,,\nonumber\\
e_{0}^{\phantom{1}i} &=& e_{a}^{\phantom{111}0}=0\,,\nonumber\\
e_{a}^{\phantom{1}i} &=&  \left(1-\frac{m_0}{pr}\right)^{-1+p/2} \left(1+\frac{m_0}{qr}\right)^{-1-q/2}\delta^{i}_a\,,
\eena
where $m_0$ is a parameter with units of mass and will be related to the physical mass of the central gravitating body in \Sec{sec:constrain-torsion}.  
The new parameters $p$ and $q$ are functions of a dimensionless parameter $\epsilon$:
\bena
\epsilon &\equiv & \frac{\kappa (c_1+c_2)}{1+\kappa (c_1+4c_2)}\,,\label{eqn:epsilondef}\\
p & \equiv & \frac{2}{1-5\epsilon}\{ [ (1-\epsilon)(1-4\epsilon)]^{1/2}-2\epsilon\}\,,\\
q &\equiv & \frac{2}{1-5\epsilon}\{ [ (1-\epsilon)(1-4\epsilon)]^{1/2}+2\epsilon\}\,.
\eena
Here $\kappa = 8\pi G$.

The line element in the static, spherically and parity symmetric field takes the exact form \cite{HS1}
\beq{eqn:sphe-line-element}
ds^2 = -\left(1-\frac{m_0}{pr}\right)^{p}\left(1+\frac{m_0}{qr}\right)^{-q}dt^2 +\left(1-\frac{m_0}{pr}\right)^{2-p} \left(1+\frac{m_0}{qr}\right)^{2+q}dx^i dx^i\,.
\eeq
In order to generalize this solution to the axisymmetric case, we transform the parallel 
vector fields into standard spherical coordinates and keep terms to first order in $m_0/r$ (the subscript ``sp'' stands for ``spherical''):

\ben
{e_{(\textrm{sp})}}\,_k^{\phantom{1}\mu} = 
\begin{array}{cl}        &   \qquad \to \mu \\
   \begin{array}{c} \downarrow \\ k \end{array} 
   & \begin{pmatrix} 1+\frac{m_0}{r} & 0 & 0 & 0 \\ 
    0 & \left[1-\frac{m_0}{r}\left(1+\frac{1}{q}-\frac{1}{p}\right) \right]\sin\theta\,\cos\phi & \frac{\cos\theta\,\cos\phi}{r} & -\frac{\csc\theta\,\sin\phi}{r} \\
    0 & \left[1-\frac{m_0}{r}\left(1+\frac{1}{q}-\frac{1}{p}\right) \right]\sin\theta\,\sin\phi & \frac{\cos\theta\,\sin\phi}{r} & \frac{\csc\theta\,\cos\phi}{r} \\
    0 & \left[1-\frac{m_0}{r}\left(1+\frac{1}{q}-\frac{1}{p}\right) \right]\cos\theta & -\frac{\sin\theta}{r} & 0 
    \end{pmatrix}
  \end{array} 
\een

A particularly interesting solution is that for the parameter choice $c_1=-c_2$ so that $\epsilon=0$ and $p=q=2$.  \Eq{eqn:sphe-line-element} shows that the resultant metric coincides with the Schwarzschild metric around an object of mass $m_0$.  The parameter $c_3$ is irrelevant here because of the static, spherically and parity symmetric field.  
When $c_1+c_2$ is small but nonzero, we have $\epsilon \ll 1$ and 
\bena
p &=& 2+\epsilon + \mathcal{O}(\epsilon^2)\,,\\
q &=& 2+9\epsilon + \mathcal{O}(\epsilon^2)\,.
\eena

By using equations (\ref{eqn:bbg2}), (\ref{eqn:conn-b}) and (\ref{eqn:tor-b}), we find that the linearized metric and torsion match our parametrization in \Sec{subsec:spher-symm}.  When $\epsilon \ll 1$, the line element is
\ben
ds^2 = -\left[1-2\frac{m_0}{r}\right]dt^2+\left[1+2(1-2\epsilon)\frac{m_0}{r}\right]dr^2+r^2 d\Omega^2\,,
\een 
and the torsion is
\beqa{eqn:torsion-HS-sp}
S_{tr}^{\phantom{01}t} &=& -\frac{m_0}{2r^2}\,,\\
S_{r\theta}^{\phantom{12}\theta}&=& S_{r\phi}^{\phantom{13}\phi} = -(1-2\epsilon)\frac{m_0}{2r^2}\,,
\eena
both to linear order in $m_0/r$.

\subsection{Solution around Earth}
\label{subsubsec:the axisym-sol2-HS}

We now investigate the field generated by a uniformly rotating spherical body to first order in $\varepsilon_a$.  
It seems reasonable to assume that to first order the metric coincides with the Kerr-like metric, \ie
\beq{eqn:kerr-like-HS}
g_{t\phi}=\mathcal{G}_0(m_0a/r)\sin ^2 \theta\,,
\eeq
around an object of specific angular momentum $a$ in the linear regime $m_0/r\ll 1$ and $a/r \ll 1$.  
Since the Kerr-like metric automatically satisfies 
$G(\{\,\})=0$ in vacuum, the vacuum field equation reduces to
\beq{eqn:reducedfield}
\nabla_{\lambda}F^{\mu\nu\lambda}+v_{\lambda}F^{\mu\nu\lambda}+H^{\mu\nu}-\frac{1}{2}g^{\mu\nu}L_2=0\,.
\eeq
We now employ our parametrization with ``mass'' in \Eq{eqn:w2-axi} replaced by $m_0$, where $m_0$ is the parameter in accordance with \Sec{subsubsec:sph-sym-HS}.  In \Sec{sec:constrain-torsion}, 
we will apply the Kerr solution $\mathcal{G}=-2$ after re-scaling $m_0$ to correspond to the physical mass.  
Imposing the no-curvature condition $R_{\mu\nu\rho\sigma}=0$, we find that this condition and \Eq{eqn:reducedfield} are satisfied to lowest order in $m_0/r$ and $a/r$ if
\beqa{eqn:torsion-HS-kerr-beta}
w_{1}^{(0)} &=& \mathcal{G}_0-\alpha_0,\,,\nonumber\\
w_{2}^{(0)} &=& -2(\mathcal{G}_0-\alpha_0)\,,\nonumber\\
w_{3}^{(0)} &=& w_{4}^{(0)}=\alpha_0\,,\nonumber\\
w_{5}^{(0)} &=& 2\alpha_0.
\eena
Here a superscript $(0)$ indicates the parametrization with $m_0$ in place of $m$. 
$\alpha_0$ is an undetermined constant and should depend on the Lagrangian parameters $c_1$, $c_2$ and $c_3$.  
This parameter has no effect on the precession of a gyroscope or on any of the other observational constraints that we consider, so its value is irrelevant to the present chapter.

The parallel vector fields that give the Kerr metric, the connection and the torsion (including the spherically symmetric part) via equations (\ref{eqn:bbg1})--(\ref{eqn:bbg2}) and (\ref{eqn:conn-b})--(\ref{eqn:tor-b}) 
take the following form to linear order:
\ben
 e_k^{\phantom{1}\mu} = {e_{(\textrm{sp})}}\,_k^{\phantom{1}\mu}
   + \begin{array}{cl}
 &  \qquad \to\mu \\ \begin{array}{c} \downarrow \\ k \end{array} & 
\begin{pmatrix} 0 & \:\:\:0\:\:\: & \:\:\:0\:\:\: & -\alpha_0 \frac{m_0a}{r^3} \\ -(\mathcal{G}_0-\alpha_0)\frac{m_0a\sin\theta\sin\phi}{r^2} & 0 & 0 & 0 \\ (\mathcal{G}_0-\alpha_0)\frac{m_0a\sin\theta\cos\phi}{r^2} & 0 & 0 & 0 \\ 0 &  0 & 0 & 0 \end{pmatrix}\end{array}
\een

\begin{table} 
\noindent 
\begin{center}
\begin{tabular}{|p{1.7cm}|p{0.5cm}|p{3cm}|p{2.2cm}||p{5.8cm}|}
\hline
\multicolumn{2}{|l|}{  } & Hayashi-Shirafuji with $m_0$ & EHS with $m_0$ & Definitions \\
\hline
metric     & $\mathcal{H}^{(0)}$ & -2 & -2 & $g_{tt}=-1-\mathcal{H}^{(0)}m_0/r+\mathcal{O}(m_0/r)^2$ \\
\cline{2-4}
parameters & $\mathcal{F}^{(0)}$ & $2(1-2\epsilon)$ & $2(1-2\tau)$ & $g_{rr}=1+\mathcal{F}^{(0)} m_0/r+\mathcal{O}(m_0/r)^2$\\
\hline
geodetic & $t_{1}^{(0)}$ & $-1$ & $-\sigma$ & anomalous, $S_{tr}^{\phantom{01}t} = t_{1}^{(0)}\,m_0/2r^2$ \\
\cline{2-4}
torsions & $t_{2}^{(0)}$ & $-(1-2\epsilon)$ & $-\sigma(1-2\tau)$ & normal, $S_{r\theta}^{\phantom{12}\theta} = S_{r\phi}^{\phantom{13}\phi} = t_{2}^{(0)}\,m_0/2r^2$\\
\hline
         & $w_{1}^{(0)}$ & $\mathcal{G}_0-\alpha_0$ & $\sigma(\mathcal{G}_0-\alpha_0)$ & $S_{r\phi}^{\phantom{0i}t} = w_{1}^{(0)}\,(m_0a/2r^2)\sin^2\theta$ \\
\cline{2-4}
frame-   & $w_{2}^{(0)}$ & $-2(\mathcal{G}_0-\alpha_0)$ & $-2\sigma(\mathcal{G}_0-\alpha_0)$ & $S_{\theta\phi}^{\phantom{0i}t} = w_{2}^{(0)}\,(m_0a/2r)\sin\theta\cos\theta$ \\
\cline{2-4}
dragging & $w_{3}^{(0)}$ & $\alpha_0$ & $\sigma\alpha_0$ & $S_{t\phi}^{\phantom{0i}r} = w_{3}^{(0)}\,(m_0a/2r^2)\sin^2\theta$ \\
\cline{2-4}
torsions & $w_{4}^{(0)}$ & $\alpha_0$ & $\sigma\alpha_0$ & $S_{t\phi}^{\phantom{0i}\theta} = w_{4}^{(0)}\,(m_0a/2r^3)\sin\theta\cos\theta$\\
\cline{2-4}
         & $w_{5}^{(0)}$ & $2\alpha_0$ & $2\sigma\alpha_0$ & $S_{tr}^{\phantom{0i}\phi} = w_{5}^{(0)} \,m_0a/2r^4$\\
\hline
\end{tabular}
\caption[Summary of metric and torsion parameters for GR, HS and EHS theories (1)]{Summary of metric and torsion parameters for General Relativity, Hayashi-Shirafuji gravity 
and Einstein-Hayashi-Shirafuji (EHS) theories.  The subscript 0 indicates all parameter values are normalized by an arbitrary constant $m_0$ (with the units of mass) that is not necessarily the physical mass of the body generating the gravity.  The parameter $\alpha_0$ in frame-dragging torsions is an undetermined constant and should depend on the Hayashi-Shirafuji Lagrangian parameters $c_1$, $c_2$ and $c_3$.  The parameter $\tau$, defined in \Eq{eqn:epsilondef} and assumed small, is an indicator of how close the emergent metric is to the Schwarzschild metric.  The values in the column of Einstein-Hayashi-Shirafuji interpolation are those in the Hayashi-Shirafuji \emph{times} the interpolation parameter $\sigma$.}
\label{table:HS1}
\end{center}
\end{table}

\section{A toy model: linear interpolation in Riemann-Cartan Space between GR and Hayashi-Shirafuji Lagrangian}
\label{subsec:linearinterpol}

We found that the Hayashi-Shirafuji Lagrangian admits both the Schwarzschild metric and (at least to linear order) the Kerr metric, but
in the Weitzenb\"ock spacetime where there is no Riemann curvature and all spacetime structure is due to torsion.
This is therefore an opposite extreme of GR, which admits these same metrics in Riemann spacetime with all curvature and no torsion.
Both of these solutions can be embedded in Riemann-Cartan spacetime, 
and we will now present a more general two-parameter family of Lagrangians that interpolates between these two extremes, always allowing the
Kerr metric and generally explaining the spacetime distortion with a combination of curvature and torsion.  
After the first version of this chapter was submitted, Flanagan and Rosenthal showed that the Einstein-Hayashi-Shirafuji Lagrangian 
has serious defects \cite{Flanagan:2007dc}, while leaving open the possibility that there may be other viable Lagrangians
in the same class (where spinning objects generate and feel propagating torsion).
This Lagrangian should therefore not be viewed as a viable physical model, but as a pedagogical toy model 
admitting both curvature and torsion, 
giving concrete illustrations of the various effects and constraints that we discuss.

This family of theories, which we will term 
Einstein-Hayashi-Shirafuji (EHS) theories, have 
an action in in Riemann-Cartan space of the form
\beq{eqn:HSlag-interp}
I_G = \int \mathrm{d}^4 x \sqrt{-g}\:[\:\frac{1}{2\kappa}R\left(\{\,\}\right)+\sigma^2\,c_1\, t^{\lambda\mu\nu}t_{\lambda\mu\nu} +\sigma^2\,c_2\, v^{\mu}v_{\mu}+\sigma^2\,c_3\, a^{\mu} a_{\mu}]\,
\eeq
where $\sigma$ is a parameter in the range $0\le\sigma\le 1$. 
Here the tensors $t_{\lambda\mu\nu}$, $v_{\mu}$ and $a_{\mu}$ are the decomposition 
(in accordance with  Eqs.\ref{eqn:decomp-torsion-t}---\ref{eqn:decomp-torsion-a})
of $\sigma^{-1}S_{\nu\mu\lambda}$,
which is independent of $\sigma$ and depends only on $e^i_{\phantom{1}\mu}$ as per \Eq{eqn:tor-b-interp}.  The function $\sigma^2$ associated with the coefficients $c_1$, $c_2$ and $c_3$ in \Eq{eqn:HSlag-interp} may be replaced by any other regular function of $\sigma$ that approaches to zero as $\sigma\to 0$.
The metric in the EHS theories is defined in \Eq{eqn:bbg2}.  
Similar to the Hayashi-Shirafuji theory, the field equation for EHS theories is obtained by varying the action with respect to the tetrad.  The resultant field equation is  identical to that
for the Hayashi-Shirafuji Lagrangian (Eq.~\ref{eqn:HSfieldeq}) 
except for the replacement $c_{1,2,3}\to \sigma^2 c_{1,2,3}$. 
Also, the $S^{\mu\sigma\rho}$ in \Eq{eqn:eom-HS-Hterm} is replaced by $\sigma^{-1}S^{\mu\sigma\rho}$.  
Thus the EHS Lagrangian admits the same solution for 
$e_k^{\phantom{1}\mu}$.  
Since the metric is independent of the parameter $\sigma$, the EHS Lagrangian admits both the spherically symmetric metric in \Eq{eqn:sphe-line-element} and the Kerr-like metric in \Eq{eqn:kerr-like-HS}, at least to the linear order.  
For the spherically symmetric metric, the parameter $\epsilon$ in Hayashi-Shirafuji theory is 
generalized to a new parameter $\tau$ in EHS theories, defined by the replacement $c_{1,2}\to \sigma^2 c_{1,2}$:
\beq{eqn:epsdefEHS}
\tau \equiv  \frac{\kappa \sigma^2(c_1+c_2)}{1+\kappa \sigma^2(c_1+4c_2)}\,.
\eeq

The torsion around Earth is linearly proportional to $\sigma$, given by the parameter $\sigma$ times the solution in \Eq{eqn:torsion-HS-sp} and (\ref{eqn:torsion-HS-kerr-beta}):
\beq{eqn:tor-b-interp}
S_{\mu\nu}^{\phantom{12}\lambda}\equiv\frac{\sigma}{2}e_k^{\phantom{1}\lambda}(\partial_{\mu}e^k_{\phantom{1}\nu}-\partial_{\nu}e^k_{\phantom{1}\mu})\,.  
\eeq
By virtue of \Eq{eqn:fullconn1} 
(the metric compatibility condition), 
it is straightforward to show that the connection is of the form
\ben
\Gamma^{\rho}_{\phantom{\rho}\mu\nu} = (1-\sigma) \left\{ \begin{array}{c} \rho \\ \mu\nu \end{array} \right\} 
+ \sigma\,e_k^{\phantom{1}\rho}\partial_{\mu}e^k_{\phantom{1}\nu}\,.
\een
EHS theory thus interpolates smoothly between metric gravity \eg\ GR $(\sigma=0)$ and the all-torsion Hayashi-Shirafuji theory ($\sigma=1$).  
If $\sigma\ne 1$, it is straightforward to verify that the curvature calculated by the full connection does not vanish.  Therefore, the EHS theories live in
neither Weitzenb\"ock space nor the Riemann space, but in the Riemann-Cartan space that admits both torsion and curvature.   

It is interesting to note that since the Lagrangian parameters $c_1$ and $c_2$ are independent of the torsion parameter $\sigma$, the effective
parameter $\tau$ is not necessarily equal to zero when $\sigma=0$ (\ie, $\sigma^2 c_1$ or $\sigma^2 c_2$ can be still finite).
In this case ($\sigma=0$ and yet $\tau \ne 0$), obviously this EHS theory is an extension to GR without adding torsion.  In
addition to the extra terms in the Lagrangian of \Eq{eqn:HSlag-interp}, the extension is subtle in the symmetry of the
Lagrangian.   In the tetrad formalism of GR, \emph{local} Lorentz transformations are symmetries in the internal space of
tetrads.  Here in this $\sigma=0, \tau\ne 0$ EHS theory, the allowed internal symmetry is \emph{global} Lorentz
transformations as in the Weitzenb\"ock spacetime, because $t_{\lambda\mu\nu}$, $v_\mu$ and $a_\mu$ contain the partial
derivatives of tetrads (see Eq.~\ref{eqn:tor-b-interp}).  So the $\sigma=0$ and $\tau \ne 0$ EHS theory is a tetrad theory in
Riemann spacetime with less gauge freedom.  

Since GR is so far consistent with all known observations, it is interesting to explore (as we will below) what observational upper limits can be placed on both $\sigma$ and $\tau$.  

\begin{table} 
\noindent 
\begin{center}
\begin{tabular}{|p{1.7cm}|p{0.3cm}|p{0.5cm}|p{3cm}|p{2.3cm}||p{4.9cm}|}
\hline
\multicolumn{2}{|l|}{  } & GR &  EHS with autoparallels & EHS with extremals & Definitions \\
\hline
mass & $m$ & $m=m_0$ & $m=(1-\sigma)m_0$ & $m=m_0$ & set by Newtonian limit \\
\hline
metric     & $\mathcal{H}$ & -2 & $-2/(1-\sigma)$ & $-2$ & $g_{tt}=-1-\mathcal{H}m/r+\mathcal{O}(m/r)^2$ \\
\cline{2-5}
parameters & $\mathcal{F}$ & 2 & $2(1-2\tau)/(1-\sigma)$ & $2(1-2\tau)$ & $g_{rr}=1+\mathcal{F}m/r+\mathcal{O}(m/r)^2$\\
\cline{2-5}
           & $\mathcal{G}$ & -2 & -2 & -2 & $g_{t\phi}=\mathcal{G}(ma/r)\sin^2\theta$\\
\hline
geodetic & $t_1$ & 0 & $-\sigma/(1-\sigma)$ & $-\sigma$ & anomalous, $S_{tr}^{\phantom{01}t} = t_1\,m/2r^2$ \\
\cline{2-5}
torsions & $t_2$ & 0 & $-\sigma(1-2\tau)/(1-\sigma)$ & $-\sigma(1-2\tau)$ & normal, $S_{r\theta}^{\phantom{12}\theta} = S_{r\phi}^{\phantom{13}\phi} = t_2\,m/2r^2$\\
\hline
         & $w_1$ & 0 & $\sigma(\mathcal{G}-\alpha)$ & $\sigma(\mathcal{G}-\alpha)$ & $S_{r\phi}^{\phantom{0i}t} = w_1\,(ma/2r^2)\sin^2\theta$ \\
\cline{2-5}
frame-   & $w_2$ & 0 & $-2\sigma(\mathcal{G}-\alpha)$ & $-2\sigma(\mathcal{G}-\alpha)$ & $S_{\theta\phi}^{\phantom{0i}t} = w_2\,(ma/2r)\sin\theta\cos\theta$ \\
\cline{2-5}
dragging & $w_3$ & 0 & $\sigma\alpha$ & $\sigma\alpha$ & $S_{t\phi}^{\phantom{0i}r} = w_3\,(ma/2r^2)\sin^2\theta$ \\
\cline{2-5}
torsions & $w_4$ & 0 & $\sigma\alpha$ & $\sigma\alpha$ & $S_{t\phi}^{\phantom{0i}\theta} = w_4\,(ma/2r^3)\sin\theta\cos\theta$\\
\cline{2-5}
         & $w_5$ & 0 & $2\sigma\alpha$ & $2\sigma\alpha$ & $S_{tr}^{\phantom{0i}\phi} = w_5 \,ma/2r^4$\\
\hline
\hline
effective  & $\mu_1$ & 0 & $-\sigma$ & $-\sigma$ & $\mu_1 = (w_1-w_2-w_3+2w_4+w_5)/(-3\mathcal{G})$ \\ \cline{2-5}
torsions   & $\mu_2$ & 0 & $-\sigma$ & $-\sigma$ & $\mu_2 = (w_1-w_3+w_5)/(-\mathcal{G})$ \\ 
\hline
\hline
bias & $b_t$ & 1 & $1-4\tau/3$ & $1-\sigma-4\tau/3$ & $b_t=(1+\mathcal{F}+2t_2+|\eta|t_1)/3$\\ \cline{2-5}
     & $b_\mu$ & 1 & $(-\mathcal{G}/2)(1-\sigma)$ & $(-\mathcal{G}/2)(1-\sigma)$ & $b_\mu = (-\mathcal{G}/2)(1+3\mu_1-2\mu_2)$ \\ \hline
\end{tabular}
\caption[Summary of metric and torsion parameters for EHS theories (2)]{Summary of metric and torsion parameters for Einstein-Hayashi-Shirafuji (EHS) theories of interpolation parameter $\sigma$ in autoparallel scheme and in extremal scheme.  All parameter values are normalized by the physical mass $m$ of the body generating the gravity.  The parameter $\mathcal{G}$ and $\alpha$ are related to $\mathcal{G}_0$ and $\alpha_0$ in Table \ref{table:HS1} by $\mathcal{G}=\mathcal{G}_0/(1-\sigma)$ and $\alpha=\alpha_0/(1-\sigma)$ in autoparallel scheme, $\mathcal{G}=\mathcal{G}_0$ and $\alpha=\alpha_0$ in extremal scheme.  The value for $\mathcal{G}$ is set to $-2$ by the Kerr metric in linear regime $m/r \ll 1$ and $a/r \ll 1$. 
}
\label{table:EHS1}
\end{center}
\end{table}

\begin{table}
\noindent 
\footnotesize{
\begin{center}
\begin{tabular}{|p{2.5cm}|p{2.5cm}|p{4cm}|p{4.5cm}|}
\hline
            & General Relativity & EHS with autoparallels & EHS with extremals \\ \hline
Averaged Geodetic Precession & $(3m/2r_0)\vec{\omega}_O\times\vec{S}_0$ & $(1-4\tau/3)(3m/2r_0)\vec{\omega}_O\times\vec{S}_0$ & $(1-\sigma-4\tau/3)(3m/2r_0)\vec{\omega}_O\times\vec{S}_0$ \\ \hline
Averaged Frame-dragging & $(I/2r_0^3)\vec{\omega}_E\times\vec{S}_0$ & $(-\mathcal{G}/2)(1-\sigma)(I/2r_0^3)\vec{\omega}_E\times\vec{S}_0$ & $(-\mathcal{G}/2)(1-\sigma)(I/2r_0^3)\vec{\omega}_E\times\vec{S}_0$ \\ \hline
Second moment $\vec{a}_2$ & $(3I\omega_E/4r_0^3)\hat{z}\times\vec{S}_0$ & $(-3\mathcal{G}I\omega_E/8r_0^3)(1-\sigma)\hat{z}\times\vec{S}_0$ & $(-3\mathcal{G}I\omega_E/8r_0^3)(1-\sigma)\hat{z}\times\vec{S}_0 -\eta \sigma m\omega_O (S_0^x\hat{z}+S_0^z\hat{x})/4r_0$ \\ \hline
Second moment $\vec{b}_2$ & $(3I\omega_E/4r_0^3)\hat{x}\times\vec{S}_0$ & $(-3\mathcal{G}I\omega_E/8r_0^3)(1-\sigma)\hat{x}\times\vec{S}_0$ & $(-3\mathcal{G}I\omega_E/8r_0^3)(1-\sigma)\hat{x}\times\vec{S}_0 -\eta \sigma m\omega_O (S_0^x\hat{x}-S_0^z\hat{z})/4r_0$ \\ \hline
\end{tabular}
\caption[Summary of the predicted precession rate for GR and EHS theories]{Summary of the predicted Fourier moments of the precession rate for General Relativity and the Einstein-Hayashi-Shirafuji (EHS) theories in autoparallel scheme and in extremal scheme.  $\eta=+1$ for extremal scheme using $S^{\mu\nu}$, and $-1$ for extremal scheme using $S^{\mu}$.  Other multiple moments vanish.  Here $m$ and $I\omega_E$ are the Earth's mass and rotational angular momentum, respectively.}
\label{table:EHS2}
\end{center}
}
\end{table}

\begin{table}
\noindent 
\footnotesize{
\begin{center}
\begin{tabular}{|p{2.9cm}||p{3.5cm}|p{2.3cm}|p{2cm}|p{2.3cm}|}
\hline
Effects  & Torsion Biases  &   EHS in autoparallel scheme &  EHS in extremal scheme & PPN biases\\ \hline
Shapiro time delay & $\Delta t/\Delta t ^{(GR)}=(\mathcal{F}-\mathcal{H})/4$ &  $1+\sigma-\tau$ & $1-\tau$ & $(1+\gamma)/2$ \\\hline
Deflection of light & $\delta/\delta^{(GR)}=(\mathcal{F}-\mathcal{H})/4$ & $1+\sigma-\tau$ & $1-\tau$ &  $(1+\gamma)/2$ \\\hline
Gravitational redshift & $(\Delta\nu/\nu)/(\Delta\nu/\nu)^{(GR)}=-\mathcal{H}/2$ & $1+\sigma$ & 1 & $1+\alpha$ \\\hline
Geodetic Precession & $\Omega_G/\Omega_G^{(GR)}=b_t$ & $1-\frac{4}{3}\tau$ & $1-\sigma-\frac{4}{3}\tau$ & $(1+2\gamma)/3$ \\ \hline
Frame-dragging & $\Omega_F/\Omega_F^{(GR)}=b_\mu$ & $1-\sigma$ & $1-\sigma$ & $(1+\gamma+\alpha_1/4)/2$ \\ \hline
\end{tabular}
\caption[Biases of EHS theories relative to GR predictions]{Summary of solar system experiments (1): the biases relative to GR predictions for the Einstein-Hayashi-Shirafuji (EHS) theories.  Both parameters $\tau$ and $\sigma$ are assumed small.  The biases in the PPN formalism are also listed for comparison, taken from \cite{Will:2005va}.}
\label{tab:solar-constraint1}
\end{center}
}
\end{table}

\begin{table}
\noindent 
\footnotesize{
\begin{center}
\begin{tabular}{|p{1.7cm}||p{3.1cm}|p{3.1cm}|p{3.1cm}|p{2cm}|}
\hline
Effects  &  PPN  &   EHS in autoparallel scheme &  EHS in extremal scheme & Remarks \\\hline
Shapiro time delay & $\gamma-1 = (2.1\pm 2.3)\times 10^{-5}$ & $\sigma-\tau = (1.1\pm 1.2)\times 10^{-5}$ &  $\tau = (-1.1\pm 1.2)\times 10^{-5}$ & Cassini tracking \cite{Bertotti:2003rm} \\\hline
Deflection of light & $\gamma-1 = (-1.7\pm 4.5)\times 10^{-4}$ & $\sigma-\tau = (-0.8\pm 2.3)\times 10^{-4}$ &  $\tau = (0.8\pm 2.3)\times 10^{-4}$ & VLBI \cite{Shapiro2004} \\\hline
Gravitational redshift & $|\alpha|<2\times 10^{-4}$ & $|\sigma|<2\times 10^{-4}$ & no constraints & Vessot-Levine rocket \cite{vessot1980} \\\hline
Geodetic Precession & $\left|\gamma-1\right|< 1.1\times 10^{-4}$ & $|\tau|<5.7\times 10^{-5}$ &  $|\sigma+4\tau/3|<7.6\times 10^{-5}$ & Gravity Probe B \\\hline
Frame-dragging & $\left|\gamma-1+\frac{1}{4}\alpha_1\right|<0.024$ & $|\sigma|<0.012$ & $|\sigma|<0.012$ & Gravity Probe B \\\hline
\end{tabular}
\caption[Constraints of the PPN and EHS parameters with solar system experiments]{Summary of solar system experiments (2): constraints on the PPN and EHS parameters.  The constraints on PPN parameters are taken from Table 4 and Page 12 of \cite{Will:2005va}.   
The full results of Gravity Probe B are yet to be released, so whether the frame dragging 
will agree with the GR prediction is not currently known.  The last two rows show the limits that would correspond to a GPB result consistent with GR,
assuming an angle accuracy of 0.5 milli-arcseconds.}
\label{tab:solar-constraint2}
\end{center}
}
\end{table}

\section{Example: testing Einstein Hayashi-Shirafuji theories with GPB and other solar system experiments}
\label{sec:constrain-torsion}

Above we calculated the observable effects that arbitrary Earth-induced torsion, if present, would have on GPB.
As a foil against which to test GR, let us now investigate the observable effects that would result for the explicit 
Einstein-Hayashi-Shirafuji class of torsion theories that we studied in \Sec{subsubsec:the axisym-sol2-HS} and \ref{subsec:linearinterpol}.

There are four parameters $c_1$, $c_2$, $c_3$ and $\sigma$ that define an EHS theory via the action in \Eq{eqn:HSlag-interp}.
We will test EHS theories with GPB and other solar system experiments.  For all these weak field experiments, only two EHS parameters ---  $\tau$ (defined in \Eq{eqn:epsdefEHS}) and $\sigma$, both assumed small --- that are functions of the said four are relevant and to be constrained below.  

The predicted EHS metric and torsion parameters, studied in \Sec{subsec:linearinterpol}, are listed in Table \ref{table:HS1}.  Below, we will test both the autoparallel and extremal calculation schemes.  In each scheme, the physical mass $m$ will be determined by the Newtonian limit.  All metric and torsion parameters are converted in accordance with $m$ and listed in Table \ref{table:EHS1}.  Then the parameter space ($\tau$, $\sigma$) will be constrained by solar system experiments.

\begin{figure}
\centering
\begin{displaymath}
\begin{array}{cc}
\includegraphics[width=0.5\textwidth]{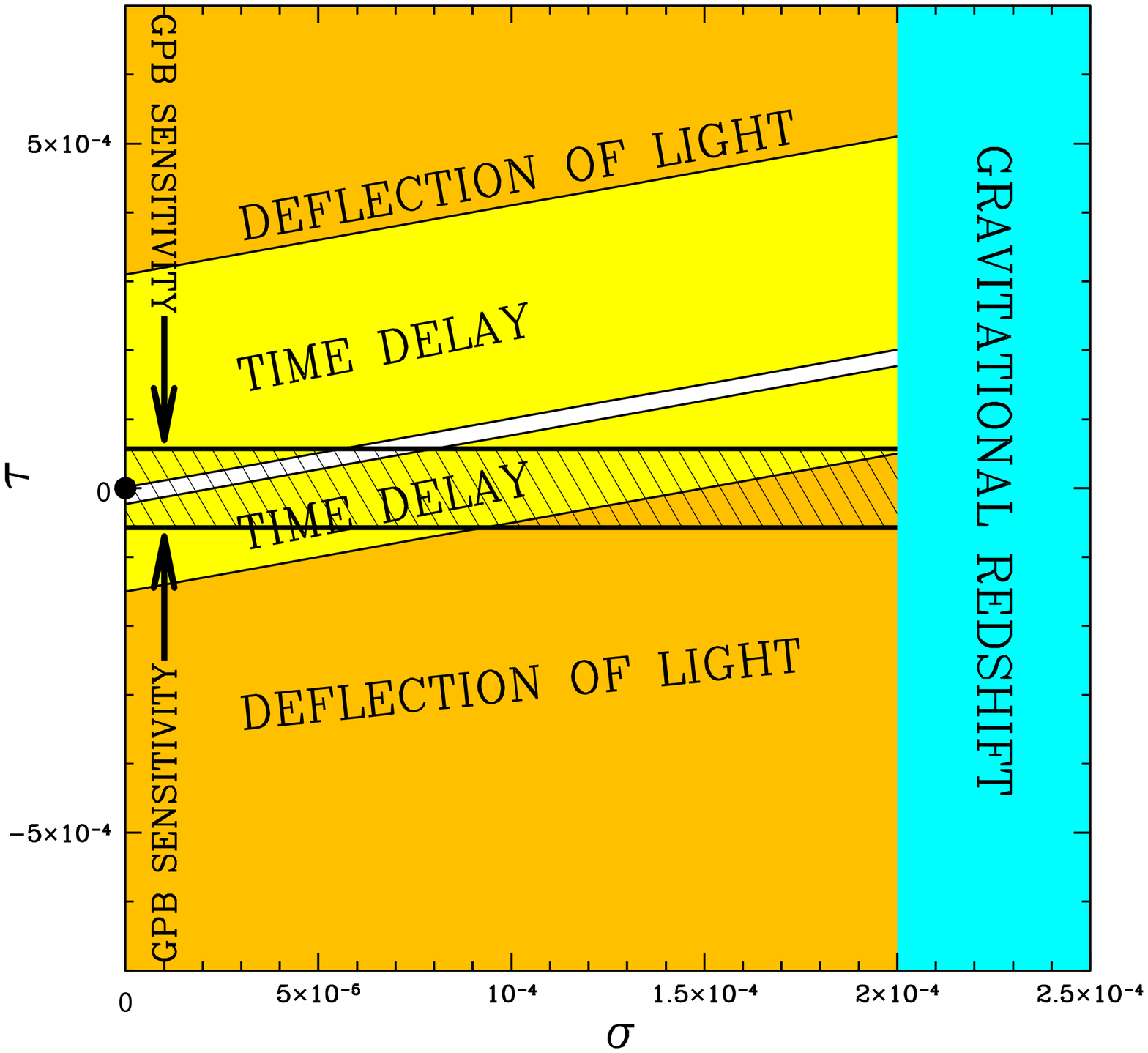} & 
\includegraphics[width=0.5\textwidth]{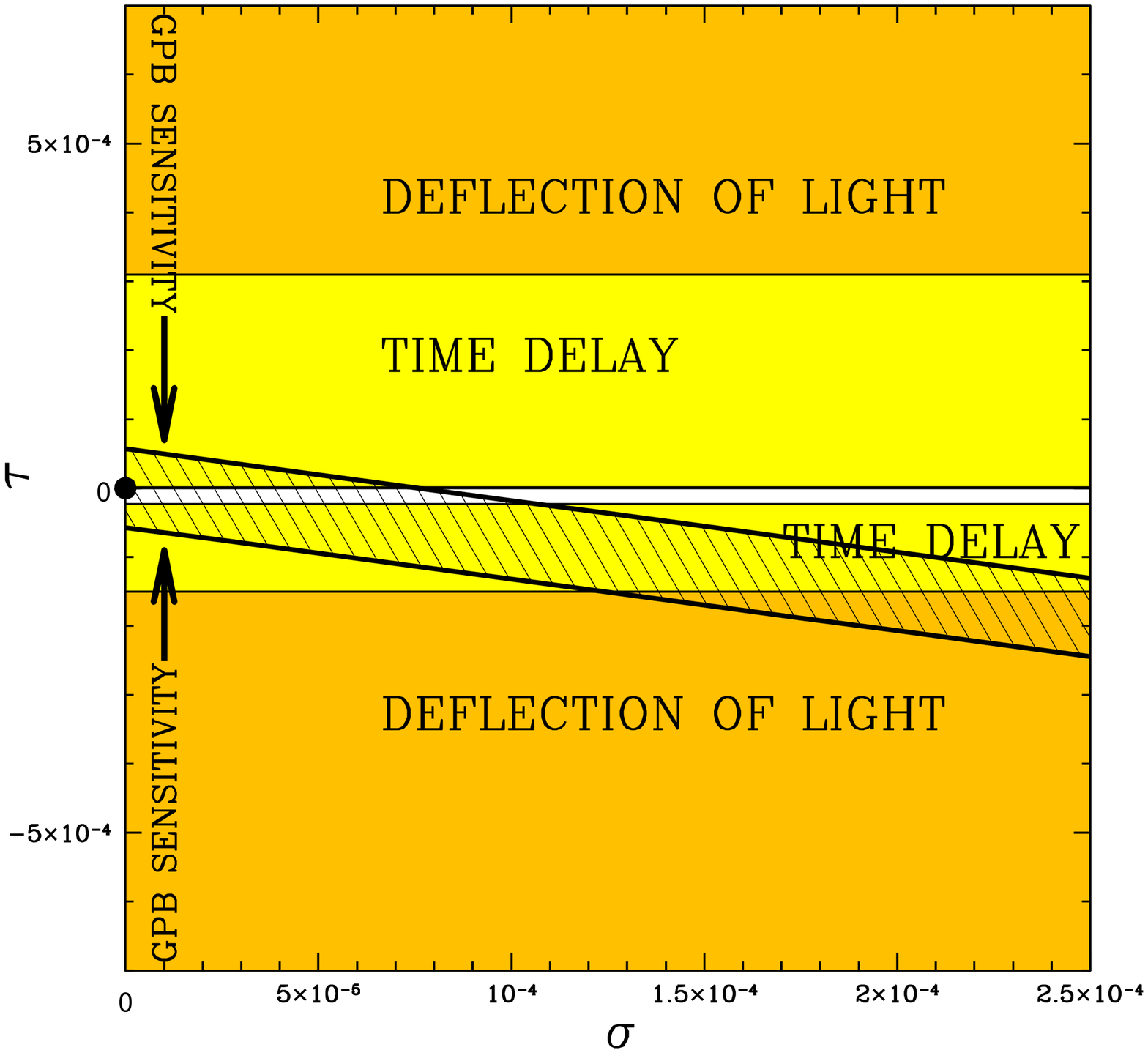} 
\end{array}
\end{displaymath}
\caption[Constraints on the EHS parameters from solar system tests]{\footnotesize%
Constraints on the EHS parameters $(\sigma,\tau)$ from solar system tests in the autoparallel scheme (left panel) and in the extremal scheme (right panel).  
General Relativity corresponds
to the black dot ($\sigma=\tau=0$). The shaded regions in the parameter space have already been ruled out by Mercury's perihelion
shift (red/dark grey), the deflection of light (orange/grey), Shapiro time delay (yellow/light grey) 
and gravitational redshift (cyan/light grey).  
If the geodetic
precession and frame-dragging measured by Gravity Probe B are consistent with GR to the target accuracy of 0.5 milli-arcseconds,
this will rule out everything outside the hatched region, implying that: (1) in the autoparallel scheme, 
$0\le \sigma<8.0\times 10^{-5}$ and $-2.3\times 10^{-5} <\tau< 5.7\times 10^{-5}$; (2) in the extremal scheme, $0\le \sigma<1.1\times 10^{-4}$ and $-2.3\times 10^{-5} <\tau< 0.1\times 10^{-5}$.
Preliminary result of Gravity Probe B have only confirmed the geodetic precession to about 1\%, thus (1) in the autoparallel scheme, bringing no further constraints beyond those from gravitational redshift, 
and (2) in the extremal scheme, implying that $\sigma < 0.01$.
}
\label{fig:all-constraints}
\end{figure}

\subsection{Autoparallel scheme}

Hayashi-Shirafuji maximal torsion theory is inconsistent with the autoparallel scheme, since $t_1-\mathcal{H}/2=0$ (see $t_1$ and
$\mathcal{H}$ in Table \ref{table:HS1}).  By \Eq{eqn:auto}, this means that $\mathrm{d}\vec{v}/\mathrm{d}t=0 + \mathcal{O}(m/r)^2$. 
The violation of Newton's law rules out the application of the autoparallel scheme to the Hayashi-Shirafuji theory.  

However, the Einstein-Hayashi-Shirafuji theories can be consistent with this scheme.  Using Table \ref{table:HS1}, the Newtonian limit can be written as 
\beq{eqn:EHSnewtonian} 
\frac{d\vec{v}}{dt}=-(1-\sigma)\frac{m_0}{r^2}\hat{e}_r\,,
\eeq
so the physical mass of the central gravitating body is 
\ben
m=(1-\sigma)m_0\,.
\een

Table \ref{table:EHS1} lists values of metric and torsion parameters in accordance with the physical mass $m$.  Using these parameters, the precession rates of gyroscopes in GPB orbit can be calculated via equations (\ref{eqn:moment2}),(\ref{eqn:btdef}),(\ref{eqn:bmudef}) and (\ref{eqn:n2moments}).  The results are listed in Table \ref{table:EHS2}.  For GPB, the average precession rates are the only experimentally accessible observables in practice.  GPB will measure the precession of gyroscopes with respect to two different axes: the orbital angular velocity $\vec{\omega}_O$ (geodetic precession) and the Earth's rotational angular velocity $\vec{\omega}_E$ (frame-dragging).  As indicated in Table \ref{table:EHS2}, the geodetic precession and frame-dragging rates are 
\bena
\Omega_G &=& (1-\frac{4}{3}\tau)\Omega_G^{(GR)}\,,\\
\Omega_F &=& \left(-\frac{\mathcal{G}}{2}\right)(1-\sigma)\Omega_F^{(GR)}\,,
\eena
where $\Omega_G^{(GR)}$ and $\Omega_F^{(GR)}$ are the geodetic precession and frame-dragging rate predicted by General Relativity, respectively.  

The existing solar system experiments, including Shapiro time delay, deflection of light, gravitational redshift, advance of Mercury's perihelion, can put constraints on the parameters $\tau$ and $\sigma$.  The derivation of these constraints essentially follow any standard textbook of General Relativity \cite{MTW-Weinberg-Islam:book} except for more general allowance of parameter values, so we leave the technical detail in Appendix \ref{appendix:solar-tests} with the results summarized in Table \ref{tab:solar-constraint1}.

It is customary that  biases of GR predictions are expressed in terms of PPN parameters on which observational constraints can be placed with solar system experiments.  In EHS theories, these biases are expressed in terms of the parameters $\tau$ and $\sigma$.  Thus we can place constraints on the EHS parameters $\tau$ and $\sigma$ by setting up the correspondence between PPN and EHS parameters via the bias expression.  Table \ref{tab:solar-constraint1} lists the biases in the PPN formalism for this purpose, and Table \ref{tab:solar-constraint2} lists the observational constraints on the EHS parameters $\tau$ and $\sigma$ with the existing solar system tests.

If GPB would see no evidence of the torsion induced precession effects, the ($\tau$,$\sigma$) parameter space can be further constrained.  Together with other solar system experiments, the observational constraints are listed in Table \ref{tab:solar-constraint2} and shown in \Fig{fig:all-constraints} (left panel).

\begin{figure}
\centering
\includegraphics[width=0.5\textwidth]{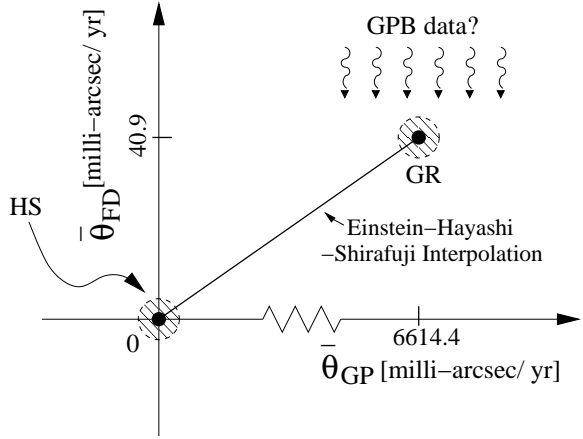}
\caption[Predictions for the \emph{average} precession rate by GR]{\footnotesize%
Predictions for the \emph{average} precession rate by General Relativity, Hayashi-Shirafuji (HS) gravity and Einstein-Hayashi-Shirafuji
theories (for the case of $\tau=0$ and the Kerr solution $\mathcal{G}=-2$) that interpolate between these two extremes, in the extremal
scheme. $\bar{\theta}_{\textrm{GP}}$ is the geodetic precession rate around the orbital angular velocity vector $\vec{\omega}_O$ and
$\bar{\theta}_{\textrm{FD}}$ is the angular frame-dragging rate around Earth's rotation axis $\vec{\omega}_E$.  The shaded areas of
about 0.5 milli-arcseconds per year in radius are the approximate forecast GPB measurement uncertainties.  The two calculation
schemes using $S^\mu$ and $S^{\mu\nu}$ with extremals for the Hayashi-Shirafuji Lagrangian (labeled ``HS'' in the figure) agree on
the predicted average rates.
The unpublished preliminary results of Gravity Probe B have confirmed the geodetic precession to better 
than 1\%, so this already rules out the Hayashi-Shirafuji Lagrangian and most EHS theories in the extremal scheme in the sense 
that $\sigma < 0.01$. 
}
\label{gpb4}
\end{figure}

\subsection{Extremal scheme}

Einstein-Hayashi-Shirafuji theories predict $\mathcal{H}=-2$ regardless of $\tau$ and $\sigma$.  By the Newtonian limit, therefore, the physical mass of the central gravitating body is just the mass parameter $m_0$, \ie $m=m_0$.  So the parameter values do not need rescaling and are re-listed in Table \ref{table:EHS1}.  By these parameters the precession rates can be calculated and listed in Table \ref{table:EHS2}.  As indicated in Table \ref{table:EHS2}, the geodetic precession and frame-dragging rates are
\bena
\Omega_G &=& (1-\sigma-\frac{4}{3}\tau)\Omega_G^{(GR)}\,,\\
\Omega_F &=& \left(-\frac{\mathcal{G}}{2}\right)(1-\sigma)\Omega_F^{(GR)}\,.
\eena

It is worth noting again that the extremal scheme is not a fully consistent framework from the theoretical point of view.  However, it serves perfectly to show the role of EHS theories as the bridge between no-torsion GR and Hayashi-Shirafuji maximal torsion theory.  \Fig{gpb4} illustrates this connectivity in terms of the
 predictions of GR, Hayashi-Shirafuji theory and the intermediate $0<\sigma<1$ EHS theories, taking $\tau=0$ and Kerr solution $\mathcal{G}=-2$, on the \emph{average} precession rate (the $\vec{a}_0$ in Table \ref{table:EHS2}).  The EHS theories are seen to connect the  extreme GR and HS cases with a straight line.  If the data released by GPB ends up falling within the shaded area corresponding to the GR prediction, the Hayashi-Shirafuji Lagrangian will thus have been ruled out with very high significance, and the GPB torsion constraints can be quantified as sharp upper limits on the $\sigma$-parameter.

More generally, Gravity Probe B will improve the constraints on the ($\tau$,$\sigma$) parameter space by its precise measurements of precession rates, in addition to the constraints put by existing solar system experiments.  These constraints are listed in Table \ref{tab:solar-constraint2} and shown in \Fig{fig:all-constraints} (right panel).  As before, the technical details are given in Appendix \ref{appendix:solar-tests}.

\subsection{Preliminary constraints from GPB's unpublished results}

In April 2007, Gravity Probe B team announced that, while they continued mining the data for the ultimately optimal accuracy, the geodetic precession was found to agree with GR at the 1\% level.  The frame-dragging yet awaits to be confirmed.  Albeit preliminary, these unpublished results, together with solar system tests, already place the first constraint on some torsion parameters to the 1\% level.  More quantitatively, $|t_2+\frac{|\eta|}{2}t_1|\lesssim 0.01$ in the model-independent framework, while  $w_1+w_2-w_3-2w_4+w_5$ is not constrained.  In the context of EHS theories, the constraint is scheme dependent.  In the autoparallel scheme, GPB's preliminary results place no better constraints than those from gravitational redshift ($\sim 10^{-4}$).  In the extremal scheme, however, the preliminary results give the constraint $\sigma < 0.01$.  The bottom line is that GPB has constrained torsion parameters to the 1\% level now and will probably reach the $10^{-4}$ level in the future.

\section{Conclusions and Outlook}
\label{sec:conclusion}

The PPN formalism has demonstrated that a great way to test GR is to embed it in a broader parametrized class of theories, 
and to constrain the corresponding parameters observationally. In this spirit, we have explored observational constraints on 
generalizations of GR including torsion. 

Using symmetry arguments, we showed that to lowest order, the torsion field around a uniformly rotating spherical mass such as Earth is
determined by merely seven dimensionless parameters.
We worked out the predictions for these seven torsion parameters for a two-parameter
Einstein-Hayashi-Shirafuji generalization of GR which includes as special cases both standard no-torsion GR ($\sigma=0$) 
and the no-curvature, all torsion ($\sigma=1$) Weitzenb\"ock spacetime.
We showed that classical solar system tests rule out a large class of these models, and that Gravity Probe B (GPB) can further improve
the constraints. 
GPB is useful here because this class of theories suggested that, depending on the Lagrangian, rotating objects can generate torsion observable with gyroscopes.
In other words, despite some claims in the literature to the contrary, the question of whether there is observable 
torsion in the solar system is one which ultimately can and should be tested experimentally.

Our results motivate further theoretical and experimental work.
On the theoretical side, it would be interesting to address in more detail the question of which Lagrangians make torsion couple to rotating objects.
A well-defined path forward would be to generalize the matched asymptotic expansion method of \cite{D'Eath1975a,D'Eath1975b} to match
two generalized EHS Kerr-like Solutions in the weak-field limit to obtain the laws of motion for two well-separated rotating objects, and determine which of the
three non-equivalent prescriptions above, if any, is correct. It would also be interesting to look for generalizations of the EHS Lagrangian that populate a large fraction 
of the seven torsion degrees of freedom that symmetry allows.
Finally, additional observational constraints can be investigated involving, \eg, binary pulsars, gravitational 
waves and cosmology.

On the experimental side, Gravity Probe B has now successfully completed its data taking phase. We have shown that the GPB data constitute a potential gold mine of information
about torsion, but that its utility for constraining torsion theories will depend crucially on how the data are analyzed and released. 
At a minimum, the average geodetic and frame dragging precessions can be compared with the predictions shown in \fig{gpb4}. However, if it is technically feasible for 
the GPB team to extract and publish also different linear combinations of the instantaneous precessions corresponding to the second moments of these precessions, 
this would enable looking for further novel effects that GR predicts should be absent.
In summary, although the nominal goal of GPB is to look for an effect that virtually everybody expects will be present (frame dragging), it also has the potential 
to either discover torsion or to build further confidence in GR by placing stringent limits on torsion theories.

We wish to thank Francis Everitt, Thomas Faulkner, Friedrich Hehl, Scott Hughes,  Erotokritos Katsavounidis, Barry Muhlfelder, Tom Murphy,  
Robyn Sanderson, Alexander Silbergleit, Molly Swanson, Takamitsu Tanaka and Martin White 
for helpful discussions and comments.  


\begin{subappendices}

\section{Parametrization of torsion in the static, spherically and parity symmetric case}
\label{appendix:spher-symm}

In this appendix, we derive a parametrization of the most general static, spherically and parity
symmetric torsion  in isotropic 
rectangular and spherical
coordinates.  The symmetry conditions are described in Section
\ref{subsubsec:general-setup1} with the quantity $\mathcal{O}$ now being the torsion tensor $S_{\mu\nu}^{\phantom{12}\rho}$.  
Note that torsion (the antisymmetric part of the connection) is a tensor under general coordinate transformations 
even though the full connection is not.

First note that time translation invariance is equivalent to the independence of torsion on time. Then consider
time reversal, under which a component of torsion flips its sign once for every temporal index.
Invariance under time reversal therefore requires that non-zero torsion components have either zero or two
temporal indices. Together with the fact that torsion is antisymmetric in its first two indices, this restricts the non-zero
components of torsion to be $S_{0i}^{\phantom{0i}0}$ and $S_{jk}^{\phantom{0i}i}$ ($i=1,2,3$).

Now consider the symmetry under (proper or improper) rotation (see \Eq{xform:spherical}). 
The orthogonality of the matrix $\R$
enables one to write
\beq{eqn:R-xform-appen}
\frac{\partial x'^i}{\partial x^j}=R^{ij}\,,\qquad \frac{\partial x^i}{\partial x'^j}=R^{ji}\,,\qquad \frac{\partial t'}{\partial t}=\frac{\partial t}{\partial t'}=1\,.
\eeq
Thus formal functional invariance means that
\beq{eqn1:form-inv2}
\begin{array}{lclcl}
S_{\phantom{1}0i}^{'\phantom{0i}0}(x') &=& R^{ij}S_{0j}^{\phantom{0i}0}(x) &=& S_{0i}^{\phantom{0i}0}(x'),\\
S_{\phantom{1}jk}^{'\phantom{jk}i}(x') &=& R^{jm}R^{kn}R^{il}S_{mn}^{\phantom{mn}l}(x) &=& S_{jk}^{\phantom{0i}i}(x').
\end{array}
\een
\Eq{eqn1:form-inv2} requires that the torsion should be built up of  $x^i$ and quantities invariant under O(3), such as scalar
functions of radius and Kronecker $\delta$-functions, since $\delta
'_{i'j'}=R^{i'i}R^{j'j}\delta_{ij}=R^{i'i}R^{j'i}=R^{i'i}(R^{-1})^{ij'}=\delta_{i'j'}\,.$  
Note that we are interested in the
parity symmetric case, whereas the Levi-Civita symbol $\epsilon_{ijk}$ is a three-dimensional \emph{pseudo}-tensor 
under orthogonal transformations, where ``pseudo'' means that $\epsilon_{ijk}$ is a tensor under SO(3) but not under O(3), 
since $\epsilon '_{i'j'k'}=R^{i'i}R^{j'j}R^{k'k}\epsilon_{ijk}=\det R \times \epsilon_{i'j'k'}\,.$  
Therefore, $\epsilon_{ijk}$ is prohibited
from entering into the construction of the torsion tensor by \Eq{eqn1:form-inv2}.

Thus using arbitrary combinations of scalar functions of radius, $x^i$ and Kronecker $\delta$-functions, 
the most general torsion tensor that can be constructed takes the form 
\bena
S_{0i}^{\phantom{0i}0} &=& t_1\frac{m}{2r^3}x^i\, ,\\
S_{jk}^{\phantom{0i}i} &=& t_2\frac{m}{2r^3}(x^j \delta_{ki}-x^k \delta_{ji})\,,\label{eqn:t2-appen}
\eena
where the combinations $t_1 m$ and $t_2 m$ are arbitrary functions of radius.
Note that in \Eq{eqn:t2-appen}, terms proportional to $x^i x^j x^k$ or $x^i\delta_{jk}$ are forbidden by the antisymmetry of the torsion.  
We will simply treat the functions $t_1(r)$ and $t_2(r)$ as constants, since GPB orbits at a fixed radius.

Transforming this result to spherical coordinates, we obtain
\begin{eqnarray*}
S_{tr}^{\phantom{01}t} &=&  S_{ti}^{\phantom{01}t} \frac{\partial x^i}{\partial r}= t_1\frac{m}{2r^2},\\
S_{r\theta}^{\phantom{12}\theta} &=& S_{jk}^{\phantom{12}i}\frac{\partial x^j}{\partial r}\frac{\partial x^k}{\partial \theta}\frac{\partial \theta}{\partial x^i}= t_2\frac{m}{2r^2}\,,\\
S_{r\phi}^{\phantom{13}\phi} &=& S_{jk}^{\phantom{12}i}\frac{\partial x^j}{\partial r}\frac{\partial x^k}{\partial \phi}\frac{\partial \phi}{\partial x^i} = t_2\frac{m}{2r^2}\, .
\end{eqnarray*}
All other components not related by the antisymmetry vanish. In the above equations, the second equalities follow from the chain rule and the facts that $\partial
x^i/\partial r=\hat{x}^i=\hat{e}_r^i$, $\partial x^i/\partial \theta=r\hat{e}_\theta^i$, and $\partial x^i/\partial
\phi=r\sin\theta\hat{e}_\phi^i$, where $\hat{e}_r^i$, $\hat{e}_\theta^i$ and $\hat{e}_\phi^i$ are the $i$th-components of the
unit vectors in spherical coordinates.  To first order in the mass $m$ of the central object, we need not distinguish between isotropic and standard spherical coordinates.
\bigskip

\section{Parametrization in stationary and spherically axisymmetric case}
\label{appendix:axisymm}

Above we considered the 0$th$ order contribution to the metric and torsion corresponding to the static, spherically and parity symmetric case
of a non-rotating spherical source.
In this appendix, we derive a parametrization of the most general 1$st$ order correction (denoted by a superscript $(1)$) to this metric and torsion that could be caused by rotation
of the source, \ie corresponding to the stationary and spherically axisymmetric case.
The symmetry conditions are described in Section \ref{subsubsec:prob-setup-axisym}, with the quantity
$\mathcal{O}$ replaced by the metric $g_{\mu\nu}^{(1)}$ for Appendix \ref{appen-subsec:metric} and by the torsion $S_{\mu\nu}^{(1)\rho}$
for Appendix \ref{appen-subsec:torsion}.

\subsection{The Metric}\label{appen-subsec:metric}

The invariance under time translation makes the metric time independent.  Under time reversal $\J\to -\J$, and a component of the metric flips its sign once for every temporal index.  Thus, the formal functional invariance equation for time reversal reads
\beq{eqn:metric-form-inv-1}
\pm g^{(1)}_{\mu\nu}(x|\J) = g^{(1)}_{\mu\nu}(x|-\J)\,.
\eeq

The plus sign in \Eq{eqn:metric-form-inv-1} is for components with even numbers of temporal indices, and minus sign for those with odd numbers.  Since only terms linear in $J/r^2=\varepsilon_m \varepsilon_a$ are concerned, the minus sign in the argument $-\J$ can be taken out as an overall factor, implying that the non-vanishing components of metric can have only one temporal index.  Thus the only nonzero first-order correction to $g_{\mu\nu}$ in rectangular coordinates is $g_{ti}^{(1)}$ (i=1,2,3). 

Now consider the transformation property under (proper or improper) rotation. 
By the orthogonality of the matrix $\R$, the vector $\x$
transforms as $\x\to\x'\equiv \R\x$ (\Eq{eqn:R-xform-appen}).  Since $\J$ is invariant under parity, formally the transformation of $\J$ writes as
\ben
\J \to \J'=(det\R) \times \R\J \,.
\een
The formal functional invariance for rotation reads
\ben
g^{(1)'}_{ti}(x'|\J) = R^{ij}g^{(1)}_{tj}(x|\J) = g^{(1)}_{ti}(x'|\J')\,.
\een
That $\J$ is a pseudo-vector under improper rotation requires that the Levi-Civita symbol $\epsilon_{ijk}$, also a pseudo-tensor, appear once and only once (because $\J$ appears only once) in the metric so as to compensate the $det\R$ factor incurred by transformation of $\J$.  Other possible elements for construction of the metric include scalar functions of radius, $x^i$, $J^i$, $\delta_{ij}$.  Having known the elements, the only possible construction is therefore
\ben
g_{ti}^{(1)}=\frac{\mathcal{G}}{r^2}\epsilon_{ijk}J^j \hat{x}^k\,,
\een
where $\hat{x}^i=x^i/r$ is the unit vector of position vector and $\mathcal{G}$ is dimensionless.  Assuming that there is no new scale other than the angular momentum $\J$ built into the 1$st$ order of torsion theory, \ie no new dimensional parameter with units of length, $\mathcal{G}(r)$ must be a constant by dimensional analysis, since the factor $J^i$ has explicitly appeared. 

In spherical polar coordinates where the $z$-axis is parallel to $\J$, this first-order correction to the metric takes the form
\ben
g_{t\phi}^{(1)} = \mathcal{G}\frac{ma}{r}\sin^2\theta\,,
\een
where $ma=J$ is the magnitude of $\J$.  All other components vanish.

\bigskip

\subsection{The Torsion}\label{appen-subsec:torsion}

We follow the same methodology as for our parametrization of the metric above. 
Given the time-independence, the property that $\J$ reverses under time-reversal requires that the non-vanishing components of torsion 
have only one temporal index, so they are $S_{\phantom{1}ij}^{(1)\,t},\; S^{(1)}_{tij}$ (i,j=1,2,3) in rectangular coordinates. 
(The antisymmetry of torsion over its first two indices excludes the possibility of three temporal indices.)  
Under (proper or improper) rotation, the formal functional invariance equation reads
\[ \begin{array}{lclcl}
S_{\phantom{(1)}ij}^{(1)\,'\phantom{0}t}(x'|\J) &=& R^{ik}R^{jl}S_{\phantom{1}kl}^{(1)\,t}(x|\J) &=&  S_{\phantom{1}ij}^{(1)\,t}(x'|\J')\,,\\
S^{(1)\,'}_{tij}(x'|\J) &=& R^{ik}R^{jl}S^{(1)}_{tkl}(x|\J) &=&  S^{(1)}_{tij}(x'|\J')\,.\\
\end{array} \]
Again, in building the torsion, one should use the Levi-Civita symbol $\epsilon_{ijk}$ once and only once to cancel the $det\R$ factor from the transformation of $\J$.
  The most general construction using scalar function of radius, $x^i$, $\delta_{ij}$, $J^i$ (also appearing once and only once) and $\epsilon_{ijk}$ is
\begin{eqnarray*}
S_{\phantom{0}ij}^{(1)\,t} &=& \frac{f_1}{2r^3}\epsilon_{ijk}J^k+\frac{f_2}{2r^3}J^k \hat{x}^l (\epsilon_{ikl}\hat{x}^j-\epsilon_{jkl}\hat{x}^i)\,,\\
S_{tij}^{(1)} &=& \frac{f_3}{2r^3}\epsilon_{ijk}J^k+\frac{f_4}{2r^3}J^k \hat{x}^l \epsilon_{ikl}\hat{x}^j+\frac{f_5}{2r^3}J^k \hat{x}^l \epsilon_{jkl}\hat{x}^i\,.
\end{eqnarray*}
By the same dimensional argument as in Appendix (\ref{appen-subsec:metric}), $f_1,\ldots,f_5$ must be dimensionless constants.

Transforming the above equations to spherical coordinates where the $z$-axis is parallel to $\J$, we obtain to first order
\begin{eqnarray*}
S_{\phantom{0}r\phi}^{(1)\phantom{0}t} &=& S_{ij}^{\phantom{0i}t}\frac{\partial x^i}{\partial r}\frac{\partial x^j}{\partial \phi} =  w_1\frac{ma}{2r^2}\sin^2\theta\,,\\
S_{\phantom{0}\theta\phi}^{(1)\phantom{0}t} &=& S_{ij}^{\phantom{0i}t}\frac{\partial x^i}{\partial \theta}\frac{\partial x^j}{\partial \phi} =  w_2\frac{ma}{2r}\sin\theta\cos\theta\,,\\
S_{\phantom{0}t\phi}^{(1)\phantom{0}r} &=& g^{rr} S_{tij}\frac{\partial x^i}{\partial \phi}\frac{\partial x^j}{\partial r} = w_3\frac{ma}{2r^2}\sin^2\theta\,, \\
S_{\phantom{0}t\phi}^{(1)\phantom{0}\theta} &=& g^{\theta\theta}S_{tij}\frac{\partial x^i}{\partial \phi}\frac{\partial x^j}{\partial \theta} = w_4\frac{ma}{2r^3}\sin\theta\cos\theta\,,\\
S_{\phantom{0}tr}^{(1)\phantom{0}\phi} &=& g^{\phi\phi} S_{tij}\frac{\partial x^i}{\partial r}\frac{\partial x^j}{\partial \phi} =w_5 \frac{ma}{2r^4}\,,\\
S_{\phantom{0}t\theta}^{(1)\phantom{0}\phi} &=& g^{\phi\phi}S_{tij}\frac{\partial x^i}{\partial \theta}\frac{\partial x^j}{\partial \phi} = -w_4\frac{ma}{2r^3}\cot\theta\,.\label{eqn:w6-axi}
\end{eqnarray*}
All other components vanish.  The constants are related by $w_1=f_1-f_2$, $w_2=f_1$, $w_3=f_4-f_3$, $w_4=-f_3$, $w_5=f_5+f_3$.

\section{Constraining torsion with solar system experiments}
\label{appendix:solar-tests}

\subsection{Shapiro time delay}
\label{appendix:shapiro}

For the electromagnetic field, if torsion is coupled to the vector potential $A_\mu$ by the ``natural'' extension, \ie, $\partial_\mu A_\nu \to \nabla_\mu A_\nu$ using the full connection, the Maxwell Lagrangian $-\frac{1}{4}F_{\mu\nu}F^{\mu\nu}$ will contain a quadratic term in $A_\mu$ that makes the photon massive and breaks gauge invariance in the conventional form.  Since the photon mass has been experimentally constrained to be $\lesssim 10^{-17}$ eV, we assume that $A_\mu$ does not couple to torsion.  Instead, we assume that the Maxwell field Lagrangian in the curved spacetime with torsion follows the extension $\partial_\mu A_\nu \to \nabla_\mu^{\{\}} A_\nu$ using the Levi-Civita connection.  Since the Levi-Civita connection depends on the metric and its derivatives only, light rays follow extremal curves (metric geodesics).  

In general, assume the line element in the field around a (physical) mass $m$ is 
\beq{eqn:metricDefappendix}
\mathrm{d}s^2 =  -\left[1+\mathcal{H}\frac{m}{r}\right]\mathrm{d}t^2 + \left[1+\mathcal{F}\frac{m}{r}\right]\mathrm{d}r^2+r^2 d\Omega^2\,.
\een
The effect of the rotation of the mass can be ignored when the rotation is slow.  

\begin{figure}
\centering
\includegraphics[width=0.5\textwidth]{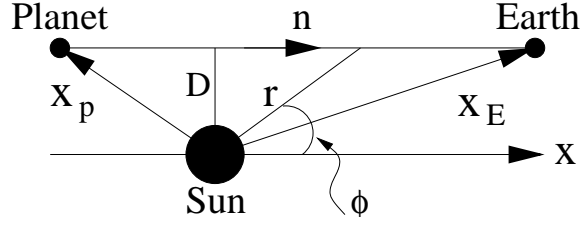}
\caption[Shapiro time delay]{\footnotesize%
Geometry of the Shapiro time delay measurement.}
\label{fig:deflection}
\end{figure}

Light deflection angle is tiny for the solar system tests we consider, so a ray can be well approximated by a straight line.  Let us use coordinates where the Sun (of mass $m$), the Earth and a planet reflecting the light ray are all in the $x$-$y$ plane ($\theta=\pi/2$) and the $x$-axis points along the ray from the planet to Earth (see \Fig{fig:deflection}).  Let $D$ be the minimal distance of the ray from the Sun.  Then $r\sin\phi=D$, or $rd\phi=-\tan\phi dr$.  Since $\mathrm{d}s^2=0$ for a light ray,
\bena
dt^2 &=& (1+\mathcal{H}\frac{m}{r})^{-1} (1+\mathcal{F}\frac{m}{r}+\tan^2\phi)dr^2\nonumber\\
     &\approx & \frac{r^2 dr^2}{r^2-D^2}[1+(\mathcal{F}-\mathcal{H})\frac{m}{r}-\mathcal{F}\frac{mD^2}{r^3}]\,, \nonumber\\
dt &\approx & \frac{r |dr|}{\sqrt{r^2-D^2}}[1+(\mathcal{F}-\mathcal{H})\frac{m}{2r}-\mathcal{F}\frac{mD^2}{2r^3}]\,.
\eena
The round-trip travel time for an electromagnetic signal bouncing between Earth and the Planet in the gravitational field of the Sun is
\bena
T &=& 2\left[ \int_{r=D_p}^{r=D} dt + \int_{r=D}^{r=D_E} dt \right] \,,\nonumber\\
  &\approx & 2{[\sqrt{D_p^2-D^2}+\sqrt{D_E^2-D^2}]}+(\mathcal{F}-\mathcal{H})m  \times\ln\left[\frac{(\sqrt{D_p^2-D^2}+D_p)(\sqrt{D_E^2-D^2}+D_E)}{D^2}\right]\nonumber\\
  & & -\mathcal{F}m\left(\frac{\sqrt{D_p^2-D^2}}{D_p}+\frac{\sqrt{D_E^2-D^2}}{D_E}\right)\,.\label{eqn:T-appC}
\eena
If $D\ll D_E$ and $D\ll D_p$, the third term in \Eq{eqn:T-appC} is negligible compared to the second one.  The excess travel time $\Delta t$ of a round-trip light ray is 
\ben
\Delta t \equiv  T - 2{[\sqrt{D_p^2-D^2}+\sqrt{D_E^2-D^2}]}\,
         \approx  \left(\frac{\mathcal{F}-\mathcal{H}}{4}\right)\Delta t^{(GR)}\,,
\een
where $\Delta t^{(GR)}$ is the excess time predicted by GR
\ben
\Delta t^{(GR)} = 4m\ln\left[\frac{(D_E+\vec{x}_E\cdot\hat{n})(D_p-\vec{x}_p\cdot\hat{n})}{D^2}\right]\,.
\een
Here $\vec{x}_E$ ($\vec{x}_p$) is the vector from the Sun to the Earth (the planet), and $\hat{n}$ is the unit vector from the planet to Earth (see \Fig{fig:deflection}).

For EHS theories in the autoparallel scheme, $(\mathcal{F}-\mathcal{H})/4=(1-\epsilon)/(1-\sigma)\approx 1+\sigma-\epsilon$, if $\sigma \ll 1$.  For EHS theories in the extremal scheme, $(\mathcal{F}-\mathcal{H})/4=1-\epsilon$.

\subsection{Deflection of light}
\label{subsec:deflec}

As discussed in Appendix \ref{appendix:shapiro}, we assume that a light ray follows an extremal curve (metric geodesic), taking the form 
\ben
\frac{D^{\{\}}u^{\mu}}{D\tau}=\frac{d^2 x^\mu}{d\tau^2}+\left\{\begin{array}{c} \mu \\ \nu\rho \end{array}\right\} \frac{dx^\nu}{d\tau}\frac{dx^\rho}{d\tau}=0\,.
\een
Here $D^{\{\}}/D\tau$ denotes the covariant differentiation using the Levi-Civita connection.

The $\mu=t$ component of the metric geodesic is 
\[ \frac{d^2 t}{d\tau^2}-\mathcal{H}\frac{m}{r^2}\frac{dt}{d\tau}\frac{dr}{d\tau}=0,\]
or, to order $\mathcal{O}(m/r)$, where $m$ is the mass of the Sun deflecting the light,
\[ \frac{d}{d\tau}\left[ (1+\mathcal{H}\frac{m}{r})\frac{dt}{d\tau}\right] = 0\,.\]
Integrating this gives a conserved quantity,
\beq{eqn:k-light}
k\equiv (1+\mathcal{H}\frac{m}{r})\frac{dt}{d\tau} = {\rm const}\,.
\een
The $\mu=\theta$ component of the metric geodesic admits the planar solution $\theta=\pi/2$.  The $\mu=\phi$ component of the metric geodesic, when $\theta=\pi/2$, is
\[ \frac{d^2 \phi}{d\tau^2}+\frac{2}{r}\frac{dr}{d\tau}\frac{d\phi}{d\tau}=0,\]
whose first integral gives another conserved quantity,
\beq{eqn:h-light}
h\equiv r^2 \frac{d\phi}{d\tau}= {\rm const}\,.
\een

For light rays in the equatorial plane $\theta=\pi/2$,
\beq{eqn:light-pi/2}
\frac{ds^2}{d\tau^2}=-\left[1+\mathcal{H}\frac{m}{r}\right]\left(\frac{dt}{d\tau}\right)^2 + \left[1+\mathcal{F}\frac{m}{r}\right]\left(\frac{dr}{d\tau}\right)^2+r^2 \left(\frac{d\phi}{d\tau}\right)^2=0\,.
\een
Note that the $\mu=r$ component of the metric geodesic is not independent of \Eq{eqn:light-pi/2}.  Rewriting $dt/d\tau$ and $d\phi/d\tau$ in terms of $k$ and $h$ via \Eq{eqn:k-light} and \Eq{eqn:h-light}, respectively, and using the fact that $dr/d\tau = (dr/d\phi)(d\phi/d\tau)$, one finds 
\ben
\frac{d^2 u}{d\phi^2}+u=\frac{3}{2}\mathcal{F}mu^2-\frac{k^2}{h^2}\frac{\mathcal{F}+\mathcal{H}}{2}m\,,
\een
where $u\equiv 1/r$.  The solution to order $\mathcal{O}(m)$ is 
\ben
u=\frac{\sin\phi}{D}+\frac{\mathcal{F}m}{2D^2}(1+C\cos\phi+\cos^2\phi)-\frac{k^2}{h^2}\frac{\mathcal{F}+\mathcal{H}}{2}m\,,
\een
where $D$ is the minimal distance of the ray to the Sun.  The $x$-axis is set up to be along the incoming direction of the ray.  $C$ is an arbitrary constant that can be determined at $\phi=\pi$ (incoming infinity).  As long as deflection angle $\delta\ll 1$,
\ben
\delta \simeq \frac{2\mathcal{F}m}{D}-\frac{k^2}{h^2}m(\mathcal{F}+\mathcal{H})D\,.
\een
Using 
\ben
\frac{h}{k}=r^2\frac{d\phi}{dt}(1-\mathcal{H}\frac{m}{r})\approx r^2\frac{d\phi}{dt}=D
\een
is the angular momentum of the light ray relative to the Sun, we finally obtain
\ben
\delta \simeq \frac{\mathcal{F}-\mathcal{H}}{4}\delta^{(GR)}\,,
\een
where $\delta^{(GR)}=4m/D$ is the deflection angle predicted by GR to lowest order.

\subsection{Gravitational Redshift}

As discussed above, we assume that the orbits of light rays are metric geodesics even when there is non-zero torsion.
Non-relativistically, the metric geodesic equation
for a test particle is
\ben
\frac{d\vec{v}}{dt}=-\frac{(-\mathcal{H})}{2}\frac{m}{r^2}\hat{e}_r\,.
\een
Effectively this introduces the gravitational potential $U$, defined by $d\vec{v}/dt=\vec{F}\equiv -\nabla U$,
\ben
U=-\frac{(-\mathcal{H})}{2}\frac{m}{r}\,.
\een
Thus the gravitational redshift of photons is 
\ben
\frac{\Delta\nu}{\nu}=\frac{(-\mathcal{H})}{2}\left(\frac{\Delta\nu}{\nu}\right)^{(GR)}\,,
\een
where $(\Delta\nu/\nu)^{(GR)}$ is the redshift predicted by GR
\ben
\left(\frac{\Delta\nu}{\nu}\right)^{(GR)}=-\frac{m}{c^2}(\frac{1}{r_1}-\frac{1}{r_2})\,.
\een
For EHS theories in the autoparallel scheme, $-\mathcal{H}/2=1/(1-\sigma)\approx 1+\sigma$ for $\sigma \ll 1$.  For EHS theories in extremal scheme,
$-\mathcal{H}/2=1$ exactly.

\subsection{Advance of Mercury's Perihelion in autoparallel scheme}
\label{subsec:advance-perih-auto}

In the autoparallel scheme, a massive test particle (\eg\ a planet in the field of the Sun) follows an autoparallel curve (\ie an affine geodesic). 
We now derive the advance of the perihelion when torsion is present. The autoparallel equation reads
\beq{eqn:auto-perih-appendix}
\frac{Du^{\mu}}{D\tau}=\frac{d^2 x^\mu}{d\tau^2}+\Gamma^\mu_{\phantom{1}\nu\rho} \frac{dx^\nu}{d\tau}\frac{dx^\rho}{d\tau}=0\,,
\een
where $D/D\tau$ is the covariant differentiation by the full connection.

The $\mu=t$ component of \Eq{eqn:auto-perih-appendix} reads
\[ \frac{d^2 t}{d\tau^2}+(t_1-\mathcal{H})\frac{m}{r^2}\frac{dt}{d\tau}\frac{dr}{d\tau}=0,\]
or, to order $\mathcal{O}(m/r)$, where $m$ is the mass of the central gravitating body (\eg\ the Sun),
\[ \frac{d}{d\tau}\left[ \left(1+(\mathcal{H}-t_1)\frac{m}{r}\right)\frac{dt}{d\tau}\right] = 0\,.\]
The integral gives a conserved quantity $k$,
\beq{eqn:k-light}
k\equiv \left(1+(\mathcal{H}-t_1)\frac{m}{r}\right)\frac{dt}{d\tau} = {\rm const}\,.
\een
The $\mu=\theta$ component of \Eq{eqn:auto-perih-appendix} admits the planar solution $\theta=\pi/2$.  
The $\mu=\phi$ component of \Eq{eqn:auto-perih-appendix}, when $\theta=\pi/2$, is
\[ \frac{d^2 \phi}{d\tau^2}+(\frac{2}{r}-t_2\frac{m}{r^2})\frac{dr}{d\tau}\frac{d\phi}{d\tau}=0,\]
whose first integral gives another conserved quantity $h$,
\beq{eqn:h-light}
h\equiv r^2 \frac{d\phi}{d\tau}(1+t_2\frac{m}{r})= {\rm const}\,.
\een
The path parameter $\tau$ can be chosen so that 
\beq{eqn:tau-time-appdenx}
ds^2/d\tau^2=g_{\mu\nu}\frac{dx^{\mu}}{d\tau}\frac{dx^{\nu}}{d\tau}=-1\,.
\een
\Eq{eqn:tau-time-appdenx} is consistent with the autoparallel scheme since $\nabla_\rho g_{\mu\nu}=0$ and $Du^\mu/D\tau=0$.  Note that the $\mu=r$ component of \Eq{eqn:auto-perih-appendix} is not independent of \Eq{eqn:tau-time-appdenx}.  For a test particle in the equatorial plane $\theta=\pi/2$, \Eq{eqn:tau-time-appdenx} reads
\beq{eqn:perih-pi/2}
-\left[1+\mathcal{H}\frac{m}{r}\right](\frac{dt}{d\tau})^2 + \left[1+\mathcal{F}\frac{m}{r}\right](\frac{dr}{d\tau})^2+r^2 (\frac{d\phi}{d\tau})^2=-1\,.
\een

Reusing the trick employed in Appendix \ref{subsec:deflec}, we find
\beq{eqn:eomdefl1}
\frac{d^2 u}{d\phi^2}+u = \frac{3}{2}\mathcal{F}mu^2+\frac{m}{2h^2}\left[k^2(-\mathcal{H}-\mathcal{F} +2t_1+2t_2)+\mathcal{F}-2t_2\right],
\een
to order $\mathcal{O}(mu)$, where $u\equiv 1/r$.  
Note that to lowest order $k\approx 1+\mathcal{O}(m,(velocity)^2)$, so the second term on the right hand side of \Eq{eqn:eomdefl1} becomes
$(t_1-\mathcal{H}/2)m/h^2$. Since $m$ \emph{is} the physical mass of the central gravitating body, the autoparallel scheme requires
$t_1-\mathcal{H}/2=1$.  Now \Eq{eqn:eomdefl1} becomes 
\ben
\frac{d^2 u}{d\phi^2}+u=\frac{m}{h^2}+\frac{3}{2}\mathcal{F}mu^2\,.
\een
Solve the equation perturbatively in the order of $\varepsilon\equiv (m/h)^2$, \ie use the ansatz $u=u_0+\varepsilon u_1$.  One finds
\bena
u_0 &=& \frac{m}{h^2}(1+e\cos\phi)\,\label{eqn:u0-appen}\\
u_1 &=& \frac{3\mathcal{F}m}{2h^2}\left[1+e\phi\sin\phi+\frac{e^2}{2}(1-\frac{1}{3}\cos 2\phi)\right]\label{eqn:u1-appen}
\eena
\Eq{eqn:u0-appen} gives the classical elliptical orbit with eccentricity $e$ and 
the semi-latus rectum $p\equiv a(1-e^2)=h^2/m$.  
The $\phi\sin\phi$ term in \Eq{eqn:u1-appen}  contributes to the advance of the perihelion, 
while the constant and $\cos 2\phi$ terms do not. Therefore
\beq{eqn:u-appen}
u\approx \frac{m}{h^2}\left\{1+e\cos\left[\phi\left(1-\frac{3\mathcal{F}m^2}{2h^2}\right)\right]\right\}\,.
\een
In \Eq{eqn:u-appen}, we used the fact that the second term inside the cosine 
is $\ll 1$.  The advance of the perihelion is now given by
\ben
\Delta\theta = \frac{2\pi}{1- \frac{3\mathcal{F}m^2}{2h^2}}-2\pi 
             = \frac{\mathcal{F}}{2}\Delta\theta^{(GR)}\,,\label{eqn:adv-perih-final}
\een
where $\Delta\theta^{(GR)}=6\pi m^2/h^2 = 6\pi m/p$ is the perihelion advance predicted by GR.

\subsection{Advance of Mercury's Perihelion in extremal scheme}
\label{subsec:advance-perih-extreme}

The extremal scheme assumes that a test particle (\eg, a planet) follows the metric geodesic even though the torsion is present.  Following the same
algebra as in Appendix \ref{subsec:advance-perih-auto}, and noting that $\mathcal{H}=-2$ for the extremal scheme, we finds that the advance of the
perihelion in the extremal scheme has the same bias factor $\mathcal{F}/2$, \ie, \Eq{eqn:adv-perih-final} holds.  

\section{Constraining torsion parameters with the upper bounds on the photon mass } 
\label{appendix:photon-mass}

In this Appendix, we derive the contraints on torsion parameters that result from assuming that 
the ``natural'' extension $\partial_\mu \to \nabla_\mu$ (using the full connection) in the electromagnetic Lagrangian.
This breaks gauge invariance, and the photon generically gains a mass via an 
additional term of the form $-\frac{1}{2}m_\gamma^2 g^{\mu\nu}A_\mu A_\nu$ in the Lagrangian as we will now show. 
The assumption gives 
\ben
F_{\mu\nu}\equiv \nabla_\mu A_\nu - \nabla_\nu A_\mu=f_{\mu\nu}-2S_{\mu\nu}^{\phantom{12}\lambda}A_\lambda\,,
\een
where $f_{\mu\nu}\equiv\partial_\mu A_\nu - \partial_\nu A_\mu$.  
The Maxwell Lagrangian therefore becomes
\bena
\mathcal{L}_{\rm EM} &=& -\frac{1}{4}g^{\mu\alpha}g^{\nu\beta}F_{\mu\nu}F_{\alpha\beta}\,,\nonumber\\
&=& 
-\frac{1}{4}g^{\mu\alpha}g^{\nu\beta}f_{\mu\nu}f_{\alpha\beta}
-K^{\mu\nu}A_\mu A_\nu + S^{\mu\nu\lambda}A_\lambda f_{\mu\nu}\,,\nonumber\\
\eena
where $K^{\mu\nu}\equiv S_{\alpha\beta}^{\phantom{12}\mu}S^{\alpha\beta\nu}$\,.  The Euler-Lagrange equation for the action $S=\int d^4 x \sqrt{-g} \mathcal{L}_{\rm EM}$ yields the following equation of motion for $A_\mu$:
\beq{eqn:cov-A-eom}
\nabla^{\Gamma}_\mu f^{\mu\nu} = 2S_{\mu\lambda}^{\phantom{12}\mu}f^{\lambda\nu} 
+ 2 K^{\lambda\nu}A_\lambda + 2\nabla^{\{\}}_\mu (S^{\mu\nu\lambda}A_\lambda)\,.
\een
Here $\nabla^{\Gamma}_\mu$ and $\nabla^{\{\}}_\mu$ are the covariant derivative w.r.t. the full connection and the Levi-Civita connection, respectively.
Both the 2nd and 3rd terms on the right hand side of \Eq{eqn:cov-A-eom} contain the coupling to $A_\mu$.  To clarify this, 
\Eq{eqn:cov-A-eom} can be rewritten non-covariantly as
\beq{eqn:noncov-A-eom}
\nabla^{\Gamma}_\mu f^{\mu\nu} = 2S_{\mu\lambda}^{\phantom{12}\mu}f^{\lambda\nu} 
+ 2 A_\lambda \left[ K^{\lambda\nu} + \partial_\mu S^{\mu\nu\lambda} 
 + \left\{ \begin{array}{c} \alpha \\ \alpha\mu \end{array} \right\} S^{\mu\nu\lambda}\right] + 2 S^{\mu\nu\lambda}\partial_\mu A_\lambda\,,
\een
in which the 2nd term on the right hand side is the direct coupling of $A_\mu$.  

The matrix $K^{\mu\nu}$ is symmetric. If it is also positive definitive up to the metric signature $(-+++)$,
the first term in the square bracket may be identified as the photon mass term. 
In the field of a non-rotating mass, using the parametrization 
(Eqs.~\ref{eqn:t1} and \ref{eqn:t2}), it can be shown that
\bena
K^{00} &=& -\frac{t_1^2 m^2}{2r^4}\,,\\
K^{0i} &=& 0\,,\\
K^{ij} &=& \frac{t_2^2 m^2}{2r^4} \left(\delta_{ij}-\frac{x^i x^j}{r^2}\right)\,.
\eena
The matrix $K$ has the eigenvalues $-\frac{t_1^2 m^2}{2r^4}$, 0 (with eigenvector $\hat{r}$) and 
$\frac{t_2^2 m^2}{2r^4}$ (with 2 degenerate eigenvectors).  Since the metric signature is $(-+++)$, 
all photon masses are positive or zero, The nonzero ones are of order
\ben
m_\gamma \simeq t \frac{m}{r^2} \,,
\een
or (with units reinserted)
\ben
m_\gamma c^2 \simeq t \frac{\hbar G}{c} \frac{m}{r^2} \,.
\een
Here $t=\max(|t_1|,|t_2|)$ and $r$ is the distance of the experiment location to the center of the 
mass $m$ that generates the torsion. For a ground-based experiment here on Earth, this gives 
\ben
t \simeq 4.64\times 10^{22} m_\gamma c^2 /(1\>{\rm eV})\,.
\een
The upper bound on the photon mass from ground-based experiments is $m_\gamma c^2 < 10^{-17}\,{\rm eV}$ \cite{Lakes:1998mi},
so the constraint that this bound places on the dimensionless torsion parameters is quite weak.

Experimentalists can also search for an anomalous electromagnetic force and translate the null results 
into photon mass bounds. To leading order, the anomalous force is 
$2\partial_\mu S^{\mu\nu\lambda} A_\lambda$, since the K-term is proportional to $S^2$, 
while the 2nd term in the square bracket of \Eq{eqn:noncov-A-eom} is proportional to $S$.  
In a field of a non-rotating mass $m$, 
\bena
(\partial_\mu S^{\mu\nu\lambda})^{00} &=& (\partial_\mu S^{\mu\nu\lambda})^{0i} = (\partial_\mu S^{\mu\nu\lambda})^{i0} =0\,,\\
(\partial_\mu S^{\mu\nu\lambda})^{ij} &=& t_2\frac{m}{2r^3}\left(-\delta_{ij}+3\frac{x^i x^j}{r^2}\right)\,,
\eena
which has eigenvalues $\frac{t_2 m}{2r^3}\times (0,-1,-1,2)$.  This 
cannot be identified as a mass term since there must be a negative ``mass squared'' regardless of 
the sign of $t_2$.  However, the anomalous electromagnetic force expressed as a photon mass 
can be estimated as
\ben
m_\gamma c^2 \simeq \sqrt{|t_2| \hbar^2 G \frac{m}{r^3}}\,,
\een
or 
\ben
\sqrt{|t_2|} \simeq 1.23\times 10^{18} m_\gamma c^2 /{\rm eV}\,.
\een
This implies that current  ground-based experimental upper bounds on the photon mass are too weak
(giving merely $|t|\simlt 10^2$, as compared to $|t|=1$ from Hayashi-Shirafuji gravity)
to place constraints on torsion parameters that are competitive with those from GPB.

\end{subappendices}
\chapter{Constraining $f(R)$ Gravity as a Scalar Tensor Theory}

\section{Introduction}


Although the emerging cosmological standard model fits measurements 
spectacularly well (see \cite{Spergel:2006hy,Tegmark:2006az} 
for recent reviews),
it raises three pressing questions: what is the physics of the 
postulated dark matter, dark energy and inflation energy?
The need to postulate the existence of as many as three new substances 
to fit the data has 
caused unease among some cosmologists 
\cite{Peebles:1999eb,Peebles:2000ay,Sellwood:2000wk,Tegmark:2001zc}
and prompted concern that these complicated dark matter flavors constitute a
modern form of epicycles.
Our only knowledge about these purported substances comes from their gravitational effects.
There have therefore been numerous suggestions that
the apparent complications
can be eliminated by modifying the laws of 
gravity to remove the need for dark matter 
\cite{Milgrom:1983ca,Bekenstein:2004ne},
dark energy 
\cite{Boisseau:2000pr,Esposito-Farese:2000ij,Carroll:2003wy}
and inflation \cite{Starobinsky:1980te}, and perhaps 
even all three together \cite{Liddle:2006qz}.
Since attempts to explain away dark matter with modified gravity have been severely challenged by recent observations, 
notably of the so-called bullet cluster \cite{Clowe:2006eq}, 
we will focus on dark energy (hereinafter ``DE'') and inflation.

There is also a second motivation for exploring alternative gravity theories: 
observational constraints on parametrized departures from general relativity (GR) have provided increasingly precise tests of 
GR and elevated confidence in its validity \cite{Will:1993ns,Will:2005va}.

\subsection{$f(R)$ gravity}

An extensively studied generalization of general relativity involves modifying 
the Einstein-Hilbert Lagrangian
in the simplest possible way, replacing $R - 2\Lambda$ by a more general 
function $f(R)$
\footnote{For the general case where $f$ depends on the full Riemann tensor $R^{\mu}_{\nu\alpha\beta}$
rather than merely on its contraction into the Ricci scalar $R$, 
this program is more complicated; a subset of these 
theories which are ghost
free can be written as $f(R,G)$, where $G = 
R^\beta_{\hphantom{\beta}\mu\nu\alpha} R_\beta^{\hphantom{\beta}\mu\nu\alpha}
- 4 R^{\mu\nu} R_{\mu\nu} + R^2$ is the Gauss-Bonnet scalar in 
4 dimensions \cite{Navarro:2005da}.
These theories lack a simple description
in terms of canonical fields; there is no so-called Einstein Frame. 
Progress has nevertheless been made along these lines,
and such Lagrangians may have more relevance to DE \cite{Navarro:2005da,Mena:2005ta,Navarro:2005gh} than ones independent of $G$.} 
\cite{Carroll:2003wy,Dick:2003dw,Dolgov:2003px,
Chiba:2003ir,Amendola:2006kh,
Capozziello:2006dj,Nojiri:2006gh,
Clifton:2005aj,Song:2006ej,Bean:2006up,
Erickcek:2006vf, Navarro:2006mw, Chiba:2006jp,PerezBergliaffa:2006ni,
Cembranos:2005fi,Olmo:2005hc,Zhang:2005vt}.
The equations of motion derived from this Lagrangian 
differ from Einstein's field equations when $f(R)$ is nonlinear, but
the theory retains the elegant property of general coordinate invariance. 
In such a theory, the acceleration of our universe may be explained 
if $f(R)$ departs from linearity at small $R$, corresponding to 
late times in cosmological evolution.
In this case it may be possible to avoid invoking a cosmological
constant to explain cosmic acceleration, although one then
replaces the problem of a small cosmological constant with the
problem of no cosmological constant. 
In such models, the effective DE is dynamic, i.e., it is not
equivalent to a cosmological constant, leading to potentially interesting
observational signatures.
We refer to these models as $f(R)$-DE theories.

In addition to potentially explaining late-time acceleration, $f(R)$
theories may be relevant to early-universe physics, particularly
if $f(R)$ is non-linear at large $R$ 
\cite{Starobinsky:1980te,Schmidt:5}.
More generally,
it is of interest to study
$f(R)$ theories
because they are arguably the simplest setting in which one can attack
the
general question of which modified theories of gravity are allowed.
By examining $f(R)$ theories, a broad class of theories
containing GR as a special case, we continue the program of testing GR
as best we can.

\subsection{The equivalence with scalar tensor gravity}


The modified Einstein field equations (and so the new Friedmann equation)
resulting from a non-linear $f(R)$ in the action can be seen
simply as the addition of a new scalar degree of freedom. 
In particular, it is well-known that 
these theories are exactly
equivalent to a scalar-tensor theory
\cite{Whitt:1984pd,
Jakubiec:1988ef}. It is therefore no surprise that 
for $f(R)$-DE theories, 
it is this scalar 
which drives the DE.
Before reviewing the mathematics of this equivalence in full detail in section \ref{sec:equiv},
we will discuss some important qualitative features below.

One can discuss the theory
in terms of the original metric $g_{\mu\nu}$,
in which case the degrees of freedom are not manifest.  Alternatively, 
by a 
conformal relabeling, one can reveal the theory to be regular gravity
$\tilde{g}_{\mu\nu}$, plus scalar field $\phi$. The former viewpoint is
referred to as the Jordan Frame (JF) and the latter as the Einstein Frame (EF).
Here
$\phi$ has the peculiar feature that in the JF, it exactly determines the
Ricci scalar $R$ and vice versa. So in the JF, the
Ricci scalar can in a sense be considered a non-canonical yet dynamical scalar field.
This feature is absent in normal general relativity,
where $R= - T/\Mpl^2 + 4 \Lambda$
is algebraically fixed by the trace of the stress energy tensor $T$.
Working in either frame
is satisfactory as long as one is careful about
what quantities are actually measurable, but we will find 
that the EF is much more useful for most of our calculations.

The coupling of the scalar field to matter is fixed in
$f(R)$ gravity, 
and is essentially of the same strength as the coupling of the graviton to
matter, except for the important case of massless conformally
invariant fields, which do not couple to $\phi$ at all.  
The dynamics of the theory are completely specified by the 
potential
$V(\phi)$ for
the scalar field in the EF, which is uniquely
determined by the functional form of $f(R)$.


After this lightning introduction to $f(R)$ theory we
are ready to summarize our main motivations for studying $f(R)$ theories
:
\begin{itemize}
\item There is recent renewed interest in this class of theories due to
their possible relevance to DE.
\item These theories may have an
interesting explanation in terms of a more complete theory of gravity.
\item Although there is an exact equivalence between $f(R)$ theories
and a class of scalar tensor theories, $f(R)$ theories may provide
a new perspective on scalar tensor theories.
For example,
a simple $f(R)$ may generate a complicated non-trivial
scalar potential $V(\phi)$ that you would not have thought
of using if just studying scalar tensor theories.
\item Exploring modified
or alternative theories is a way to test general relativity.
\end{itemize}

\subsection{The $R-\mu^4/R$ example}


Such a scalar field is not without observational consequence for
solar system tests of gravity, especially for $f(R)$-DE models.
For \emph{any} scalar field driving DE, we can come to the following
conclusions:
First, the field value $\phi$ must vary on a time scale of order 
the Hubble time $H_0^{-1}$, if the DE is distinguishable
from a cosmological constant
(for longer time scale,  
the DE looks like a cosmological constant; for shorter time scales, 
we no longer get acceleration). 
On general grounds, such a scalar field must have a mass
of order $m_\phi^2 \sim H_0^2$. Second, the Compton wavelength of this scalar
field is on the order of the Hubble distance, so it will
mediate an attractive fifth force which is distinguishable from gravity
by the absence of any coupling to light.
Unless the coupling to matter is tiny compared to that
of gravity, many solar system based tests
of gravity would fail, such as measurements of the bending of light 
around the Sun \cite{Will:1993ns,Will:2005va}.

The archetypal example of $f(R)$-DE suffers from problems
such as these.
This model invokes the function 
\cite{Carroll:2001bv},
\begin{equation}
\label{eq:arch}
f(R) = R - \frac{\mu^4}{R}
\end{equation}
for $\mu \approx H_0$. This 
gives a $V(\phi)$ in the EF with a runaway exponential potential
at late times: 
$V(\phi) \sim H_0^2 \Mpl^2 \exp(-(3/\sqrt{6})\phi/\Mpl)$ (here large $\phi$ 
means small $R$ which means late times.)
With no matter, 
such a potential in the JF 
gives rise to an accelerating
universe with the equation of state parameter $w_X = -2/3$
\cite{Carroll:2001bv}.
This model, however, is riddled with problems.
First, the theory does not pass solar system tests 
\cite{Chiba:2003ir,Dick:2003dw,Dolgov:2003px,Erickcek:2006vf},
and second, the cosmology is inconsistent with observation
when non-relativistic matter is present 
\cite{Amendola:2006kh}. Both problems can be understood
in the dual scalar tensor theory. 

For cosmology, during
the matter dominated phase but at high redshifts,
the influence on the dynamics of $\phi$ from
the potential $V$ is small compared to the influence
from the coupling 
to matter, which manifests itself in terms of an effective potential
for $\phi$ of the form
\begin{equation}
V_{\rm{eff}}(\phi) = 
V(\phi) + \bar{\rho}_{\rm{NR}} \exp\left( - \frac{
\phi}{ \sqrt{6} \Mpl} \right),
\end{equation}
where $\bar{\rho}_{NR}$ is the energy density of non-relativistic (NR) matter.
(More
details of the exact form of this potential will be presented in 
the next section.) The second term dominates because $H_0^2 \Mpl^2 \ll
\bar{\rho}_{\rm{NR}}$, and 
$\phi$ then rolls down the potential generated by $\bar{\rho}_{\rm{NR}}$ and not
$V$. The result is that the universe is driven away from
the expected matter dominated era (MDE) into a
radiation dominated expansion in the JF with $H^2 \propto a^{-4}$,
after which it crosses directly into the accelerating phase, with
expansion driven by DE with an effective equation of state parameter $w=-2/3$.
This special radiation-dominated-like 
phase (which is not driven by radiation) was
dubbed the $\phi$MDE by \cite{Amendola:2006kh}, where it
was made clear that 
this phase is inconsistent with observation. We say that this
potential $V$ is unstable to large cosmological non-relativistic densities.

For the solar system tests, the potential $V(\phi)$ is also negligible, so 
the theory behaves exactly like a scalar-tensor theory with
no potential. Because the coupling to matter has the same strength
as that to gravity, the scalar field mediates a long-range fifth force,
and the theory is ruled out by solar system tests.
In particular,
\cite{Chiba:2003ir} found that $\gamma=1/2$ in the PPN framework,
which is in gross violation of the experimental bound.

The above solar system tests also seem to rule
out more general classes of $f(R)$-DE models 
\cite{Olmo:2005hc,Chiba:2006jp,Erickcek:2006vf,Navarro:2005gh,Navarro:2006mw}.
However on the cosmology front, 
it seems that one \emph{can} cook up 
examples of $f(R)$ consistent with some dynamical dark energy
\cite{Capozziello:2006dj,Nojiri:2006gh,Song:2006ej}: by demanding that
the cosmological expansion $a(t)$ take a certain form, one can integrate
a differential equation for the function $f$ that by design 
gives a universe with any desired expansion history $a(t)$.
In this way, one gets around the cosmological instability of the 
archetypal model mentioned above.
However, these functions are arguably very contrived,
and further investigation of solar system predictions is required to determine 
whether these models are viable.

\subsection{What $f(R)$-theories are allowed?}


We now try to find viable $f(R)$ theories by examining what \emph{is}
acceptable on the scalar tensor
side. We focus
on theories that pass solar system tests.
Because 
the coupling of the scalar field 
to matter is fixed in $f(R)$ theories, and the only freedom we have is with the potential
$V$, we must choose $V$ in such a way as 
to hide the scalar field
from the solar system tests that caused problems
for the models described above. We are aware of only two ways to do this.
The first is the Chameleon scalar field, which
uses non-linear effects from a very specific singular 
form of potential
to hide the scalar field from current tests of gravity 
\cite{Khoury:2003aq,Khoury:2003rn}.
The second is simply to give the scalar field a quadratic potential
with mass $m_\phi \gtrsim 10^{-3} \rm{eV}$,
so that the fifth force has an extent less than $0.2\rm{mm}$ 
\footnote{At the time I finished this thesis, I was pointed out that the limits deviations on the gravitational inverse square law is down to 56 micrometers \cite{Kapner:2006si}.  }
and so cannot be currently measured by laboratory searches for a fifth
force \cite{Hoyle:2000cv}.

We will find simple $f(R)$ models which reproduce these two types
of potentials and so by design pass solar system tests.
Finding functions $f$ which give \emph{exactly} these potentials
will simply generate models which are
indistinguishable from their scalar-tensor equivalent. However, if
we search for simple choices of $f$ that reproduce these
potentials in a certain limit, then these theories
will not be exactly equivalent and might have distinguishable
features.


The Chameleon type $f(R)$ model seems to be the most plausible
model for attacking DE, as at first glance it seems to get
around the general problems mentioned above.
Indeed, 
one Chameleon model will arise quite naturally from a simple choice
of $f$. However,
we will show that the solar system constraints on this model
preclude any possible interesting late-time cosmological
behavior: the acceleration is observationally indistinguishable from a cosmological
constant.
In particular, for all the relevant physical situations 
this Chameleon model is the same as has been
considered before with no distinguishing features. However,
this model might provide clues in a search for
viable $f(R)$ theories that pass solar system tests and that may
give interesting late-time behavior. 

In an independent recent analysis, \cite{Navarro:2006mw,Brookfield:2006mq} 
also discussed
the Chameleon effect in $f(R)$ theories. They focus on a slightly 
different set of Chameleon potentials and come to similar
conclusions. Their results and ours together suggest
that the Chameleon effect may be generic
to $f(R)$ theories. 


We now turn from attempts to explain DE in $f(R)$ models to an arguably more
plausible scenario, which is simply to give the scalar
field a large mass.  These models
have no 
relevance for dynamic DE, but they do
have interesting consequences for
early universe cosmology.
The most theoretically best motivated functions, namely polynomials in $R$,
fit this class of $f(R)$ theories. The aim of this investigation 
is to explore
what we can possibly know about the function $f$. Because
this
question is very general, we will restrict our attention to a sub-class
of plausible $f(R)$ models. 

For these polynomial models, we will investigate possible
inflationary scenarios where the scalar partner $\phi$ is
the inflaton. We find the relevant model parameters which seed the
fluctuations of the CMB in accordance with experiment. We then
investigate general constraints on the model parameters
where $\phi$ is not an inflaton. We use solar system tests, 
nucleosynthesis constraints
and finally
an instability which is present in these theories when another
slow roll inflaton $\psi$ is invoked to explain CMB fluctuations.
This instability is analogous to that of the $\phi$MDE 
described above.


The rest of this chapter is organized as follows.
In section \ref{sec:equiv} we review the equivalence of $f(R)$ theories with
scalar tensor theories, elucidating all the essential points we
will need to proceed. Then in sections \ref{sec:cham} and \ref{sec:mass}
we explore the Chameleon model and massive theories, respectively, focusing on observational constraints.
We summarize our conclusions in section \ref{sec:concl}.

\section{$f(R)$ duality with Scalar Tensor theories \label{sec:equiv}}

We study the ``modified'' gravity theory defined by the action
\begin{equation}
\label{eq:theaction}
S_{JF} = \int d^4x \sqrt{-g} \frac{\Mpl^2}{2} f(R) +
S_M\left[g_{\mu \nu}, \Psi, A_\alpha, \ldots\right]
\end{equation}

Where for example $ \Psi, A_\alpha, \ldots$ label the matter fields
of the Standard Model.
Here we present a run
down of the 
map to the scalar tensor theory, displaying the most
important points needed to proceed.
See for example 
\cite{Jakubiec:1988ef,Hindawi:1995cu,Carroll:2003wy} for more details
of the equivalence with scalar tensor theories. 

We \emph{choose} to 
fix the connection in $R$ as the Christoffel symbols and not
an independent field, as opposed to the Palatini
formalism, which results in a very different theory
\cite{Vollick:2003aw,Flanagan:2003rb,
Flanagan:2003iw,Carroll:2006jn,Amarzguioui:2005zq}.  

If one simply varies the action Eq.~(\ref{eq:theaction}) with respect
to the metric $g_{\mu\nu}$, then a fourth order equation for the
metric results. One can argue (using general coordinate invariance)
that the degrees of freedom in the field $g_{\mu\nu}$ can be split
into a massless spin-2 field $\tilde{g}_{\mu\nu}$ and
a massive scalar field $\phi$ with second order equations of motion.
This split is easily revealed at the level of the action. 
Following for example \cite{Hindawi:1995cu} we introduce
a new auxiliary scalar field $Q$ (a Lagrange multiplier).  The
gravity part of Eq.~(\ref{eq:theaction}) may be written as
\begin{equation}
\label{eq:gravaction1}
S_{\rm{grav}} = \int d^4x \sqrt{-g} \frac{\Mpl^2}{2}
\left( f'(Q) (R - Q) + f(Q) \right)
\end{equation}
As long as $f''(Q) \neq 0$, the equation of motion ($\delta / \delta Q$) gives
$Q=R$ and Eq.~(\ref{eq:gravaction1}) becomes the original gravity action.
This may be written in the more suggestive form 
\begin{equation}
S_{\rm{grav}} = \int d^4x \sqrt{-g} \left(
\frac{\Mpl^2}{2}
\chi R  - \chi^2 V(\chi) \right)
\end{equation}
by relabeling
$f'(Q) \equiv \chi$. 
This is a scalar tensor theory of gravity 
with Brans Dicke parameter $\omega_{BD}=0$ \cite{Brans:1961sx} 
and potential \cite{Hindawi:1995cu}
\begin{equation}
\label{eq:potential}
V(\chi) = \frac{\Mpl^2}{2 \chi^2} \left[ Q(\chi) \chi
- f(Q(\chi)) \right]
\end{equation}
Here $Q(\chi)$ solves $\chi = f'(Q)$. Finally a rescaling of the
metric (which should be thought of as a field relabeling) 
\begin{equation}
\label{eq:conf}
\tilde{g}_{\mu\nu} = \chi g_{\mu\nu} = e^{(2/\sqrt{6}) \phi/\Mpl}
g_{\mu\nu}
\end{equation}
reveals the kinetic terms for the scalar field:
\ben
\label{eq:efaction}
S_{EF} = \int d^4x \sqrt{-\tilde{g}} \left( 
\frac{\Mpl^{2}}{2}\tilde{R}
- \frac{1}{2}\tilde{g}^{\mu\nu} \partial_\mu \phi \partial_\nu \phi
- V(\phi) \right)  + S_M [\tilde{g}_{\mu\nu} e^{-\sqrt{\frac{2}{3}}
\frac{\phi}{M_{\rm{pl}} } }, \Psi, A_\alpha, \ldots]\,,
\een
where the new canonical scalar field $\phi$ is related to $\chi, Q, R$
through
\begin{equation}
\label{eq:scalars}
f'(R) = f'(Q) = \chi = \exp\left(\sqrt{2/3}\,\phi/M_{\rm{pl}}\right)\,.
\end{equation}
As the kinetic terms for $\tilde{g}_{\mu\nu}$ and $\phi$ are
now both canonical, we see that these are the true degrees
of freedom of $f(R)$ gravity.
This demonstrates that the theories defined by $S_{JF}$ (
the Jordan Frame) and
$S_{EF}$ (the Einstein Frame) 
are completely equivalent when $f''(Q) \neq 0$. We choose
to analyze the theory in the Einstein Frame 
as things are much simpler here. It is,
however, important to be careful to interpret results correctly,
making reference to what is observed. In particular, matter
is defined in the Jordan Frame, and hence it will be most sensible to talk about
JF observables. We will give a simple example of this when we 
have introduced some matter.

The equations of motion for $\phi$ 
resulting from Eq.~(\ref{eq:efaction}) are 
\begin{equation}
\label{eq:scalar-eom}
-\tilde{\Box} \phi = - \frac{d V}{d \phi} 
- \frac{\tilde{T}^M}{\sqrt{6} M_{\rm{pl}} }\,,
\end{equation}
and for the metric $\tilde{g}_{\mu\nu}$,
\begin{equation}
\tilde{R}_{\mu\nu} - \frac{1}{2} \tilde{g}_{\mu\nu} \tilde{R}
= M_{\rm{pl}}^{-2} \left( \tilde{T}^M_{\mu\nu} + \tilde{T}^\phi_{\mu\nu}
\right) \end{equation}
with the energy momentum tensors
\ben
\label{eq:em-einstein}
\tilde{T}^M_{\mu\nu} =  \chi^{-1} 
T^M_{\mu\nu} \left(\chi^{-1} \tilde{g}_{\mu\nu} \ldots \right) 
\een
\ben
\label{eq:em-phi}
\tilde{T}^\phi_{\mu\nu} = \partial_{\mu} \phi \partial_\nu \phi
+ \tilde{g}_{\mu\nu} \left( - \frac{1}{2} \tilde{g}^{\alpha\beta}
\partial_\alpha \phi \partial_\beta \phi
+ V(\phi)
\right)
\een
Note that only the combination 
$\tilde{T}^M_{\mu\nu} + \tilde{T}^\phi_{\mu\nu}$
is conserved in the EF.

There are two important observations to be made about Eq.~(\ref{eq:efaction})
relating to the extra coupling to matter. First, the 
$\tilde{T}^M/\Mpl \sqrt{6}$ term in Eq.~(\ref{eq:scalar-eom}) 
represents an additional density dependent
``force'' on the scalar field, and for special cases where
we can solve for the functional form of the $\phi$ dependence of 
$\tilde{T}^M /\Mpl \sqrt{6}$ explicitly, as in
\cite{Khoury:2003rn}, we can think of the
scalar field living in an effective potential. We will see
two examples where this force is important, the most dramatic 
being the Chameleon effect.

Second, $\phi$ couples to matter as strongly as conventional
gravity ($\tilde{g}_{\mu\nu}$) does.
Hence,
as was already mentioned, $\phi$ will mediate a 
detectable fifth force 
for solar system tests
unless we do something dramatic to hide it. Finding theories which
hide $\phi$ from solar system tests is the focus of this chapter.

\subsection{Matter and Cosmology in $f(R)$ theories}

Let us first consider the coupling to standard model fields,
assuming that they are defined in the JF.
This is important for understanding how $\phi$ may decay. 
Massless scalar fields conformally coupled to gravity and 
massless gauge bosons
behave the same in the two frames and so do not couple to $\phi$.
However, a minimally coupled (real) scalar field $\Phi$ and a Dirac field $\Psi$
have extra interactions with $\phi$ in the EF:
\begin{eqnarray}
S_{\Phi} &=& \int d^4x \sqrt{-\tilde{g}} \left\{
-\frac{1}{2} \left(\partial \tilde{\Phi}\right)^2
- \frac{1}{2} m_\Phi^2 \chi^{-1} \tilde{\Phi}^2  
 - \frac{1}{12 \Mpl^2} \tilde \Phi^2 \tilde{g}^{\mu\nu}
\partial_\mu \phi
\partial_\nu \phi 
- \frac{1}{\sqrt{6} \Mpl} \tilde \Phi \tilde{g}^{\mu\nu}
 \partial_\mu \tilde{\Phi} \partial_\nu \phi \right\} \nonumber \\
\label{eq:einstein-dirac}
S_{\Psi} &=& \int d^4x \sqrt{-\tilde{g}} \bar{\tilde{\Psi}}
\left( i \tilde{\gamma}^\mu \tilde{D}_\mu
- m_\Psi \chi^{-1/2} \right)
\tilde{\Psi}\,,
\end{eqnarray}
where the JF fields have been rescaled as $\tilde \Phi = \chi^{-1/2} \Phi$
and $\tilde \Psi = \chi^{-3/4} \Psi$. Note 
that the cosmologically evolving field $\phi = \bar{\phi}(t)$
will change the masses of the standard model particles in the EF as
\begin{equation}
\label{eq:masses-scale}
\tilde{m} = m \chi^{-1/2} = m \exp\left( -(\sqrt{1/6}) \bar{\phi}(t)/\Mpl
\right)
\end{equation}
and small excitations $\delta \phi$ 
around the average value $\bar{\phi}(t)$ will roughly speaking
interact via the vertices defined by the interaction Lagrangian,
\begin{equation}
\label{eq:vertices}
\frac{1}{\sqrt{6} \Mpl} \left(
\tilde{m}_\Phi^2  \delta \phi \tilde{\Phi}^2
- \tilde \Phi \tilde{g}^{\mu\nu}
\partial_\mu \tilde{\Phi} \partial_\nu \delta \phi 
+ \tilde{m}_\Psi \delta \phi \bar{\tilde{\Psi}}
\tilde{\Psi} \right)
\end{equation}
to lowest order in $1/\Mpl$.
The mass shift in Eq.~(\ref{eq:masses-scale}) has an interesting
consequence in the EF; it shifts the frequency of the
absorption and emission lines by a factor of $\chi^{-1/2}$.
This effect will be indistinguishable from the normal cosmological
redshift due to expansion, and our effective redshift will be the combination of
both cosmological expansion and mass shift: $(1+z)^{-1} = \tilde{a}
\chi^{-1/2}$, where $\tilde a$ is the scale
factor in the EF normalized equal to unity today. This combination 
turns out to be the Jordan frame scale factor $a$ (see below), so
our redshift measurements coincide in both frames as expected.
These ideas were recently discussed in the context of conformal
cosmology \cite{Behnke:2001nw,Blaschke:2004wa}, 
where the observed redshifts
are explained completely in terms of an evolving scalar field.

Perfect fluids are best examined in the JF, because it is here
that their energy momentum tensor is conserved. For a general
JF metric one can solve for the flow of the fluid using conservation
of $T^M_{\mu\nu}$ and number flux
$n U^\mu$ (or other relevant physical principles)
and then map into the EF via Eq.~(\ref{eq:em-einstein}).

\subsubsection{The homogeneous and isotropic case}

For example, consider a homogeneous isotropic cosmology,
\begin{eqnarray}
{\rm (JF)} \quad ds^2 = d t^2 - a(t)^2 d \vec{x}^2 \,,
&\quad & U^\mu = (\partial_t)^\mu \,, \\
{\rm (EF)} \quad d\tilde{s}^2 = d \tilde{t}^2 - 
\tilde{a}(\tilde{t})^2 d \vec{x}^2 \,,
&\quad & \tilde{U}^\mu = \left(\partial_{\tilde{t}}\right)^\mu \,,
\end{eqnarray}
where $U^\mu$ and $\tilde{U}^\mu$ are the local fluid velocities
in the two frames. The quantities above are related by 
\begin{equation}
\tilde{a} = \chi^{\frac{1}{2}} a\,, \quad 
d\tilde{t} = \chi^{\frac{1}{2}} dt\,, \quad 
\tilde{U}^\mu = \chi^{-1/2} U^\mu\,.
\end{equation}
These relations imply that the Hubble parameters in the two frames
are related by
\begin{equation}
\label{eq:hubble}
H = \chi^{1/2} \left( \tilde H - \frac{\tilde{\partial}_t \phi}{\sqrt{6} \Mpl}
\right)\,.
\end{equation}
For example, applying the principles of entropy ``conservation''
and number conservation in the JF
(one may also need to demand thermal and chemical
equilibrium as relevant to the early universe) results
in known functions $\rho(a)$ and $p(a)$ such that the EF energy
momentum tensor may be written as
\begin{equation}
\tilde{T}^M_{\mu\nu} = \tilde{\rho} \tilde{U}_\mu \tilde{U}_\nu + \tilde{p}
\left( \tilde{U}_\mu \tilde{U}_\nu + \tilde{g}_{\mu\nu} \right)\,,
\end{equation}
where 
\begin{equation}
\label{eq:densitymap}
\tilde{\rho} = \chi^{-2} \rho\left( \tilde a \chi^{-1/2}\right)\,,
\quad
\tilde{p} = \chi^{-2} p\left( \tilde a \chi^{-1/2}\right)\,.
\end{equation}
The cosmological equations of motion are
\begin{equation}
\label{eq:einstein-cosmo}
3 \tilde{H}^2 \Mpl^2 =  \tilde{\rho}
+ \frac{1}{2} \left( \tilde{\partial}_t \phi \right)^2 + V(\phi)\,,
\end{equation}
\vspace{-12pt}
\begin{equation}
\label{eq:scalar-cosmo}
\tilde{\partial}_t^2 \phi \!+\! 3\tilde{H} \tilde{\partial}_t \phi
= - \frac{\partial V_{\rm{eff}}(\phi,\tilde a)}{\partial \phi}
= - \frac{d V_{E}}{d\phi}  - \frac{\tilde{T}^M}{\sqrt{6} \Mpl}\,.
\end{equation}
The effective potential for the scalar field coupled to 
homogeneous and isotropic matter is 
\begin{equation}
\label{eq:raweffective}
V_{\rm{eff}} (\phi,\tilde a) = V(\phi) + \tilde \rho = 
V(\phi) + \chi^{-2} \rho\left( \tilde a \chi^{-1/2}\right)\,.
\end{equation}
For the special case where the only density
components present are non-relativistic 
($\rho = \rho_{\rm{NR}} \propto a^{-3}$)
and ultra-relativistic ($\rho = \rho_{\rm{R}} \propto a^{-4}$) fluids, 
the effective potential is
\begin{equation}
\label{eq:effective}
V_{\rm{eff}}(\phi) = V(\phi) +  \bar{\rho}_{\rm{NR}}(\tilde a)
e^{-\frac{\phi}{\Mpl \sqrt{6}}}
 + \bar{\rho}_{\rm{R}}(\tilde{a})
\end{equation}
where for convenience we define 
$\bar{\rho}_{\rm{NR}}(\tilde a) \equiv \chi^{-3/2} \rho_{\rm{NR}}
(\tilde a \chi^{-1/2}) \propto \tilde a^{-3}$ and 
$\bar{\rho}_{\rm{R}}(\tilde a) \equiv \chi^{-2} 
\rho_{\rm{R}} (\tilde a \chi^{-1/2})\propto \tilde{a}^{-4}$.
These expressions are now independent of $\phi$: all the $\phi$ dependence
is explicitly seen in Eq.~(\ref{eq:effective}).
Note that relativistic particles provide
no force on $\phi$ because $\tilde T$ vanishes, or
equivalently because $\bar{\rho}_R(\tilde a)$ appears simply as an additive
constant to the potential in Eq.~(\ref{eq:effective}).

\subsubsection{The spherically symmetric case}

We now turn to the case of a spherically symmetric distribution of 
non-relativistic matter $\rho_{\rm{NR}}(r)$ in the JF,
for which we aim to solve for the metric $g_{\mu\nu}$. We
wish to consider this problem in the EF, where $\phi$
will take a spherically symmetric form and gravity
behaves like GR coupled to $\tilde{\rho} = \chi^{-2} \rho_{\rm{NR}}$.
In the
weak field limit, we write the metrics in the two frames as
\begin{subequations}
\begin{eqnarray}
{\rm (JF)}\quad ds^2 &=& - (1- 2 A(r)) d t^2 
+ (1 + 2 B(r) ) d r^2  + r^2 d\Omega^2 \\
\label{eq:EFmetric}
{\rm (EF)}\quad d\tilde{s}^2 &=& - (1- 2 \tilde{A}(\tilde r)) 
d t^2 
+ (1 + 2 \tilde{B}(\tilde r) ) d \tilde{r}^2 +  \tilde{r}^2 d \Omega^2 
\end{eqnarray}
\end{subequations}
where $\tilde r = \chi^{1/2} r$ and for small $\phi/\Mpl$,
the gravitational potentials are related by
\begin{subequations}
\label{eq:potentials}
\begin{eqnarray}
A(r) &\approx& \tilde{A}(\tilde r) + \frac{\phi(\tilde r)}{\sqrt{6} \Mpl}\,,
\\
B(r) &\approx& \tilde{B}(\tilde r) + 
\frac{1}{\sqrt{6} \Mpl} \frac{ d \phi(\tilde r)}{d \ln \tilde r}\,.
\end{eqnarray}
\end{subequations}
Following \cite{Khoury:2003rn} we define a 
non-relativistic energy density
$\bar{\rho}_{\rm{NR}}(\tilde r)  
= \chi^{-3/2} \rho (r)$ in the EF
which is conserved there and is analogous
to $\bar{\rho}_{\rm{NR}}(\tilde a)$ defined above for cosmology. 
Ignoring the back reaction of the metric on $\phi$, we take
$\tilde{g}_{\mu\nu} \approx \eta_{\mu\nu}$
in Eq.~(\ref{eq:scalar-eom}) and find as in \cite{Khoury:2003rn} that
\begin{equation}
\label{eq:spherical}
\frac{1}{\tilde r^2} \frac{d}{d \tilde r} \left( \tilde r^2
\frac{d \phi}{ d\tilde r}\right) = V'(\phi) - 
\chi^{-\frac{1}{2}} \frac{ \bar{\rho}_{\rm{NR}}(\tilde r)}{ \sqrt{6} \Mpl}
= \frac{ \partial V_{\rm{eff}}(\phi , \tilde r)}{ \partial \phi}\,,
\end{equation}
where again the effective potential is 
\begin{equation}
\label{eq:effective2}
V_{\rm{eff}} 
= V (\phi) + \chi^{-1/2} \bar{\rho}_{\rm{NR}}(\tilde r)\,.
\end{equation}
Solving Eq.~(\ref{eq:spherical}) for $\phi$ then allows us to find the
metric in the JF via Eq.~(\ref{eq:potentials}).

As an instructive example, consider the quadratic 
potential $V(\phi) = m_\phi^2 \phi^2/2$
and a uniform sphere of mass $M_c$ and radius $R_c$.
The solution external to 
the sphere is given by a Yukawa potential
\begin{equation}
\label{eq:yukawa}
\frac{\phi(r)}{\Mpl} = \frac{1}{\sqrt{6}} \frac{ M_c e^{-m_\phi r}}
{ 4 \pi \Mpl^2 r}
\end{equation}
assuming that $m_\phi R_c \ll 1$ and $\phi/\Mpl \ll 1$ so that
$\tilde r \approx r$.
If we ignore the energy density of the profile $\phi(r)$,
then outside the object there is vacuum.
The metric in the EF is then simply the Schwarzschild solution for
mass $M_c$. In other words, the potentials 
in Eq.~(\ref{eq:EFmetric}) are given by
$\tilde A(\tilde r)  = \tilde B(\tilde r) = M_c / 8 \pi \Mpl^2 \tilde r $
in the weak field limit $|\tilde A|,\, |\tilde B| \ll 1$.
In the JF using Eq.~(\ref{eq:potentials}), one finds the corresponding
potentials
\begin{subequations}
\begin{eqnarray}
A(r) &\approx& \tilde{A}(r) \left[ 1 + \frac{1}{3} e^{-m_\phi r} \right]\,, \\
B(r) &\approx& \tilde{A}(r) \left[1 - \frac{1}{3} e^{-m_\phi r}
(1 + m_\phi r ) \right]\,.
\end{eqnarray}
\end{subequations}
For $r \ll m_\phi^{-1}$ we find that the PPN parameter $\gamma=1/2$,
a well known result for a Brans Dicke theory \cite{Brans:1961sx}
with $\omega_{BD} = 0$ \cite{Will:1993ns}. 

The key feature here is the effective potential
from Eqs.~(\ref{eq:effective}) and (\ref{eq:effective2}). 
We have now seen that it makes a crucial difference
in two situations, and it will
play an important role in the next two sections as well.

\section{An $f(R)$ Chameleon \label{sec:cham}}

In this section, we consider $f(R)$ theories that are able
to pass solar system tests of gravity because of the 
so-called ``Chameleon'' effect.
We first present a theory that
is by design very similar to the original Chameleon 
model presented in \cite{Khoury:2003aq}.
We will give a brief description of how this model
evades solar system constraints,
and then move on to the cosmology of these $f(R)$ theories,
concentrating in particular on their relation to DE. 
Throughout this discussion we refer the
reader to the original work \cite{Khoury:2003aq,Brax:2004qh,Khoury:2003rn,
Mota:2006ed},
highlighting the differences between the original
and $f(R)$ Chameleons.

The Chameleon model belongs to the following general class of models,
\begin{equation}
f(R) = R - (1-m) \mu^2 \left(\frac{R}{\mu^2}\right)^m - 2 \Lambda.
\end{equation}
The sign of the second factor is important to
reproduce the Chameleon,
and the $(1-m)$ factor ensures that the theory is equivalent
to GR as $m\rightarrow 1$.
These models have been considered before in the literature
\cite{Carroll:2003wy,Amendola:2006kh}; 
in particular, this class contains the original DE $f(R)$ of
Eq.~(\ref{eq:arch}) 
when $m=-1$, $\Lambda=0$ and $2 \mu^4 \rightarrow \mu^4$.

The potential for $\phi$ in the EF is
\begin{equation}
\label{eq:expansion}
V(\phi) = \frac{\Mpl^2 \mu^2}{2 \chi^2} (m-1)^2 \left(
\frac{\chi-1}{m^2-m} \right)^{\frac{m}{m-1}} + \frac{\Mpl^2 \Lambda}{\chi^2}.
\end{equation}
where $\chi = \exp( \sqrt{2/3} \phi/\Mpl)$ as usual.
For $0<m<1$ and for $|\phi/\Mpl| \ll 1$, this reduces to
\begin{equation}
\label{eq:champot}
V(\phi) = M^{4+n} \left(-\phi\right)^{-n} + \Mpl^2 \Lambda,
\end{equation}
defined for $\phi < 0$, 
where the old parameters $\mu,m$ are related to the new
parameters $M,n$ through
\begin{equation}
m = \frac{n}{1+n} ,\quad
\mu^2 = 
\frac{ \left( 2 (1+n)^2 \right)^{1+n} }{\left( \sqrt{6} n \right)^n}
\frac{ M^{4+n} }{\Mpl^{n+2}}\,.
\end{equation}
The preferred values used in \cite{Khoury:2003aq}
are $M \sim 10^{-3} \rm{eV}$ and $n \sim  1$.
In the $f(R)$ theory, these values
give $m \sim 1/2$ and $\mu \sim 10^{-50} \rm{eV}$, i.e. 
much smaller than the Hubble scale today.

For small $|\phi|/\Mpl$, this
singular potential is equivalent
to the potential considered in \cite{Khoury:2003aq} 
for the Chameleon scalar field, albeit
with $\phi \rightarrow -\phi$. The coupling to matter, which
is a very important feature of this model, is also
very similar.  In \cite{Khoury:2003aq}, a species of particles $i$ is
assumed to have its own Jordan Frame metric $g^{(i)}_{\mu\nu}$,
with respect to which it is defined,
and a conformal coupling to the metric in the EF 
\begin{equation}
\label{eq:beta}
g^{(i)}_{\mu\nu} = e^{2\beta_i \phi/\Mpl}\tilde{g}_{\mu\nu}.
\end{equation}
Comparing this to Eq.~(\ref{eq:conf}),
the $f(R)$ Chameleon has $\beta_i = - 1/\sqrt{6}$ for all
matter species, so that all the Jordan Frame metrics coincide.

In the original Chameleon model, the $\beta_i$ were specifically
chosen to be different so that $\phi$ would show up in tests
of the weak equivalence principle (WEP). The $f(R)$ Chameleon does
not show up in tests of the WEP, so the
solar system constraints will be less stringent here.

\begin{figure}
\centering
\includegraphics[width=0.7\textwidth]{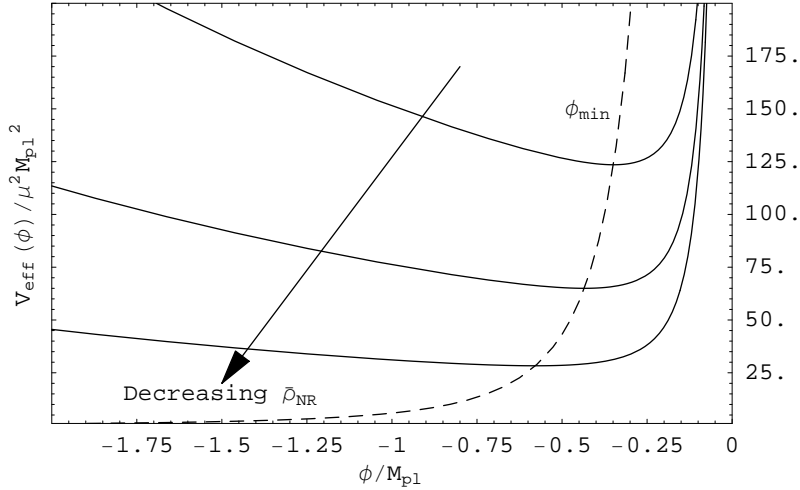}
\caption[Effective potential for the Chameleon model]{Effective potential for the Chameleon model Eq.~(\ref{eq:expansion})
with decreasing $\bar{\rho}_{\rm{NR}}/\mu^2 \Mpl^2 = 100,50,20
$ and $m=1$. Note that $\phi_{\rm{min}}$ and the mass $m_\phi^2$
(the curvature of the minimum)
are very sensitive to the background energy density
$\bar{\rho}_{\rm NR}$.\label{fig:champot}}
\end{figure}

This coupling to matter, along with the singular potential 
Eq.~(\ref{eq:champot}),
are the defining features of this $f(R)$ that make it a Chameleon 
theory. 
The effective potential $V_{\rm{eff}}$, discussed
in the previous section (see for example Eq.~ \ref{eq:effective}),
is then a balance between two forces;
$V$ pushing $\phi$ toward more negative values and the 
density-dependent term pushing $\phi$ toward more positive values.
So although the singular potential Eq.~(\ref{eq:champot})
has no minimum and hence no stable ``vacuum'', the effective potential
Eq.~(\ref{eq:effective}) including the coupling to matter does have a minimum. 
In fact,  
the density dependent term pushes the scalar field $\phi$ up
against the potential wall created by the singularity in
$V$ at $\phi=0$. 
Indeed, the field value $\phi_{\rm min}$ at the minimum
of the effective potential $V_{\rm{eff}}$ and the 
mass $m_\phi$ of $\phi$'s excitation around that 
given minimum are both highly
sensitive increasing functions of 
the background density $\bar{\rho}_{\rm NR}$,
as illustrated in Fig.\ \ref{fig:champot}. Using
Eq.~(\ref{eq:effective2}) for small $|\phi|/\Mpl$, the 
field value at the minimum 
and the curvature of the minimum are, respectively,
\begin{eqnarray}
\label{eq:minsolar}
- \frac{\phi_{\rm min}}{\sqrt{6} \Mpl}
&=& \frac{ m(1-m)}{2} \left( \frac{ \Mpl^2 \mu^2}
{ \bar{\rho}_{\rm{NR}}}
\right)^{1-m} \, ,  \\
\label{eq:masssolar}
m_\phi^2 &=& \frac{2}{3(1-m)} \frac{\bar{\rho}_{\rm NR}}{\Mpl^2}
\left( -\frac{\sqrt{6} \Mpl}{\phi_{\rm min}} \right)\,.
\end{eqnarray}
It is plausible that a scalar field $\phi$ which is very light for
cosmological densities is heavy for solar system densities
and hence currently undetectable. 
However, as we will now see, the actual mechanism that
``hides'' $\phi$ from solar system tests is a bit more complicated
than this.

\subsection{Solar System Tests}

In this section, we will derive solar system and laboratory
constraints on the parameters $(\mu,m)$, summarized
in Figure \ref{fig:cham}.
The profile of $\phi(\tilde r)$ in the solar system (around the Earth,
around the Sun, etc.) is of interest for solar system tests
of gravity: it determines the size of the 
fifth force and the post-Newtonian
parameter $\gamma$.
Because the effective potential for $\phi$ changes in different
density environments, the differential equation governing the
profile $\phi(\tilde r)$ in
Eq.~(\ref{eq:spherical}) is highly non-linear, and
the standard Yukawa profile of Eq.~(\ref{eq:yukawa}) does not always arise. 
These non-linear
features have been studied in \cite{Khoury:2003rn}, where
it was found that
for a spherically symmetric object of 
mass $M_c$ and radius $R_c$ surrounded by a gas of asymptotic density 
$\rho_\infty$, the profile is governed by the so-called 
``thin-shell'' parameter,
\begin{equation}
\label{eq:thinshell}
\Delta = \frac{| \phi_{\rm{min}}^\infty - \phi_{\rm{min}}^c |}{\sqrt{6} \Mpl} 
\frac{24 \pi \Mpl^2 R_c}{M_c},
\end{equation}
where $\phi_{\rm min}^\infty$ and 
$\phi_{\rm min}^c$ are the minima of the effective potential
in the presence of the asymptotic energy density 
$\bar{\rho}_{\rm{NR}} = \rho_\infty$ 
and central energy densities $\bar{\rho}_{\rm{NR}} = \rho_c$
respectively, see Eq.~(\ref{eq:minsolar}).
If $\Delta$ is large, then the external profile of $\phi$ is the
usual Yukawa profile Eq.~(\ref{eq:yukawa})
with mass $m^2_\infty=m^2_\phi~$,
the curvature of the effective potential in the 
presence of the asymptotic density $\bar{\rho}_{\rm{NR}} = \rho_\infty$; 
see Eq.~(\ref{eq:masssolar}).
If $\Delta$ is small, then the Yukawa profile is suppressed
by a factor of $\Delta$. The term
``thin shell'' comes from the fact that only a portion
of such a ``thin shell'' object contributes to the external Yukawa profile, 
the thickness of the shell being roughly $ (\Delta R_c)$.
We simply treat $\Delta$ as a parameter that suppresses this profile
if $\Delta \ll 1$.

For example, let us consider the profile
$\phi$ around the Sun, with $M_c = M_{\rm Sun}$ and 
$R_c = R_{\rm Sun}$. Assuming that we are in the 
thin shell regime ($\Delta\ll 1$), the Yukawa
profile of Eq.~(\ref{eq:yukawa})
suppressed by a factor $\Delta$ becomes,
\begin{equation}
\phi(r) = \frac{\Delta}{\sqrt{6}}  
\frac{M_{\rm Sun} e^{- m_\infty r} }{4 \pi \Mpl r} + \phi_{\rm min}^\infty \,.
\end{equation}
As in \cite{Khoury:2003aq}, we take the asymptotic 
density used to find $\phi_\infty$ and $m_\infty$
as that of the local homogeneous density of dark and baryonic matter
in our Galaxy: $\rho_\infty \approx  10^{-24} \rm{g}/\rm{cm}^3$.
Following the discussion in Section \ref{sec:equiv},
the metric in  the EF external to the Sun is just the Schwarzschild metric
(in the weak field limit) with
Newtonian potential $\tilde{A}(r) \approx M_{\rm Sun}/(8\pi\Mpl^2 r)$.
Using Eq.~(\ref{eq:potentials}) to map this metric 
into the JF metric 
$g_{\mu\nu} 
= \chi^{-1} \tilde{g}_{\mu\nu}$, we find
\beq{eq:metric-JF}
d s^2 = - \left[1 - 2 \tilde{A}(r) \left( 1 + \frac{\Delta}{3} e^{-m_\infty r}
) \right) \right] d t^2 
+ r^2 d\Omega^2  
 + \left[1 + 2 \tilde{A}(r) \left(1- \frac{\Delta}{3} e^{-m_\infty r} (
1+ m_\infty r) \right) \right] d r^2\,.
\een
Assuming that the Compton wavelength $m_\infty^{-1}$ is much
larger than solar system scales (we will confirm this later),
we obtain within the PPN formalism \cite{Will:1993ns} that
\begin{equation}
\gamma = \frac{3-\Delta}{3+\Delta}
\approx 1 - (2/3) \Delta
\end{equation}
There are several observational constraints
on $|\gamma-1|$, including ones from the deflection of light
and from Shapiro time delay. The tightest solar system
constraint comes from Cassini 
tracking, giving $|\gamma-1| \lesssim 2.3 \times 10^{-5}$ \cite{Will:2005va}.
Thus the  ``thin shell'' parameter satisfies
$\Delta \lesssim 3.5 \times 10^{-5}$. We note that 
$|\phi_{\rm min}^c| \ll |\phi_{\rm min}^\infty|$ because of the
sensitive dependence of $\phi_{\rm min}$ on the local density, so
the definition of $\Delta$ in Eq.~(\ref{eq:thinshell}) becomes
\begin{equation}
\Delta \approx 3 |\phi_{\rm min}^\infty|/ \sqrt{6} \Mpl \tilde{A}(R_{\rm Sun})
\end{equation}
where $\tilde{A}(R_c = R_{\rm{Sun}})
\approx 10^{-6}$ is the Newtonian potential at the surface
of the Sun. Using Eq.~(\ref{eq:minsolar})
with 
$\bar{\rho}_{\rm{NR}} = \rho_\infty \approx  10^{-24} \rm{g}/\rm{cm}^3$ 
gives the constraint
\begin{equation}
\label{eq:sol}
\frac{\mu^2}{H_0^2} \lesssim  3 \left( \frac{2}{m(1-m)} \right)^{\frac{1}{1-m}}
10^{\frac{-6 -5 m}{1-m} }
\end{equation}
on the theory parameters $\mu$ and $m$.
For theories which fail this bound, we find that the 
Compton wavelength of
$\phi$ for the asymptotic background density of our galaxy 
satisfies
$m^{-1}_\infty \gtrsim 10^{10} \rm{AU}$. This confirms
the assumption that $m^{-1}_\infty$ is large compared
to solar system scales, which was used to derive this bound.

As was already noted, the solar system constraints
derived in \cite{Khoury:2003rn} are more restrictive. This is
because they demanded that the couplings ($\beta_i$) to different
species of particles in equation (\ref{eq:beta}) be different.
This
gives violations of the weak equivalence principle
on Earth-based experiments unless the Earth and
atmosphere have a thin shell.  
However, in the $f(R)$ Chameleon model, all the $\beta_i$
are the same, so there will be no 
weak equivalence principle violations. 

The $f(R)$ Chameleon may still show up in searches
for a fifth force, in particular in tests of the inverse
square law. 
The strongest comes from Earth-based
laboratory tests of gravity such as in the E\"{o}t-Wash 
experiments \cite{Hoyle:2000cv}. 
By demanding that the test masses acquire thin shells, 
\cite{Khoury:2003rn} 
found constraints on the parameters $(M,n)$ which map
into the following bound on the
$f(R)$-parameters $(\mu,m)$:
\begin{equation}
\label{eq:ssconstraints}
\frac{\mu^2}{H_0^2} \lesssim
(1-m) \left(\frac{2}{m(1-m)}\right)^{\frac{m}{1-m}} 
10^{\frac{-4 -24 m}{1-m}}
\end{equation}

\subsection{Cosmology}

We now turn to
the cosmology of the Chameleon scalar field, which
was studied in \cite{Brax:2004qh}.  It was found there and already commented
on in \cite{Khoury:2003rn} that the mass of $\phi$ on cosmological
scales is not small enough to give any interesting DE behavior
for $M \approx 10^{-3} \rm{eV}$ and $n \sim 1$. We will
revisit this question in the $f(R)$ context: do any
\emph{allowed} parameters ($\mu,m$) in 
Eqs.~(\ref{eq:sol}-\ref{eq:ssconstraints}) give non-vanilla DE?
Will there be any cosmologically observable differences between
this $f(R)$ Chameleon and the original model (which
is in principle possible because higher
order terms in the expansion of $V$ in Eq.~(\ref{eq:expansion})
may become important)? 
We will see that the answer to both of these questions
is \emph{no} for the same reason: solar system tests
preclude the minimum of the effective potential from 
lying beyond $\phi \lesssim - \Mpl$ on cosmological scales today.

Let us try to understand this by looking at the details
of Chameleon cosmology. We first note that, as opposed to
\cite{Brax:2004qh}, we do \emph{not} fix $\Lambda \Mpl^2 = M^4$,
so we are less restricted as to what $M$ or $\mu$ can be. 
The essence
of the argument, however, is the same as in \cite{Brax:2004qh}.
Working in the
EF, for a large
set of initial conditions in the early universe, $\phi$ is attracted 
to the minimum of the effective
potential given by Eq.~(\ref{eq:effective}). 
The scalar field tracks the minimum, which shifts 
$\phi(\tilde a) \equiv \phi_{\rm min}$ as the universe expands. The 
energy density in coherent oscillations around this minimum
are negligible and so there is no ``moduli problem''. (
In contrast, this \emph{may} be a problem
for the case considered below in Section \ref{sec:mass}.)  

We will see that the condition for such a tracking solution
to be valid is that the minimum satisfies
\begin{equation}
\label{eq:less}
-\phi(\tilde a)/\Mpl \ll 1,
\end{equation}
so we consistently make this assumption to derive properties
of the tracking minimum. 
After matter-radiation equality we have the tracking solution
\begin{equation}
- \frac{\phi(\tilde a)}{\sqrt{6} \Mpl}
= \frac{ m(1-m)}{2} \left( \frac{ \Mpl^2 \mu^2}
{ \bar{\rho}_{\rm{NR}}(\tilde a) + 4 V(\phi(\tilde a)) }
\right)^{1-m} .
\end{equation}
Along this tracking solution, the curvature (mass)
around the minimum and the speed of the minimum
are, respectively
\begin{equation} 
\label{eq:mass}
\frac{m_\phi^2(\tilde a)}{\tilde{H}^2}
= \frac{2}{1-m} \left(\frac{\sqrt{6} \Mpl}{-\phi(\tilde a)}\right)
\left( \frac{ \bar{\rho}_{\rm{NR}}(\tilde a) + 4 V(\phi(\tilde a)) }
{\bar{\rho}_{\rm{NR}}(\tilde a) + V(\phi(\tilde a)) } \right)\,,
\end{equation}
\begin{equation}
\frac{- 1}{\Mpl \tilde H} \frac{ d\phi(\tilde a)}{d \tilde t}
= -3 \left( \frac{- \phi(\tilde a)}{\Mpl} \right)
\frac{(1-m) \bar{\rho}_{\rm{NR}}(\tilde a)}{ \bar{\rho}_{\rm{NR}}(\tilde a) + 4 V(\phi(\tilde a)) }
\end{equation}
Since $\phi$ will track the minimum while $m_\phi(\tilde a) \gg \tilde H$,
Eq.~(\ref{eq:mass}) shows that the assumption of Eq.~(\ref{eq:less}) 
is indeed consistent. 

\begin{figure}[h!]
\centering
\includegraphics[width=0.7\textwidth]{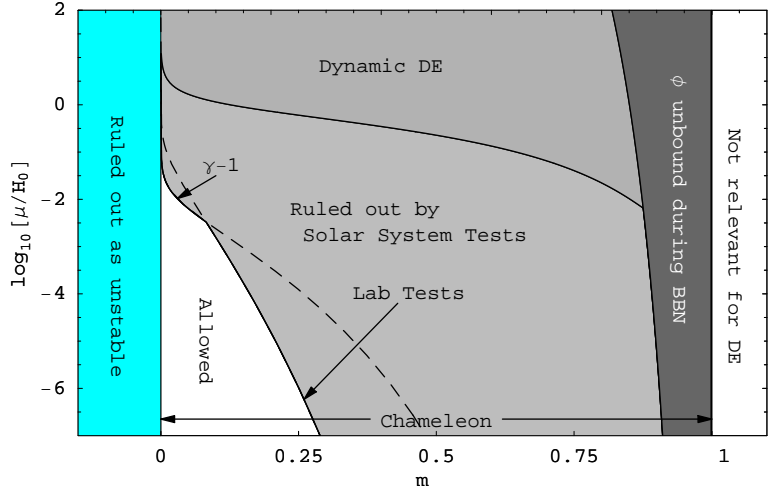}
\caption[Solar system constraints on the $f(R)$ Chameleon]{Solar system constraints on the $f(R)$ Chameleon are seen to exclude 
all models where the ``dark energy'' is observationally distinguishable from a cosmological constant
(labeled ``dynamic DE''). 
The two different solar system constraint curves come
from Eq.~(\ref{eq:sol}) and Eq.~(\ref{eq:ssconstraints}). Although it is not
clear from the plot, the limits $m\rightarrow 0$, $m\rightarrow1$
and $\mu \rightarrow 0$ are all acceptable and yet give
no dynamical DE. Indeed these are exactly the limits in which
we recover standard GR.}
\label{fig:cham}
\end{figure}

Also, during radiation domination one can show that $m_\phi^2(\tilde a)/\tilde{H}^2 \sim 
(-\Mpl/\phi(\tilde a)) \, \tilde a / \tilde a_{\rm{MR}}$, where $\tilde
a_{\rm MR}$ is the scale factor at matter-radiation equality,
so 
it is possible that at early times the scalar field is unbound.
We know the expansion history
and the effective value of Newton's constant $G_N$ quite well \cite{Carroll:2001bv,Copi:2003xd}
around big-bang nucleosynthesis (BBN); 
if $\phi$ is 
unbound, we have no reason to believe that $G_N$, which varies
as $\phi$ varies, is near
today's value.
Requiring that it is bound before the beginning of BBN gives a 
constraint that we have included in Figure \ref{fig:cham}.

Returning to the matter-dominated era, Eq.~(\ref{eq:less})
implies that the 
expansion history in the JF may be written as
\begin{equation}
\label{eq:freed}
3 \Mpl^2 H^2 = \rho_{\rm{NR}}(a) + V(\phi(a_0)) + 
\mathcal{O}(\frac{\phi}{\Mpl})\,,
\end{equation}
For $|\phi (a_0)|/\Mpl \ll 1$ 
today, this is just the usual Friedmann equation
with a cosmological constant,
where in accordance with experiment 
we are forced to identify $V(\phi(a_0))$ with
$\rho_X(0)$, the current dark energy
density. 
Note that the parameter
$\Lambda$ in $V$, which we have not fixed,
allows us to make this choice independent
of any values of $\mu$ and $m$. For $m$ not small, $
V(\phi(a_0)) \approx \Lambda \Mpl^2$ so $\Lambda$ is fixed
at $\Lambda \approx \rho_X(0) / \Mpl^2$;
however, for small $m$ we will see later that the situation
will be slightly different.

This implies that the only
way to get interesting late-time cosmological behavior
is to not have $|\phi (a_0)/\Mpl| \ll 1$ but
rather $|\phi (a_0)/\Mpl| \sim 1$
today. In this case the tracking solution above is not valid; the scalar
field is no longer stuck at the minimum, and we might not have
to invoke a constant  
$\Lambda$ in $V$ to explain todays accelerated expansion.  
Rather the acceleration would be driven by 
a quintessence type phase.

However, one can show that given the solar system
constraints,
$|\phi(a_0)/\Mpl| \sim 1$ is not possible.
In fact, as we will now show, a stronger statement can be
made: even if we continue
to assume Eq.~(\ref{eq:less}), so that the tracker solution is still
valid, the solutions that are consistent with solar system
tests always give DE behavior that is ``vanilla,'' i.e.,
indistinguishable from a cosmological constant.

In these models, the
effective dark energy density is
\beq{eq:effcc}
\rho_X(a) \approx V(\phi(a)) + \left(\frac{- \phi(a)}{\sqrt{6}\Mpl}\right)
\left( \rho_{\rm{NR}}(a) + \rho_X(0) \right)  
\times \left( 2 + \frac{ 6 \rho_{\rm{NR}} (a) (1-m)}
{\rho_{\rm{NR}}(a) + 4 \rho_X(0)} \right),
\eeq
where $V(\phi(a)) - \rho_X(0)= \mathcal{O}(\phi/\Mpl)$. 
If we expected Eq.~(\ref{eq:effcc}) to give interesting behavior
in the allowed region of parameter space, we would fit 
the Friedmann equation with $\rho_X(a)$ to the combined knowledge
of the expansion history and find the allowed values of
$(\mu,m)$. We will instead adopt a simpler
approach, \emph{defining} ``non-vanilla
DE'' through the effective equation of state parameter,
\begin{equation}
w_X = -\frac{1}{3} \frac{d \ln \rho_X(a) }{d \ln a} -1.
\end{equation}
This is the relevant equation of state that one would
measure from the expansion history (that is \emph{not} 
$p_\phi / \rho_\phi$).
We say that the DE is non-vanilla if
$|w_X+1|>.01$, which is quite optimistic
as to future observational capabilities 
\cite{Bock:2006yf}.  However, because
our result is
null the exact criterion is not important. 

The resulting constraint
on $\mu$ and $m$ is shown in Figure \ref{fig:cham} along with
the solar system constraints. As the Figure shows, all models
consistent with solar system tests are ``vanilla'' -- that is, 
indistinguishable from a cosmological constant.


The most interesting
part of this parameter space is the limit $m\rightarrow 0$,
which is one of the limits in which we
should recover general relativity.  The theory then becomes
\begin{equation}
f(R) \approx R - (\mu^2 + 2 \Lambda ) + \mu^2 m \ln \left( R /\mu^2 \right) 
\end{equation}
with the Chameleon-like (singular at $\phi=0$) potential 
\begin{equation}
V (\phi) \approx 
\frac{\Mpl^2}{2} e^{-\frac{4}{\sqrt{6}} \frac{\phi}{\Mpl} }
\left( \mu^2 + 2 \Lambda -  m \mu^2 \ln\left(1 - e^{\sqrt{\frac{2}{3}} 
\frac{\phi}{\Mpl}} \right) \right)
\end{equation}
In this limit we 
are forced to fix $\Lambda = \rho_X(0)/\Mpl^2 - \mu^2/2$.
The DE energy equation of state parameter is
$w_X = -1 -0.05 m \mu^2 / H_0^2$.
The tightest solar system constraint on $\mu^2$ in this limit
is from $|\gamma-1|$ in Eq.~(\ref{eq:sol}) which gives 
$m \mu^2 \lesssim 6 \times 10^{-6} H_0^2$.   The equation
of state parameter for DE is then constrained to be $|w_X+1|
\lesssim 0.3 \times 10^{-6}$ which is definitely unobservable. 

Finally we note that the ultimate fate of the $f(R)$ chameleon
is different from that of the original model. This is because
$V(\phi)$ actually does have a minimum relevant for
cosmological energy densities. This is due to the
$\phi$ dependence of the
$\Lambda \Mpl^2 \chi^{-2}$ term in 
Eq.~(\ref{eq:expansion}), which is absent in the
original models. Eventually $\phi$ will settle into this minimum
and the universe will enter an inflating de Sitter phase, much like
the fate of a universe with a simple cosmological constant.
The original model on the other hand \emph{eventually} enters
a quintessence like expansion. However, this distinction
is unobservable today.

In conclusion to this section, we 
have found a previously unstudied class of $f(R)$ theories
that gives acceptable local gravity by exploiting the
Chameleon effect. For the allowed
parameters of this model, there is no interesting late-time
cosmological behavior (observably dynamic DE). 
That is not to say that these
models have no interesting physics --- there may 
indeed be some interesting
effects of such models for future solar system tests
\cite{Khoury:2003rn} or on large scale structure \cite{Brax:2005ew},
and this might warrant further study in the context of $f(R)$ models.
We also noted that the $f(R)$ model is
subtly different from the original Chameleon model.
It does not violate the weak equivalence principle
, so solar system constraints are less
stringent and the ultimate fate of the universe is now simply an 
inflating de Sitter spacetime.

This mechanism might also be a starting point for constructing
working modified gravity models which do give non-vanilla DE,
somehow exploiting this mechanism more effectively and  bridging
the gap in Figure \ref{fig:cham}
between solar system constraints and non-vanilla DE.
We suspect they will not be as simple as the one presented. This
mechanism may also be relevant for attempting to understand
the Newtonian limit of the artificially constructed $f(R)$ models
mentioned earlier
that reproduce an exact expansion history. We make this claim because
an important property of the model presented in \cite{Song:2006ej}
is that the parameter $B \propto f''(R)$ is a rapidly growing function of
the scale factor $a$. 
For small $f''(R)$, one can show that 
the mass curvature of $V$ 
is $m_\phi^2 \sim 1/f''(R)$.
Hence, in this theory the mass of the 
scalar field during cosmological evolution 
is large at early times and small at late times, as in
the Chameleon models. A more detailed
analysis, beyond the scope of this chapter, is required
to see whether non-linear effects play a part in the Newtonian limit
of these theories.

\section{Massive $f(R)$ theories \label{sec:mass}}

We now consider arguably
more realistic $f(R)$ theories, namely
polynomials 
$f(R) =-2 \Lambda + R + a R^2 + b R^3  \ldots$. 
These theories have been extensively studied, especially
for quadratic $f(R)$; see \cite{Schmidt:5} and references therein.
They are more natural from the point of view
of renormalization and effective field theories: a high energy
completion of gravity would allow us to find these higher order terms.
However, common wisdom would have the higher order terms suppressed by 
inverse powers of $\Mpl$ and would force us to include other
terms of the same mass dimension such as $R^{\mu\nu} R_{\mu\nu}$. 
Despite this, we wish to explore the phenomenology
of such polynomial $f(R)$ theories and hence constrain them
with cosmological observations. In doing so, we will 
explore the full range of values for the coefficients ($a,b,...$) 
of the higher order terms to
be conservative rather than assume that they are order unity in Planck units. 

This class of theories can only match 
the currently observed cosmic acceleration via an explicit cosmological constant term 
$f(0) = -2 \Lambda$, giving the identification
$\Lambda = \rho_X(0)/\Mpl^2 = 3 H_0^2 \Omega_\Lambda$,
so there is 
no hope of dynamical DE.
Rather, these theories are more relevant to
very early universe cosmology where $R$ is large, and hence some of our results will be
quite speculative. 

\begin{figure}[h!]
\centering
\includegraphics[width=0.7\textwidth]{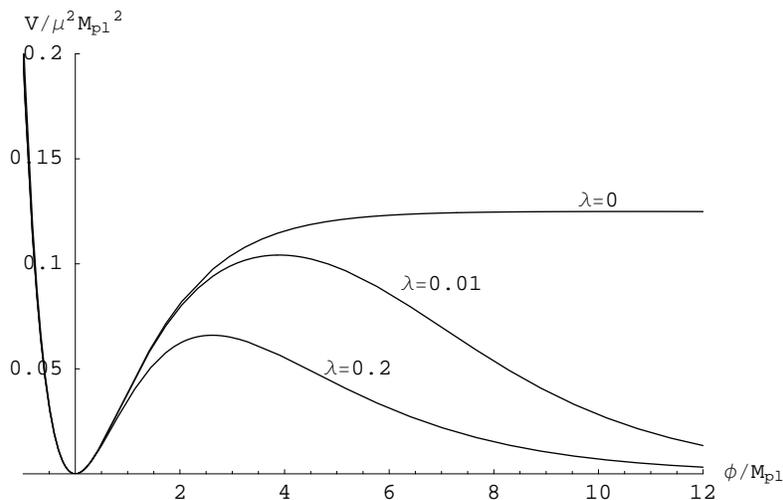}
\caption[Potential for the $f(R)$ model]{Potential for the $f(R)$ model in
Eq.~(\ref{eq:polymodel}) with various values of $\lambda$. Notice that
the $\lambda=0$ case has an asymptotically flat potential as
$\phi\rightarrow\infty$.
\label{fig:polymodel}}
\end{figure}

Consider for simplicity the two-parameter model
\begin{equation}
\label{eq:polymodel}
f(R) = R + R \left(\frac{R}{\mu^2} \right) +  \lambda R
\left(\frac{R}{\mu^2} \right)^2
\end{equation}
We restrict to the parameter range $\mu^2 >0$ and $0 < \lambda < 1/3$,
so that the resulting potential $V$ has a stable quadratic minimum 
and is defined for all $\phi$.
The Einstein frame potential for $\phi$ or $\chi$ is given by
\begin{equation}
\label{eq:polypot}
V_{E}(\chi) = \frac{ \Mpl^2 \mu^2}{2 \chi^2} q^2 \left( 1 + 2 \lambda q\right),
\end{equation}
where 
\begin{equation}
\label{eq:qDefEq}
q \equiv {1\over 3\lambda}\left[\sqrt{1 - 3\lambda(1-\chi)}- 1\right]
\end{equation}
is the larger of the two roots of $1-\chi + 2q + 3 \lambda q^2$  
(this ensures that the
resulting potential has a stable minimum). 
We plot this
potential for various $\lambda$ in Figure \ref{fig:polymodel}.

We will first explore the possibility that $\phi$ is the inflaton,
then discuss other constraints from our knowledge about
the early universe. Figure~\ref{fig:polycons}
summarizes our constraints.

\subsection{$f(R)$ inflation}

\begin{figure}[h!]
\centering
\includegraphics[width=0.7\textwidth]{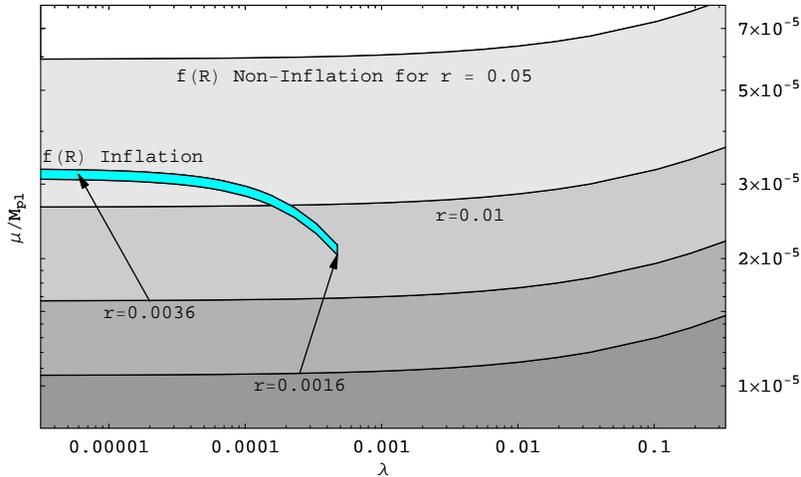}
\caption[Constraints on the cubic $f(R)$ model]{Constraints on the cubic $f(R)$ model. The thin
blue/grey sliver corresponds to observationally allowed
$f(R)$ inflationary scenarios. 
Shaded are regions
we may rule out given a measurement of the tensor to scalar
ratio $r$ and the assumption that they were generated
by a period of slow roll
inflation in the early universe. The $r=0.05$ and $r=0.01$ are the
most realistic curve, in the sense that future experiments
are sensitive to such values as low as $r=0.01$ \cite{Bock:2006yf}.
\label{fig:polycons}}
\end{figure}

The possibility that higher order corrections
to the gravitational Lagrangian might be responsible for a de Sitter inflationary
period was examined thoroughly early on in the inflationary game 
\cite{Starobinsky:1980te,Starobinsky:1983}.
For $\lambda = 0$, the potential $V(\phi)$ is  very flat
for large $\phi$, which is perfect for inflation.
This model and other related ideas were extensively studied in 
\cite{Mijic:1986iv,Mukhanov:1989rq,Hwang:2001pu,Kalara:1990ar,Ellis:1998gf,
Barrow:1988xh},
which also confirmed the existence of a viable inflationary model. 
We now search for
possible inflationary scenarios with $\lambda \neq 0$ that are consistent with
current observations. This
question was already considered in \cite{Berkin:1990nu}, which
found $\lambda \ll 1$; however, we wish to be more quantitative
in light of the latest CMB measurements. 

As usual in these models, it is important to keep careful
track of whether we are working in the EF or the JF: recall
that the potential
$V$ is defined in the EF, while matter is most naturally 
considered in the JF. Nonetheless,
we will argue that the inflationary
predictions are exactly the same as those of general relativity 
plus a normal
slow rolling scalar field with potential $V(\phi)$. The argument
goes as follows. Slow roll inflation works normally in the EF where
the graviton and scalar field have canonical actions. In
particular, the EF is where one \emph{should} calculate the spectrum
of tensor and scalar mode fluctuations. Re-heating
and the transformation of fluctuations in $\phi$ to 
adiabatic density fluctuations also works as usual in the EF, because at this time
the cosmic fluid is relativistic and hence governed by the same 
equations of motion in both frames. After reheating, $\phi$
is frozen out at the minimum of $V$, 
with $\phi = 0$ and $\chi=1$, so 
there is no longer any distinction
between the JF and the EF ($\tilde{g}_{\mu\nu} = g_{\mu\nu}$). 
Calculations for $\lambda=0$ were performed both as above
and in the JF in \cite{Hwang:2001pu}, and the results were
found to be consistent as expected.

Using this idea, calculating the inflationary predictions is
straightforward.  Using Eq.~(\ref{eq:vertices}), we can estimate the reheating
temperature as $T_{\rm{RH}} \approx 1.3\times10^{-2} g_*^{-1/4} \left( N_s
\mu^3 / \Mpl \right)^{1/2}$, where $N_s$ is the number of minimally
coupled scalar fields into which $\phi$ decays (it decays most strongly
into these fields). Then the scale factor (normalized to $a=1$ today) 
is 
\begin{equation}
a_{\rm{end}} = 7.5 \times 10^{-32} \left(\frac{\mu}{\Mpl}\right)^{-1/6}
g_*^{-1/12} N_s^{1/6}
\end{equation}
at the end of inflation.
Integrating the slow roll equations of motion,
$\phi' = - V'(\phi)/3\tilde H$, and assuming 
$\lambda \ll 1$, 
the number of $e$-foldings of inflation for a mode $k$ is
\begin{equation}
\label{eq:efold}
N_k \approx  \frac{3\, \mbox{arctanh}(\sqrt{\lambda} q )}{2\sqrt{\lambda}}
- \frac{3}{4} \ln\left( 1+2 q_k \right) 
+ N_0(\lambda).
\end{equation}
Here $N_0$ is a small number defined such that $N_k(q_{\rm end}) = 0$
at the end of inflation, where $q = q_{\rm end}\approx 1/\sqrt{3}$,
and $q_k$ is related to the conformal factor 
$\chi_k = 1 + 2 q_k + 3 \lambda q_k^2$ when the mode $k$ crosses
the horizon:
\begin{equation}
\label{eq:crossing}
\tilde{H}_k \approx \mu/\sqrt{24} = k e^{N_k} /a_{\rm end}
\end{equation}
This particular mode will have a scalar fluctuation amplitude
(also referred to as $\delta_H^2$ in the literature)
\begin{equation}
\label{eq:amp}
Q_k^2 = \frac{1}{ 1200 \pi^2 \epsilon_k} \left(\frac{\mu^2}{\Mpl^2} \right)
\end{equation}
where the slow roll parameters (using the definitions
in \cite{Bassett:2005xm}) are
\begin{equation}
\epsilon_k \approx \frac{ (1- \lambda q_k^2)^2}{3 q_k^2}
\,, \quad 
\eta_k \approx - \frac{ 2(1 + \lambda q_k^2)}{3 q_k}.
\end{equation}
We then use these to find the the scalar spectral index $n_s=1-6\epsilon+2\eta$, 
the ratio of 
tensor to scalar modes $r=16\epsilon$ etc.
Using the combined WMAP+SDSS measurements \cite{Tegmark:2006az} 
$Q = 1.945\pm0.05 \times 10^{-5}$ for modes $k=0.002 / \rm{Mpc}$ 
we can use Eq.~(\ref{eq:efold}-\ref{eq:amp}) together to fix $\mu$.
For $\lambda\rightarrow0$ the result is 
\begin{eqnarray}
\mu&\approx&(3.2 \pm 0.1) \times 10^{-5} \Mpl,\\
n_s&\approx&0.964,\\
r  &\approx&0.0036,
\end{eqnarray}
which is consistent with the both the theoretical results of 
\cite{Hwang:2001pu,Mukhanov:1989rq} and recent observational constraints
\cite{Spergel:2006hy,Tegmark:2006az}.

In addition, $n_s$ is sensitive to the value of $\lambda$. The observational
constraint 
$0.937 < n_s <  0.969$ (68\% 
C.L.) from \cite{Tegmark:2006az},
translates into a strong upper bound on $\lambda$:
\begin{equation}
\lambda < 4.7 \times 10^{-4}.
\end{equation}
This is an example of the usual fine tuning that is needed for observationally allowed
inflationary potentials and is consistent with the findings
of \cite{Berkin:1990nu}.
More precisely the values of $\mu,\lambda$ appropriate for
inflation are shown in Fig \ref{fig:polycons}.

\subsection{Other constraints}


Above we explored the possibility that $\phi$ was the inflaton.
Let us now turn to the alternative possibility that $\phi$ is not the inflaton,
and compute miscellaneous constraints on the parameters $\mu$ and $\lambda$ when they are varied freely.
We will first consider the
fifth force mediated by $\phi$, then investigate how the scalar
field behaves dynamically in the early universe, where the
most interesting effect comes from considering
a period of slow-roll inflation driven by some \emph{other}
scalar field.
As noted in Section \ref{sec:equiv},
the dynamics of $\phi$ is still governed by an effective potential
Eq.~(\ref{eq:effective}) which is important when there is a
component of matter whose energy-momentum tensor has nonzero 
trace.

To begin with, we ignore any effect that such a term may have on
the minimum of $V_{\rm{eff}}$ for these polynomial models, 
which is a good approximation if $|\tilde{T}^{\mu}_{\mu}|\ll
\mu^2 \Mpl^2$. We will see that for the first few 
constraints that we derive, this
will indeed be the case. Then we will return to the question of where
this is a bad approximation, which will naturally lead
to our discussion of slow-roll inflation by some other scalar
field.

\subsubsection{Fifth force constraints}


The minimum
of the effective potential lies at $\chi=1$, $\phi = 0$.
The curvature of this minimum is $m^{2}_{\phi} = \mu^2 /6$. Hence
we can get around solar system constraints simply by making
$\mu$ large enough so that the range of the fifth
force will be small. Clearly it must have an range smaller
than the solar system, otherwise, as was discussed above,
it will violate the bound on the PPN parameter $\gamma$. (Recall that
there is no Chameleon effect here, so $\Delta=1$ 
in Eq.~(\ref{eq:metric-JF}) and $\gamma=1/2$.)
For smaller scales, we consider searches for a fifth force
via deviations from the inverse square law.
The profile for a quadratic potential, i.e,
Eq.~(\ref{eq:yukawa}), gives a Yukawa potential between two
tests masses $m_1$ and $m_2$:
\begin{equation}
\label{eq:yukawa_potential}
V(r)= - \alpha {m_1 m_2 \over 8 \pi \Mpl^2} {e^{- m_\phi r}\over r},
\end{equation}
where $\alpha = 1/3$. For this $\alpha$-value, a fifth force
is ruled out for any Compton wavelength $m_\phi^{-1}$ ranging from
solar system scales down to $0.2 \rm{mm}$, where the lower
bound comes from the E\"{o}t-Wash experiments \cite{Hoyle:2000cv}.
This bound translates to 
\begin{equation}
\mu \gtrsim 1.0 \times 10^{-3} \rm{eV}.
\end{equation}
This implies 
$V(\phi) \sim \mu^2 \Mpl^2 \gg \rho_{\rm solar}$, a typical
solar system density, so 
for this constraint we were justified in ignoring any
effects of the density-dependent term on the
minimum of $V_{\rm{eff}}$.

\subsubsection{Nucleosynthesis constraints}


Given this preliminary constraint from local gravity tests, let
us now consider the cosmology of $\phi$ in the EF. We may approximate
the potential around the minimum by a quadratic potential
$V_{\rm{eff}}(\phi) \approx (\mu^2/12) \phi^2$, which is 
valid for $| \phi | \lesssim \Mpl$. The interesting
behavior will come during the radiation dominated epoch,
so in Eq.~(\ref{eq:einstein-cosmo}) we take
$\tilde\rho(\tilde a) \approx \bar{\rho}_R(\tilde a) \propto \tilde a^{-4}$, 
and  we ignore the $\tilde{T}^\mu_\mu$ term in Eq.~(\ref{eq:scalar-cosmo}) 
to find the cosmological equations of motion
\begin{eqnarray}
\label{eq:einstein-new}
3 \tilde{H}^2 \Mpl^2 &=& \bar{\rho}_R(\tilde a)
+ \frac{\mu^2}{12} \phi^2 + \frac{1}{2} 
(\phi')^2, \\
\phi'' &+& 3 \tilde{H} \phi' + \frac{\mu^2}{6} \phi =0,
\label{eq:cosmo-new}
\end{eqnarray}
where the primes denote $d/d\tilde t$.
There are two interesting limiting behaviors, corresponding to $\tilde H\gg \mu$ and $\tilde H\ll \mu$, which we will now explore in turn.

For $\tilde H \gg \mu$, the friction term in
Eq.~(\ref{eq:cosmo-new}) dominates, and $\phi$ is frozen out
at some value $\phi_*$ with $d \phi/ d \tilde t = 0 $. 
The energy density of $\phi$ is
subdominant in Eq.~(\ref{eq:einstein-new}).  Therefore, in the EF we
have the usual radiation dominated expansion, and in the JF using 
Eq.~(\ref{eq:densitymap}) and Eq.~(\ref{eq:hubble}) we have the same
Friedmann-Robertson-Walker (FRW) expansion with a different effective Newton's constant $G_N^*$:
$3 H^2 = 8 \pi G_N^* \rho(a) \propto a^{-4}$, where
\begin{equation}
\label{eq:newt}
G_N^* = \frac{1}{8\pi \Mpl^2} \exp\left(-\sqrt{\frac{2}{3}} \frac{\phi_*}{\Mpl}
\right)
\end{equation}

For $ \tilde H \ll \mu$, on the other hand, assuming $\phi_* < \Mpl$,
the field $\phi$ starts to oscillate with frequency $\mu/\sqrt{6}$ and
an amplitude that redshifts as $\tilde a^{-3/2}$.
Hence in the EF, the
energy density of $\phi$ in Eq.~(\ref{eq:einstein-new}) from these
zero momentum field oscillations is
$\rho_\phi = (\mu^2/12) \phi^2 +  \phi'^2/2 \approx 
\rho^*_\phi (\tilde{a}/\tilde{a}_*)^{-3}$ 
, where
$\rho^*_\phi \approx (\mu^2/12) \phi_*^{2}$. Mapping back into the JF, 
and averaging
over a cycle of this oscillation, we obtain the Friedmann equation
\begin{equation}
3 H^2 \Mpl^2 = \rho_R(a) + \frac{3}{2} \rho^*_\phi (a/a_*)^{-3},
\end{equation}
where the unusual factor
of $3/2$ comes from the averaging of the oscillations of $G_N^*$
in Eq.~(\ref{eq:newt}), as is discussed in more depth in \cite{Accetta:1990yb}.

The crossover between these two behaviors occurs when
$\tilde H$ is comparable to $\mu$, and given the laboratory tests of 
gravity above we can say that this must occur 
when the universe has at least the temperature 
$T_* \gtrsim 1 \rm{TeV}$. We were therefore justified in assuming radiation
domination in our calculation.

Let us examine further the zero momentum oscillations of 
$\phi$ that give this extra non-relativistic energy density. In the
absence of some mechanism (such as an extra period of low scale inflation
\cite{Randall:1994fr}), we expect the initial 
amplitude of oscillations to be of
the order $\Mpl$. 
This is because the potential in Figure \ref{fig:polypot}
varies on the scale of $\Mpl$ independently of the height of
$V$. Hence in the absence of any other scale, the initial amplitude must be 
around this size.  
Recall that at the onset of oscillations, $\tilde H \sim \mu$,
so the initial energy density of these oscillations is 
\begin{equation}
\rho^{*}_\phi \sim \Mpl^2 \mu^2 \sim \tilde H^2 \Mpl^2 \sim \rho_R(a_*).
\end{equation}
This energy density 
subsequently grows relative to the radiation density component,
quickly forcing the universe into a matter dominated period of expansion.
This is unacceptable if this component does not decay before the onset of
BBN, because  
then at the time of BBN the expansion would
be much faster than the normal radiation dominated expansion,
which would be inconsistent with observed primordial
abundances \cite{Copi:2003xd}.

The fact that $\phi$ interacts weakly with other particles
(the vertices in Eq.~(\ref{eq:vertices}) are suppressed by $1/\Mpl$)
so that $\phi$ decays too slowly
is exactly what is known as the cosmological \emph{moduli problem}. 
To be more
precise, we can use Eq.~(\ref{eq:vertices}) to estimate the decay
rate of zero momentum modes into other massive particles:
\ben
\Gamma_\phi \approx \sum_s \left( \frac{m_s^4}{m_\phi \Mpl^2 96 \pi} 
- \frac{m_\phi m_s^2}{ 96 \pi \Mpl^2} + 
\frac{m^3_\phi}{ \Mpl^2 384 \pi} \right) 
+ \sum_f \frac{ m_f^2 m_\phi}{\Mpl^2 12\pi},
\een
where the sums are over minimally coupled scalar particles 
and fermions with masses $2 m_s, \, 2m_f < m_\phi$. The
requirement $\Gamma_\phi > H_{\rm{BBN}}$ translates into the constraint
$\mu \gtrsim 100 \,\rm{TeV}$ for the Standard Model. One would
expect the bound on $\mu$ to be slightly smaller if one includes
other particles that have not been detected yet with mass smaller
than $100\, \rm{TeV}$.
This constraint should not be taken too seriously, however, because
the moduli problem may hypothetically be resolved by electroweak scale inflation
\cite{Knox:1992iy} or even by
a brief second period of inflation at the electroweak scale
\cite{Randall:1994fr}.

\subsubsection{Density dependent forces}


We now consider how the extra density dependent term in $V_{\rm eff}$
may effect cosmology.
In other words, when can we not neglect
the forcing term $\tilde{T}^{\mu}_{\mu}$ of Eq.~(\ref{eq:scalar-cosmo})?
After $\phi$ enters the oscillating phase when $\mu \gg
\tilde H$, the extra term has little effect
on the minimum since then it is small compared to the size of the potential
itself ($V \sim \mu^2 \Mpl^2$).  As a result, $\phi$ simply oscillates as
expected. Before the crossover, when $\phi$ is frozen, we showed that the
universe must be radiation dominated so that in particular,
as $\tilde{T}^{\mu}_\mu \ll \tilde \rho = 3 \tilde H^2 \Mpl^2$ 
during this phase, the Hubble friction will 
dominate compared to the force term of $\tilde{T}^{\mu}_\mu$
in Eq.~(\ref{eq:scalar-cosmo}), and we were justified 
in claiming that $\phi$ is frozen out.
The cosmology here does not suffer from the instability
that plagues Eq.~(\ref{eq:arch}).

There are, however, some exceptions that might
lead to interesting constraints.
First, consider a 
relativistic component $i$ of the cosmological plasma that becomes
non-relativistic and dumps its energy into the other
relativistic components. In this case,
$-\tilde{T}_\mu^\mu \sim (g_i/g_*)
\tilde{\rho}$  for a period
of about one e-folding, so $\phi$ receives a kick and
is displaced by an amount $\Delta \phi \approx (g_i / g_*) \Mpl/\sqrt{6}$
\cite{Brax:2004qh}. 
This might lead to an interesting effect such as
$\phi$ being kicked out of the basin of attraction
of $V$. The extreme case would be that $\phi$ does not  
end up oscillating around the minimum as expected when
$\tilde H \sim \mu$, 
but instead ends up rolling down the tail of $V$, an effect
which is clearly only possible for $\lambda\neq 0$.
In principle, such kicks could
even invalidate the predictions of BBN: near the onset
of BBN, $e^{\pm}$ annihilation occurs, displacing $\phi$
and consequently changing $G_N^*$ significantly as per Eq.~(\ref{eq:newt}).
However, we have already shown that $\phi$
must be in the oscillatory phase long before the onset of BBN,
and we have argued that 
these kicks have no effect while $\phi$ is in the oscillator phase, so
in fact this effect is unlikely to have relevance for BBN.
Such kicks may effect
other important cosmological dynamics
at temperatures
higher than $T > 1 \rm{TeV}$, such as baryogenesis.  However,
the effects 
are extremely model dependent, and it is hard to say anything 
definitive at this point.

\subsubsection{Non-inflation}



Another situation when we cannot ignore the density
dependent force on $\phi$ is during inflation.
Here $\tilde{T}^\mu_\mu$
is large for many e-foldings. Remember that in this section, we are 
not considering $\phi$ as our inflaton;
instead we consider a slow roll inflationary period driven by some
other scalar field $\psi$ defined in the JF. 
We wish to examine the 
effect a modified gravity Lagrangian such as Eq.~(\ref{eq:polymodel}) has
on the inflationary scenario.  In particular,
we will be interested in situations where inflation by
the field $\psi$
does not work, being effectively sabotaged by $\phi$. We will discuss the
generality of these assumptions at the end.

Such models have been considered before in the context
of both the $\lambda = 0$ models \cite{PhysRevD.43.2510,Kofman:1985aw,
Cardenas:2003tg} and other generalized gravity
models \cite{Berkin:1991nm}. There the goal was generally
to make the inflationary predictions 
more successful, focusing on working models.

\begin{figure}[h!]
\centering
\includegraphics[width=0.7\textwidth]{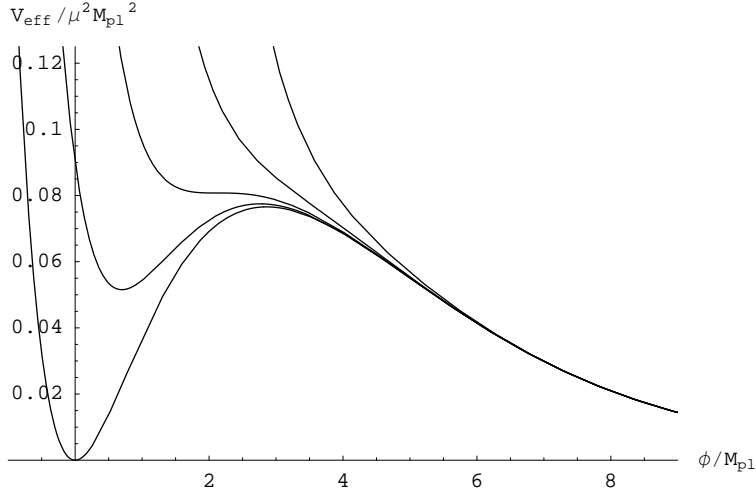}
\caption[Effective potential for the polynomial model]{Effective potential for the polynomial model Eq.~(\ref{eq:polymodel})
with $\lambda=0.1$.
for various JF inflationary energy densities $U(\psi)$ 
\label{fig:polypot}}
\end{figure}

In the JF, consider a scalar field $\psi$ with a potential
$U(\psi)$. We assume that $\psi$ is slow rolling;
$d \psi/dt \approx - U'(\psi) / 3 H(t)$. 
This is the assumption that
\begin{equation}
\label{eq:slowroll}
\frac{d^2 \psi}{ d t^2} \ll U'(\psi) \, , \quad
\left(\frac{d \psi}{d t}\right)^2 \ll U(\psi)
\end{equation}
which must be checked for self-consistency 
once we have solved for $H(t)$.  We can now easily calculate
$H(t)$  by first working in the EF and mapping back to the JF.
The equations of motion in the EF, 
Eqs.~(\ref{eq:einstein-cosmo}-\ref{eq:scalar-cosmo}),
become
\begin{equation}
\label{eq:infeom}
3 \tilde{H}^2 \Mpl^2 = \frac{1}{2} \phi'^2
+ V_{\rm{eff}}(\phi) \,,\quad
\phi'' + 3 \tilde{H} \phi'
 = - V'_{\rm{eff}}(\phi) 
\end{equation}
\vspace{-12pt}
\begin{equation}
\label{eq:infeffective}
V_{\rm{eff}}(\phi) = V(\phi) + U(\psi) \chi^{-2}
\end{equation}
It is interesting that a constant vacuum term in the JF does
not translate into to a constant term in the EF. See Figure \ref{fig:polypot}
for some examples of the effective potential $V_{\rm{eff}}$; we see that
for large enough $U(\psi) \gg \mu^2 \Mpl^2$, the minimum vanishes. One
finds that there is no minimum of the effective potential for
\begin{equation}
\label{eq:zeroconst}
U(\psi) > \frac{\mu^2 \Mpl^2}{18 \sqrt{3 \lambda}}.
\end{equation}
In particular, there is always a minimum for $\lambda = 0$.

The resulting behavior of the inflaton $\psi$ depends
on the size of $U(\psi)$ compared to $\mu^2 \Mpl^2$.
For small $U(\psi) \ll \mu^2 \Mpl^2$, it is clear that there is a stable minimum
around which $\phi$ will oscillate. 
In this situation, the effective potential has a minimum at
$\phi \approx 0$ with value
approximately $V_{\rm{eff}} (\phi \approx 0) \approx U(\psi)$, so after the
energy density of $\phi$ oscillations redshift away, we are left
with an exponentially expanding universe with $\chi\approx 1$, 
$3 \tilde{H}^2 \Mpl^2 \approx U(\psi)$ and $\tilde H \approx H$. 
Hence in the JF,
gravity behaves as it normally would in
general relativity: for a flat potential, the slow roll
conditions are satisfied, and inflation driven by $\psi$ 
works as it normally would. This is the
expected situation, and it will happen for $\mu \approx \Mpl$.


On the other hand, we now show that when $U(\psi) \gg \mu^2 \Mpl^2$ 
and when there is no minimum of the effective potential ($\lambda
\neq 0$),
we get a contradiction to the assumption that $\psi$
was slow rolling. Hence we show that it is not possible
for $\psi$ to drive slow-roll inflation.
For large $U(\psi) \gg \mu^2 \Mpl^2$,
the potential may be approximated as 
$V_{\rm{eff}} \approx U(\psi) \chi^{-2} $.
We treat $U(\psi)$ as a constant and find
that there is an exact attractor solution to Eq.~(\ref{eq:infeom})
of the form $ \chi \sim \tilde t$ and $\tilde a \sim \tilde t^{3/4}$.
Mapping this into the EF, we find the behavior
$a \sim t^{1/2}$, i.e.,  a period of radiation dominated
expansion analogous to the $\phi$MDE of \cite{Amendola:2006kh}
More specifically, we find
\begin{equation}
3 \Mpl^2 H^2 \approx U(\psi) a^{-4}
\end{equation}
This is clearly not an inflating universe. So
the slow roll assumptions of Eq.~(\ref{eq:slowroll})
are not consistent in this case.  We therefore
conclude that it is not possible
for $\psi$ to drive slow-roll inflation.

Instead, $\psi$ dumps most of its energy $U(\psi_0)$ into radiation, and
as before, $\phi$ is left frozen at some point $\phi_*$
until $U(\psi_0) \tilde a^{-4} \sim \mu^2 \Mpl^2$. 
After this, $\phi$ can either drive an inflationary period
itself as in the original discussion of
$f(R)$ inflation, or if $\phi_*$ 
is not in the basin of attraction of $V_{\rm{eff}}$, it will roll down
the tail of $V_{\rm{eff}}$. In neither situation has $\psi$ inflated
our universe. 
From this combination of inflaton $\psi$ plus $f(R)$ gravity
with $U(\psi) \gg \mu^2 \Mpl^2$, we only get
satisfactory inflation 
if $(\mu,\lambda)$ lies in the region of parameter space
appropriate
for $f(R)$ inflation (the blue/grey sliver in Figure \ref{fig:polycons}) 
and if $\phi_*$ sits
at a point which allows for the required number of e-foldings.

\subsubsection{Gravitational wave constraints}


It is well-known that inflation produces horizon-scale gravitational waves of amplitude 
$Q_t\sim H/\Mpl$, so that the energy scale of inflation can be bounded from above by the
current observational upper limit 
$Q_t\simlt 0.6\times 10^{-5}$ \cite{Spergel:2006hy,Tegmark:2006az} and perhaps 
measured by a detection of the gravitational wave signal with future CMB experiments \cite{Bock:2006yf}. 
Using such a detection one might try to constrain $\mu^2$ by the arguments
of the previous section. Specifically, by demanding that during
inflation there is a minimum of the effective potential one can find
a constraint by invoking
Eq.~(\ref{eq:zeroconst})
with $U(\psi)$ (incorrectly) 
replaced by the measured energy density of inflation.

However, because of the EF-JF duality, one needs to carefully define what one means by 
``the energy scale of inflation''. 
The bound from the above argument simply
precludes inflatons with a given energy density
$U(\psi)$ in the JF, but 
$U(\psi)$ is merely a parameter which
does not necessarily set the energy scale of inflation. 
In addition to this problem, we cannot use Eq.~(\ref{eq:zeroconst})
to derive a constraint 
for $\lambda = 0$, because
in this case there is always a minimum in the effective potential
and it is always possible for $\psi$ to slow roll
(this situation is described in 
greater depth in \cite{Kofman:1985aw,PhysRevD.43.2510}).

To make these ideas more concrete and resolve both
of these ambiguities, we will operationally define the energy scale of inflation 
to be the one that makes the standard GR formula for the gravitational wave amplitude valid.
It is clear that the amplitude of gravitational waves
should be calculated in the EF where
the metric has a canonical action. The result is then
passed trivially into the JF after inflation and when $\phi = 0$.
The Hubble scale $\tilde H$
then sets the size of the fluctuations,
but it is a complicated
model dependent calculation to find exactly when the relevant
fluctuations are generated. However, there is a limit
to the size of $\tilde H$ for which the EF is approximately
inflating, and so gravitational waves are being generated. 
Following the discussion above of non-working
inflatons, we demand that $\phi$ must
be slow rolling down the effective potential
$V_{\rm eff}$ defined in Eq.~(\ref{eq:infeffective})
for both frames to be inflating.
In this situation, both scalar and gravity modes are being generated.

The procedure is thus to find the maximum value of $\tilde H$ 
(that is, from Eq.~(\ref{eq:infeom}), the maximum
value of $V_{\rm{eff}}$) such that $\phi$ is slow rolling.
We then maximize this $\tilde H$ with respect to the parameter 
$U(\psi)$ to find the largest amplitude of gravitational waves
that can possibly be produced. At each step in this 
procedure, we wish to be as conservative as possible; for
example, we define slow roll through the slow roll parameter constraints
$\epsilon < 1$ and $|\eta| <2$ to allows
for the possibility of power law inflation. Where again we
use the standard defintion of $\eta$ and $\epsilon$ from \cite{Bassett:2005xm}.

As an example, consider the $\lambda=0$ case. Here it is possible
to show that for $\phi$ to be slow rolling, it must satisfy
\begin{equation}
\label{eq:phisr}
\phi > \phi_{\rm sr}\equiv \sqrt{\frac{3}{2}} \Mpl \ln \left(
\frac{2}{3} + \sqrt{\frac{7}{9} + \frac{8 U(\psi)}{3 \Mpl^2 \mu^2}} 
\right),
\end{equation}
where $\phi_{\rm sr}$ always lies to the left of the minimum
of $V_{\rm eff}$.
The maximum Hubble scale in the EF for
a given $U(\psi)$ is then 
$\tilde{H}^2 < \mbox{max}\{ V_{\rm eff}(\phi_{\rm sr})/3 \Mpl^2\, ,\,\,
\mu^2/24 \}$. This is maximized for large $U(\psi)/\mu^2\Mpl^2$,
with the result that $\tilde{H}^2< \mu^2/6$ where we
have used Eq.~(\ref{eq:phisr}). This translates into a constraint
on the maximum gravitational wave 
amplitude that can be produced,
\begin{equation}
Q_t^{\rm{MAX}} \approx 0.04 \frac{\mu}{\Mpl} \,
\Longrightarrow \, r^{\rm{MAX}} \approx 5 \times 10^{6} \frac{\mu^2}{\Mpl^2}
\end{equation}
Given a measurement of 
the tensor to scalar ratio $r$, this places a limit on $\mu$:
\begin{equation}
\mu \gtrsim  3 \times 10^{-4} r^{1/2} \Mpl
\end{equation}
Numerically, we find similar
results for non-zero $\lambda$. We plot examples of this 
constraint in Figure~\ref{fig:polycons}, combined with the
already discussed working $f(R)$ inflationary models. Note that
for a given $r$, it is important that this constraint lies 
below the corresponding working $f(R)$ inflationary model (the blue/grey
thin sliver of Fig.~\ref{fig:polycons}) with the
same $r$; fortunately, as is indicated by the arrows in
this figure, it does.

If gravitational waves are not detected, 
then this argument gives no lower bound on $\mu$.  In particular, it is possible
that inflation occurred at the electroweak scale, in which case 
the constraint $\mu\gtrsim 2\times 10^{-3} \rm{eV}$ is
the best we can do.

Note that we completely ignore the production of scalar fluctuation modes for
this argument. This is because the scalar modes are much more difficult
to calculate, since there are two scalar fields in the mix, $\psi$
and $\phi$, which are canonically defined in different frames. But the
scalar modes are also model dependent and one should generally be able to
fine tune $U(\psi)$ to give the correct amplitude and spectral
index without affecting the above argument. This more complicated
problem was considered for chaotic inflation with
$R^2$ gravity in \cite{Cardenas:2003tg}.


This constraint applies only to slow-roll inflation models.
There are
classes of fast-roll inflation, but these models
have problems of their own and generally fail to reproduce
the required scale invariance (see \cite{Linde:2001ae} for 
a review).

Finally, let us discuss some inflaton models that might
circumvent this constraint. It is possible to add
an inflaton in the EF.  However, this theory is then not conformally
equivalent to an $f(R)$ theory: the two scalar fields $\tilde{\psi}$ 
and $\phi$ get mixed up. Hence it is not in the class of models we set
out to constrain. Another possibility is to add
an inflaton which is conformally coupled to gravity and has
a $V \propto \psi^4$ potential.  This does not change from
frame to frame and so inflation might be expected to work. 
However, it was shown by \cite{Komatsu:1997hv} that non-minimally
coupled scalar fields cannot drive inflation. 

In any case, if gravitational waves are
found, then this constraint must be thought about seriously when
using such $f(R)$ models in other astrophysical or local
gravity situations.

\section{Conclusions}
\label{sec:concl}

We have searched for viable $f(R)$ theories 
using the wealth of knowledge on scalar tensor theories
to which $f(R)$ theories are equivalent. 
We studied two classes of models:
the $f(R)$ Chameleon and massive $f(R)$ theories, which may well be the only classes of models that can be
made consistent with local gravity observations.

The $f(R)$ Chameleon that was studied is a special kind of scalar field which
hides itself from solar system tests of gravity
using non-linear effects associated with the 
all-important density-dependent effective potential. 
It was shown that, despite this
Chameleon behavior, solar system tests still preclude the possibility
of observably dynamical DE; the best we could do was $|w_X - 1| 
\lesssim 0.3 \times 10^{-6}$ for the effective DE equation of
state parameter $w_X$ relevant for the dynamics of the expansion.
There are of course interesting effects of the 
Chameleon both
for local gravity \cite{Khoury:2003aq} and on cosmological density perturbations
\cite{Brax:2005ew},
and these may be worth future studies 
in the context of $f(R)$ theories. 

The massive theories were found to be more relevant for very high
energy cosmology, so the conclusions were more speculative. First,
the scalar field may be the inflaton, in which case we found the required
polynomial $f(R)$ to be quite fine tuned as is usual for inflationary
potentials. 
If the scalar field was not the inflaton, then we saw that possible instabilities could spoil both inflation
and Big Bang nucleosynthesis, giving interesting constraints on the shape of $f(R)$. 
If primordial gravitational waves are detected using the CMB, then the 
most naive models of inflation have serious problems unless the
mass of the $f(R)$-scalar is very large; a measured 
scalar to tensor ratio of $r=0.05$ requires $\mu\gtrsim 7 \times 10^{-5} \Mpl$.
If gravitational waves are not found, then the best we can say comes from the 
E\"{o}t-Wash laboratory experiments constraining the extent of a 
5th force: $\mu \gtrsim 2 \times 10^{-3} \rm{eV}$.


General relativity adorned with nothing but a cosmological constant, 
i.e., $f(R)=R-2\Lambda$, is a remarkable successful theory.
As we have discussed, a host of observational data probing scales from $10^{-2}$m to $10^{26}$m not only 
agree beautifully with GR, but also place sharp constraints on the parametrized departures from 
GR that we have explored.
In particular, both viable classes of $f(R)$ theories that we studied
were found to have no relevance for dynamic dark energy that is 
observationally distinguishable from ``vanilla'' dark energy, i.e., a cosmological constant. 
Since we have no good reason to believe that there are additional viable classes of $f(R)$-theories,
it appears likely that no viable $f(R)$ theories can produce the sort of interesting non-vanilla dark energy
that many observers are hoping to find. However, without a much larger
study of the parameter space (which is of course incredibly large)
we shy away from making a stronger statement here.

We would like to thank 
Serkan Cabi, Alan Guth, Robert Wagoner and Matias Zaldarriaga
for helpful discussion.  

\chapter{Constraining cosmological parameters with 21cm tomography}

\section{Introduction}
\label{sec:intro}

\begin{figure}[h!]
\centering
\includegraphics[width=0.7\textwidth]{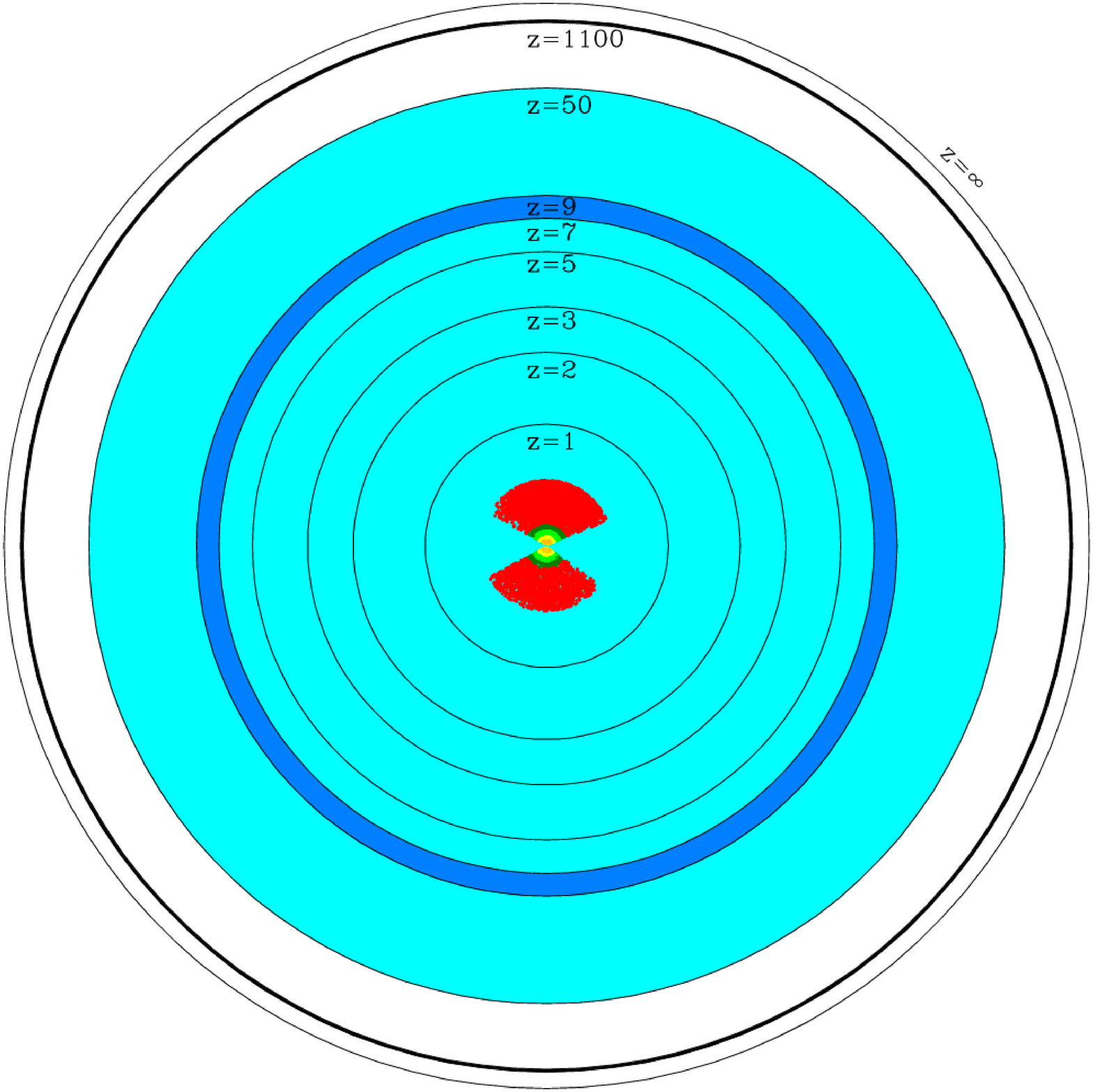}
\caption[21 cm tomography]{21 cm tomography can potentially map most of our observable
universe (light blue/light grey), whereas the CMB probes mainly a thin shell
at $z\sim 10^3$ and current large-scale structure maps (here
exemplified by the Sloan Digital Sky Survey and its luminous red
galaxies) map only small volumes near the center.  This chapter focuses
on the convenient $7\la z \la 9$ region (dark blue/dark grey).
\label{fig:spheres}
}
\end{figure}

Three-dimensional mapping of our Universe using the redshifted 21 cm
hydrogen line has recently emerged as a promising cosmological probe,
with arguably greater long-term potential than the cosmic microwave
background (CMB).  The information garnered about cosmological
parameters grows with the volume mapped, so the ultimate goal for the
cosmology community is to map our entire horizon volume, the region
from which light has had time to reach us during the 14 billion years
since our Big Bang.  Figure~\ref{fig:spheres} illustrates that whereas
the CMB mainly probes a thin shell from $z \sim 1000$, and current
large-scale structure probes (like galaxy clustering, gravitational
lensing, type Ia supernovae and the Lyman $\alpha$ forest) only map
small volume fractions nearby, neutral hydrogen tomography is able
to map most of our horizon volume.

\begin{table}
\caption{Factors that affect the cosmological parameter measurement accuracy.} \label{tab:methodology}
\begin{minipage}{\textwidth}
\footnotesize{
\begin{center}
\begin{tabular}{p{1.8cm}|p{2.8cm}|p{2.9cm}|p{2.9cm}|p{2.9cm}}
\hline\hline
\multicolumn{2}{c|}{Assumptions}  &  Pessimistic & Middle & Optimistic \\ \hline
Power modeling & Ionization power spectrum modeling &  
  Marginalize over arbitrary $\Pmuzero$ and $\Pmutwo$ & 
  Marginalize over constants that parametrize $\sPxx(k)$ and $\sPxd(k)$ & 
  No ionization power spectrum, $\sPdd(k) \propto P_{\Delta T}(\bfk)$.  \\ \cline{2-5}

 &Non-linear cut-off scale $\kmax$ & $1\,\perMpc$ & $2\,\perMpc$  &$ 4\,\perMpc$ \\ \cline{2-5}

 & Non-Gaussianity of ionization signals & Doubles sample variance & \multicolumn{2}{c}{Negligible} \\  \hline

Cosmological & Reionization history & 
  \multicolumn{2}{p{5.8cm}|}{Gradual reionization over wide range of redshifts} & 
  Abrupt reionization at $z \lesssim 7$
   \\ \cline{2-5}
 &Redshift range  &7.3-8.2&  	$6.8 - 8.2$    &6.8 - 10\\ \cline{2-5}
 & Parameter space & 
Vanilla model plus optional parameters & 
       &
Vanilla model parameters\\ \hline   
   
Experimental & 
 Data & MWA, LOFAR, 21CMA & Intermediate case  &  SKA, FFTT \\ \cline{2-5}
 & Array configuration \footnote{For the FFTT, we consider only the case where all dipoles are in a giant core.}
 & $\eta=0.15$ & $\eta=0.8$, $n=2$ & Giant core \\ \cline{2-5}
 & Collecting area \footnote{See designed or assumed values of $A_e$ in Table \ref{tab:spec}.}
  &  $0.5\>\times$ design values & Design values & $2\>\times$ Design values \\ \cline{2-5}
 & Observation time \footnote{Assumes observation of two places in the sky.}
  & $1000\,{\rm hours}$  & $4000\,{\rm hours}$ & $16000\,{\rm hours}$ \\ \cline{2-5}
 & System temperature & $2\times T_{\rm sys}$ in \cite{bowman05cr}   &  $T_{\rm sys}$ given in \cite{bowman05cr}   & $0.5\times T_{\rm sys}$ in \cite{bowman05cr}   \\ \cline{2-5}
  \hline

Astrophysical & 
Residual foregrounds cut-off scale $\kmin$\footnote{It is hard to predict the level of the residual foregrounds after the removal procedure. 
To quantify contributions from other factors, we take the approximation that there is no residual foregrounds at $k>\kmin$.  Here in the table, $yB$ is the comoving (l.o.s.) distance width of a single $z$-bin.}  
 & $4\pi / yB$ & $2\pi / yB$ & $\pi / yB$ \\ \hline\hline
\end{tabular}	
\end{center}
}
\end{minipage}
\end{table}

\begin{sidewaystable}
\centering
\caption[The dependence of cosmological constraints on the full range of assumptions]{The dependence of cosmological constraints on the full range of assumptions.  
We assume the fiducial values given in Section \ref{sec:cosmology}, and employ the Fisher matrix 
formalism to forecast the $\onesig$ accuracy of 21cm tomography measurements.   
Unless otherwise noted, errors are computed by marginalizing over all other parameters in the 
first ten columns (which we refer to as the ``vanilla'' parameters). 
In ``All OPT/MID/PESS'', we use the assumptions of the right, middle and left column of Table~\ref{tab:methodology}, respectively.
We assume that the total observing time is split between two sky regions, each for an amount in Table \ref{tab:methodology},  
using a giant/quasi-giant/small core array configuration where
100\%/80\%/15\%  of the antennae in the inner core are compactly laid at the array center while the rest 0\%/20\%/85\% of antennae 
fall off in density as $\rho\sim r^{-2}$ outside the compact core.}
\label{tab:omp}
\footnotesize{
\begin{tabular}{llp{1cm}p{1.5cm}p{1.5cm}p{1cm}p{1cm}p{1cm}p{1cm}p{1cm}p{1cm}p{1cm}p{1cm}p{1cm}p{1cm}}

\hline\hline
    &      & \multicolumn{10}{c}{\it Vanilla Alone} &  & &   \\ \cline{3-12}
		&  		    & $\Delta\Ol$ & 
 $\Delta\ln(\Omega_m h^2)$   & $\Delta\ln(\Omega_b h^2)$ & 
 $\Delta\ns$		       & $\Delta\ln\As$ & 
 $\Delta\tau$	               & $\Delta\xH(7.0)$ \footnote{$\xH(z)$ refers to the mean neutral fraction at redshift $z$.} & 
 $\Delta\xH(7.5)$ & 
 $\Delta\xH(8.0)$            & $\Delta\xH(9.2)$ & $\Delta\Ok$ &
 $\Delta\mnu$ [\eV]	       & $\Delta\al$ \\ \hline
Planck    &          & 0.0070&0.0081&0.0059&0.0033&0.0088&0.0043&...&... &...&... & 0.025  & 0.23  & 0.0026 \\ \hline
          & All OPT  &0.0044&0.0052&0.0051&0.0018&0.0087&0.0042&0.0063&0.0063&0.0063&0.0063&0.0022&0.023&0.00073 \\
\:+LOFAR  & All MID  &0.0070&0.0081&0.0059&0.0032&0.0088&0.0043&0.18&0.26&0.23& ... &0.018&0.22&0.0026 \\
	  & All PESS &0.0070&0.0081&0.0059&0.0033&0.0088&0.0043& ... &51&49& ... &0.025&0.23&0.0026 \\ \hline
	  & All OPT  &0.0063&0.0074&0.0055&0.0024&0.0087&0.0043&0.0062&0.0062&0.0062&0.0062&0.0056&0.017&0.00054 \\
\:+MWA	  & All MID  &0.0061&0.0070&0.0056&0.0030&0.0087&0.0043&0.32&0.22&0.29 & ... &0.021&0.19&0.0026 \\
	  & All PESS &0.0070&0.0081&0.0059&0.0033&0.0088&0.0043& ... &29&30& ...&0.025&0.23&0.0026 \\ \hline
	  & All OPT  &0.00052&0.0018&0.0040&0.00039&0.0087&0.0042&0.0059&0.0059&0.0059&0.0059&0.0011&0.010&0.00027 \\
\:+SKA	  & All MID  &0.0036&0.0040&0.0044&0.0025&0.0087&0.0043&0.0094&0.014&0.011 & ... &0.0039&0.056&0.0022 \\
	  & All PESS &0.0070&0.0081&0.0059&0.0033&0.0088&0.0043& ...&1.1&1.0& ...&0.025&0.23&0.0026  \\ \hline
	  & All OPT  &0.00010&0.0010&0.0029&0.000088&0.0086&0.0042&0.0051&0.0051&0.0051&0.0051&0.00020&0.0018&0.000054 \\
\:+FFTT\footnote{FFTT stands for Fast Fourier Transform Telescope, a future square kilometer array optimized for 21 cm tomography as described in \cite{FFTT}.    Dipoles in FFTT are all in a giant core, and this configuration does not vary.} 
  &All MID & 0.00038&0.00034&0.00059&0.00033&0.0086&0.0042&0.0013&0.0022&0.0031&... &0.00023&0.0066&0.00017 \\
          & All PESS & 0.0070&0.0081&0.0059&0.0033&0.0088&0.0043&...&0.0043&0.0047& ...&0.025&0.11&0.0024 \\	

\hline\hline
\end{tabular} 
}
\end{sidewaystable}

Several recent studies have forecast the precision with which such 21
cm tomography can constrain cosmological parameters, both by mapping
diffuse hydrogen before and during the reionization epoch
\cite{McQuinn:2005hk,Bowman:2005hj,Santos:2006fp} and by mapping
neutral hydrogen in galactic halos after reionization
\cite{Wyithe:2007rq}.  These studies find that constraints based on
the cosmic microwave background measurements can be significantly
improved.  However, all of these papers make various assumptions, and
it is important to quantify to what extent their forecasts depend on
these assumptions.  This issue is timely because 21 cm experiments
(like LOFAR \cite{LOFAR}, 21CMA \cite{21CMA}, MWA \cite{MWA} and SKA
\cite{SKA}) are still largely in their planning, design or
construction phases.  These experiments will be described in detail 
in Section \ref{sec:experiments}.  
In order to maximize their scientific ``bang for
the buck'', it is therefore important to quantify how various design
tradeoffs affect their sensitivity to cosmological parameters.

The reason that neutral hydrogen allows mapping in three rather than
two dimensions is that the redshift of the $21$ cm line provides the
radial coordinate along the line-of-sight (l.o.s.).  This signal can
be observed from the so-called dark ages \cite{shapiro05,Lewis:2007kz} 
before any stars had formed,
through the epoch of reionization (EoR), and even to the current epoch
(where most of the neutral hydrogen is confined within galaxies).  We
focus in this study on the $21$ cm signal from $6 < z < 20$ -- the end
of the dark ages through the EoR.  This is the redshift
range at which the synchrotron foregrounds are smallest, and
consequently is the range most assessable for all planned $21$ cm
arrays. 

There are three position-dependent quantities that imprint signatures
on the 21 cm signal: the hydrogen density, the neutral fraction, and
the spin temperature. For cosmological parameter measurements, only
the first quantity is of interest, and the last two are nuisances.
(For some astronomical questions, the situation is reversed.)  The 21
cm spin-flip transition of neutral hydrogen can be observed in the
form of either an absorption line or an emission line against the CMB
blackbody spectrum, depending on whether the spin temperature is lower
or higher than the CMB temperature.

During the epoch of reionization, the spin temperature is likely coupled
to the gas temperature through Ly$\alpha$ photons via the
Wouthuysen-Field Effect \cite{wout,field58}, and the gas in the IGM has been
heated by X-ray photons to hundreds of Kelvin from the first stars
\cite{Pritchard:2007}.  If this is true, the 21cm signal will only
depend on the hydrogen density and the neutral fraction.  However,
astrophysical uncertainties prevent a precise prediction for exactly
when the gas is heated to well above the CMB temperature and is
coupled to the spin temperature.  In this chapter, we follow
\cite{McQuinn:2005hk,Bowman:2005hj} and focus entirely on the regime
when the spin temperature is much larger than the CMB temperature
\cite{zald04,furlanetto04a,santos05}, such that the observed signal
depends only on fluctuations in density and/or the neutral fraction.
Specifically, we focus on the time interval from when this
approximation becomes valid (around the beginning of the reionization
\cite{zald04,furlanetto04a,santos05}) until most hydrogen has become
ionized, illustrated by the darkest region in
Figure~\ref{fig:spheres}.  Despite this simplification, the methods
that we apply to model the ionization fluctuations almost certainly can be
applied to model spin temperature fluctuations with minimal
additional free parameters.

In Table \ref{tab:methodology}, we list all the assumptions that affect the accuracy of cosmological parameter measurements, including ones about power
modeling, cosmology, experimental design, and astrophysical foregrounds. 
For each case, we provide three categories of assumptions:
one pessimistic (PESS), one middle-of-the-road (MID) 
and one optimistic (OPT). Since we wish to span the entire range of
uncertainties, we have made both the PESS and OPT models rather extreme.
The MID model is intended to be fairly realistic, but somewhat
on the conservative (pessimistic) side.

Before describing these assumptions in detail in the next section, it is important to note that taken together, they make a huge difference.
Table~\ref{tab:omp} illustrates this by showing the cosmological parameter constraints resulting from
using all the OPT assumptions, all the MID assumptions or all the PESS assumptions,  respectively.  
For example, combining CMB data from Planck and 21 cm data from FFTT,
the $\onesig$ uncertainty differs by a factor of 125 for $\Ok$ and by
a factor of 78 for $\mnu$ depending on assumptions.  It is therefore
important to sort out which of the assumptions contribute the most to
these big discrepancies, and which assumptions do not matter
much. This is a key goal of our chapter.

The rest of this chapter is organized as follows.  In Section
\ref{sec:assumptions}, we explain in detail the assumptions in the
same order as in Table \ref{tab:methodology}, and also present a new
method for modeling the ionization power spectra.  In Section
\ref{sec:results}, we quantify how the cosmological parameter
measurement accuracy depends on each assumption, and we derive simple
analytic approximations of these relations.  In Section
\ref{sec:conclusion}, we conclude with a discussion of the relative
importance of these assumptions, and implications for experimental
design.

\section{Forecasting Methods \& Assumptions}\label{sec:assumptions}

\subsection{Fundamentals of 21cm cosmology}
\subsubsection{Power spectrum of 21 cm radiation}\label{sec:review}

We review the basics of the 21 cm radiation temperature and power spectrum only briefly here, and refer the
interested reader to \cite{Furlanetto:2006jb} for a more comprehensive discussion of the relevant physics.
The difference between the observed 21 cm brightness temperature at the
redshifted frequency $\nu$ and the CMB temperature \Tcmb\ is \cite{field59a}
\beq{eqn:tb1}
T_b(\bfx) = \frac{3 c^3 h A_{10} n_H(\bfx) [\Ts(\bfx)-\Tcmb]}{32\pi k_B \nu_0^2 \Ts(\bfx) (1+z)^2 (dv_{\parallel}/dr)}\,,
\een
where \Ts\ is the spin temperature, $n_H$ is the number density of the neutral hydrogen gas, and
$A_{10}\approx 2.85\times 10^{-15}$s$^{-1}$ is the spontaneous decay rate of 21cm transition. The factor
$dv_{\parallel}/dr$ is the gradient of the physical velocity along the line of sight ($r$ is the
comoving distance), which is $H(z)/(1+z)$  on average (\ie for no peculiar velocity).  Here $H(z)$ is the
Hubble parameter at redshift $z$. The spatially averaged brightness temperature at redshift $z$ is (in units of $\milliK $)
\ben
\bar{T_b} \approx 23.88 \xH \left(\frac{\bar{\Ts} -\Tcmb}{\bar{\Ts}}\right) 
            \left(\frac{\Ob h^2}{0.02}\right) \left(\frac{0.15}{\Om h^2} \frac{1+z}{10} \right)^{1/2}\,,
\een
where $\xH$ is the mean neutral fraction, and $\bar{\Ts}$ is the averaged spin temperature.
If $\Ts \gg \Tcmb$ in the EoR, the 21cm emission should therefore be observed at the level of milli-Kelvins.  

To calculate the fluctuations, we rewrite \Eq{eqn:tb1} in terms of
$\delta$ (the hydrogen mass density fluctuation), $\delta_x$ (the
fluctuation in the ionized fraction), and the gradient of the peculiar
velocity $\partial v_r / \partial r$ along the line of sight, using
the fact that $dv_{\parallel}/dr = H(z)/(1+z) + \partial v_r /
\partial r$: 
\ben
T_b(\bfx) = \tilde{T}_b \left[
1-\xI(1+\delta_x)\right] (1+\delta) \left(1-\frac{1}{Ha}\frac{\partial
v_r }{ \partial r }\right)  \times \left(\frac{\bar{\Ts} -\Tcmb}{\bar{\Ts}}\right) \,. \label{eqn:tb2}
\een 
Here $\xI\equiv 1-\xH$ is the mean ionized fraction, and we have
defined $\tilde{T}_b\equiv \bar{T}_b/\xH\times [\bar{\Ts}/(\bar{\Ts}
-\Tcmb)]$.  We write $\delta_v \equiv (Ha)^{-1} \partial v_r /
\partial r$. In Fourier space, it is straightforward to show that, as
long as $\delta\ll 1$ so that linear perturbation theory is valid,
$\delta_v (\bfk) = -\mu^2 \delta$, where $\mu=\hat{\bfk}\cdot\hat{{\bf
n}}$ is the cosine of the angle between the Fourier vector $\bfk$ and
the line of sight.  In this chapter, we restrict our attention to the
linear regime.  We will also throughout this chapter assume $\Ts \gg
\Tcmb$ during the EoR, making the last factor in \Eq{eqn:tb2} unity
for the reasons detailed in Section \ref{sec:intro}.

In Fourier space, the power spectrum $\PDT(\bfk)$ of the 21cm fluctuations is defined by $\langle \Delta T_b^{\,*}(\bfk)  \Delta T_b(\bfk ') \rangle
\equiv  (2\pi)^3\delta^3 (\bfk-\bfk ') \PDT(\bfk)$, where $\Delta T_b$ is the deviation from the mean brightness
temperature.  It is straightforward to show from \Eq{eqn:tb2} that, to leading order,
\ben
 \PDT(\bfk)  =  \tilde{T}_b^2 \left\{ [\xH^2 \Pdd -2\xH \Pxd + \Pxx]  + 2\mu^2 [\xH^2 \Pdd - \xH \Pxd ] + \mu^4 \xH^2 \Pdd \right\}\,. \label{eqn:21cmpower}
\een
Here $\Pxx = \xI^2 P_{\delta_x \delta_x}$ and $\Pxd = \xI P_{\delta_x \delta}$ are the ionization power spectrum and the density-ionization power spectrum respectively.   
For convenience, we
define $\sPdd(k)\equiv \tilde{T}_b^2 \xH^2 \Pdd(k)$, $\sPxd(k) \equiv \tilde{T}_b^2 \xH \Pxd(k)$ and $\sPxx(k) \equiv \tilde{T}_b^2
\Pxx(k)$, so the total 21 cm power spectrum can be written as
three terms with different angular dependence:
\beq{eqn:fourmoments}
\PDT(\bfk) = \Pmuzero(k) + \Pmutwo(k) \mu^2 + \Pmufour(k) \mu^4,
\een
where 
\bena
\Pmuzero&=&\sPdd -2 \sPxd + \sPxx,\label{eqn:pmu0}\\ 
\Pmutwo&=&2(\sPdd - \sPxd),\label{eqn:pmu2}\\ 
\Pmufour&=&\sPdd.\label{eqn:pmu4}
\eena
Since $\Pmufour$ involves only the matter
power spectrum that depends only on cosmology, 
Barkana and Loeb \cite{barkana04a} argued that in principle, one can separate cosmology from astrophysical
``contaminants'' such as $\sPxx$ and $\sPxd$ whose physics is hitherto far from known. 
We will quantify the accuracy of this conservative approach (which corresponds to our PESS scenario for ionization
power spectrum modeling below) in Section \ref{sec:results}.

\subsubsection{From ${\bf u}$ to ${\bf k}$}

The power spectrum $\PDT(\bfk)$ and the comoving 
 vector ${\bf k}$ (the Fourier dual of the comoving position vector ${\bf r}$) are not directly measured by 21cm experiments.  
An experiment cannot directly determine which position vector ${\bf r}$ a signal is coming from, but instead which 
vector ${\bf \Theta} \equiv \theta_x \hat{e}_x +\theta_y \hat{e}_y + \Delta f \hat{e}_z$ it is coming from, where 
$(\theta_x,\theta_y)$ give the angular location on the sky plane, and $\Delta f$ is the frequency difference from the central redshift of a $z$-bin.
For simplicity, we assume that the sky volume observed is small enough that we can linearize 
the relation between ${\bf \Theta}$ and ${\bf r}$.
Specifically, we assume that the sky patch observed is much less than a radian across, 
so that we can approximate the sky as flat 
\footnote{The FFTT is designed for all-sky mapping (i.e. the field of view is of order $2\pi$).  However, 
since the angular scales from which we get essentially all our cosmological information are much smaller than a radian 
(with most information being on arcminute scales), the flat-sky approximation is accurate as long as
the data is analyzed separately in many small patches and the constraints are subsequently combined.}, 
and that separations in frequency near the mean redshift $z_*$ are approximately proportional to separations in comoving distance.
In these approximations, if there are no peculiar velocities, 
\bena
{\bf\Theta}_\perp	&=&\frac{{\bf r}_\perp}{d_A(z_*)} \,\label{eqn:thetaperp},\\ 
\Delta f		 &=&\frac{\Delta r_\parallel}{y(z_*)}.\label{eqn:thetaz}
\eena
Here ``$\perp$'' denotes the vector component perpendicular to the line of sight, 
{\it i.e.}, in the $(x,y)$-plane, and
$d_A$ is the comoving angular diameter distance given by \cite{KolbTurnerBook}
\ben\label{dAEq}
d_A(z)={c\over H_0}|\Omega_k|^{-1/2}S\left[|\Omega_k|^{1/2}\int_0^z \frac{dz'}{E(z')}\right],
\eeq
where 
\beq{EdefEq}
E(z)\equiv \frac{H(z)}{H_0} = \sqrt{\Om(1+z)^3 + \Ok(1+z)^2+\Ol},
\eeq
is the relative cosmic expansion rate and
the function $S(x)$ equals $\sin (x)$ if $\Ok<0$, $x$ if $\Ok=0$, and $\sinh x$ if $\Ok>0$. 
The conversion factor between comoving distances intervals and frequency intervals is 
\ben\label{yDefEq} 
y(z) = \frac{\lambda_{21}(1+z)^2}{H_0 E(z)},
\een
where $\lambda_{21}\approx 21$ cm is the rest-frame wavelength of the 21 cm line.  

We write the Fourier dual of ${\bf \Theta}$ as $\bfu \equiv u_x \hat{e}_x + u_y \hat{e}_y  + \upar \hat{e}_z$ ($\upar$ has units of time).  
The relation between $\bfu$ and $\bfk$ is therefore
\bena
\uperp&=&d_A {\bf k}_\perp\,,\label{eqn:uperp}\\ 
\upar &=&y\,\kpar\,.\label{eqn:uz}
\eena
In $\bfu$-space, the power spectrum $\PDT(\bfu)$ of 21cm signals is defined by 
$\langle \Delta \tilde{T}^{*}_b(\bfu) \Delta \tilde{T}_b(\bfu ')\rangle=(2\pi)^3 \delta^{(3)}(\bfu-\bfu')  \PDT(\bfu)  $,
and is therefore related to $\PDT(\bfk)$ by 
\beq{PuPkEq}
\PDT(\bfu)= {1\over d_A^2 y} \PDT(\bfk)\,.
\een

Note that cosmological parameters affect $\PDT(\bfu)$ in two ways: 
they both change $\PDT(\bfk)$ and alter
the geometric projection from ${\bf k}$-space to ${\bf u}$-space.
If $d_A$ and $y$ changed while $\PDT(\bfk)$ remained fixed, the observable power spectrum
$\PDT(\bfu)$ would be dilated in both the $\uperp$ and $\upar$ directions and rescaled in amplitude,
while retaining its shape.
Since both $d_A$ and $y$ depend on the three parameters  
$(\Ok,\Ol,h)$, and the Hubble parameter is in turn given by the parameters in Table~\ref{tab:omp}
via the identity $h=\sqrt{\Omega_m h^2/(1-\Ol-\Ok)}$, we see that these geometric effects provide information only
about our parameters $(\Ok,\Ol,\Omega_m h^2)$. 
Baryon acoustic oscillations in the power spectrum provide a powerful ``standard ruler'',
and the equations above show that if one generalizes to the dark energy to make $\Ol$ an arbitrary function of $z$, then 
the cosmic expansion history $H(z)$ can be measured separately at each 
redshift bin, as explored in \cite{Wyithe:2007rq,Mao:2007ti,Chang:2007xk}.
21 cm tomography information on our other cosmological
parameters ($\ns$, $\As$, $\Omega_b h^2$, $m_\nu$, $\alpha$, etc.) 
thus comes only from their direct effect on $\PDT(\bfk)$. 
Also note that $(\Ok,\Ol)$ affect $\PDT(\bfk)$ only by modulating the rate of linear perturbation growth,
so they alter only the amplitude and not the shape of $\PDT(\bfk)$.

If we were to use \Eq{PuPkEq} to infer $\PDT(\bfk)$ from the measured power spectrum $\PDT(\bfu)$
while assuming incorrect cosmological parameter values, then this geometric scaling
would cause the inferred $\PDT(\bfk)$ to be distorted by the so-called Alcock-Paczy\'nski (AP) effect
\cite{nusser04,barkanaAP} and not take the simple form of
Eqns.(\ref{eqn:fourmoments})-(\ref{eqn:pmu4}).
To avoid this complication, we therefore perform our Fisher matrix analysis directly in terms of 
$\PDT(\bfu)$, since this quantity is directly measurable without any cosmological assumptions.

The above transformations between ${\bf u}$-space and ${\bf r}$-space are valid when there are no peculiar velocities.
The radial peculiar velocities $v_r$ that are present in the real world induce the familiar redshift space distortions
that were discussed in Section \ref{sec:review}, causing $\mu^2$ and $\mu^4$ power spectrum anisotropies that were described there.

\subsection{Assumptions about \sPxx\ and \sPxd }
\label{sec:ion-model}

During the EoR, ionized bubbles (HII regions) in the IGM grow and
eventually merge with one another.  Consequently, \sPxx(k) and
\sPxd(k) contribute significantly to the total 21cm power spectrum.
The study of the forms of these two ionization power spectra has made
rapid progress recently, particularly through the semi-analytical
calculations \cite{furlanetto04a,zald04,mcquinn05,Zahn:2005fn} and radiative
transfer simulations \cite{McQuinn:2007dy,Zahn:2006sg}.  However, these models
depend on theoretically presumed parameters whose values cannot
currently be calculated from first principles.  From the experimental
point of view, it is therefore important to develop data analysis
methods that depend only on the most generic features of the
ionization power spectra. In this chapter, we consider three methods ---
our OPT, MID and PESS models --- that model $\sPxx$ and $\sPxd$ as
follows: 
\beq{eqn:allP0} {\rm (OPT)} \qquad\qquad \left\{
\begin{array}{lcl}
\sPxx(k) & = & 0 \\ 
\sPxd(k) & = & 0
\end{array}
\right.
\een

\beq{eqn:parn_pxxd} 
{\rm (MID)} \qquad\qquad 
\left\{ 
\begin{array}{lcl}
\sPxx(k) & = & \bsqxx \left[ 1+\alxx(k\,\Rxx) + \,(k\,\Rxx)^2\right]^{- {\gaxx \over 2}} \sPdd^{\rm (fid)}\\ 
\sPxd(k) & = & \bsqxd \,\exp{\left[-\alxd (k\,\Rxd)-(k\,\Rxd)^2\right]} 		 \sPdd^{\rm (fid)}
\end{array}
\right.
\een

\ben
{\rm (PESS)} \qquad\qquad
\left\{ 
\begin{array}{lcl}
\sPxx(k) & = & {\rm arbitrary} \\ 
\sPxd(k) & = & {\rm arbitrary}
\end{array}
\right.
\een
In the next three subsections, we explain these models in turn.

\subsubsection{OPT model}

It is likely that before reionization started (while $\xH=1$ and $\sPxx=\sPxd=0$), 
hydrogen gas had already been sufficiently heated that $\Ts \gg \Tcmb$. 
In this regime, \Eq{eqn:allP0} holds. This OPT scenario
is clearly the simplest model, since the total 21cm power spectrum is simply proportional to $\sPdd$:
$\PDT(\bfk)=\sPdd(k) (1+\mu^2)^2$.  To forecast the $\onesig$ error, we use the Fisher matrix
formalism \cite{tegmark97b}.  
Repeating the derivation in \cite{galfisher},
the Fisher matrix for cosmological parameters $\lambda_a$ ($a=1,\ldots,N_p$) is 
\beq{FisherIntegralEq} 
{\bf F}_{ab} = \frac{1}{2}\int 
\left(\frac{\partial\ln\PDT^{\rm tot}(\bfu)}{\partial\lambda_a}\right)  
\left(\frac{\partial\ln\PDT^{\rm tot}(\bfu)}{\partial\lambda_b}\right) 
V_{\Theta} \frac{d^3 u}{(2\pi)^3},
\eeq
where the integral is taken over the observed part of ${\bf u}$-space, 
and $\PDT^{\rm tot}(\bf u)$ denotes the combined power spectrum from cosmological 
signal and all forms of noise.  Here $V_{\Theta} = \Omega \times B$ is the volume of
 the ${\bf \Theta}$-space where $\Omega$ is the solid angle within the field of view (f.o.v.) and $B$ is the frequency size of a $z$-bin.  
The Fisher matrix determines the parameter errors as $\Delta\lambda_a = \sqrt{({\bf F}^{-1})_{aa}}$. 

For computational convenience, we subdivide $u$-space into pixels so small that the 
power spectrum is roughly constant in each one,
obtaining 
\beq{eqn:FM4cp} 
{\bf F}_{ab}\approx\sum_{\rm pixels} \frac{1}{[\delta \PDT(\bfu)]^2}\left(\frac{\partial \PDT(\bfu)}{\partial \lambda_a}\right)  \left(\frac{\partial \PDT(\bfu)}{\partial \lambda_b}\right),
\een
where the power spectrum measurement error in a pixel at $\bfu$ is 
\beq{OPT_P_err_eq}
\delta\PDT(\bfu)=\frac{\PDT^{\rm tot}(\bfu)}{N_c^{1/2}} =  \frac{\PDT(\bfu)+P_N(u_\perp)}{N_c^{1/2}}.
\eeq
Here $P_N(u_\perp)$ is the noise power spectrum and will be discussed in detail in Section \ref{sec:noise};
note that it is independent of $u_\parallel$ and depends only on $u_\perp$ through the
baseline distribution of the antenna array.
\beq{NcDefEq}
N_c=2\pi k^2\sin\theta \Delta k \Delta\theta \times {\rm Vol}/(2\pi)^3
\eeq
is the number of independent cells in an annulus summing over
the azimuthal angle.  We have the factor $\sqrt{1/N_c}$ in $\delta\PDT$ instead of the normal $\sqrt{2/N_c}$ because we only
sum over half the sphere.

\begin{table}
\centering
\footnotesize{
\caption[Fiducial values of ionization parameters]{\label{tab:ri-fid} Fiducial values of ionization parameters adopted for Figure \ref{fig:ri}.  $\Rxx$ and $\Rxd$ are in units of $\Mpc$, while other parameters are unitless. }
\begin{tabular}{lcccccccc}
\hline\hline
$z\quad\quad$     & \xH  & \bsqxx   & \Rxx     &  \alxx   & 
\gaxx    & \bsqxd  & \Rxd     &  \alxd   \\ \hline

9.2  & 0.9 & 0.208 & 1.24 & -1.63 & 0.38 & 0.45 & 0.56  & -0.4 \\
8.0  & 0.7 & 2.12  & 1.63 & -0.1  & 1.35 & 1.47 & 0.62  & 0.46 \\
7.5  & 0.5 & 9.9   & 1.3  & 1.6   & 2.3  & 3.1  & 0.58  & 2.   \\
7.0  & 0.3 & 77.   & 3.0  & 4.5   & 2.05 & 8.2  & 0.143 & 28.  \\ 
\hline\hline

\end{tabular}
}
\end{table}

\begin{figure}[ht]
\centering
\begin{displaymath}
\begin{array}{cccc} 
  \includegraphics[width=0.24\textwidth]{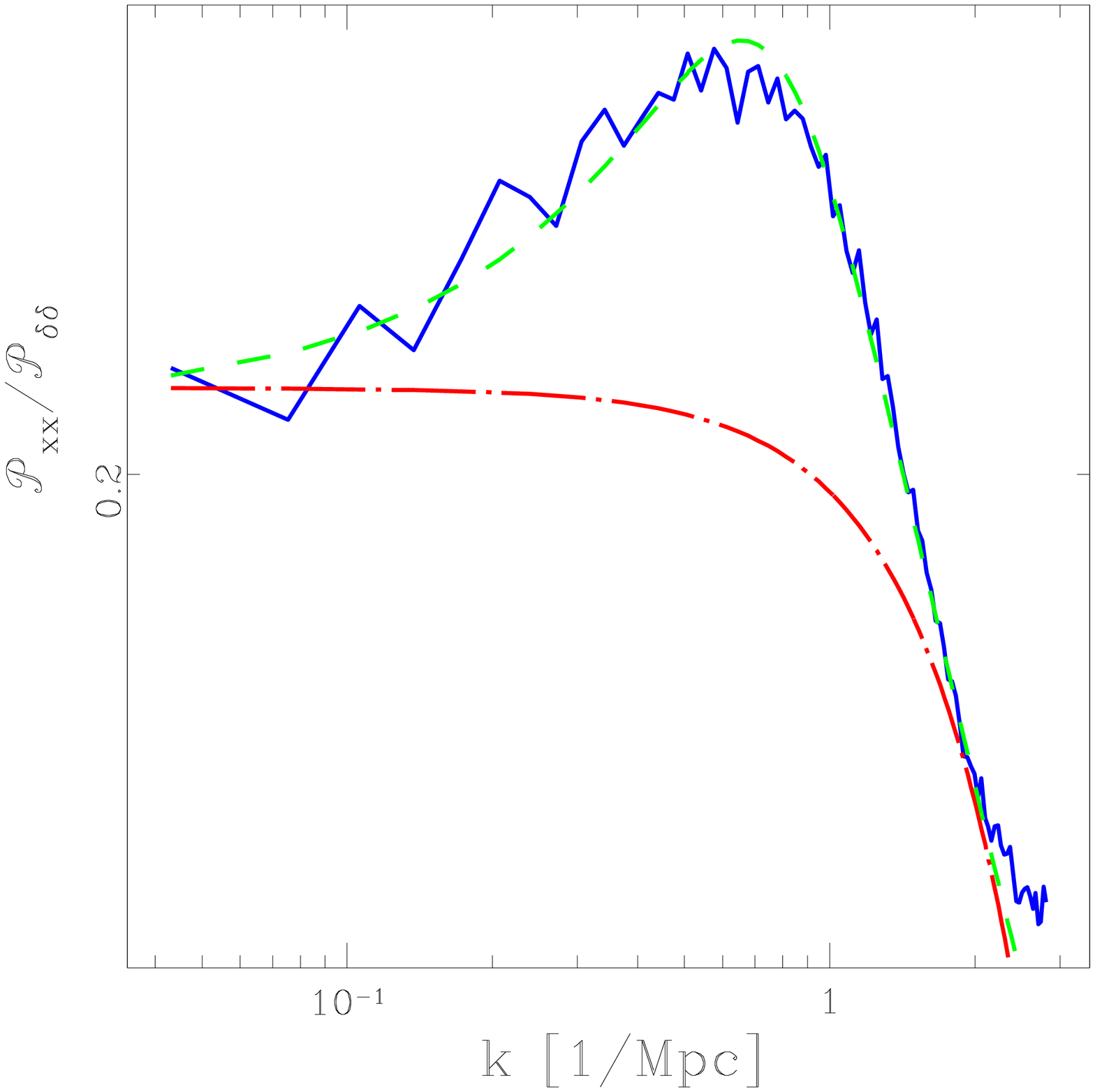} & 
  \includegraphics[width=0.24\textwidth]{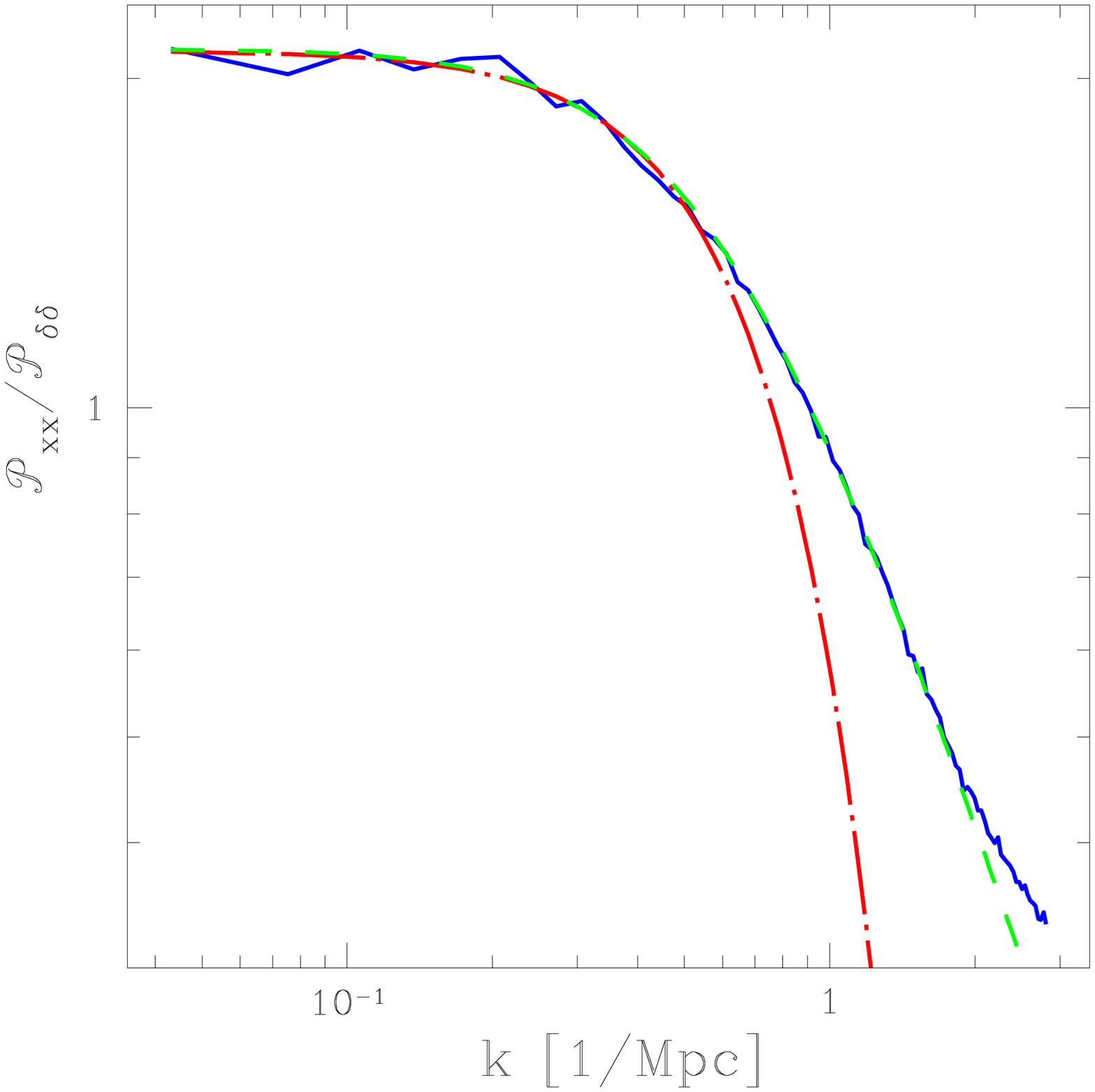} & 
  \includegraphics[width=0.24\textwidth]{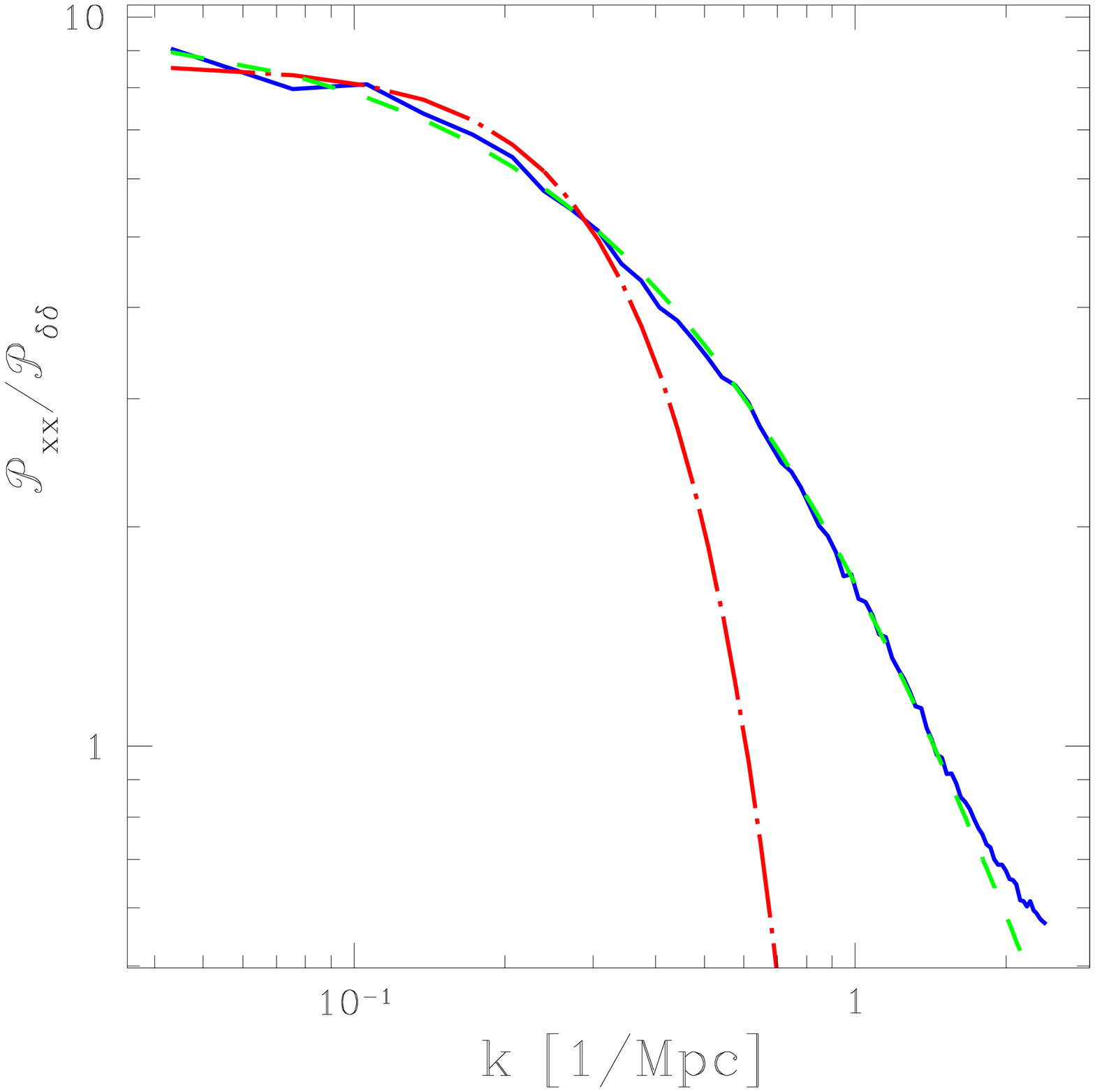} & 
  \includegraphics[width=0.24\textwidth]{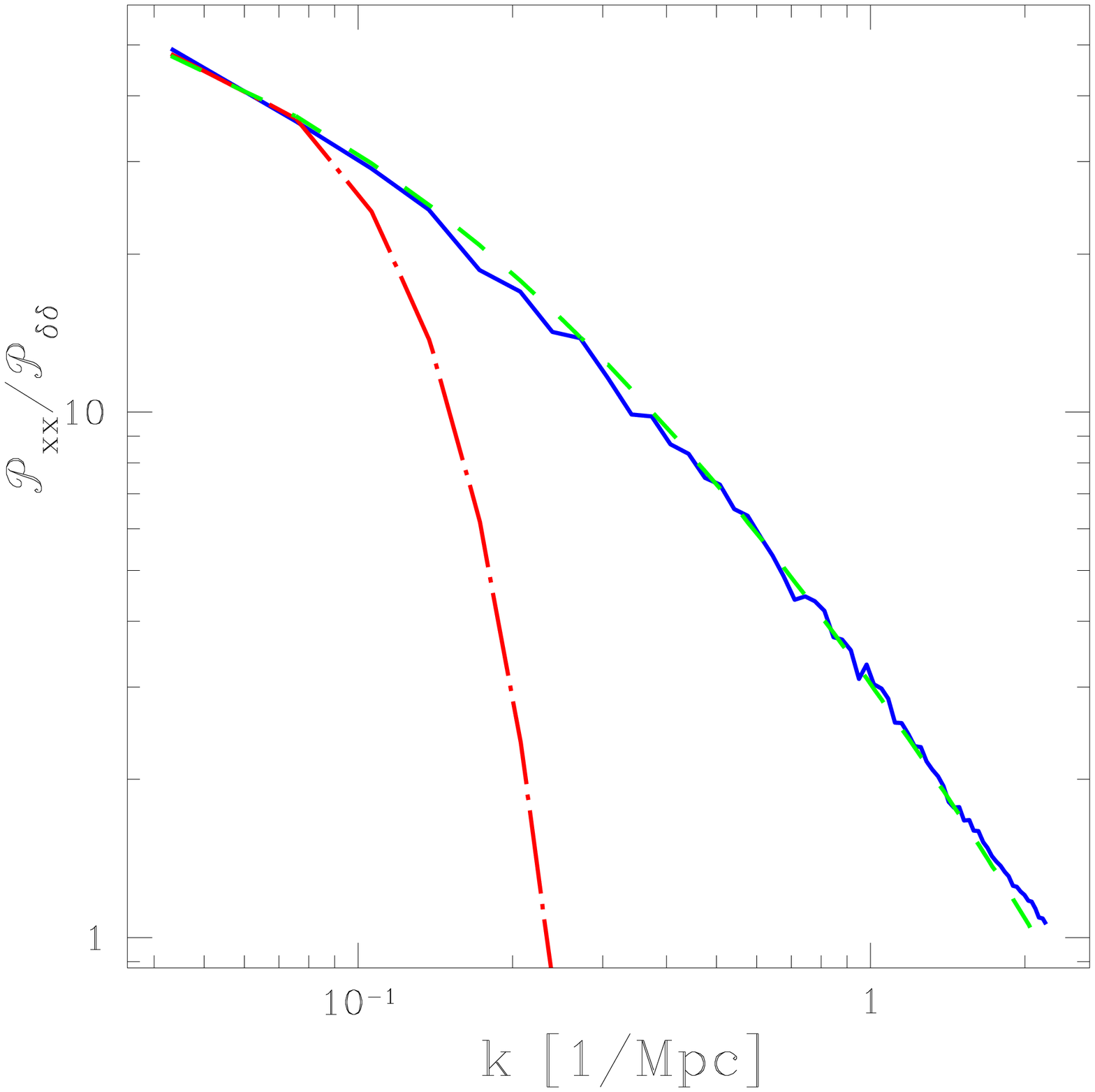} \\
  \includegraphics[width=0.24\textwidth]{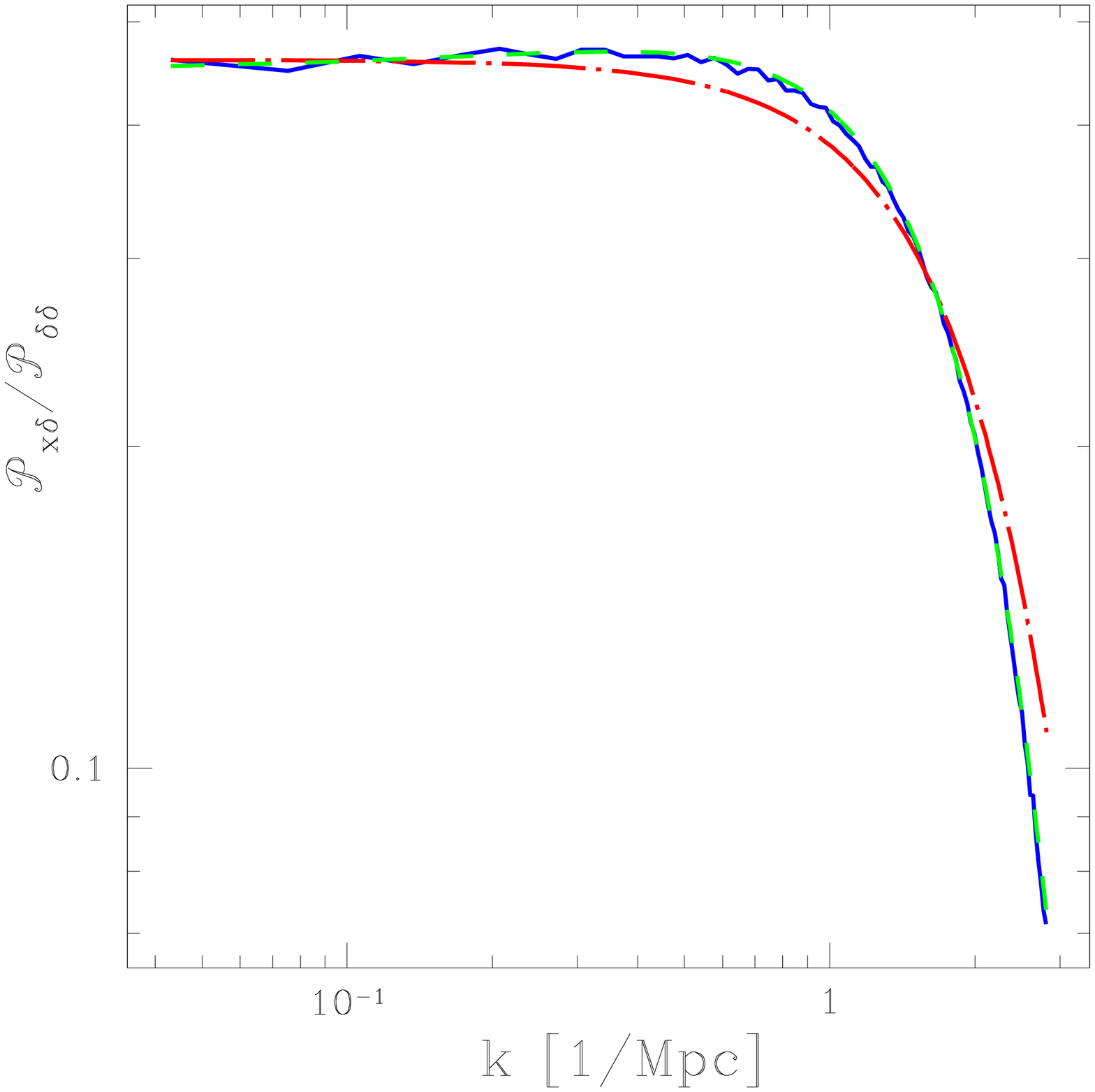} &  
  \includegraphics[width=0.24\textwidth]{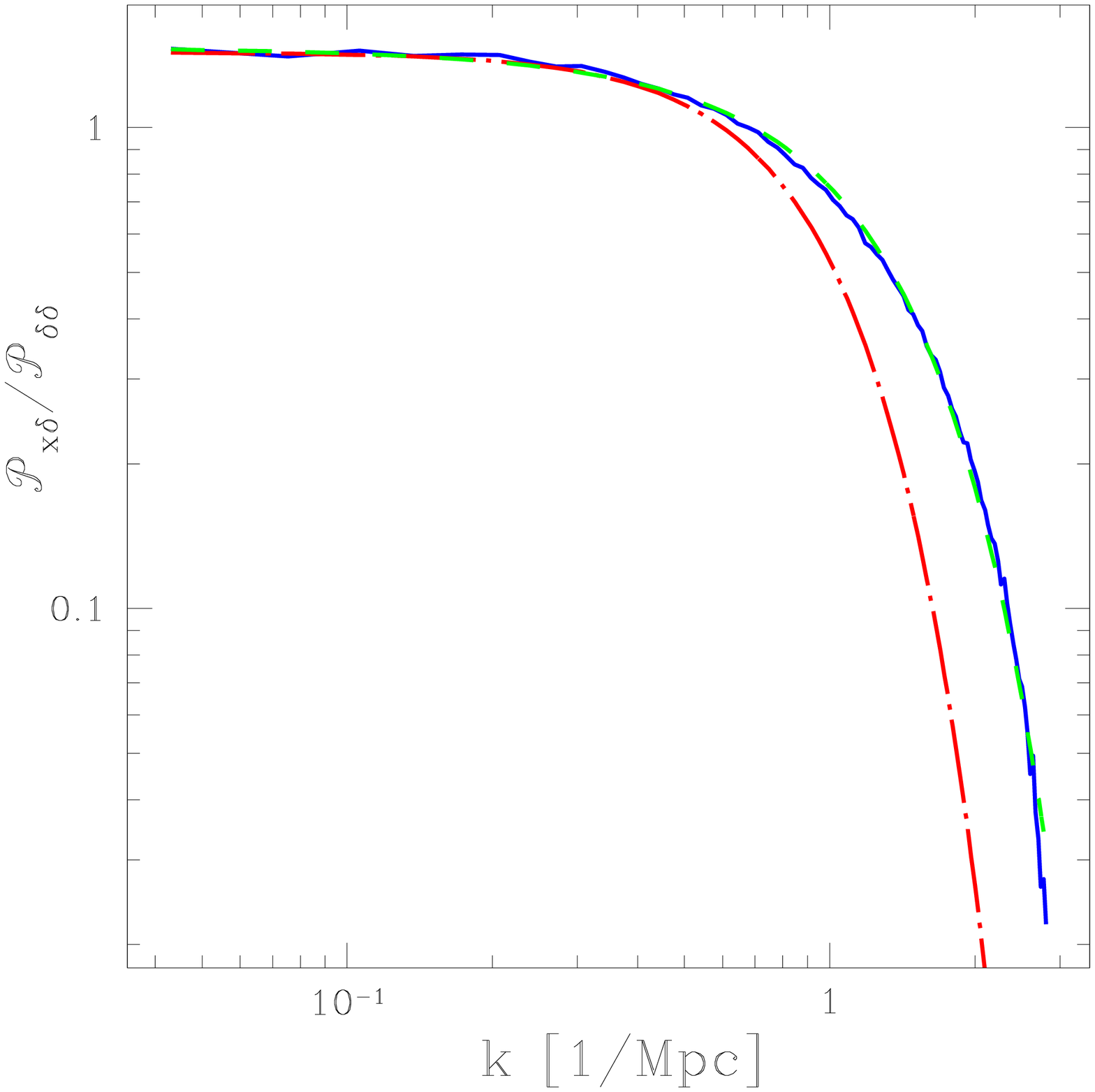} &  
  \includegraphics[width=0.24\textwidth]{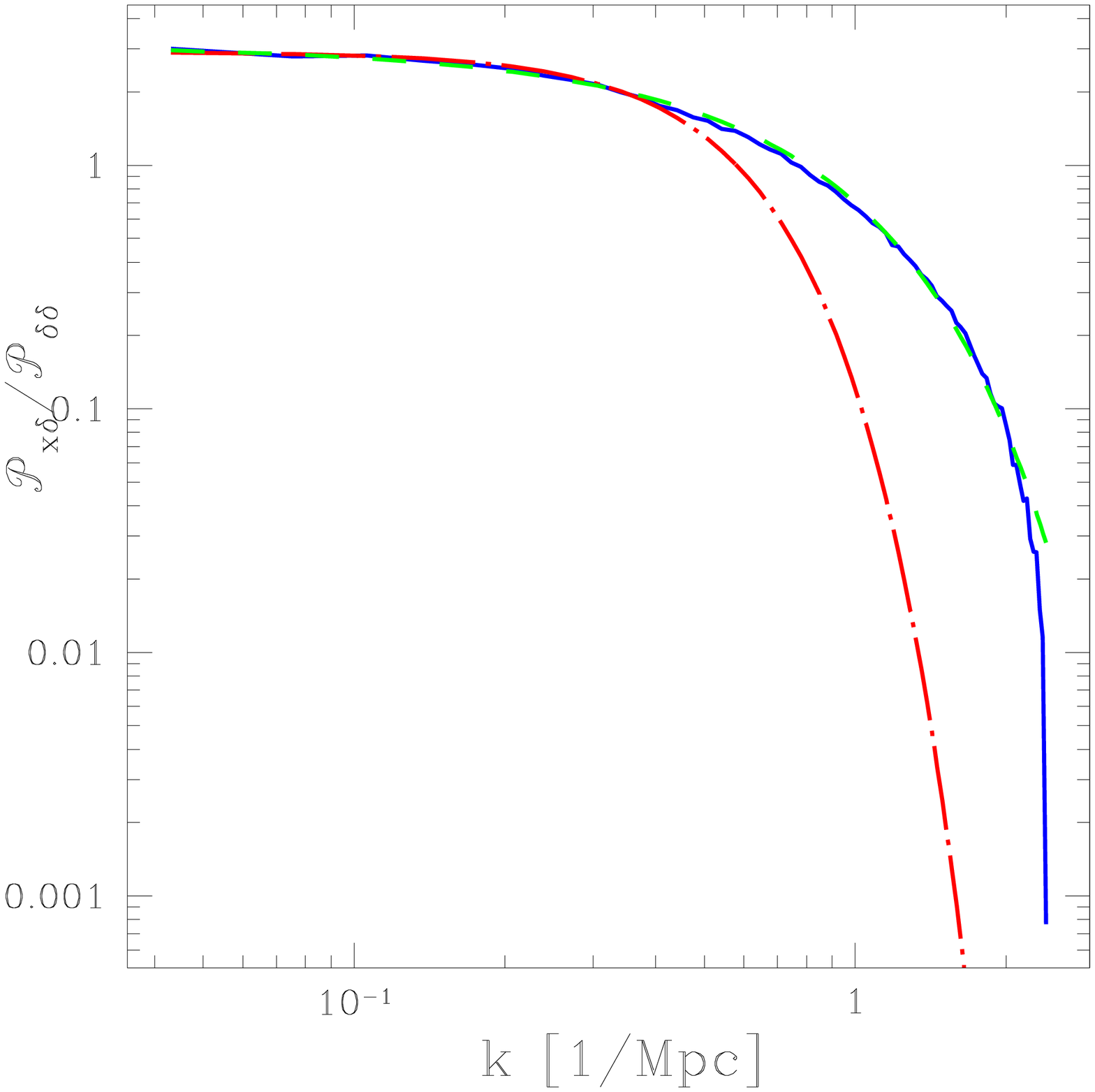} &  
  \includegraphics[width=0.24\textwidth]{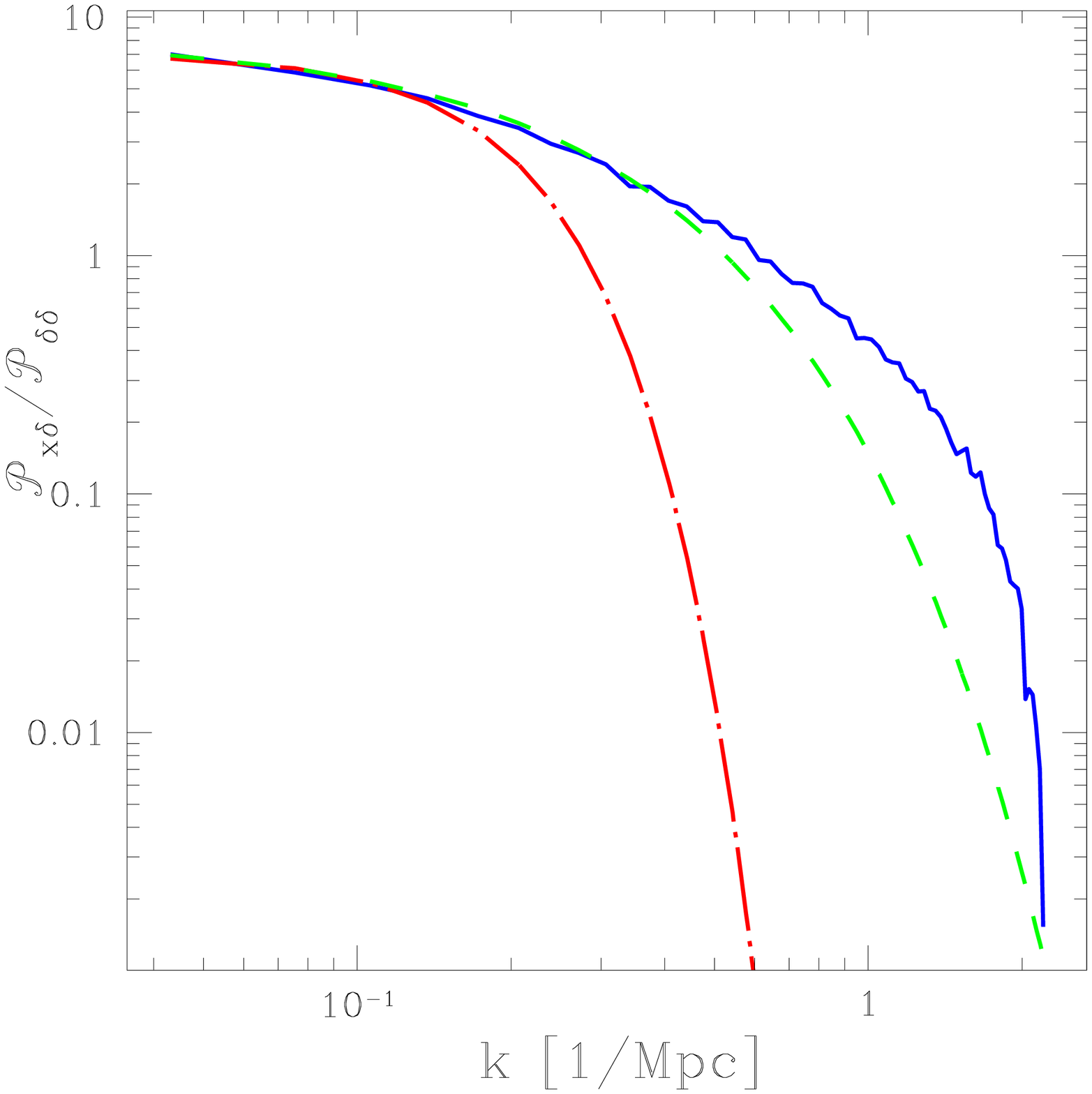} 
\end{array}
\end{displaymath}
\caption[Parametrization of ionization power spectra and the fit to simulation results]{ Fits to the ionization power spectra at several redshifts.
Solid (blue) lines are the results of the radiative transfer
simulation in Model I of the McQuinn \etal paper
\cite{McQuinn:2007dy}.  Dashed (green) lines are fitting curves of our
parametrization. Dot-dashed (red) lines are best fits using the
parametrization suggested by Santos and Cooray \cite{Santos:2006fp} .
Top panels: $ \sPxx/\sPdd=\Pxx/(\xH^{2} \Pdd)$.  Bottom panels: $
\sPxd/\sPdd=\Pxd/(\xH \Pdd)$.  From left to right:
$z=9.2,\,8.0,\,7.5,\,7.0$ ($\xI=0.10,\, 0.30,\, 0.50,\, 0.70$
respectively).}
\label{fig:ri}
\end{figure}

\subsubsection{MID model}\label{sec:asmp-mid}

After reionization starts, both ionization power spectra $\sPxx$ and
$\sPxd$ make significant contribution to the total 21cm power
spectrum.  We explore two different analysis methods --- our MID and
PESS models --- for separating the cosmological signal from these
astrophysical contaminants (\ie $\sPxx$ and $\sPxd$).

Our MID model assumes that both ionization power spectra $\sPxx(k)$
and $\sPxd(k)$ are smooth functions of $k$ which can be parametrized
by a small number of nuisance parameters $\beta_1,\ldots,\beta_{n_{\rm
ion}}$ related to reionization physics.  Combining these ionization
parameters with our cosmological ones $\lambda_a$ into a larger
parameter set $p_\alpha$ ($\alpha=1,\ldots,N_p+n_{\rm ion}$), we can
jointly constrain them by measuring $\PDT(\bfu)$.

In Appendix \ref{chi2} we will describe a $\chi^2$ goodness-of-fit
test for quantifying whether this parametrization is valid.  The
Fisher matrix for measuring $p_\alpha$ is simply \beq{MIDfisherEq}
F_{\alpha\beta} = \sum_{\rm pixels}
\frac{1}{[\delta\PDT(\bfu)]^2}\frac{\partial \PDT(\bfu)}{\partial
p_\alpha}\frac{\partial \PDT(\bfu)}{\partial p_\beta} \,.  \eeq This
Fisher matrix $F_{\alpha\beta}$ is not block diagonal, {\it i.e.},
there are correlations between the cosmological and ionization
parameters, reflecting the fact that both affect $\sPxx(k)$ and
$\sPxd(k)$.  The inversion of the Fisher matrix therefore leads to the
degradation of the constraints of cosmological parameters.  However,
the total 21cm power spectrum is usually smaller in magnitude in the
MID model than in the OPT model (see \Eq{eqn:21cmpower}), giving less
sample variance.  This means that as long as noise in a 21cm
experiment dominates over sample variance, the MID model will give
weaker constraints than the OPT model, because of the degeneracies.
For future experiments with very low noise, however, it is possible to
have the opposite situation, if the reduction in sample variance
dominates over the increase in degeneracy.  This does of course not
mean that the MID model is more optimistic than the OPT model, merely
that the OPT model is assuming an unrealistic power spectrum.

Having set up the general formalism, we now propose the specific
parametrization specified by \Eq{eqn:parn_pxxd}, with fiducial values
of ionization parameters given in Table \ref{tab:ri-fid}.  This
parametrization was designed to match the results of the radiative
transfer simulations in Model I of \cite{McQuinn:2007dy}, and
Figure~\ref{fig:ri} shows that the fit is rather good in the range
$k=0.1-2 \; \Mpc^{-1}$ to which the 21cm experiments we consider are most
sensitive.

The radiative transfer simulations implemented in
\cite{McQuinn:2007dy} are post processed on top of a $1024^3$ N-body
simulation in a box of size $186\,\Mpc$. Three models for the
reionization history are considered in \cite{McQuinn:2007dy}:
\begin{enumerate}
\item In Model I, all dark matter halos above $m_{\rm cool}$
(corresponding to the minimum mass galaxy in which the gas can cool by
atomic transitions and form stars, \eg $m_{\rm cool}\approx
10^8\,M_{\odot}$ at $z=8$) contribute ionizing photons at a rate that
is proportional to their mass.
\item In Model II, the ionizing luminosity of the sources scales as halo mass to
the $5/3$ power, \ie more massive halos dominate the production of ionizing photons than in Model I.
\item In Model III, which has the same source parametrization as in
Model I except for doubled luminosity, minihalos with $m>10^5 M_{\odot}$ absorb incident ionizing photons out to their virial radius unless they are
photo-evaporated (but do not contribute ionizing photons).
\end{enumerate}

It appears to be a generic feature in the simulation results that the ratios of functions at large $k$ fall off like a power law for
$\sPxx(k)/\sPdd(k)$, and exponentially for $\sPxd(k)/\sPdd(k)$.  At small $k$, $\sPxx(k)/\sPdd(k)$ can either increase or decrease
approximately linearly as $k$ increases, while $\sPxd(k)/\sPdd(k)$ is asymptotically constant.  Our parametrization in
\Eq{eqn:parn_pxxd} captures these features: at large $k$, $\sPxx(k)/\sPdd(k) \propto k^{-\gaxx}$ and $\sPxd(k)/\sPdd(k) \propto \exp{(-(k\,\Rxd)^2)}$; at small $k$, $\sPxx(k)/\sPdd(k) \propto \left(1- (\gaxx \, \alxx\Rxx/2)\,k\right)$, and $\sPxd(k)/\sPdd(k)
\propto \left(1- \alxd\,\Rxd\,k \right)$ (both $\alxx$ and $\alxd$ can be either positive or negative).
Figure~\ref{fig:ri} also shows that for $\Pxx(k)$ and also for $\Pxd(k)$ at large $k$, 
our parametrization further improves over the 
parametrization $P(k)/\Pdd = b^2 e^{-(k\,R)^2}$
suggested by Santos and Cooray \cite{Santos:2006fp}, which works well for $\Pxd(k)$ at small $k$.

To be conservative, we discard cosmological information from $\sPxd(k)$ and $\sPxx(k)$ in our Fisher matrix analysis by using the fiducial power spectrum $\sPdd(k)^{\rm (fid)}$ rather than the actual one $\sPdd(k)$ in \Eq{eqn:parn_pxxd}.
This means that the derivatives of $\sPxd(k)$ and $\sPxx(k)$ with respect to the cosmological parameters vanish in \Eq{MIDfisherEq}.
It is likely that we can do better in the future: once  
the relation between the ionization power spectra and the matter power spectrum can be reliably calculated either analytically or numerically, 
the ionization power spectra can contribute to further constraining cosmology.    

In addition to the fit of Model I shown in Figure \ref{fig:ri}, we also fit our model (with different fiducial values from those listed in Table \ref{tab:ri-fid}) to the simulations using Model II and III in \cite{McQuinn:2007dy}, and find that the parametrization is flexible enough to provide good fits to all three simulations,  
suggesting that the parametrization in
\Eq{eqn:parn_pxxd} may be generically valid and independent of models.  
Note, however, that at low redshifts ($\xI \gtrsim 0.7$), our parametrization of $\sPxd/\sPdd$  does not work well at large $k$, in that the
simulation falls off less  rapidly than exponentially.  This may be because when HII regions dominate the IGM, the ionized bubbles  overlap in complicated
patterns and correlate extremely non-linearly at small scales.  This partial incompatibility indicates that our parametrization (\ie Eq.\ref{eqn:parn_pxxd}) is only accurate for small
$\xI$, \ie before non-linear ionization patterns come into play.

In the remainder of this chapter, we will adopt the values in Table \ref{tab:ri-fid} as  fiducial values of the ionization parameters.

\subsubsection{PESS model}

By parametrizing the ionization power spectra with a small number of
constants, the MID model rests on our understanding of the physics of
reionization.  From the point of view of a maximally cautious
experimentalist, however, constraints on cosmological parameters
should not depend on how well one models reionization. In this spirit,
Barkana and Loeb \cite{barkana04a} proposed what we adopt as our
``PESS'' model for separating the physics $\sPdd(k)$ from the
``gastrophysics'' $\sPxx(k)$ and $\sPxd(k)$. Instead of assuming a
specific parametrization, the PESS model
makes no {\it a priori} assumptions about the ionization power spectra.
In each $k$-bin that contains more than three pixels in ${\bf u}$-space, one can in principle separate $\Pmufour(k)=\sPdd(k)$ from the other two moments.  
The PESS model essentially only constrains cosmology from the $\Pmufour$ term and therefore loses all information in $\Pmuzero$ and $\Pmutwo$.  
We now set up the Fisher matrix formalism for the PESS model that takes advantage of the anisotropy in $\PDT(\bfk)$ arising from the velocity field effect.  Numerical
evaluations will be performed in Section \ref{sec:result-power}.  

The observable in 21cm tomography is the brightness temperature $T_b(\bfx)$. In Fourier space, the covariance matrix between two pixels $\bfk_i$ and $\bfk_j$ is
${\bf C}_{ij}=\delta_{ij}[\PDT(\bfk_i)+P_N(k_{\perp})]$, assuming that the measurements in two different pixels are uncorrelated\footnote{We ignore here a 
$\delta$-function centered at the origin since 21cm experiments will not measure any $k=0$ modes.}.     
The total 21cm power spectrum is $\PDT(\bfk) = \Pmuzero(k) + \Pmutwo(k)
\mu^2 + \Pmufour(k) \mu^4$.  For convenience, we use the shorthand notation $P_A$, where $P_1\equiv\Pmuzero$, 
$P_2\equiv\Pmutwo$ and $P_3\equiv \Pmufour$ and define
the $a_A=0,2,4$ for $A=1,2,3$, respectively.  Thus the power spectrum can be rewritten as $\PDT=\sum_{A=1}^{3} P_A \mu^{a_A}$.  
Treating $P_A(k)$ at each $k$-bin as
parameters, the derivatives of the covariance matrix are simply $\partial {\bf C}_{ij}/ \partial P_A(k) = \delta_{ij} \mu^{a_A}$, 
where $|\bfk_i|$ resides in the shell of radius
$k$ with width $\Delta k$. 
Since the different $k$-bins all decouple, the Fisher matrix 
for measuring the moments $P_A(k)$ is simply a separate $3\times 3$-matrix for each $k$-bin:
\bena
F_{AA'} (k) & = & \frac{1}{2}\hbox{tr}\left[{\bf C}^{-1}\frac{\partial {\bf C}}{\partial P_A(k)}{\bf C}^{-1}\frac{\partial {\bf C}}{\partial P_{A'}(k)}\right] \nonumber\\
& = &\sum_{\textrm{upper half-shell}} \frac{\mu^{a_A + a_{A'}}}{[\delta \PDT(\bfk)]^2}\,, \label{eqn:Fpess}
\eena
where $\delta \PDT(\bfk) = N_c^{-1/2}\left[ \PDT(\bfk) + P_N(k_{\perp})\right]$.  
Here $P_N(k_{\perp})$ is related to $P_N(u_\perp)$ by \Eq{PuPkEq}.  
Again the sum is over the upper half of the spherical shell $k<|\bfk|<k+\Delta k$.  
The $\onesig$ error of $P_3=\Pmufour $ is $\delta P_3(k) = \sqrt{F^{-1}_{\phantom{-1}33}(k)}$. 
Once $\sPdd=\Pmufour$ is separated from other moments, $\sPdd$ can be used to constrain cosmological parameters $\lambda_a$ with the Fisher matrix as given in \Eq{eqn:FM4cp}.

We have hitherto discussed the anisotropy in $\PDT(\bfk)$ that arises from the velocity field effect.
However, the AP-effect may further contribute to the anisotropy in that it
creates a $\mu^6$-dependence and modifies the $\mu^4$ term \cite{nusser04,barkanaAP}.  
The AP-effect can be distinguished from the velocity field effect since the $\Pmusix$ term is unique to the AP-effect. 
Thus, one can constrain cosmological parameters from $\Pmufour$ and $\Pmusix$ \cite{McQuinn:2005hk}, 
involving the inversion of a $4\times 4$ matrix which loses even more information and therefore further weakens constraints.  
Therefore, the PESS Fisher matrix that we have derived without taking AP-effect into account can be viewed as an upper bound on how 
well the PESS approach can do in terms of cosmological parameter constraints.
However, this maximally conservative $4\times 4$ matrix approach may be inappropriately pessimistic, since the AP-induced clustering anisotropy is typically very small within the observationally allowed
cosmological parameter range, whereas the velocity-induced anisotropies can be of order unity.

\subsection{Assumptions about Linearity}

To avoid fitting to modes where $\delta_k$ is non-linear and physical
modeling is less reliable, we impose a sharp cut-off at $\kmax$ and
exclude all information for $k>\kmax$. We take $\kmax = 2\,\perMpc$
for our MID model, and investigate the $\kmax$-dependence of
cosmological parameter constraints in Section \ref{sec:kmax}.  

\subsection{Assumptions about non-Gaussianity}
\label{sec:non-gauss}

Non-Gaussianity of ionization signals generically becomes important at
high $\xI$. With a large volume, high resolution simulations of cosmic
reionization, Lidz \etal \cite{Lidz:2006vj} and Santos \etal
\cite{Santos:2007dn} found non-negligible (a factor of 1.5)
differences in the full power spectrum at high $\xI$ ($\xI \gtrsim
0.35)$).  To get a rough sense of the impact of non-Gaussianity on
cosmological parameter constraints, we simply model it as increasing
the sample variance by a factor $\xi$.  We thus write the total power
spectrum as \beq{xiDefEq}
\delta\PDT(\bfu)=N_c^{-1/2}\,[\,\xi\PDT(\bfu)+P_N(u_\perp)]\,, \eeq
where $\xi$ is the factor by which the the sample variance is
increased.  The parameter $\xi$ should take the value $\xi \approx 1$
(Gaussian) at epochs with low $\xI$ and $1 < \xi \lesssim 2$
(non-Gaussian) at high $\xI$.

\subsection{Assumptions about reionization history and redshift range}
\label{sec:asmp-history}

21cm tomography can probe a wide range of redshifts, as illustrated in
Figure~\ref{fig:spheres}.  However, one clearly cannot simply measure
a single power spectrum for the entire volume, as the clustering
evolves with cosmic time: The matter power spectrum changes gradually
due to the linear growth of perturbations \cite{tegmark05}.  More importantly, the
ionization power spectra vary dramatically with redshift through the
epoch of reionization.  We incorporate these complications by
performing our analysis separately in a series of redshift slices,
each chosen to be narrow enough that the matter and ionization power
spectra can be approximated as constraint within each slice.  This
dictates that for a given assumed reionization history, thinner
redshift slices must be used around redshifts where $\xH$ varies
dramatically.

In this chapter, we will consider two rather opposite toy models in Section \ref{sec:results}:
\begin{itemize}
\item OPT: A sharp reionization that begins and
finishes at one redshift (say $z\lesssim 7$).
\item MID/PESS: A gradual reionization that spanning a range of redshifts,
assuming the ionization parameter values that fit Model I simulation of  the McQuinn \etal paper \cite{McQuinn:2007dy}
\end{itemize}
For the latter scenario, the ionization fraction $\xH$ is not a linear function of redshift. For example, in 
in the McQuinn \etal \cite{McQuinn:2007dy} simulation, $\xH=$0.9, 0.7, 0.5 and 0.3 correspond to redshifts 
$z=9.2$, 8.0, 7.5 and 7.0, respectively.
For our different scenarios, we therefore adopt the redshift ranges $6.8<z<10$ that are divided into four redshift slices centered at the above redshifts (OPT),
$6.8<z<8.2$ split into three bins centered at $z$=7.0, 7.5 and 8.0 (MID), 
$7.3<z<8.2$ split into two slices centered at $z=7.5$ and 8.0. 

\subsection{Assumptions about cosmological parameter space}\label{sec:cosmology}

Since the impact of the choice of cosmological parameter space and related degeneracies has been extensively studied in the literature,
we will perform only a basic analysis of this here.
We work within the context of standard inflationary cosmology with adiabatic perturbations, and parametrize cosmological models in terms of 12
parameters (see, \eg, Table 2 in \cite{Tegmark:2006az} for explicit definitions) whose fiducial values are assumed as follows: 
$\Ok = 0$ (spatial curvature), $\Ol = 0.7$ (dark energy density), $\Ob = 0.046$ (baryon density),
$h = 0.7$ (Hubble parameter $H_0 \equiv 100 h \,{\rm km}\,{\rm s}^{-1}\,{\rm Mpc}^{-1}$), $\tau = 0.1$ (reionization optical depth), $\On=0.0175$
(massive neutrino density), $\ns=0.95$ (scalar spectral index), $A_s = 0.83$ (scalar fluctuation amplitude), $r=0$ (tensor-to-scalar ratio), $\al=0$
(running of spectral index), $n_t=0$ (tensor spectral index) and $w=-1$ (dark energy equation of state).  
We will frequently use the term ``vanilla'' to refer to the minimal model space parametrized by $(\Ol,\om,\ob,\ns,\As,\tau)$ combined with $\xH(z)$ and
ionization parameters at all observed $z$-bins, setting $\Ok,\on,r,\al,\nt$, and $w$ fixed at their fiducial values.  

\begin{table}
\centering
\footnotesize{
\caption{Specifications for 21cm interferometers\label{tab:spec}}
\begin{tabular}{ccccc}
\hline\hline
Experiment & $N_{\rm ant}$ & Min. baseline (m) & f.o.v. (${\rm deg}^2$) & $A_e$ (m$^2$) at z=6/8/12\footnote{We 
assume that the effective collecting area is proportional to $\lambda^2$ 
such that the
sensitivity ($A_e/T_{\rm sys}$ in ${\rm m}^2 {\rm K}^{-1}$) meets the design specification.}  \\ \hline
MWA   & 500 & 4 & $\pi\,16^2$ & 9/14/18 \\ 
SKA   & 7000 & 10 & $\pi\,8.6^2$ & 30/50/104 \\
LOFAR & 77 & 100 & $2\times\pi\,2.4^2$ & 397/656/1369 \\
FFTT  & $10^6$ & 1 & $2\pi$ & 1/1/1 \\
\hline\hline
\end{tabular}
}
\end{table}

\subsection{Assumptions about Data}\label{sec:experiments}

The MWA, LOFAR, SKA and FFTT instruments are still in their planning/design/development stages. 
In this chapter, we adopt the key design parameters from \cite{bowman05cr}
for MWA, \cite{Schilizzi07} and www.skatelescope.org for SKA, www.lofar.org for LOFAR, and \cite{FFTT} for FFTT unless explicitly stated.  

\subsubsection{Interferometers} 

We assume that MWA will have 500  
correlated $4{\rm m} \times 4{\rm m}$ antenna tiles, each with 16
dipoles.  Each individual tile will have an effective collecting area
of $14 \,{\rm m}^2$ at $z=8$ and $18\, {\rm m}^2$ at $z\gtrsim 12$.
LOFAR will have 77 large (diameter $\sim 100 \,{\rm m}\,) $ stations,
each with thousands of dipole antennae such that it has the collecting
area nearly 50 times larger than each antenna tile of MWA.  Each
station can simultaneously image $N$ regions in the sky.  We set $N=2$
in this chapter but this number may be larger for the real array.  The
design of SKA has not been finalized.  We assume the ``smaller
antennae'' version of SKA, in which SKA will have 7000 small antennae,
much like MWA, but each panel with much larger collecting area.  FFTT
stands for Fast Fourier Transform Telescope, a future square kilometer
array optimized for 21 cm tomography as described in \cite{FFTT}.
Unlike the other interferometers we consider, which add in phase the
dipoles in each panel or station, FFTT correlates all of its dipoles,
resulting in more information.  We evaluate the case where FFTT contains
a million $1{\rm m} \times 1{\rm m}$ dipole antennae in a contiguous
core subtending a square kilometer, providing a field-of-view of
$2\pi$ steradians.

For all interferometers, we assume that the collecting area $A_e \propto \lambda^2$, like a simple dipole, except that $A_e$ is saturated at $z\sim 12$ in MWA
since the wavelength $\lambda=21(1+z)\,{\rm cm}$ exceeds the physical radius of an MWA antenna panel.  The summary of the detailed specifications adopted in this
chapter is listed in Table \ref{tab:spec}. 

\subsubsection{Configuration}
\label{configuration}

The planned configurations of the above-mentioned interferometers are
quite varied. However, all involve some combination of the following
elements, which we will explore in out calculations:
\begin{enumerate}
\item A {\it nucleus} of radius $R_0$ within which the area coverage fraction is close to 100\%.
\item A {\it core} extending from radius $R_0$ our to $R_{\rm in}$ where there coverage density drops like some power law $r^{-n}$.
\item An {\it annulus} extending from $R_{\rm in}$ to $R_{\rm out}$ where the coverage density is low but rather uniform.
\end{enumerate}
In its currently planned design, the MWA will have a small ($R_0\sim
20{\rm m}$) nucleus, while the core density falls off as $r^{-2}$
until a sharp cutoff at $R_{\rm in}=750$m.  For LOFAR we assume 
32 stations in a core of radius $R_{\rm in}\sim 1\,{\rm km}$,
and another 32 stations in an outer annulus out to radius $R_{\rm out}\sim
6\,{\rm km}$.  For SKA we assume 20\% in the core, and 30\% in the
annulus out to radius $R_{\rm out}\sim 5\,{\rm km}$. We ignore the
measurements from any dilute distribution of antenna panels outside
$R_{\rm out}$. For LOFAR and SKA, we assume a uniform distribution of
antennae in the annulus, but with an inner core profile like that of
the MWA, i.e., a nucleus of radius $R_0=285/189\,{\rm m}$ (LOFAR/SKA) 
and an $r^{-2}$
fall-off outside this compact core.  We assume an azimuthally
symmetric distribution of baselines in all arrays.

For an array with $N_{\rm in}$ antennae within $R_{\rm in}$, we can define a
quantity
\ben
R_0^{\rm max} \equiv \sqrt{\frac{N_{\rm in}}{\rho_0 \pi}}\,,
\een
where $\rho_0$ is the area density of the nucleus.
$R_0^{\rm max}$ is the maximal radius of the nucleus, corresponding to the case where there it 
contains all the $N_{\rm in}$ antennae and there is no core.

It is also convenient to parametrize the distribution of these $N_{\rm in}$ antennae within $R_{\rm in}$ by two numbers: 
the fraction $\eta$ that are in the nucleus, and the fall-off index $n$ of the core.  
It is straightforward to show that $R_0$ and $R_{\rm in}$ are related to $\eta$ and $n$ by 
\ben
R_0 = \sqrt{\eta} R_0^{\rm max}\,,
\een
\beq{eqn:Rin-R0}
R_{\rm in} = R_0 \left( \frac{2-n(1-\eta)}{2\eta} \right)^{\frac{1}{2-n}}
\een
if $n\ne 2$. The analytic relation for $n=2$ is $R_{\rm in} = R_0 \exp{[(1-\eta)/(2\eta)]}$, which can be well approximated in numerical calculation by by taking $n=2+\epsilon$ in \Eq{eqn:Rin-R0} with $\epsilon\sim 10^{-10}$.  

\begin{figure}[h!]
\centering
\includegraphics[width=1.0\textwidth]{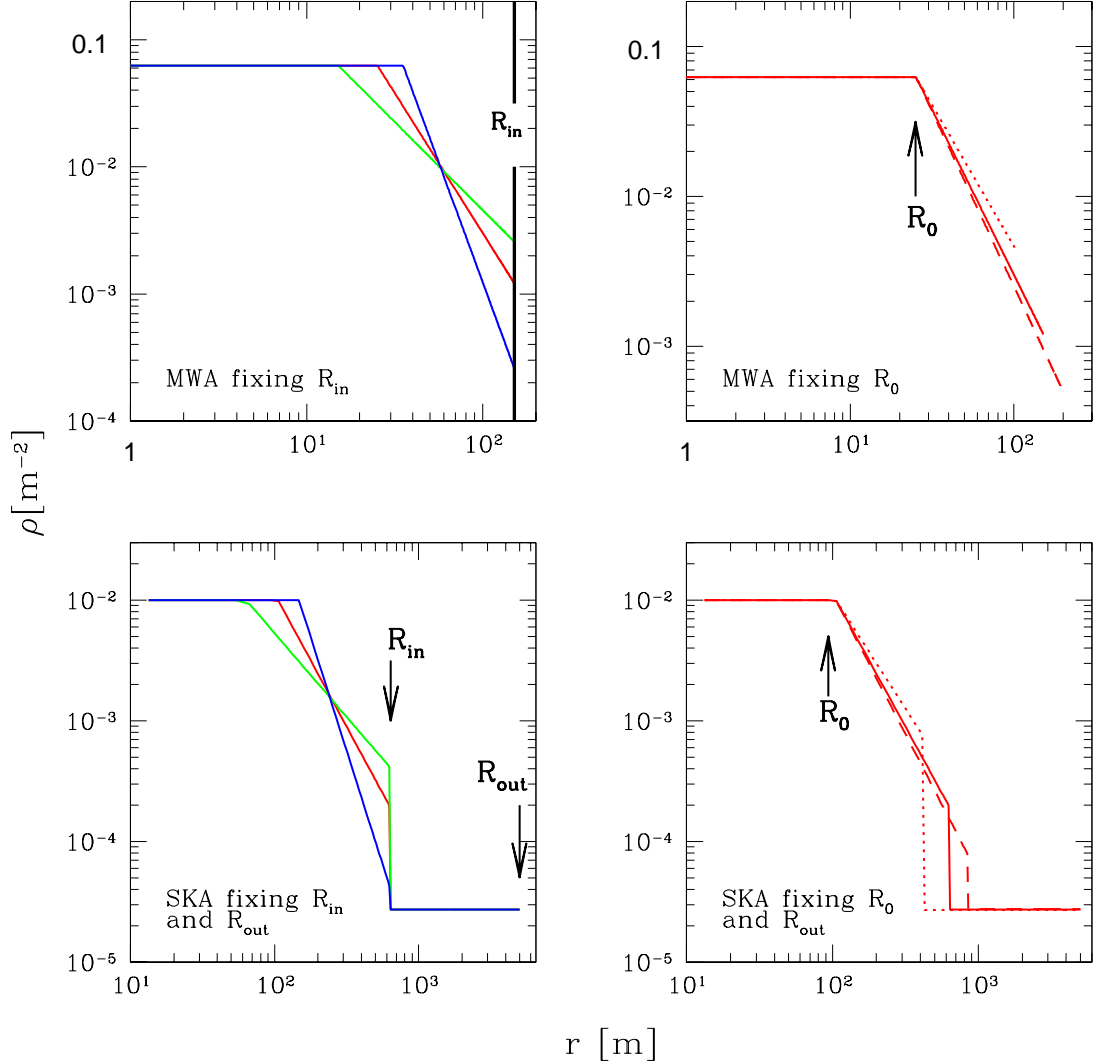} 
\caption[Examples of array configuration changes]{Examples of array configuration changes.  For MWA (upper panels), antennae are uniformly distributed inside the nucleus radius $R_0$, and the density $\rho$ falls off like a power law for $R_0<r<R_{in}$ where $R_{in}$ is the core radius.  For SKA (lower panels) and similarly for LOFAR, there is in addition a uniform yet dilute distribution of antennae in the annulus $R_{in}< r< R_{out}$, where $R_{out}$ is the outer annulus radius.  When $R_0$ is decreased ($R_0=0.7/0.5/0.3\times R_0^{\rm max}$) with $R_{in} = 3.0\times R_0^{\rm max}$ fixed (left panels), the density in the core falls off slower (blue/red/green curves).  When $R_{in}$ is decreased ($R_{in}=4.0/3.0/2.0\times R_0^{\rm max}$) with $R_0=0.5\times R_0^{\rm max}$ fixed (right panels), the density in the core also falls off less steep (dashed/solid/dotted curves).
}
\label{fig:layout}
\end{figure}

In Section \ref{sec:oc-results}, we will scan almost all possible
design configurations and find the optimal one for constraining
cosmology.  
There are two independent ways to vary array configurations, 
as illustrated by Figure \ref{fig:layout}: by varying $R_0$ with $R_{in}$ fixed, and by 
varying $R_{in}$ with $R_0$ fixed.  Contributions from antennae in 
the annulus are negligible compared to the core, so varying $R_{out}$ is 
not interesting.  

In other parts of Section \ref{sec:results}, we will
assume the intermediate configuration $\eta=0.8$ and $n=2$ 
(except for FFTT which is purely in a giant core) 
with the planned number of antennae in the core and annulus.  
Note that this configuration is optimized from the currently planned design.

\subsubsection{Detector noise} \label{sec:noise}  

21cm radio interferometers measure visibility $\bfV$.  The visibility
for a pair of antennae is defined as \cite{Morales:2004} \ben
\bfV(u_x,u_y,\Delta f) = \int dxdy \Delta T_b(x,y,\Delta f) e^{-i(u_x
x+u_y y)}\,, \een where $(u_x,u_y)$ are the number of wavelengths
between the antennae.  The hydrogen 3D map is the Fourier transform in
the frequency direction $\tilde{I}(\bfu)\equiv \int d\Delta f
\bfV(u_x,u_y,\Delta f) \exp{(-i\Delta fu_\parallel)}$ where $\bfu =
u_x\hat{e}_x+u_y\hat{e}_y+u_\parallel\hat{e}_z$.  The detector noise
covariance matrix for an interferometer is
\cite{Morales:2005,McQuinn:2005hk} \ben C^N(\bfu_i,\bfu_j) = \left(
\frac{\lambda^2 B T_{\rm sys}}{A_e}\right)^2 \frac{\delta_{ij}}{B
t_{\bfu_i}} \,, \een where $B$ is the frequency bin size, $T_{\rm
sys}$ is system temperature, and $t_{\bfu} \approx (A_e
t_0/\lambda^2)n(u_\perp)$ is the observation time for the visibility
at $|\bfu_\perp| = d_A|\bfk|\sin\theta$.  Here $t_0$ is the total
observation time, and $n$ is the number of baselines in an observing
cell.

The covariance matrix of the 21cm signal $\tilde{I}(\bfu)$ is related to the power spectrum $\PDT(\bfk)$ by \cite{McQuinn:2005hk}
\bena
C^{SV}(\bfu_i,\bfu_j) & \equiv & \langle \tilde{I}^{*}(\bfu_i) \tilde{I}(\bfu_j) \rangle\nonumber\\
 & = & \PDT(\bfu_i) \frac{\lambda^2 B}{A_e} \delta_{ij}\,.
\eena
Therefore, the noise in the power spectrum is 
\beq{eqn:PNoise}
P^{N}(u_\perp)=\left( \frac{\lambda^2 T_{\rm sys}}{A_e}\right)^2 \frac{1}{t_0 n(u_\perp)} \,.
\een

For all interferometers, the system temperature is dominated by sky temperature $T_{\rm sky} \approx 60(\lambda/1\,{\rm m})^{2.55}\,{\rm K}$ due to synchrotron radiation
in reasonable clean parts of the sky.  Following \cite{bowman05cr}, we set $T_{\rm sys} = 440\,K$ at $z=8$ and $T_{\rm sys} = 690\,K$ at $z=10$.

\subsection{Assumptions about Residual Foregrounds}\label{sec:foregrounds}

There have been a number of papers discussing foreground removal for 21 cm tomography (\eg  \cite{Wang:2005zj,DiMatteo:2004dt,Oh:2003jy,DiMatteo:2001gg} and references therein),  
and much work remains to be done on this important subject, as the 
the amplitudes of residual foregrounds depend strongly depends on cleaning techniques and assumptions, 
and can have potentially dominate the cosmological signal.
The work of Wang \etal \cite{Wang:2005zj} and McQuinn \etal \cite{McQuinn:2005hk} suggested that after fitting out a low-order polynomial from the frequency dependence in each pixel,
the residual foregrounds were negligible for $k> 2\pi/yB$ where $yB$ is the comoving width of a $z$-bin.  
To obtain a crude indication of the impact of residual foregrounds, there therefore we adopt the rough approximation that all data below some cutoff value $\kmin$ is destroyed by foregrounds while
the remainder has negligible contamination. We choose $\kmin=(1/2/4)\times \pi/yB$ for the OPT/MID/PESS scenarios, and also explore wider ranges below.

\section{Results and discussion}\label{sec:results}

In this section, we numerically evaluate how the accuracy of
cosmological parameter constraints depend on the various assumptions
listed above.  Where possible, we attempt to provide intuition for
these dependences with simple analytical approximations.  In most
cases, we explore the dependence on one assumption at a time by
evaluating the PESS, MID and OPT scenario for this assumption while
keeping all other assumptions fixed to the baseline MID scenario.

\subsection{Varying ionization power spectrum modeling and reionization histories} \label{sec:result-power}

\begin{sidewaystable}
\centering
\footnotesize{
\caption[How cosmological constraints depend on the ionization power spectrum modeling and reionization history]{\label{tab:power} How cosmological constraints depend on the ionization power spectrum modeling and reionization history.  We assume observations of 4000 hours on two places in the sky in the range of $z=6.8 - 8.2$ that is divided into three $z$-bins centered at $z=7.0$, $7.5$ and $8.0$ respectively, $\kmax = 2\perMpc$, $\kmin = 2\pi/yB$ and a quasi-giant core configuration (except for FFTT that is a giant core).  
$\onesig$ errors of ionization parameters in the MID model, marginalized over other vanilla parameters, are listed separately in \Tab{tab:mid2}.}

\begin{tabular}{llcccccccccccc}
\hline\hline
    &      & \multicolumn{9}{c}{\it Vanilla Alone} &  & &   \\ \cline{3-11}
		&   Model    & $\Delta\Ol$ & 
 $\Delta\ln(\Omega_m h^2)$   & $\Delta\ln(\Omega_b h^2)$ & 
 $\Delta\ns$		       & $\Delta\ln\As$ & 
 $\Delta\tau$	               & $\Delta\xH(7.0)$ \footnote{$\xH(z)$ denotes the mean neutral fraction at the central redshift $z$.  
$\xH(z)$'s and $A_s$ are completely degenerate from the 21cm measurement alone.  For this reason, the errors shown for $\ln\As$ from 21cm data alone is really not marginalized over $\xH(z)$'s.} & 
 $\Delta\xH(7.5)$ & 
 $\Delta\xH(8.0)$            &  $\Delta\Ok$ &
 $\Delta\mnu$ [\eV]	       & $\Delta\al$ \\ \hline

LOFAR & OPT  &0.025&0.27&0.44&0.063&0.89&...&...&...&...&0.14&0.87&0.027   \\
      & MID  &0.13&0.083&0.15&0.36&0.80&...&...&...&...&0.35&12&0.17   \\ \cline{2-14}

MWA   & OPT &0.046&0.11&0.19&0.022&0.37&...&...&...&...&0.056&0.38&0.013  \\
      & MID &0.22&0.017&0.029&0.097&0.76&...&...&...&...&0.13&9.6&0.074  \\ \cline{2-14}
      
SKA   & OPT  &0.0038&0.044&0.083&0.0079&0.16&...&...&...&...&0.023&0.12&0.0040  \\
      & MID  &0.014&0.0049&0.0081&0.012&0.037&...&...&...&...&0.043&0.36&0.0060  \\ \cline{2-14}

      & OPT  &0.00015&0.0032&0.0083&0.00040&0.015&...&...&...&...&0.00098&0.011&0.00034 \\
FFTT  & MID  &0.00041&0.00038&0.00062&0.00036&0.0013&...&...&...&...&0.0037&0.0078&0.00017 \\
      & PESS &1.1&0.017&0.037&0.010&0.19&...&...&...&...&... &0.20&0.0058 \\	     

\hline

Planck &     & 0.0070 & 0.0081 & 0.0059 & 0.0033 & 0.0088 & 0.0043 & \nodata & \nodata & \nodata & 0.025  & 0.23  & 0.0026 \\ \cline{2-14}
         & OPT  &0.0066&0.0077&0.0058&0.0031&0.0088&0.0043&0.0077&0.0084&0.0093&0.0051&0.060&0.0022 \\
\:+LOFAR & MID  &0.0070&0.0081&0.0059&0.0032&0.0088&0.0043&0.18&0.26&0.23&0.018&0.22&0.0026 \\
         & PESS &0.0070&0.0081&0.0059&0.0033&0.0088&0.0043&0.54&0.31&0.24&0.025&0.23&0.0026   \\  \cline{2-14}

         & OPT  &0.0067&0.0079&0.0057&0.0031&0.0088&0.0043&0.0065&0.0067&0.0069&0.0079&0.027&0.0014  \\
\:+MWA   & MID  &0.0061&0.0070&0.0056&0.0030&0.0087&0.0043&0.32&0.22&0.29&0.021&0.19&0.0026  \\
         & PESS &0.0070&0.0081&0.0059&0.0033&0.0088&0.0043&3.8&0.87&0.53&0.025&0.23&0.0026   \\  \cline{2-14}
      
         & OPT  &0.0031&0.0038&0.0046&0.0013&0.0087&0.0042&0.0060&0.0060&0.0060&0.0017&0.017&0.00064   \\
\:+SKA   & MID  &0.0036&0.0040&0.0044&0.0025&0.0087&0.0043&0.0094&0.014&0.011&0.0039&0.056&0.0022 \\
         & PESS &0.0070&0.0081&0.0059&0.0033&0.0088&0.0043&0.061&0.024&0.012&0.025&0.21&0.0026   \\  \cline{2-14}

         & OPT  &0.00015&0.0015&0.0036&0.00021&0.0087&0.0042&0.0056&0.0056&0.0056&0.00032&0.0031&0.000094 \\
\:+FFTT  & MID  &0.00038&0.00034&0.00059&0.00033&0.0086&0.0042&0.0013&0.0022&0.0031&0.00023&0.0066&0.00017 \\
         & PESS &0.0055&0.0064&0.0051&0.0030&0.0087&0.0043&0.0024&0.0029&0.0040&0.025&0.020&0.0010 \\     

\hline\hline
\end{tabular}
}
\end{sidewaystable}

\begin{table}
\centering
\footnotesize{
\caption[$\onesig$ marginalized errors for the ionization parameters in the MID model]{\label{tab:mid2} $\onesig$ marginalized errors for the ionization parameters in the MID model.  Assumptions are made the same as in \Tab{tab:power}.  $\Rxx$ and $\Rxd$ are in units of $\Mpc$ and other parameters are unitless. } 
\begin{tabular}{clccccccc}
\hline\hline
$z$ &   &  $\Delta\bsqxx$	 &  $\Delta\Rxx$ &  $\Delta\alxx$ &  $\Delta\gaxx$   &  $\Delta\bsqxd$	 &  $\Delta\Rxd$ &  $\Delta\alxd$ \\ \hline
        & Values& 77.   & 3.0  & 4.5   & 2.05 & 8.2  & 0.143 & 28.  \\ \cline{2-9}
        & LOFAR	&94&140&130&27&5.1&49&9600 \\
$7.0$   & MWA	&20&43&43&8.3&2.6&16&3200   \\
        & SKA   &9.1&9.8&8.7&2.0&0.49&2.6&520 \\ 
        & FFTT  &0.59&0.47&0.39&0.098&0.027&0.088&17 \\ \hline	
	
        & Values& 9.9   & 1.3  & 1.6   & 2.3  & 3.1  & 0.58  & 2.   \\	\cline{2-9}
        & LOFAR	&2.2&55&18&73&1.4&5.7&24  \\
$7.5$   & MWA	&4.3&16&4.9&22&1.8&1.8&8.1\\
        & SKA   &0.18&1.7&0.71&2.1&0.076&0.17&0.78 \\   
        & FFTT  &0.0072&0.027&0.015&0.030&0.0023&0.0021&0.012 \\ \hline	
	
        & Values& 2.12  & 1.63 & -0.1  & 1.35 & 1.47 & 0.62 & 0.46 \\ \cline{2-9}
        & LOFAR	&1.6&20&2.1&34&1.2&3.4&6.9 \\
$8.0$   & MWA	&2.7&13&4.2&24&1.5&1.6&2.8 \\
        & SKA   &0.085&0.60&0.090&0.90&0.057&0.095&0.24 \\ 
        & FFTT  &0.0017&0.013&0.0026&0.017&0.0013&0.0014&0.0030 \\
\hline\hline	
\end{tabular}
}
\end{table}

\subsubsection{Basic results}

We start with testing assumptions in the ionization power modeling of
$\Pxx$ and $\Pxd$.  
In Table \ref{tab:power} we show the accuracy with which the 21cm
power spectrum can place constraints on the cosmological parameters
from three $z$-bins ranging from $z=6.8 - 8.2$.  We fix the
assumptions concerning $\kmax$, the foreground removal, and the array
layout and specifications, but vary the sophistication with which we
model the ionization power.

Our results agree with those of previous studies
\cite{McQuinn:2005hk,Bowman:2005hj}, \ie 21cm data alone (except for
the optimized FFTT) cannot place constraints comparable with those
from Planck CMB data.  However, if 21cm data are combined with CMB
data, the parameter degeneracies can be broken, yielding stringent
constraints on $\Ok$, $\mnu$ and $\al$.  For example, in the OPT model,
from LOFAR/MWA/SKA/FFTT combined with Planck, the curvature density
$\Ok$ can be measured 5/3/15/78 times better, to a precision
$\Delta\Ok=0.005/0.008/0.002/0.0003$, the neutrino mass $\mnu$ can be constrained
4/9/14/74 times better to accuracy $\Delta\mnu=0.06/0.03/0.02/0.003$, and running
of the scalar spectral index $\al$ can be done 1/2/4/28 times better, to
$\Delta\al=0.002/0.001/0.0006/0.0001$. The more realistic MID model
yields weaker yet still impressive constraints: from SKA/FFTT
combined with Planck, $\Ok$ can be measured 6/109 times better, to
$\Delta\Ok=0.004/0.0002$, $\mnu$ 4/35 times better, to
$\Delta\mnu=0.06/0.007$, and $\al$ 1/15 times better, to
$\Delta\al=0.002/0.0002$. The improved measurements of $\Ok$ and $\al$
enable further precision tests of inflation, 
since generically $\Ok$ is predicted to vanish down to the $10^{-5}$ level, while 
the simplest inflation models (with a single slow-rolling scalar field) 
predict $\al\sim (1-\ns)^2\sim 10^{-3}$. 
For example, the inflaton potential $V(\phi)\propto\phi^2$ predicts 
$\alpha\approx -0.0007$, while $V(\phi)\propto\phi^4$ predicts $\alpha=0.008$.
In addition, 21cm data combined with CMB data
from Planck can make accurate measurements in the mean neutral
fraction $\xH(z)$ at separate redshifts, outlining the full path of
reionization, \eg at the $\Delta\xH(z)\sim 0.01/0.003$ level from
SKA/FFTT data combined with Planck data.

\subsubsection{OPT and MID models}

For most 21cm experiments, the OPT model yields stronger
constraints than the MID model.  The reason is as follows.  By
assuming $\Pxx=\Pxd=0$, there are essentially no neutral fraction
fluctuations in the OPT model.  This means that this model is an ideal
model in which the 21cm power spectrum encodes cosmological
information per se, since $\PDT(\bfk)\propto \sPdd(k)$ at each pixel
in the Fourier space.  In the more realistic MID model, however, the
nuisance ionization parameters has correlations with cosmological
parameters.  Mathematically, the inversion of a correlated matrix
multiplies each error by a degradation factor.

An exception is the FFTT, where the situation is reversed.  As mentioned in Section \ref{sec:asmp-mid}, the sample
variance $\PDT$ in the MID model is smaller than that in the OPT model 
because of two reasons: (i) the MID model assumes non-zero $\Pxx$ and $\Pxd$, and 
$\Pxd$ has negative contribution to the total power spectrum (see Eqs.\ref{eqn:pmu0} and \ref{eqn:pmu2});
(ii) the OPT model assumes $\xH=1$, but $\xH$ takes realistic values (less than 1) in the MID model, decreasing the overall amplitude.  
In a signal-dominated experiment, reduced sample variance can be more important than the degradation from correlations.  

\begin{figure}[ht]
\centering
\includegraphics[width=0.8\textwidth]{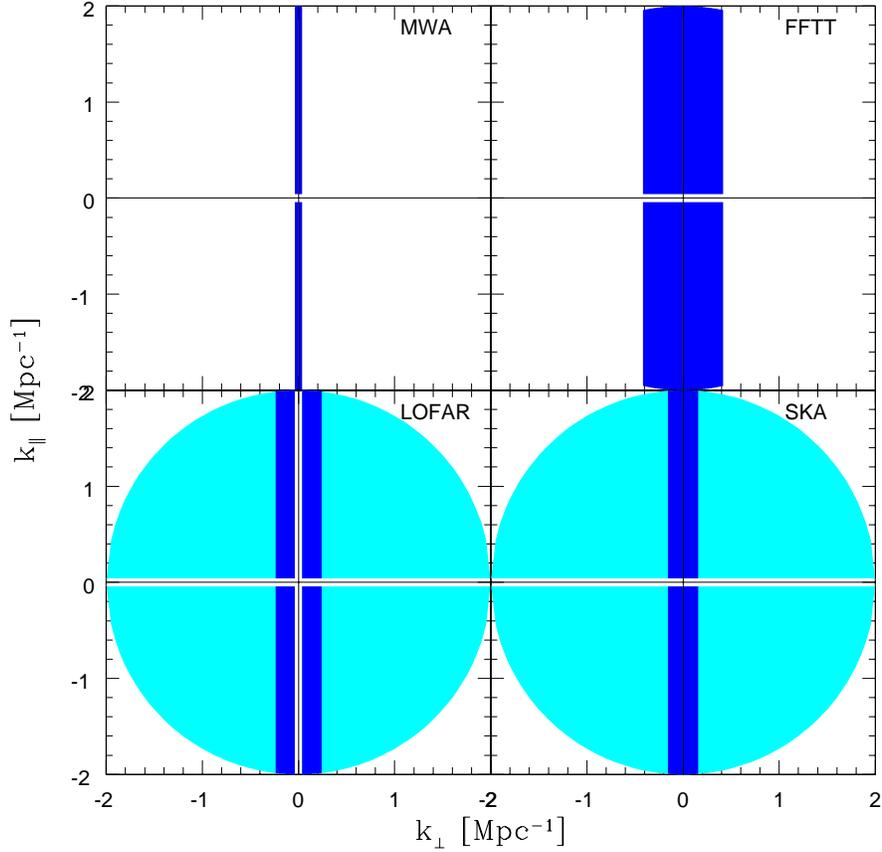} 
\caption[Available $(k_\perp,k_\parallel)$ pixels from upcoming 21cm experiments]{Available $(k_\perp,k_\parallel)$ pixels from MWA (upper left), FFTT (upper right), LOFAR (lower
left) and SKA (lower right), 
evaluated at $z=8$.  The blue/grey regions can be measured with good signal-to-noise from the nucleus and 
core of an array, while the cyan/light-grey regions are measured only with the annulus and have so poor signal-to-noise that they 
hardly contribute to cosmological parameter constraints. 
}
\label{fig:datacube}
\end{figure}

\begin{figure}[h!]
\centering
\includegraphics[width=0.65\textwidth]{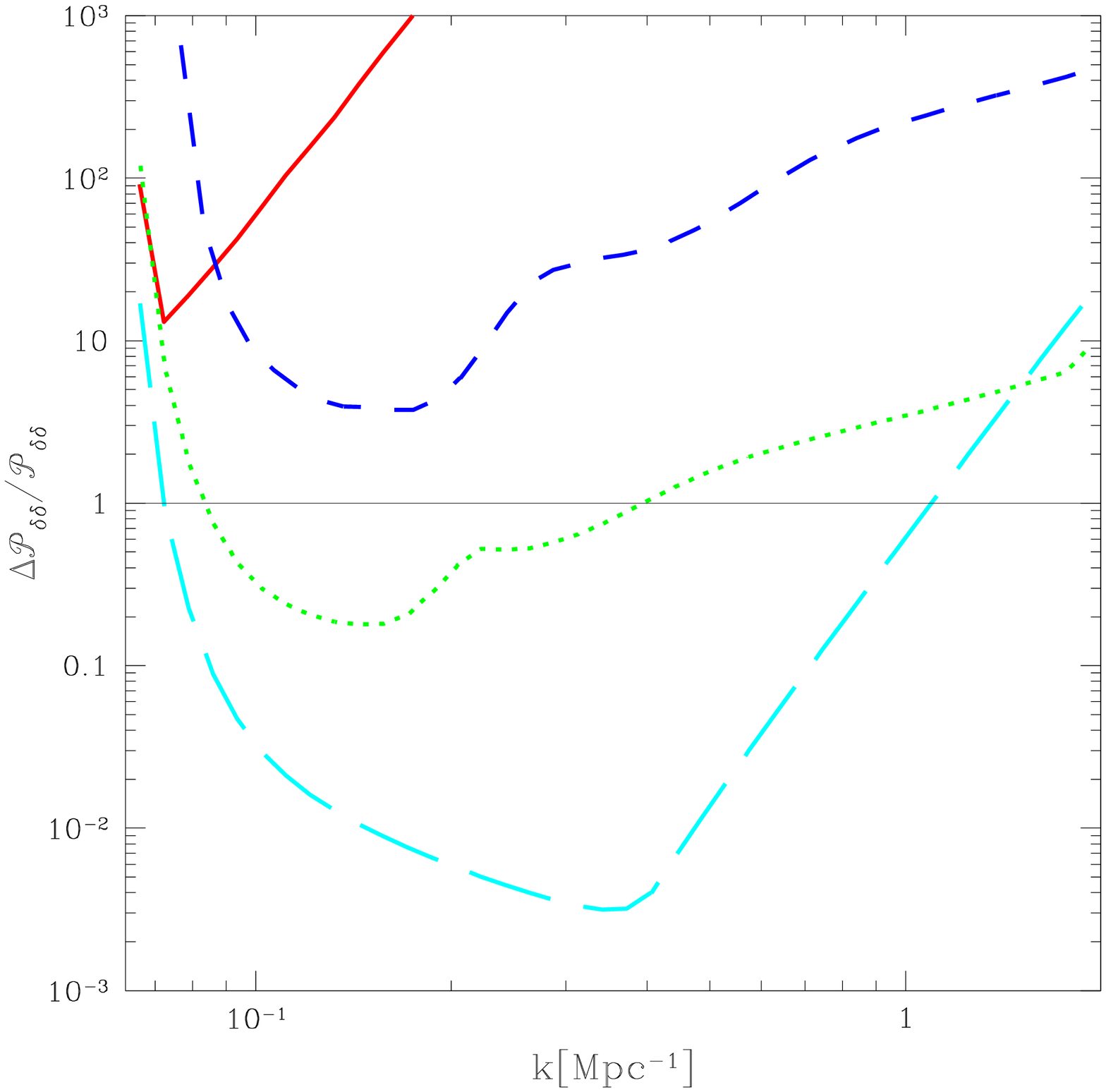} 
\caption[Relative 1$\sigma$ error for measuring $\sPdd(k)$ with the PESS model]{Relative 1$\sigma$ error for measuring $\sPdd(k)$ with the PESS model by observing a 6MHz band that is centered at $z=8$ with MWA (red/solid), LOFAR (blue/short-dashed), SKA (green/dotted) and FFTT (cyan/long-dashed).  The step size is $\Delta \ln k \approx 0.10$.}
\label{fig:Pkerr}
\end{figure}

\subsubsection{PESS model}

Our results show that even combined with CMB data from Planck, the
 21cm data using the PESS model cannot significantly improve constraints.  
There are two reasons for this failure.  Firstly, the PESS model
 essentially uses only $\Pmufour(k)$ to constrain cosmology, by
 marginalizing over $\Pmuzero$ and $\Pmutwo$.  This loses a great deal
 of cosmological information in the contaminated $\Pmuzero$ and
 $\Pmutwo$, in contrast to the situation in the OPT and MID models.
 Secondly, to effectively separate $\Pmufour(k)$ from other two
 moments, the available Fourier pixels should span a large range in
 $\mu$.  Figure \ref{fig:datacube} shows that in MWA and FFTT, the
 data set is a thin cylinder instead of a sphere.  The limitation in
 $\mu$-range will give large degradation factors during the inversion
 of Fisher matrix.  (In the limit that there is
 only one $\mu$ for each shell, then the Fisher matrix is singular and the
 degradation factor is infinite.)  These two factors work together
 with the noise level to shrink the useful $k$-modes into a rather
 narrow range: as shown in Figure \ref{fig:Pkerr}, $\Delta\sPdd <
 \sPdd$ only for $k=0.09-0.4 \; \perMpc$ for SKA, $k=0.07-1 \; \perMpc$ for
 FFTT and over zero modes for LOFAR and MWA.

\subsection{Varying $\kmax$} \label{sec:kmax}

\begin{figure}[ht]
\centering
\includegraphics[width=0.65\textwidth]{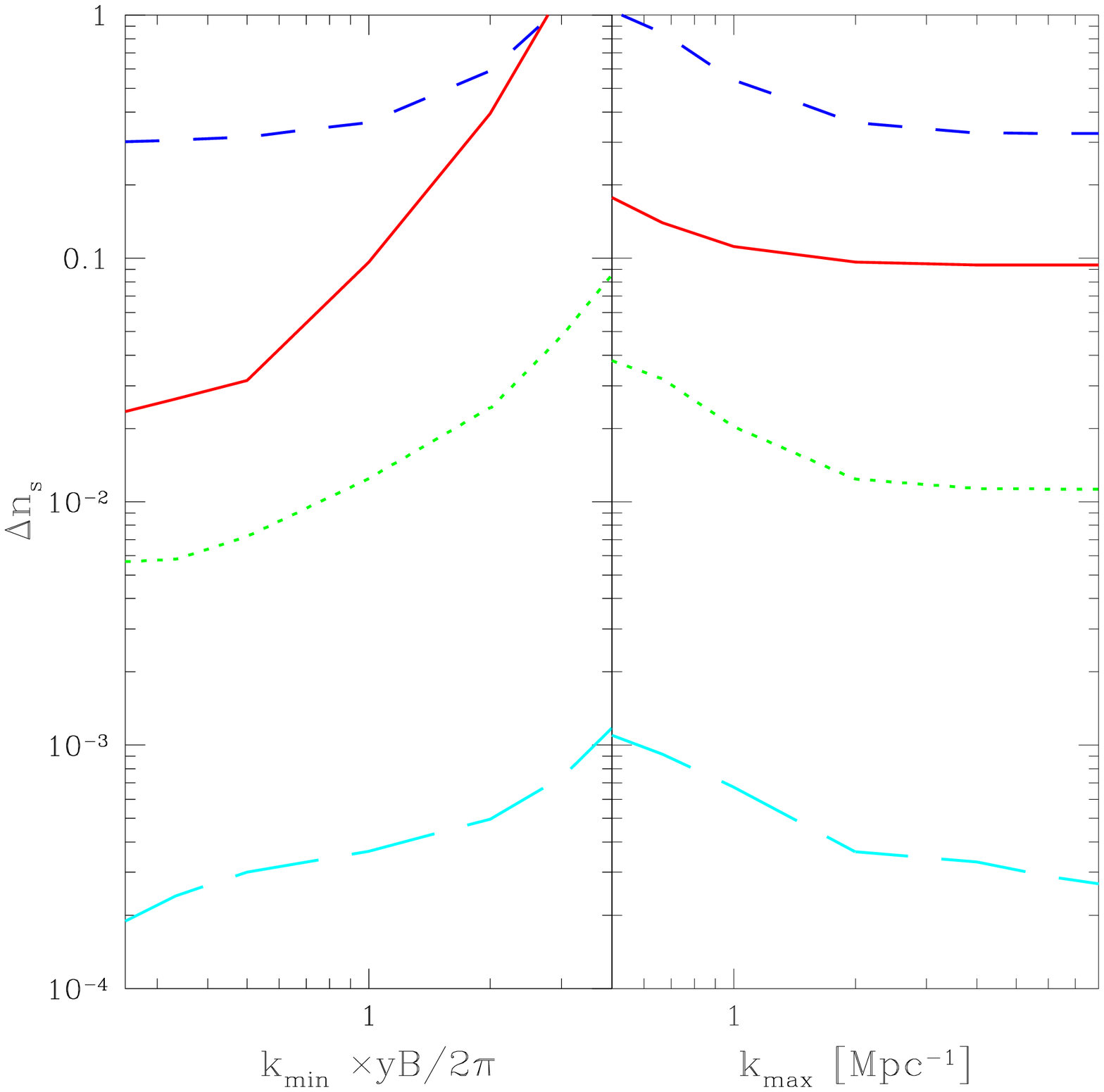} 
\caption[How cosmological constraints $\Delta n_s$ depend on $\kmin$ and $\kmax$]{How cosmological constraints $\Delta n_s$ depend on $\kmin$ (left panel) and $\kmax$ (right panel) in the MID model with the 21cm experiments MWA (red/solid), LOFAR (blue/short-dashed), SKA (green/dotted) and FFTT (cyan/long-dashed).  
We plot $\Delta n_s$ in this example because it has the strongest dependence on
$\kmin$ and $\kmax$ of all cosmological parameters.  
The quantity $2\pi/yB$ varies between different $z$-bins, so as the horizontal axis of the left panel, we
use the overall scale $\kappa_{\rm min}\equiv \kmin \times (yB/2\pi)$ which is equal for all $z$-bins,    
}
\label{fig:kminkmax}
\end{figure}

We test how varying $\kmax$ affects constraints in this section.  The
cutoff $\kmax$ depends on the scale at which non-linear physics, \eg
the non-linear clustering of density perturbations or the
irregularities of ionized bubbles, enter the power spectrum.  
It is illustrated in the right panel of Figure \ref{fig:kminkmax} that
generically cosmological constraints asymptotically approach a value
as $\kmax$ increases above $\sim 2 \; \perMpc$ (this typical scale can
be larger for cosmic variance-limit experiments such as FFTT).  Not
much cosmological information is garnered from these high-$k$ modes
because detector noise becomes increasingly important with $k$. 
The upshot is that the accuracy only weakly depends on $\kmax$.

\subsection{Varying the non-Gaussianity parameter $\xi$}
\label{sec:nongauss_sec}

Table~\ref{tab:omp} shows the effect of changing the non-Gaussianity
parameter $\xi$ in Section~\ref{sec:non-gauss} from the $\xi=1$
(Gaussian) case to $\xi=2$ in the PESS scenario, 
along with changing other assumptions.  
However, there is no need to perform
extensive numerical investigation of the the impact of $\xi$, since it
is readily estimated analytically.  Because $\onesig$ errors $\Delta
p_i$ in cosmological parameters are $\sqrt{(F^{-1})_{ii}}$, it follows
directly from \eq{xiDefEq} that $\Delta p$ does not appreciably depend
on $\xi$ for noise dominated experiments like MWA and LOFAR, whereas
$\Delta p \propto \xi^\sigma$ with $\sigma\la 1$ for (nearly) signal
dominated experiments like SKA and FFTT.  Compared with the other
effects that we discuss in this section, this (no more
than linear) dependence on the non-Gaussianity parameter $\xi$ is not
among the most important factors.

\subsection{Varying redshift ranges}
\label{sec:zbin_z}

\begin{sidewaystable}
\centering
\footnotesize{
\caption[How cosmological constraints depend on the redshift range in OPT model]{\label{tab:z1} How cosmological constraints depend on the redshift range in OPT model.  Same assumptions as in Table \ref{tab:power} but for different redshift ranges and assume only OPT model.   } 
\begin{tabular}{llp{1cm}p{1.5cm}p{1.5cm}p{1cm}p{1cm}p{1cm}p{1cm}p{1cm}p{1cm}p{1cm}p{1cm}p{1cm}p{1cm}}
\hline\hline
    &      & \multicolumn{10}{c}{\it Vanilla Alone} &  & &   \\ \cline{3-12}
		&   z range    & $\Delta\Ol$ & 
 $\Delta\ln(\Omega_m h^2)$   & $\Delta\ln(\Omega_b h^2)$ & 
 $\Delta\ns$		       & $\Delta\ln\As$ & 
 $\Delta\tau$	               & $\Delta\xH(7.0)$ & $\Delta\xH(7.5)$ & 
 $\Delta\xH(8.0)$            & $\Delta\xH(9.2)$            & $\Delta\Ok$ &
 $\Delta\mnu$ [\eV]	       & $\Delta\al$ \\ \hline
 
      & 6.8-10 &0.021&0.20&0.34&0.049&0.67&...&...&...&...&...&0.086&0.75&0.023  \\
LOFAR & 6.8-8.2&0.025&0.27&0.44&0.063&0.89&...&...&...&...&...&0.14&0.87&0.027   \\
      & 7.3-8.2&0.036&0.38&0.61&0.090&1.2 &...&...&...&...&...&0.24&1.3&0.038    \\ \cline{2-15}

      & 6.8-10 &0.037&0.072&0.14&0.016&0.25&...&...&...&...&...&0.031&0.31&0.011 \\
MWA   & 6.8-8.2&0.046&0.11&0.19&0.022&0.37 &...&...&...&...&...&0.056&0.38&0.013 \\
      & 7.3-8.2&0.070&0.15&0.27&0.032&0.51 &...&...&...&...&...&0.097&0.53&0.018 \\ \cline{2-15}
      
      & 6.8-10 &0.0032&0.031&0.061&0.0058&0.12&...&...&...&...&...&0.012&0.096&0.0032  \\
SKA   & 6.8-8.2&0.0038&0.044&0.083&0.0079&0.16&...&...&...&...&...&0.023&0.12&0.0040   \\
      & 7.3-8.2&0.0053&0.059&0.11&0.011&0.21 &...&...&... &...&...&0.042&0.17&0.0054   \\ \cline{2-15}
     
      & 6.8-10 &0.00012&0.0023&0.0058&0.00030&0.011&...&...&...&...&...&0.00045&0.0073&0.00023 \\
FFTT  & 6.8-8.2&0.00015&0.0032&0.0083&0.00040&0.015&...&...&...&...&...&0.00098&0.011&0.00034  \\
      & 7.3-8.2&0.00021&0.0042&0.011&0.00052&0.019 &...&...&...&...&...&0.0021&0.014&0.00043   \\ 
\hline

Planck &     & 0.0070 & 0.0081 & 0.0059 & 0.0033 & 0.0088 & 0.0043 & \nodata & \nodata & \nodata& \nodata & 0.025  & 0.23  & 0.0026 \\ \cline{2-15}
         & 6.8-10 &0.0065&0.0076&0.0057&0.0031&0.0088&0.0043&0.0077&0.0084&0.0082&0.0090&0.0046&0.051&0.0021   \\
\:+LOFAR & 6.8-8.2&0.0066&0.0077&0.0058&0.0031&0.0088&0.0043&0.0077&0.0084&0.0093&...   &0.0051&0.060&0.0022  \\
         & 7.3-8.2&0.0068&0.0079&0.0058&0.0032&0.0088&0.0043&...   &0.0085&0.0093&...   &0.0072&0.081&0.0024   \\ \cline{2-15}

         & 6.8-10 &0.0065&0.0076&0.0056&0.0031&0.0088&0.0043&0.0065&0.0067&0.0066&0.0067&0.0066&0.023&0.0013 \\
\:+MWA   & 6.8-8.2&0.0067&0.0079&0.0057&0.0031&0.0088&0.0043&0.0065&0.0067&0.0069&\nodata&0.0079&0.027&0.0014\\ 	     
         & 7.3-8.2&0.0068&0.0080&0.0058&0.0032&0.0088&0.0043&\nodata&0.0067&0.0069&\nodata&0.011&0.036&0.0017\\ \cline{2-15} 
      
         & 6.8-10 &0.0027&0.0035&0.0045&0.0012&0.0087&0.0042&0.0060&0.0060&0.0060&0.0060&0.0016&0.015&0.00061   \\
\:+SKA   & 6.8-8.2&0.0031&0.0038&0.0046&0.0013&0.0087&0.0042&0.0060&0.0060&0.0060&\nodata&0.0017&0.017&0.00064  \\	       
         & 7.3-8.2&0.0039&0.0047&0.0049&0.0017&0.0087&0.0042&\nodata&0.0060&0.0060&\nodata&0.0020&0.019&0.00075 \\ \cline{2-15}
     
         & 6.8-10 &0.00013&0.0014&0.0033&0.00019&0.0087&0.0042&0.0054&0.0054&0.0054&0.0054&0.00026&0.0025&0.000078 \\
\:+FFTT  & 6.8-8.2&0.00015&0.0015&0.0036&0.00021&0.0087&0.0042&0.0056&0.0056&0.0056&\nodata&0.00032&0.0031&0.000094 \\  
         & 7.3-8.2&0.00020&0.0016&0.0038&0.00023&0.0087&0.0042&\nodata&0.0057&0.0057&\nodata&0.00040&0.0038&0.00011\\ 
\hline\hline
\end{tabular}
}
\end{sidewaystable}

\begin{sidewaystable}
\centering
\footnotesize{
\caption[How cosmological constraints depend on the redshift range in MID model]{\label{tab:z2} How cosmological constraints depend on the redshift range in MID model. Same assumptions as in Table \ref{tab:power} but for different redshift ranges and assume only MID model.}
\begin{tabular}{llp{1cm}p{1.5cm}p{1.5cm}p{1cm}p{1cm}p{1cm}p{1cm}p{1cm}p{1cm}p{1cm}p{1cm}p{1cm}p{1cm}}
\hline\hline
    &      & \multicolumn{10}{c}{\it Vanilla Alone} &  & &   \\ \cline{3-12}
		&   z range    & $\Delta\Ol$ & 
 $\Delta\ln(\Omega_m h^2)$   & $\Delta\ln(\Omega_b h^2)$ & 
 $\Delta\ns$		       & $\Delta\ln\As$ & 
 $\Delta\tau$	               & $\Delta\xH(7.0)$ & $\Delta\xH(7.5)$ & 
 $\Delta\xH(8.0)$            & $\Delta\xH(9.2)$            & $\Delta\Ok$ &
 $\Delta\mnu$ [\eV]	       & $\Delta\al$ \\ \hline
 
      & 6.8-10  &0.090&0.055&0.093&0.18&0.43&...&...&...&...&...&0.22&5.7&0.080 \\
LOFAR & 6.8-8.2 &0.13&0.083&0.15&0.36&0.80  &...&...&...&...&...&0.35&12&0.17         \\
      & 7.3-8.2 &0.21&0.099&0.15&0.42&0.81  &...&...&...&...&...&0.62&15&0.18             \\ \cline{2-15}

      & 6.8-10  &0.15&0.012&0.020&0.031&0.46&...&...&...&...&...&0.092&4.4&0.025  \\
MWA   & 6.8-8.2 &0.22&0.017&0.029&0.097&0.76&...&...&...&...&...&0.13&9.6&0.074 	  \\
      & 7.3-8.2 &0.40&0.018&0.030&0.099&1.0&...&...&...&...&...&0.32&18&0.083 		  \\ \cline{2-15}
      
      & 6.8-10  &0.010&0.0031&0.0056&0.0073&0.023&...&...&...&...&...&0.031&0.23&0.0032  \\
SKA   & 6.8-8.2 &0.014&0.0049&0.0081&0.012&0.037&...&...&...&...&...&0.043&0.36&0.0060	 \\
      & 7.3-8.2 &0.018&0.0050&0.0081&0.013&0.039&...&...&...&...&...&0.072&0.41&0.0063  	 \\ \cline{2-15}
     
      & 6.8-10  &0.00029&0.00021&0.00043&0.00025&0.00097&...&...&...&...&...&0.0020&0.0055&0.00011 \\
FFTT  & 6.8-8.2 &0.00041&0.00038&0.00062&0.00036&0.0013&...&...&...&...&...&0.0037&0.0078&0.00017 	   \\
      & 7.3-8.2 &0.00050&0.00039&0.00062&0.00037&0.0013&...&...&...&...&...&0.0058&0.0083&0.00018 	   \\
\hline

Planck &     & 0.0070 & 0.0081 & 0.0059 & 0.0033 & 0.0088 & 0.0043 & \nodata & \nodata & \nodata& \nodata & 0.025  & 0.23  & 0.0026 \\ \cline{2-15}
         & 6.8-10  &0.0069&0.0080&0.0058&0.0032&0.0088&0.0043&0.18&0.26&0.15&0.23&0.017&0.22&0.0026 \\
\:+LOFAR & 6.8-8.2 &0.0070&0.0081&0.0059&0.0032&0.0088&0.0043&0.18&0.26&0.23&...&0.018&0.22&0.0026      \\
         & 7.3-8.2 &0.0070&0.0081&0.0059&0.0032&0.0088&0.0043&...&0.27&0.23&...&0.023&0.22&0.0026 	    \\ \cline{2-15}

         & 6.8-10  &0.0056&0.0065&0.0054&0.0029&0.0087&0.0043&0.32&0.22&0.091&0.36&0.020&0.11&0.0025 \\
\:+MWA   & 6.8-8.2 &0.0061&0.0070&0.0056&0.0030&0.0087&0.0043&0.32&0.22&0.29&...&0.021&0.19&0.0026       \\
         & 7.3-8.2 &0.0061&0.0071&0.0056&0.0030&0.0087&0.0043&...&0.25&0.29&...&0.024&0.19&0.0026 	     \\ \cline{2-15}
      
         & 6.8-10  &0.0025&0.0027&0.0038&0.0023&0.0087&0.0042&0.0094&0.014&0.0075&0.024&0.0032&0.033&0.0020 \\
\:+SKA   & 6.8-8.2 &0.0036&0.0040&0.0044&0.0025&0.0087&0.0043&0.0094&0.014&0.011&...&0.0039&0.056&0.0022	    \\
         & 7.3-8.2 &0.0036&0.0041&0.0044&0.0025&0.0087&0.0043&...&0.015&0.011&...&0.0053&0.056&0.0023		    \\ \cline{2-15}
     
         & 6.8-10  &0.00033&0.00021&0.00043&0.00024&0.0086&0.0042&0.0013&0.0022&0.0030&0.0040&0.00020&0.0052&0.00011 \\
\:+FFTT  & 6.8-8.2 &0.00038&0.00034&0.00059&0.00033&0.0086&0.0042&0.0013&0.0022&0.0031&...&0.00023&0.0066&0.00017 	     \\
         & 7.3-8.2 &0.00041&0.00035&0.00059&0.00033&0.0086&0.0042&...&0.0022&0.0031&...&0.00024&0.0070&0.00017  	     \\
\hline\hline
\end{tabular}
}
\end{sidewaystable}

We now test how accuracies depend on the redshift ranges.  In Table
\ref{tab:z1} (OPT model) and \ref{tab:z2} (MID model), we consider the
optimistic/middle/pessimistic ranges, $z=6.8-10$ / $6.8-8.2$ /
$7.3-8.2$ which is divided by $n_z=4/3/2$ $z$-bins.  The results show
that, from 21cm data alone, the constraints from the extreme ranges
differ significantly (a factor of 5 for $\Delta\Ok$).
Therefore, the sensitivity of a 21cm telescope depends strongly on the
frequency range over which it can observe the signal.

\subsection{Optimal configuration: varying array layout}\label{sec:oc-results}

\begin{figure}[ht]
\centering
\begin{displaymath}
\begin{array}{ccc} 
  \includegraphics[width=0.33\textwidth]{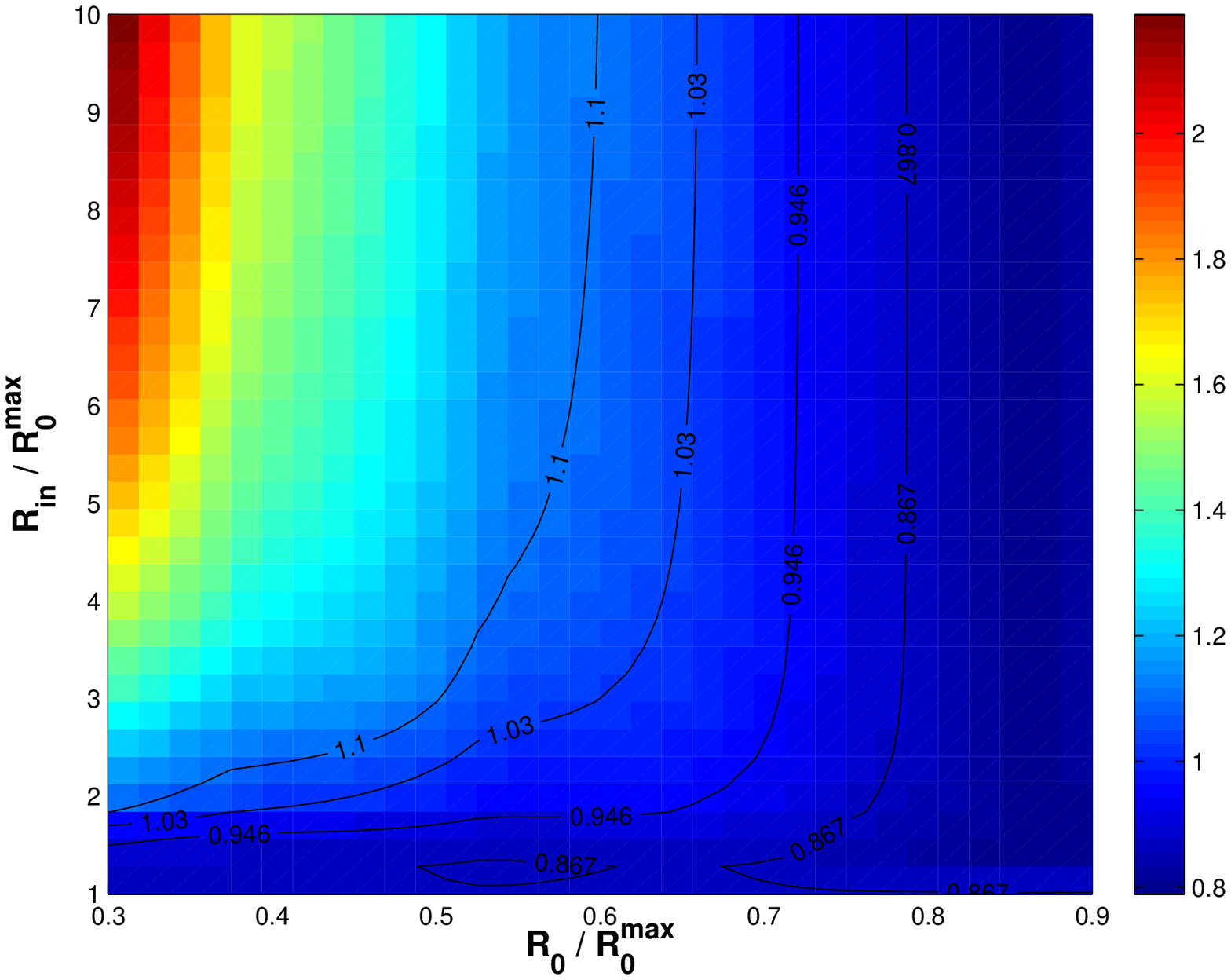} & 
  \includegraphics[width=0.33\textwidth]{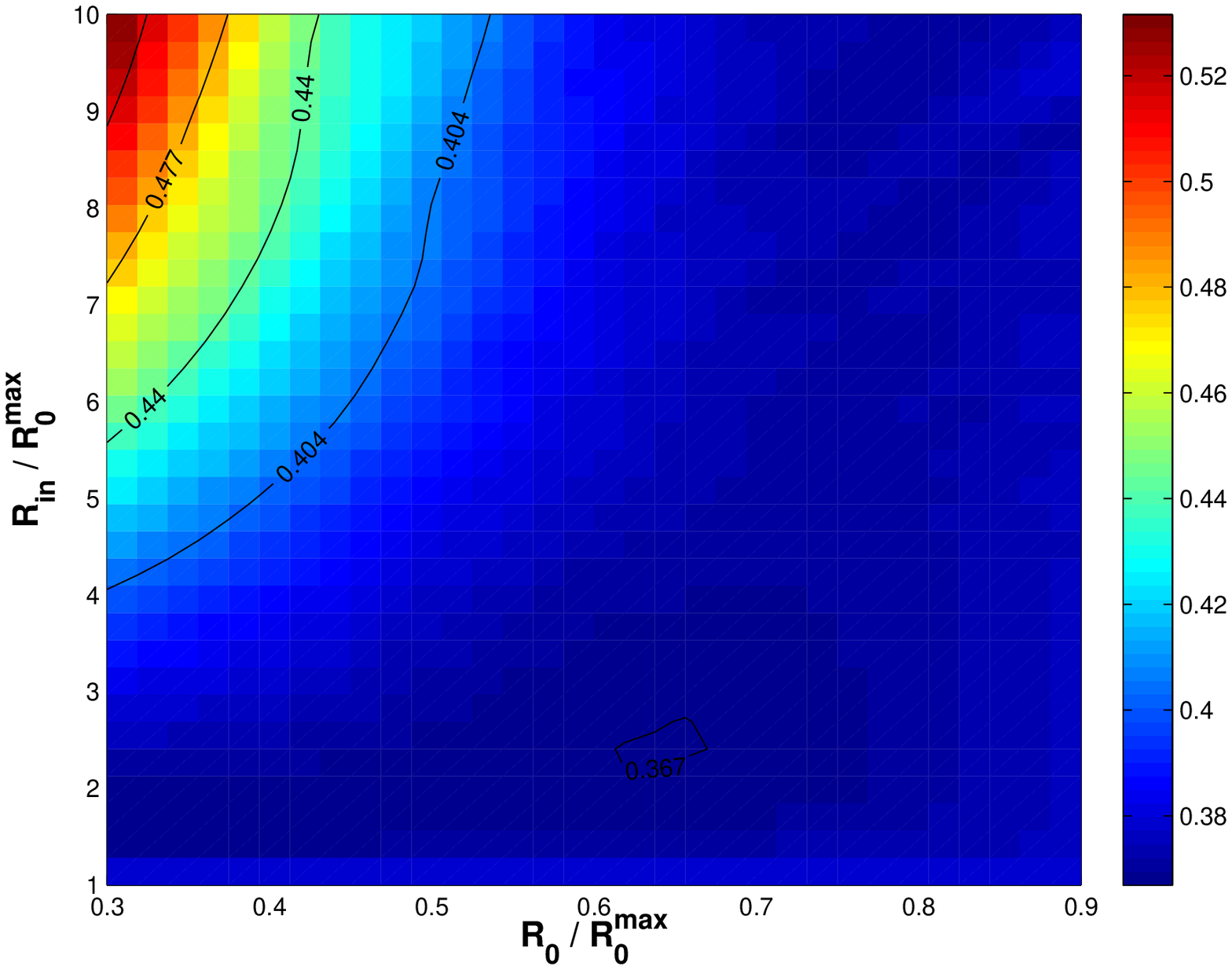} &
  \includegraphics[width=0.33\textwidth]{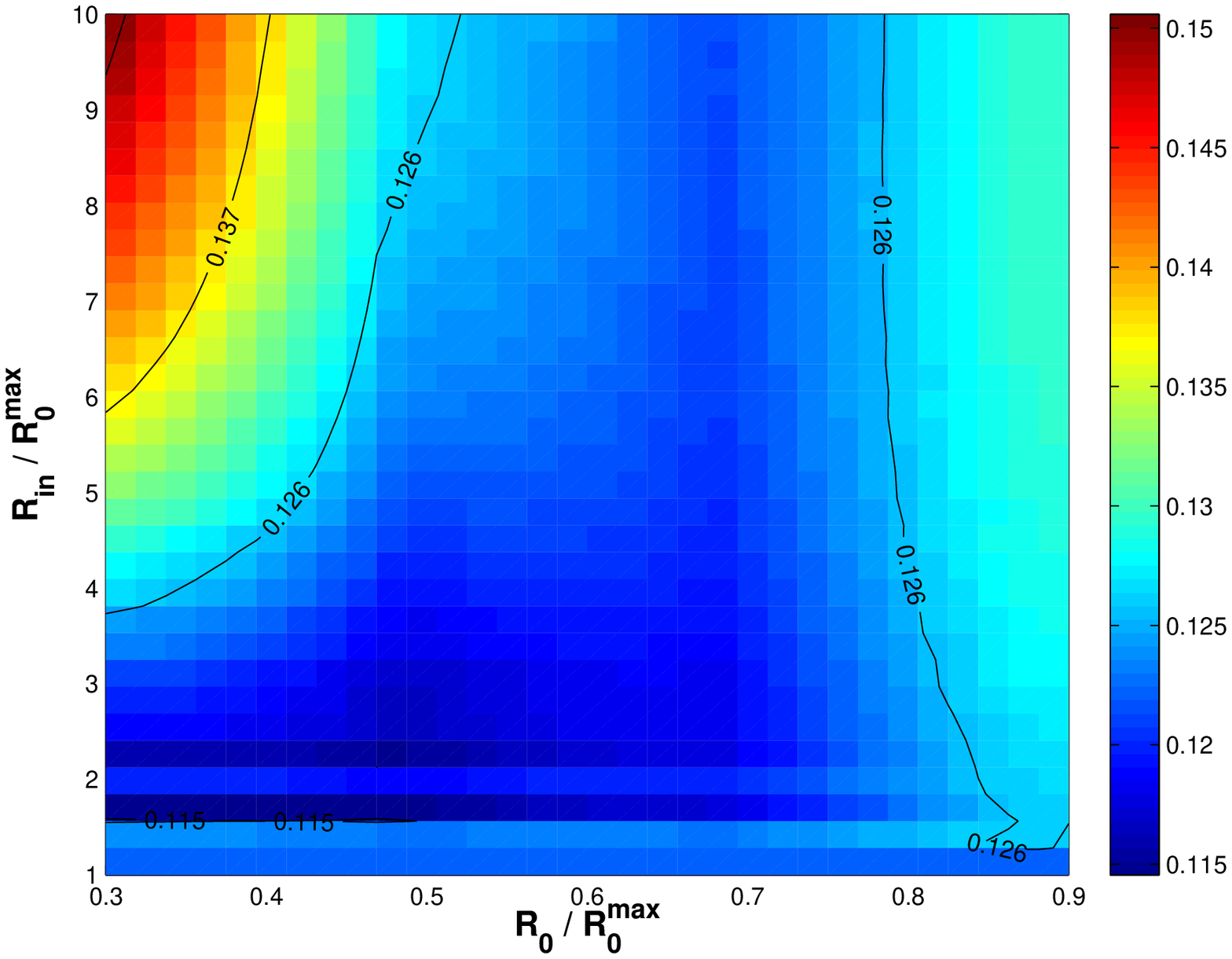} \\
\end{array}
\end{displaymath}
\caption[1$\sigma$ error for various configuration $(R_0,\, R)$ of upcoming 21cm experiments in the OPT model]{ 1$\sigma$ error for $\mnu$ marginalized over vanilla parameters for various configuration $(R_0,\, R)$ of LOFAR(left panel), MWA(middle panel) and SKA(right panel).  We made the same assumptions here as in Table \ref{tab:power} but assume only OPT model.}
\label{fig:OC1}
\end{figure}

\begin{figure}[ht]
\centering
\begin{displaymath}
\begin{array}{ccc} 
  \includegraphics[width=0.33\textwidth]{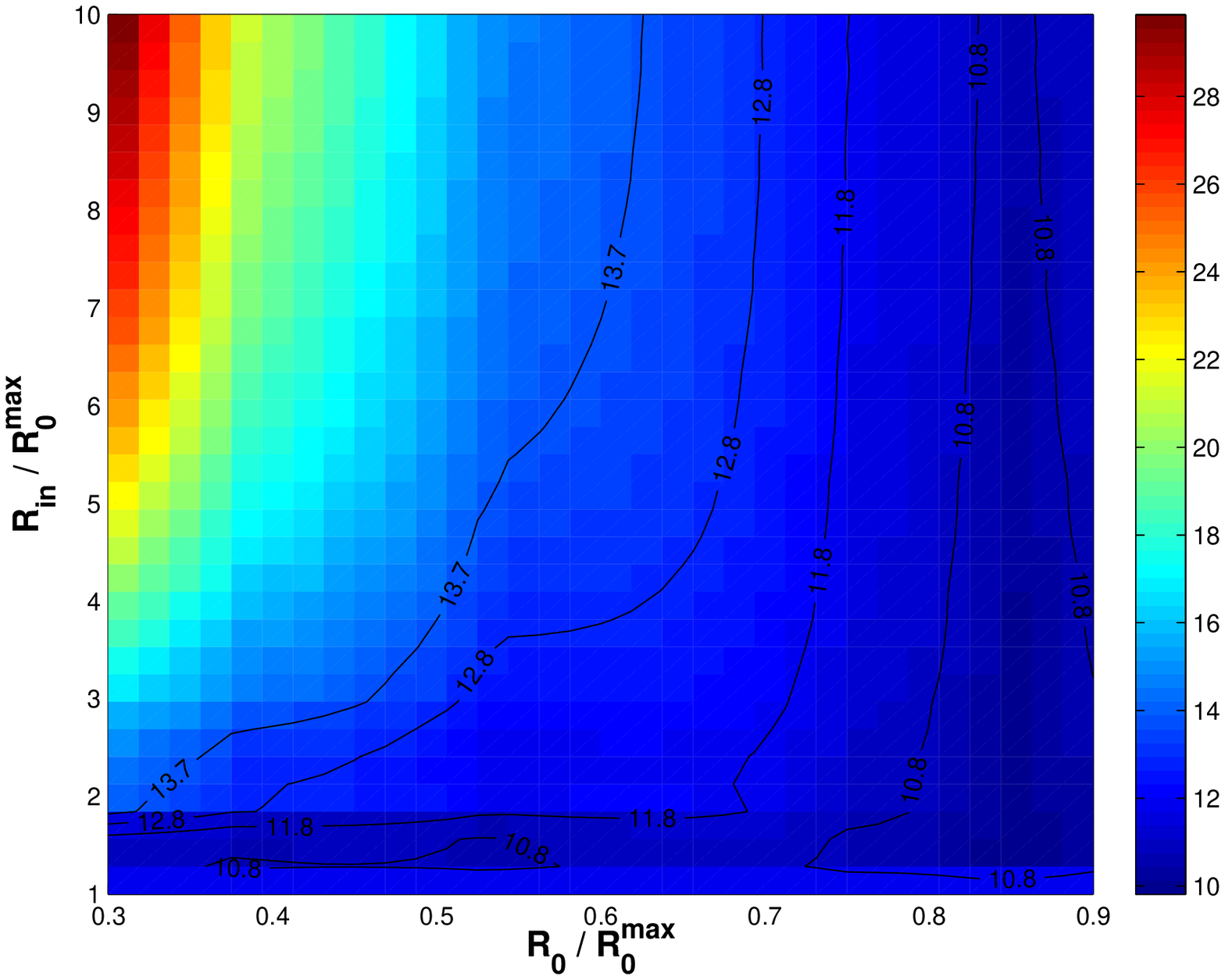} & 
  \includegraphics[width=0.33\textwidth]{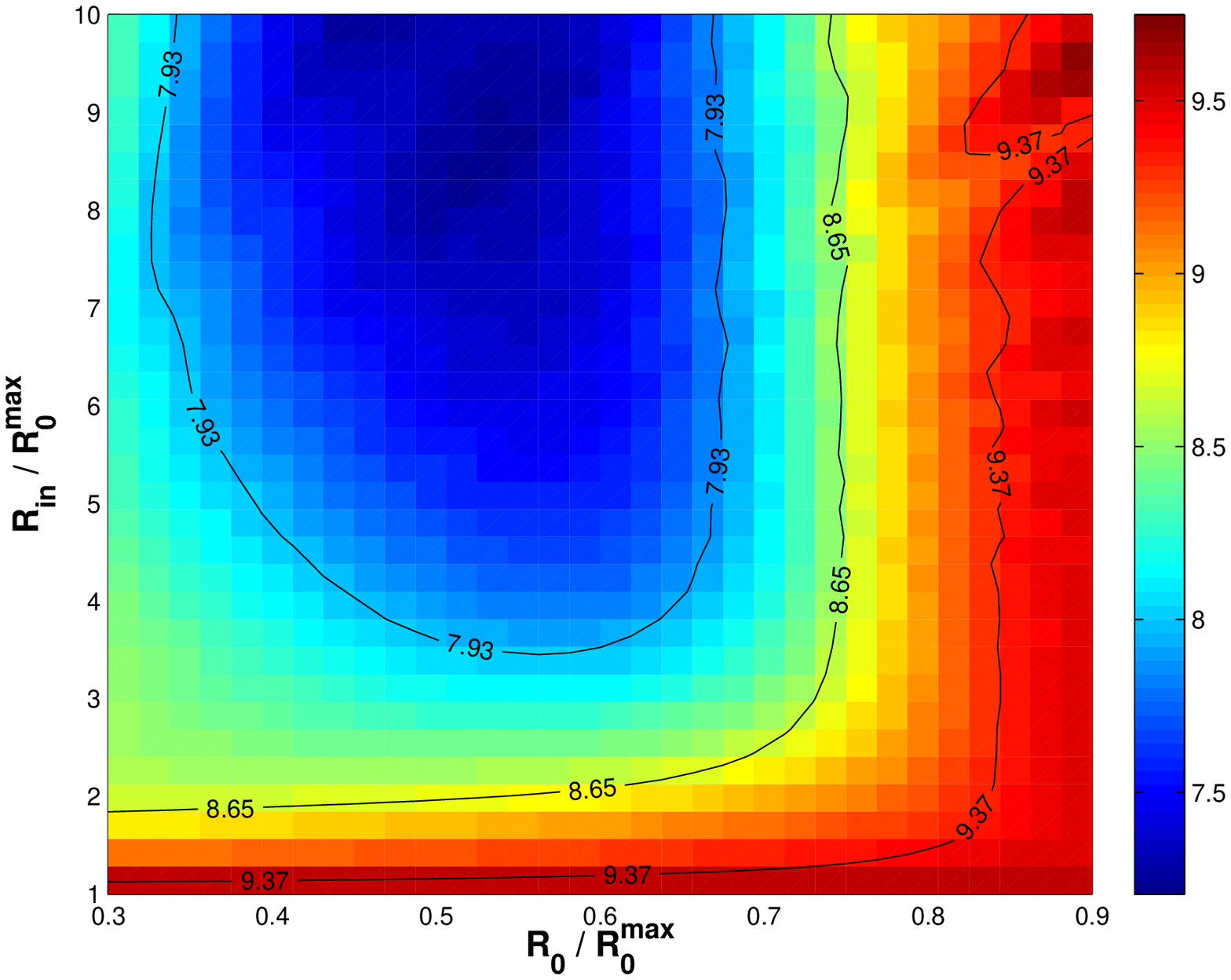} &  
  \includegraphics[width=0.33\textwidth]{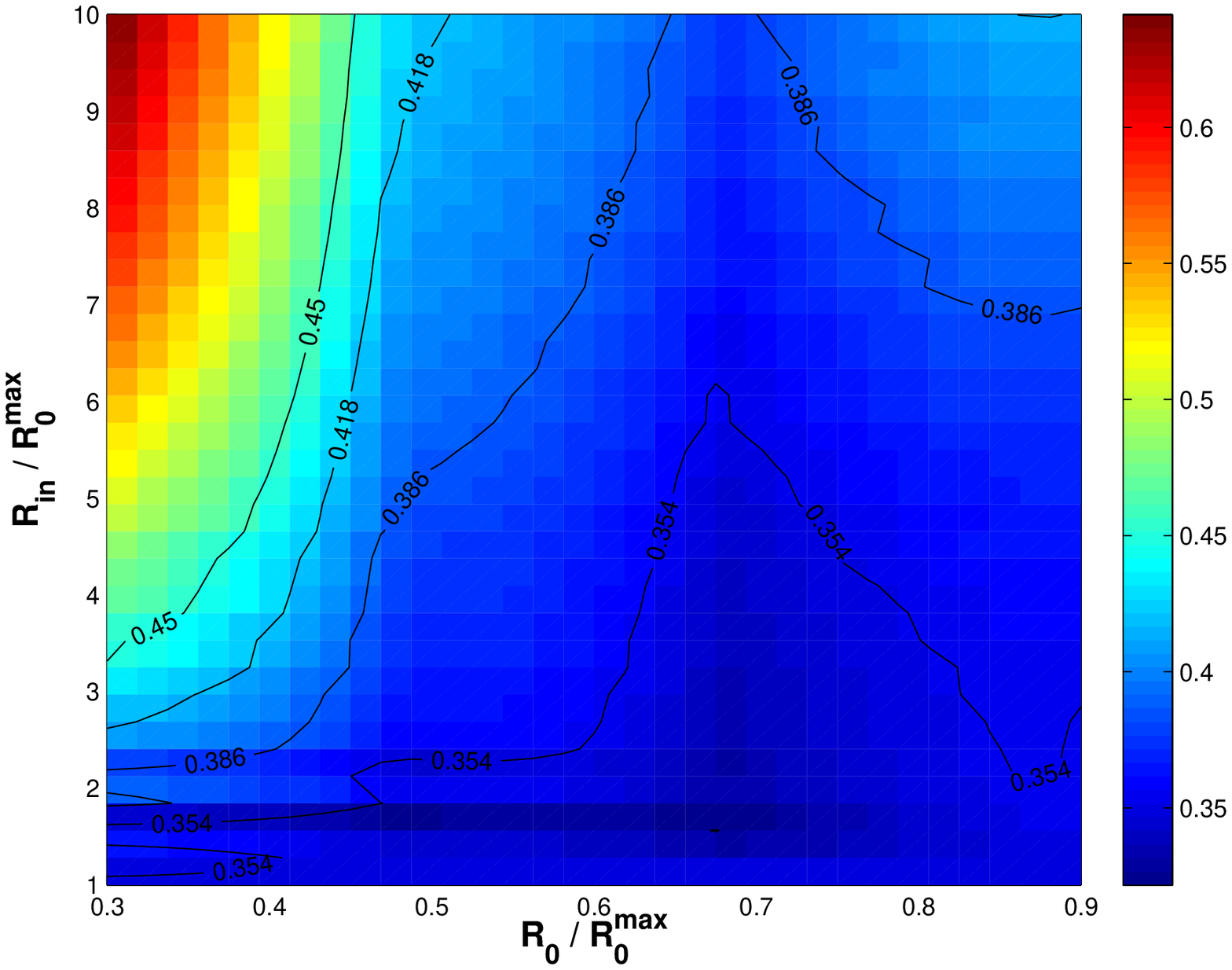} \\
\end{array}
\end{displaymath}
\caption[1$\sigma$ error for various configuration $(R_0,\, R)$ of upcoming 21cm experiments in the MID model]{Same as Figure \ref{fig:OC1} but for MID model.  Figures are for LOFAR(left panel), MWA(middle panel) and SKA(right panel).}
\label{fig:OC2}
\end{figure}

\begin{table}
\centering
\begin{minipage}{\textwidth}
\centering
\footnotesize{
\caption[Optimal configuration for various 21cm interferometer arrays]{\label{tab:op}  Optimal configuration for various 21cm interferometer arrays.  Same assumptions as in Table \ref{tab:power} but for different array layout.  $R_{\rm in}^{\rm prop}$ is the previously proposed inner core radius.  $\eta$ is the ratio of the number of antennae in the nucleus to the total number inside the core.  $n$ is the fall-off index by which $\rho \propto r^{-n}$ outside the nucleus.}
\begin{tabular}{p{1cm}p{1.5cm}p{1cm}p{1.2cm}p{1.2cm}p{1.1cm}p{0.5cm}p{0.5cm}p{3.5cm}}
\hline\hline
                         &    Experiment          &    
$R_0^{\rm max}$ (m) & 
$R_0$ $(\times R_0^{\rm max})$	    &	 $R_{\rm in}$ $(\times R_0^{\rm max})$	 &     
$R_{\rm in}^{\rm prop}$ (m) \footnote{Note that for LOFAR and SKA there is an outer core with the radius 6 km and 5 km respectively.  So for them $R_{\rm in}$ is not the size of total array.}   & 
$\eta$         &    $n$ 		& 
Comments      \\   \hline
       & LOFAR   & 319 & 0.84 & 1.28 & 1000 & 0.71 & 6.0 &  Almost a giant core      \\
OPT    & MWA     & 50  & 0.64 & 2.41 & 750  & 0.41 & 3.0 &  Close to a giant core  \\
       & SKA     & 211 & 0.30 & 1.56 & 1000 & 0.09 & 0.83 &  Almost a giant core   \\ \hline
       & LOFAR   & 319 & 0.84 & 1.28 & 1000 & 0.71 & 6.0  &  Almost a giant core \\
MID    & MWA     & 50  & 0.45 & 10   & 750  &  0.20 & 2.3 & {Both a large nucleus and a wide-spread core} \\
       & SKA     & 211 & 0.68 & 1.57 & 1000 & 0.46 & 2.9 &  Almost a giant core \\
\hline\hline	
\end{tabular}
}
\end{minipage}
\end{table}

In this section we first investigate how array layout affects the
sensitivity to cosmological parameters.  Next, we investigate the
optimal array configuration for fixed antennae number. Our
parametrization of the array configuration is discussed in Section
\ref{configuration}.

We map the constraint in $\mnu$ on the $R_0$--$R_{\rm in}$ plane in
Figure \ref{fig:OC1} (OPT model) and Figure \ref{fig:OC2} (MID model).  $R_0$
is the radius of the compact core, and $R_{\rm in}$ the radius of
inner core, both in the unit of $R_0^{\rm max}\equiv \sqrt{N_{\rm
in}/\rho_0 \pi}$.  Note that if $R_0=R_0^{\rm max}$, then $R_{\rm
in}=R_0^{\rm max}$ --- this is the case of a ``giant
core'', in which all antennae are compactly laid down with a physical
covering fraction close to unity, and is represented by the $x$-axis
in the $R_0$--$R_{\rm in}$ plane (the value of $R_0$ is meaningless if
$R_{\rm in}=R_0^{\rm max}$).  
In Table \ref{tab:op}, we list the
optimal configuration that is indicated by Figure \ref{fig:OC1} and
\ref{fig:OC2}.  The compactness of an array is represented by $R_{\rm
in}/R_0^{\rm max}$, since $R_0^{\rm max}$ is the minimum of $R_{\rm
in}$.  In comparison, $R_0/R_0^{\rm max}$ does not indicate the
compactness, since a slow fall-off configuration with a small $R_0$ is
effectively very close to a giant core.  Rather, $R_0$ is a transition
point from a flat compact core to the fall-off region.
Note that we have three configuration parameters $R_0$, $R_{\rm in}$ and $R_{\rm out}$.  
We find the annulus for SKA and LOFAR to make almost no difference
to the cosmological constraints, 
and therefore focus on how to optimize only the remaining two parameters $R_0$ and $R_{\rm in}$.

Table \ref{tab:op} shows that the optimal layout for OPT model is
close to a giant core, with the inner core much smaller than the
previously proposed.  For MID model, LOFAR and SKA still favors
the quasi-giant-core layout, but MWA favors a large core whose radius
is about the size that was previously proposed.  The accuracies in
$\mnu$ varies in the OPT model by a factor of 3 for LOFAR, 1.4-1.5 for
MWA and SKA, and in the MID model by a factor of 3 for LOFAR, 1.3 for
MWA and 2.2 for SKA.  This means that an optimal
configuration can improve the constraints by a factor up to 3 in noise
dominated experiments, and up to 2 times in signal dominated
experiments.

The plots have three interesting features.  First, the configuration
of a quasi-giant core is generically favored.  The reason for this is
that the noise on the temperature in an observing cell with $u_\perp$
is inversely proportional to the square root of the number of
baselines that probe this $u_\perp$.  A compact array increases
the number of baselines that probe small $u_\perp$, reducing the
overall noise level on these modes.  Second, a couple of the upcoming
$21$ cm experiments favor the configuration that is close but not
identical to a giant core.  The reason for this is because arrays become
sample variance-limited once they have a certain number of baselines that
probe a given $u_\perp$.  A simple estimate on the signal-to-noise
ratio for a compact MWA shows that on average $\sPdd/\bar{P}_N \approx
5 $ at the $k\sim 0.1 \; \perMpc$ and $\sPdd/\bar{P}_N \approx 1/40$ at
the $k\sim 0.7 \; \perMpc$.  
Although moving more antennae to the center can increase the
signal-to-noise, the error cannot be reduced as much if modes are
already dominated by signal.  Third, in the MID model, MWA favors a
less compact core.  This fact is due to the mixing between
cosmological and ionization parameters.  Remember that the
off-diagonal elements in the Fisher matrix are proportional to the
magnitude of ionization power spectra --- the smaller the magnitude,
the smaller degradation factor and the more accurate is the
cosmological parameter measurement.  Figure \ref{fig:ri} illustrates
that the ionization power spectrum generically falls off at large $k$
such that a relatively large core, which is more sensitive to these
large $k$, may actually improve parameter constraints.  This factor
appears to be important for MWA because, as Figure \ref{fig:datacube}
shows, a compactified MWA only occupies a rather narrow band in
$k$-space.  This means that MWA has to expand significantly in order
to use much more large $k$ modes.

It came to our attention that Lidz \etal \cite{Lidz:2007az} performed
an analysis of the optimal configuration for MWA.  Lidz \etal
\cite{Lidz:2007az} concludes that the optimal layout for MWA is a
giant core.  This conclusion is slight different than ours; we find a
compact but not exactly a giant core is optimal for MWA.  The work in
\cite{Lidz:2007az} defines the optimal configuration to be the
configuration that maximizes the total signal-to-noise, while our
definition is based on parameter constraints. In addition, the
conclusion in \cite{Lidz:2007az} is based on the comparison of a giant
core array configuration to one without a giant core, while we
investigate a range of plausible configurations. It should be pointed
out that both approaches should be tested with detailed simulations.

\subsection{Varying collecting area}

\begin{table}
\centering
\begin{minipage}{\textwidth}
\centering
\footnotesize{
\caption[How cosmological constraints depend on collecting areas in the OPT model]{\label{tab:Ae1}  How cosmological constraints depend on collecting areas in the OPT model.  Same assumptions as in Table \ref{tab:power} but for different collecting areas $A_e$ and assume only OPT model.  The exponent $\beta$ tells the rule of thumb of the $A_e$-dependence of marginalized errors $\Delta p$, assuming $\Delta p\propto (A_e)^\beta$. } 
\begin{tabular}{llccccc}
\hline\hline
 	      &  $A_e/A_e^{\rm fid}$ \footnote{$A_e^{\rm fid}$ refers to the fiducial values assumed in Table \ref{tab:spec} and are not the same for different arrays.}  & 
 $\Delta\Ol$                &$\Delta\ln(\Omega_m h^2)$ & 
 $\Delta\ln(\Omega_b h^2)$  &$\Delta\ns$	    & 
 $\Delta\ln\As$ \\ \hline
         & $ 2.0 \quad$  &0.020&0.24&0.40&0.048&0.80 \\
LOFAR    & $ 1   \quad$  &0.025&0.27&0.44&0.063&0.89 \\
         & $ 0.5 \quad$  &0.039&0.40&0.62&0.10&1.3 \\ \cline{2-7}
	 & $\beta$       & -0.48 & -0.37  & -0.32 & -0.53 & -0.35 \\ \hline
         & $ 2.0 \quad$  &0.057&0.11&0.22&0.021&0.41\\
MWA      & $ 1   \quad$  &0.046&0.11&0.19&0.022&0.37\\
         & $ 0.5 \quad$  &0.042&0.11&0.19&0.027&0.37\\ \cline{2-7}
	 & $\beta$       & 0.22  & 0 & 0.11 & -0.18 & 0.07 \\ \hline
         & $ 2.0 \quad$  &0.0027&0.048&0.099&0.0077&0.19 \\
SKA      & $ 1   \quad$  &0.0038&0.044&0.083&0.0079&0.16 \\
         & $ 0.5 \quad$  &0.0043&0.043&0.076&0.0089&0.15 \\ \cline{2-7}
	 & $\beta$       & -0.34 & 0.08 & 0.19 & -0.10 & 0.17 \\ \hline
         & $ 2.0 \quad$  &0.00014&0.0031&0.0082&0.00037&0.015\\
FFTT   & $ 1   \quad$    &0.00015&0.0032&0.0084&0.00040&0.015\\
         & $ 0.5 \quad$  &0.00017&0.0035&0.0086&0.00046&0.016\\ \cline{2-7}
	 & $\beta$       & -0.14 & -0.09  & -0.03  & -0.16 & -0.05 \\ 
\hline\hline
\end{tabular}
}
\end{minipage}
\end{table}

\begin{table}
\centering
\footnotesize{
\caption[How cosmological constraints depend on collecting areas in the MID model]{\label{tab:Ae2} How cosmological constraints depend on collecting areas in the MID model.  Same assumptions as in Table \ref{tab:power} but for different collecting areas $A_e$ and assume only MID model.  The exponent $\beta$ tells the rule of thumb of the $A_e$-dependence of marginalized errors $\Delta p$, assuming $\Delta p\propto (A_e)^\beta$.} 
\begin{tabular}{llccccc}
\hline\hline
 	      &  $A_e/A_e^{\rm fid}$   & 
 $\Delta\Ol$                &$\Delta\ln(\Omega_m h^2)$ & 
 $\Delta\ln(\Omega_b h^2)$  &$\Delta\ns$	    & 
 $\Delta\ln\As$ \\ \hline
         & $ 2.0 \quad$  &0.086&0.044&0.072&0.15&0.35 \\
LOFAR    & $ 1   \quad$  &0.13&0.083&0.15&0.36&0.80 \\
         & $ 0.5 \quad$  &0.26&0.17&0.35&0.92&2.0 \\ \cline{2-7}
         & $\beta$	 & -0.80 & -0.98 & -1.1 & -1.3 & -1.3 \\ \hline
         & $ 2.0 \quad$  &0.21&0.015&0.025&0.073&0.61\\
MWA      & $ 1   \quad$  &0.22&0.017&0.029&0.097&0.76\\
         & $ 0.5 \quad$  &0.26&0.026&0.045&0.16&1.3\\ \cline{2-7}
	 & $\beta$       & -0.15  & -0.40 & -0.42 & -0.57 & -0.55 \\ \hline
         & $ 2.0 \quad$  &0.013&0.0049&0.0079&0.0092&0.032 \\
SKA      & $ 1   \quad$  &0.014&0.0049&0.0081&0.012&0.037 \\
         & $ 0.5 \quad$  &0.016&0.0063&0.011&0.022&0.053 \\ \cline{2-7}
	 & $\beta$       & -0.15 & -0.18 & -0.24 & -0.63 & -0.36 \\ \hline
         & $ 2.0 \quad$  &0.00036&0.00037&0.00061&0.00032&0.0012\\
FFTT     & $ 1   \quad$  &0.00041&0.00038&0.00062&0.00036&0.0013\\
         & $ 0.5 \quad$  &0.00052&0.00041&0.00066&0.00046&0.0016\\ \cline{2-7}
	 & $\beta$       & -0.27 & -0.07 & -0.06  & -0.26 & -0.21 \\
\hline\hline
\end{tabular}
}
\end{table}

The survey volume and the noise per pixel are both affected by
changing the collecting area $A_e$ because the solid angle a survey
observes is $\Omega \approx \lambda^2/A_e$ and $P^N \propto 1/A_e^2$
(\Eq{eqn:PNoise}).  For noise-dominated experiments, $\delta \PDT
/\PDT \propto P^N/\sqrt{N_c} \propto A_e^{-2} /
\sqrt{A_e^{-1}}=A_e^{-3/2}$, and, for signal-dominated experiments,
$\delta \PDT /\PDT \propto 1/\sqrt{N_c} \propto A_e^{1/2}$.  If we
parametrize the scaling of the error on a cosmological parameter as
$\Delta p\propto (A_e)^\beta$, we have $-1.5<\beta<0.5$.  A caveat is
FFTT which has fixed $\Omega=2\pi$, so $\delta \PDT /\PDT \propto
A_e^0$ (signal dominated) or $\delta \PDT /\PDT \propto 1/A_e^2$
(noise dominated).  Since nearly signal dominated, $\beta\lesssim 0$
for FFTT.

We show how collecting area affects the accuracy in Table
\ref{tab:Ae1} (OPT model) and \ref{tab:Ae2} (MID model).  In the OPT
model, it appears that $\beta\approx -0.4$ for LOFAR, $|\beta|
\lesssim 0.2$ for MWA, $|\beta| \lesssim 0.3$ for SKA, and $\beta\sim
-0.1$ for FFTT.  In the MID model, it appears that $\beta\sim -1.3$
for LOFAR, $\beta\sim -0.5$ for MWA, $\beta\sim -0.6$ for SKA,
$\beta\sim -0.3$ for FFTT.  These exponents are compatible with the
above arguments.  The upshot is that varying $A_e$ does not
significantly affect parameter constraints.

\subsection{Varying observation time and system temperature}

\begin{table}
\centering
\footnotesize{
\caption[How cosmological constraints depend on observation time in the OPT model]{\label{tab:t0-1}  How cosmological constraints depend on observation time in the OPT model.  Same assumptions as in Table \ref{tab:power} but for different observation time $t_0$ and assume only OPT model.  The exponent $\epsilon$ tells the rule of thumb of the $t_0$-dependence of marginalized errors $\Delta p$, assuming $\Delta p\propto (t_0)^{-\epsilon}$.  $t_0$ is in units of 4000 hours.} 
\begin{tabular}{llccccc}
\hline\hline
 	      &  $t_0$   & 
 $\Delta\Ol$                &$\Delta\ln(\Omega_m h^2)$ & 
 $\Delta\ln(\Omega_b h^2)$  &$\Delta\ns$	    & 
 $\Delta\ln\As$ \\ \hline
         & $ 4.0 \quad$  &0.014&0.17&0.28&0.034&0.56\\
LOFAR    & $ 1   \quad$  &0.025&0.27&0.44&0.063&0.89\\
         & $ 0.25\quad$  &0.055&0.56&0.88&0.14&1.8\\ \cline{2-7}
	 & $\epsilon$    & 0.49 & 0.43 & 0.41 & 0.51 & 0.42 \\ \hline
         & $ 4.0 \quad$  &0.040&0.081&0.16&0.015&0.29\\			  
MWA      & $ 1   \quad$  &0.046&0.11&0.19&0.022&0.37\\			  
         & $ 0.25\quad$  &0.059&0.15&0.27&0.038&0.52\\ \cline{2-7} 	  
	 & $\epsilon$	 & 0.14 & 0.22 & 0.19 & 0.34 & 0.21 \\ \hline 
         & $ 4.0 \quad$  &0.0019&0.034&0.070&0.0054&0.13\\			  
SKA      & $ 1   \quad$  &0.0038&0.044&0.083&0.0079&0.16\\			  
         & $ 0.25\quad$  &0.0060&0.061&0.11&0.013&0.21\\ \cline{2-7} 	  
	 & $\epsilon$	 & 0.41 & 0.21 & 0.16 & 0.32 & 0.17 \\ \hline 
         & $ 4.0 \quad$  &0.00014&0.0031&0.0082&0.00037&0.015\\			  
FFTT     & $ 1   \quad$  &0.00015&0.0032&0.0084&0.00040&0.015\\			  
         & $ 0.25\quad$  &0.00017&0.0035&0.0086&0.00046&0.016\\ \cline{2-7} 	  
	 & $\epsilon$	 & 0.07 & 0.04 & 0.02 & 0.08 & 0.02 \\ 
\hline\hline
\end{tabular}
}
\end{table}

\begin{table}
\centering
\footnotesize{
\caption[How cosmological constraints depend on observation time in the MID model]{\label{tab:t0-2}  How cosmological constraints depend on observation time in the MID model. Same assumptions as in Table \ref{tab:power} but for different observation time $t_0$ and assume only MID model. The exponent $\epsilon$ tells the rule of thumb of the $t_0$-dependence of marginalized errors $\Delta p$, assuming $\Delta p\propto (t_0)^{-\epsilon}$.  $t_0$ is in units of 4000 hours. } 
\begin{tabular}{llccccc}
\hline\hline
 	      &  $t_0$   & 
 $\Delta\Ol$                &$\Delta\ln(\Omega_m h^2)$ & 
 $\Delta\ln(\Omega_b h^2)$  &$\Delta\ns$	    & 
 $\Delta\ln\As$ \\ \hline
         & $ 4.0 \quad$  &0.061&0.031&0.051&0.11&0.25\\
LOFAR    & $ 1   \quad$  &0.13&0.083&0.15&0.36&0.80\\
         & $ 0.25\quad$  &0.36&0.24&0.50&1.3&2.9\\ \cline{2-7}
	 & $\epsilon$    & 0.64 & 0.74 & 0.82 & 0.89 & 0.88 \\ \hline
         & $ 4.0 \quad$  &0.15&0.010&0.017&0.052&0.43\\			  
MWA      & $ 1   \quad$  &0.22&0.017&0.029&0.097&0.76\\			  
         & $ 0.25\quad$  &0.36&0.037&0.064&0.23&1.8\\ \cline{2-7} 	  
	 & $\epsilon$	 & 0.32 & 0.47 & 0.48 & 0.54 & 0.52 \\ \hline 
         & $ 4.0 \quad$  &0.0089&0.0035&0.0056&0.0065&0.022\\			  
SKA      & $ 1   \quad$  &0.014&0.0049&0.0081&0.012&0.037\\			  
         & $ 0.25\quad$  &0.023&0.0090&0.015&0.031&0.075\\ \cline{2-7} 	  
	 & $\epsilon$	 & 0.34 & 0.34 & 0.36 & 0.56 & 0.44 \\ \hline 
         & $ 4.0 \quad$  &0.00036&0.00037&0.00061&0.00032&0.0012\\			  
FFTT     & $ 1   \quad$  &0.00041&0.00038&0.00062&0.00036&0.0013\\			  
         & $ 0.25\quad$  &0.00052&0.00041&0.00066&0.00046&0.0016\\ \cline{2-7} 	  
	 & $\epsilon$	 & 0.13 & 0.04 & 0.03 & 0.13 & 0.10 \\ 
\hline\hline
\end{tabular}
}
\end{table}

The detector noise is affected by changing the observation time and system temperature.  From \Eq{eqn:PNoise}, the noise $P^N \propto T_{\rm sys}^2/t_0$.  Therefore, for noise dominated experiments, $\delta \PDT /\PDT \propto P^N/\sqrt{N_c} \propto T_{\rm sys}^2/t_0$, and for signal dominated experiments, $\delta \PDT /\PDT \propto 1/\sqrt{N_c} \propto (T_{\rm sys}^2/t_0)^0$.  Assuming that errors in cosmological parameter $\Delta p \propto (T_{\rm sys}^2/t_0)^\epsilon$, we have $0<\epsilon<1$.  

Since $T_{\rm sys}^2$ and $t_0^{-1}$ shares the same exponent, we
evaluate the $\epsilon$ by varying only $t_0$ in Table \ref{tab:t0-1}
(OPT model) and \ref{tab:t0-2} (MID model).  It appears that in
average $\epsilon\sim 0.5$ for LOFAR, $\epsilon\sim 0.3$ for MWA,
$\epsilon\sim 0.3$ for SKA, $\epsilon < 0.1$ for FFTT in the OPT
model, and $\epsilon\sim 0.8$ for LOFAR, $\epsilon\sim 0.5$ for MWA,
$\epsilon\sim 0.4$ for SKA, $\epsilon \lesssim 0.1$ for FFTT in the
MID model.  These exponents are compatible with the expected
$0<\epsilon<1$ from the above argument.  The upshot is that the order
unity changes in $T_{\rm sys}$ and $t_0$ play a marginal role in the
accuracy for future signal-dominated experiments.

\subsection{Varying foreground cutoff scale $\kmin$}

Finally, we test how accuracy is affected by varying $\kmin$ 
above which foregrounds can be cleaned from the signal.  
One expect that the constraints tend to approach 
asymptotic values at small enough $\kmin$.  However, 
the most effectively constrained modes are at
small $k$ ($k \sim 0.1 ~{\rm Mpc}^{-1}$) for noise dominated
experiments, while the contributions from larger $k$ modes are more
important for cosmic variance-limit experiments.  This means that $\kmin$ 
affects the noise dominated experiments most.  
Left panel of Figure \ref{fig:kminkmax} illustrates this by plotting 
cosmological constraints as a function of the relative minimum cutoff 
$\kappa_{\rm min} \equiv \kmin \times y(z)B(z)/2\pi$ which is a constant scale 
factor for all $z$-bins by definition.  The slopes at 
$\kappa_{\rm min} = 1$ are rather large for MWA (varying from $\kappa_{\rm min} = 0.5$ 
to $2$, $\Delta\ns=0.032$ to 0.39, about 10 times larger).  
For a signal dominated experiment like SKA, the constraints can be off by a factor of 3, 
or FFTT by a factor of 1.6.  This suggests that in general $\kmin$ is among top factors to affect cosmological constraints.

\section{Conclusion \& outlook}\label{sec:conclusion}

\subsection{Which assumptions matter most?}

\begin{figure}[ht]
\centering
\includegraphics[width=1.0\textwidth]{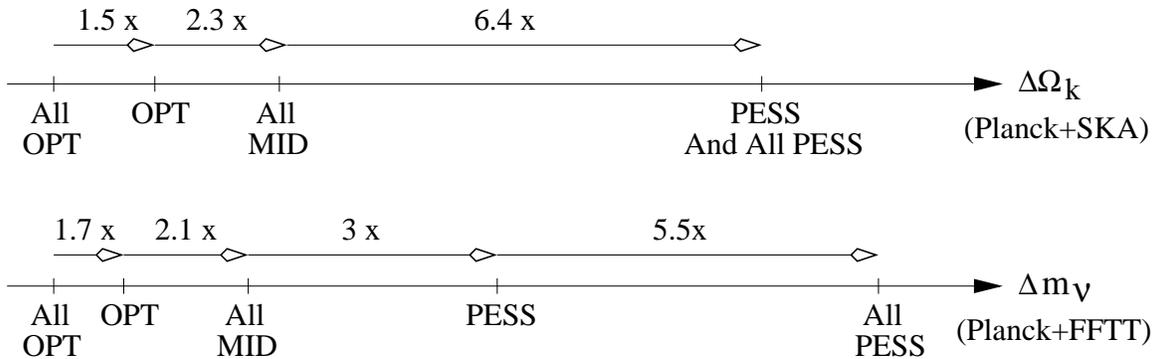} 
\caption[Cartoon showing how cosmological parameter measurement accuracy depends on various assumptions]{Cartoon showing how cosmological parameter measurement accuracy depends on various assumptions. The cases labeled merely ``PESS'' or ``OPT'' 
have the PESS/OPT ionization power spectrum modeling with MID assumptions for everything else.
}
\label{fig:lineup}
\end{figure}

In Section \ref{sec:results}, we have quantified how cosmological
parameter measurement accuracy depends on assumptions about ionization
power modeling, reionization history, redshift range, experimental
specifications such as the array configuration, and astrophysical
foregrounds.  We now return to the overarching question from
Section~\ref{sec:intro} that motivated our study: among these
assumptions, which make the most and least difference?

To quantify this, we consider two of the parameters for which 21cm
tomography has the most potential for improving on Planck CMB
constraints based on our estimates: $\Ok$ and $\mnu$.  Figure
\ref{fig:lineup} shows $\Delta\Ok$ based on data from Planck plus SKA
as well as $\Delta\mnu$ from Planck plus FFTT.  Varying
the ionization power modeling from PESS to OPT models improves the
constraints on these two parameters by a factor of 6--15.
From 21cm data alone in the OPT model, the optimal array configuration
can affect accuracies up to a factor 3 (Figure~\ref{fig:OC1}),
redshift ranges affect it by up to a factor of 5 (Table~\ref{tab:z1}),
and residual foregrounds affect it by up to a factor of 10
(Figure~\ref{fig:kminkmax}, left panel).  In summary, the assumptions
can be crudely ordered by importance as ionization power modeling
$\gg$ foregrounds $\sim$ redshift ranges $\sim$ array layout $ > A_e
\sim T_{\rm sys} \sim t_0 \sim\kmax \sim$ non-Gaussianity.



\subsection{Outlook}

We have investigated how the measurement of cosmological parameters
from 21 cm tomography depends on various assumptions.  We have found
that the assumptions about how well the reionization process can be
modeled are the most important, followed by the assumptions pertaining
to array layout, IGM evolution, and foreground removal.

Our results motivate further theoretical and experimental work. On the
theoretical side, it will be valuable to develop improved EoR data
analysis techniques.  The OPT approach is restricted to when neutral
fraction fluctuations are not important, which is not an accurate
approximation during the EOR.  On the other hand, although the PESS
approach is in principle insensitive to our poor understanding of
reionization by marginalizing over it, in practice this approach
destroys too large a fraction of the cosmological information to be
useful.  Hopefully more detailed EoR simulations will
enable our MID approach to be further improved into a phenomenological
parametrization of our ignorance that is robust enough to be reliable,
yet minimizes the loss of cosmological information.
\footnote{
It is also possible to constrain cosmological parameters using lensing of 21cm 
fluctuations \cite{Zahn:2005ap,Metcalf:2006ji,Metcalf:2008gq,Zhang:2006fc}.}

On the experimental side, there are numerous complications that are
beyond the scope of this chapter, but that are important enough to
deserve detailed investigation in future work.  To what extent can
radio-frequency interference be mitigated, and to what extent does it
degrade cosmological parameter accuracy? This is particularly
important for instruments in densely populated parts of the world,
such as LOFAR.  To what extent is the subtraction of the foreground
point sources hampered the complicated off-center frequency scaling of
the synthesized beam?  To what extent does the dramatic variation of
the synchrotron brightness temperature across the sky affect our
results and optimal array design?  Performing a realistic end-to-end
simulation of possible experiments (from sky signal to volts and back)
should be able to settle all of these issues.

These are difficult questions, but worthwhile because the potential
for probing fundamental physics with 21 cm tomography is impressive: a
future square kilometer array optimized for 21 cm tomography could
improve the sensitivity of the Planck CMB satellite to spatial
curvature and neutrino masses by up to two orders of magnitude, to
$\Delta\Omega_k\approx 0.0002$ and $\Delta m_\nu\approx 0.007$ eV, and
detect at $4\sigma$ the running of the spectral index predicted by the
simplest inflation models.

We wish to thank Judd Bowman, Jacqueline Hewitt and Miguel Morales 
for helpful discussions and comments.  
YM thanks Yi Zheng for technical help.  

\begin{subappendices}
\section{$\chi^2$ goodness of fit in the MID model}
\label{chi2}

In this appendix, we elucidate some issues in separating cosmological information from astrophysics in the MID model, and give the $\chi^2$ goodness-of-fit test.

The parametrization of ionization power spectra is based on the assumption that these power spectra are smooth functions of $k$, and therefore can be parametrized with as many parameters as necessary to fit the data at some accuracy.  However, the separation of cosmology from astrophysics implicitly depends on another assumption that the shapes of ionization power spectra are distinguishable from that of matter power spectrum, since one can only measure the \emph{total} 21cm power spectrum.  Albeit sometimes the shape may be similar at small $k$ (see the plateaus in the ratios of power spectra in Figure \ref{fig:ri}), the slope and amplitude of ionization power spectrum at the fall-off region can in principle distinguish nuisance functions from the matter power spectrum, determine the overall amplitude, and in return use the data at small $k$ to further constrain the nuisance 
parameters that correspond to the amplitudes.  

There are standard statistical methods for testing whether the parametrization is successful.  We now give a compact description of the $\chi^2$ goodness-of-fit test, and refer interested readers to \cite{PDG} for a useful review on the statistics.  We want to test the hypothesis $H_0$ that the parametrization with fitting parameter values is an accurate account of the ionization power spectra.  The parameter vector to be fitted is $\Theta\equiv \left( \lambda_i\,(i=1,\ldots,N_p),\beta_\al\,(\al=1,\ldots,n_{\rm ion})\right)$, where $N_p$ and $n_{\rm ion}$ are the number of cosmological and ionization parameters, respectively.  The observed data vector is $\bfy\equiv (y_1,\ldots,y_N)$ where $y_i \equiv \PDT(\bfk_i)$ at each pixel $\bfk_i$ labeled by $i=1,\ldots, N$, where $N$ is the total number of pixels.  Assuming the Gaussian statistic in the measurements, the corresponding vector $\bfF$ for the expected value is  
$F(\bfk_i;\Theta)=(\sPdd-2\sPxd+\sPxx)+2(\sPdd-\sPxd)\mu^2+\sPdd\mu^4$, and the variance is $ \sigma_i^2 \equiv (\delta \PDT(\bfk_i))^2 = \frac{1}{N_c}[\PDT(\bfk_i)+P_N(k_{i\,\perp})]^2$.  
We can now compute $\chi^2$: 
\beq{eq:chi2}
\chi^2(\Theta)=(\bfy-\bfF(\Theta))^T C^{-1} (\bfy-\bfF(\Theta))\,, 
\een 
where $C$ is the covariance matrix.  If each measurement $y_i$ is independent, then $C$ becomes diagonal with $C_{ii}=\sigma_i^2$.  Then \Eq{eq:chi2} is simplified to be 
\ben
\chi^2(\Theta)=\sum_{i=1}^{N} \frac{ [y_i - F(\bfk_i;\Theta)]^2}{\sigma_i^2}\,.
\een
We can define the $p$-value as the probability, under the assumption of the hypothesis $H_0$, of obtaining data at least as incompatible with $H_0$ as the data actually observed.  So 
\ben
p = \int_{\chi^2(\Theta)}^{\infty} f(z;n_d) dz\,,
\een
where $f(z;n_d)$ is the $\chi^2$ probability density function (p.d.f.) with $n_d$ degrees of freedom $n_d=N-(N_p+n_{\rm ion})$.  Values of the $\chi^2$ p.d.f. can be obtained from the CERNLIB routine PROB \cite{prob}. 
To set the criterion, a fit is good if $p \ge 0.95$, \ie the real data fit the parametrization better than the 95\% confidence level.

\end{subappendices}
\chapter{Conclusions}

We set out in this thesis to use the avalanche of new astrophysical data to shed light on the fundamental laws of physics involving gravitation and cosmology.  

\section{Summary of results}

We have generalized tests of GR to allow the testing of assumptions that are normally not questioned, for example whether a type of space-time distortion known as torsion exists, and whether the gravitational Lagrangian contains extra terms that are general functions of the Ricci scalar and could affect cosmic expansion and structure formation.  
Specifically, using symmetry arguments, we have generalized the Parametrized Post-Newtonian formalism by parametrizing any torsion field around a uniformly rotating spherical mass with seven dimensionless parameters that provide a concrete framework for further testing GR.  Using the fact that torsion could in principle affect precession of a gyroscope in Earth orbit, we have shown that the ongoing satellite experiment Gravity Probe B can in principle measure the values of torsion parameters to an unprecedented accuracy of one part in ten thousand.  

Testing gravity in a separate direction, we have searched for viable theories of $f(R)$ gravity, and find that models can be made consistent with solar system constraints either by giving the emergent scalar a high mass or by exploiting the chameleon effect.  Furthermore we have explored observational constraints from the  late-time cosmic acceleration, big bang nucleosynthesis and inflation.  

In looking for precision tests of cosmological models, 
we have demonstrated that twenty-one-centimeter tomography has the potential to become one of the most promising cosmological probes.  Upcoming experiments such as MWA, LOFAR, 21CMA and SKA will map neutral hydrogen throughout the universe in 3D by measuring the 21 cm radio waves that neutral hydrogen atoms emit.  To help optimize such observations, we have quantified how the precision with which cosmological parameters can be measured depends on a broad range of assumptions, enabling experimentalists to exploit design tradeoffs to maximize the scientific bang for the buck.  We have also presented an accurate yet robust method for measuring cosmological
parameters in which the ionization power spectra can be accurately fit by seven phenomenological parameters.  
We find that a future square kilometer array optimized for 21 cm tomography has great potential,
improving the sensitivity to spatial curvature and neutrino masses by up to two orders of magnitude, to $\Delta\Omega_k\approx 0.0002$ and
$\Delta m_\nu\approx 0.007$ eV, and giving a $4\sigma$ detection of the spectral index running predicted by the simplest inflation models.

\section{Looking ahead}

One particularly interesting direction for future gravitational studies is to 
constrain parametrized departures from the GR-based standard cosmological model by testing GR on the scale of the cosmos.  
This is timely, with current and upcoming precision cosmological experiments involving 
CMB anisotropy and polarization, large scale structure surveys, etc.  
It would be valuable to develop a solid and general cosmological Parametrized Post-Friedmannian (PPF) framework, and to constrain the values of PPF parameters with cosmological experiments.  

As a rising star in precision cosmology, 21cm tomography raises many important open questions worth pursuing.  For example, it will be useful to develop a new data analysis method that enhances the signal-to-noise at the epoch of reionization, by optimally extracting cosmologically dependent information from the total 21cm power spectrum that is contaminated by ionized hydrogen bubbles during the EOR.  
Foregrounds generated by synchrotron radiation and other sources are a serious challenge to 21cm observation.  Unpolarized foregrounds have been quantified by de Oliveira-Costa et al. \cite{deOliveiraCosta:2008pb}, and can hopefully be adequately removed by exploiting the smooth dependence of the foreground power spectrum on frequency.  Polarized foregrounds should be non-Gaussian, which can hopefully be used to further improve foreground removal.  Improving the technique of optimal foreground removal will remain at the frontier of 21cm observations. 
In the long run, assuming that solutions to technical difficulties can be found, 21cm observations will carry unique information pertaining to structure formation and the dark ages, and will therefore reveal much about particle physics and gravitation.  Thus, efforts will be well-rewarded to improve foreground removal techniques and thermal noise reduction, since foreground and noise are among the most serious impediments to present observations of 21cm signal from the dark ages.

As a complement to 21cm tomography, further theoretical study should be devoted to the numerical modeling of reionization.  Larger N-body hydrodynamic simulations are a powerful tool in the search for the signature that galaxy formation imprints on patchy reionization.  In particular, simulations need to trace how ionizing photons (either ultraviolet photons from stars or x-rays from black holes) propagate through the surrounding gas, a process that is critical to the forecasting of what can be seen in observations.

In conclusion, astrophysics can link extraterrestrial observations to fundamental physics.  The results in this thesis suggest that this exciting link can be made even stronger in the future.

\begin{singlespace}

\bibliographystyle{plain}

\end{singlespace}

\end{document}